\documentclass[aps,pra,reprint,superscriptaddress,longbibliography]{revtex4-1}
\usepackage{graphicx,color}
\usepackage{amsmath, amssymb}
\usepackage{longtable}
\usepackage{natbib}
\usepackage{bm}

\usepackage{multirow}

\UseRawInputEncoding
\begin{document}

\title{Jahn-Teller effect in cubic fullerides A$_{3}$C$_{60}$}

\author{Zhishuo Huang}
\affiliation{Theory of Nanomaterials Group, KU Leuven, Celestijnenlaan 200F, B-3001 Leuven, Belgium}
\author{Munirah D. Albaqami}
\affiliation{Chemistry Department, College of Science, King Saud University, P.O. Box 2455, Riyadh 11451, Saudi Arabia}
\author{Tohru Sato}
\affiliation{Fukui Institute for Fundamental Chemistry, Kyoto University, Takano Nishihiraki-cho 34-4, Sakyo-ku, Kyoto 6068103, Japan}
\author{Naoya Iwahara}
\email[]{naoya.iwahara@gmail.com}
\affiliation{Department of Chemistry, National University of Singapore, Block S8 Level 3, 3 Science Drive 3, 117543, Singapore}
\affiliation{Theory of Nanomaterials Group, KU Leuven, Celestijnenlaan 200F, B-3001 Leuven, Belgium}
 \affiliation{Graduate School of Engineering, Chiba University, 1-33 Yayoi-cho, Inage-ku, Chiba 263-8522, Japan} 
\author{Liviu F. Chibotaru}
\email[]{liviu.chibotaru@kuleuven.be}
\affiliation{Theory of Nanomaterials Group, KU Leuven, Celestijnenlaan 200F, B-3001 Leuven, Belgium}
\date{\today}

\begin{abstract}
Compared to isolated C$_{60}^{3-}$ ions, characterized by a three-dimensional equipotential trough at the bottom of the lowest adiabatic potential energy surface (APES), the Jahn-Teller (JT) effect in cubic fullerides is additionally influenced by the interaction of JT distortions at C$_{60}$ sites with vibrational modes of the lattice. This leads to modification of JT stabilization energy and to the warping of the trough at each fullerene site, as well as to the interaction of JT distortions at different sites. Here we investigate these effects in three fcc fullerides with A=K,Rb,Cs and in Cs$_3$C$_{60}$ with bcc (A15) structure. DFT calculations of orbital vibronic coupling constants at C$_{60}$ sites and of phonon spectra have been done for fully ordered lattices (1 C$_{60}$/u.c.). Based on them the elastic response function for local JT distortions has been evaluated and the lowest APES investigated. To this end an expression for the latter in function of trough coordinates of all sites has been derived. The results show that the JT stabilization energy slightly increases compared to an isolated C$_{60}^{3-}$ and a warping of the trough of few meV occurs. The interaction of JT distortions on nearest- and next-nearest-neighbor fullerene sites is of similar order of magnitude. These effects arise first of all due to the interaction of C$_{60}$ sites with the displacements of neighbor alkali atoms and are more pronounced in fcc fullerides than in the A15 compound. The results of this study support the picture of weakly hindered independent rotations of JT deformations at C$_{60}$ sites in cubic A$_3$C$_{60}$.           
\end{abstract}

\maketitle
\section{Introduction}
Cubic akali-doped fullerides A$_{3}$C$_{60}$ (A = K, Rb, Cs) have attracted much attention due to a large variety of unusual electronic properties, such as the superconductivity 
 in equilibrium
\cite{Gunnarsson1997a, Structure_nature_K_Rb, Ganin2008a, Takabayashi2009, Ganin2010a} 
and in nonequilibrium
\cite{Mitrano2016, Cantaluppi2017}
states
and the metal-insulator transition controlled by the carrier concentration $n$ (number of electrons in the LUMO band) \cite{Gunnarsson1997a, Yildirim1996, Kerkoud1996}, type of alkali atom A and applied pressure \cite{Ganin2008a, Takabayashi2009, Ganin2010a, Ihara2010}. 
The electronic properties of these materials are strongly influenced by the Jahn-Teller (JT) effect on fullerene sites, notably the superconductivity, for which it represents the main mechanism of Cooper pairing. One should stress that A$_{3}$C$_{60}$ superconductors are remarkable in several respects. They are high-$T_c$ superconductors with a highest critical temperature ($T_c =38$ K for A = Cs) among organic superconductors. Recently a transient superconductivity with a critical temperature exceeding many times its value at equilibrium was found in K$_{3}$C$_{60}$ \cite{Mitrano2016, Cantaluppi2017, Nava2018cooling}. 
In addition these are the only superconducting materials with JT mechanism of pairing [comment]. 
Finally, the JT effect in these fullerides is dynamic due to a threefold orbital degeneracy of fullerene sites C$_{60}^{3-}$ imposed by their cubic site symmetry, which persists also in metallic compounds \cite{Zadik2015, StructureK3C60}.

The Jahn-Teller effect in fullerene ions was thoroughly investigated during the last decades. With increased accuracy of DFT methods and well resolved photoemission bands, the value of Jahn-Teller stabilization energy in an isolated anion C$_{60}^{-}$ was firmly established \cite{Iwahara2010a}.  Recently, by applying a novel exchange-correlation functional, the theoretical reproduction of the magnitude of vibronic coupling constants for active vibrational modes $n$H$_g$, $n$=1-8, became possible \cite{Zhishuo2020}. At the same time, the Jahn-Teller effect on fullerene sites in fulleride materials was not investigated yet. 


Compared to isolated fullerene molecules, the JT effect on C$_{60}$ sites in fullerides is expected to be more complex. Thus, although the strength of vibronic coupling to intra fullerene vibrational modes is not expected to be significantly modified compared to isolated molecules (this is confirmed by the present calculations), the fivefold degenerate active vibrational modes themselves split into threefold and twofold degenerate modes each, $H_g =T_g +E_g$ due to the symmetry reduction from I$_{\text{h}}$ to T$_{\text{h}}$ group \cite{Altmann1994} when the fullerene is placed in a cubic crystal. This effect gives a contribution to the warping of the two-dimensional rotational trough of the adiabatic potential of the corresponding $T\times h$ vibronic problem (actually of a three-dimensional rotational trough in the case of C$_{60}^{3-}$). Second, the nearest environment of a C$_{60}$ molecule in a cubic crystal (nearest-neighbour alkali atoms and fullerene molecules) provides additional JT active nuclear modes which can modify the local Jahn-Teller effect on sites. Finally, the active JT modes at neighbor fullerene sites interact with each other through common phonon modes thus hindering or, eventually blocking the free rotation of Jahn-Teller deformations on individual fullerene sites. Despite the importance, the strength of these three effects was never estimated in fullerides. On the other hand the quantitative knowledge of these effects will allow to conclude on the character of JT effect on fullerene sites in A$_{3}$C$_{60}$. 

Despite intensive experimental research \cite{Klupp2012, Kamaras2013, Potocnik2014}, it is proved to be hardly conclusive on the details of intrasite vibronic interactions due to the superposition of additional effects influencing the spectroscopic bands in fullerides. On the contrary, given the success of DFT investigation of JT effect in individual fullerene ions, we expect that a similar methodology could shed light on its manifestation in fullerides. 
In the present work, the the Jahn-Teller effect on fullerene sites in A$_3$C$_{60}$ is thoroughly investigated fully taking into account the effects mentioned above. 


%
\begin{figure}[tb]
\includegraphics [width=0.5\textwidth]{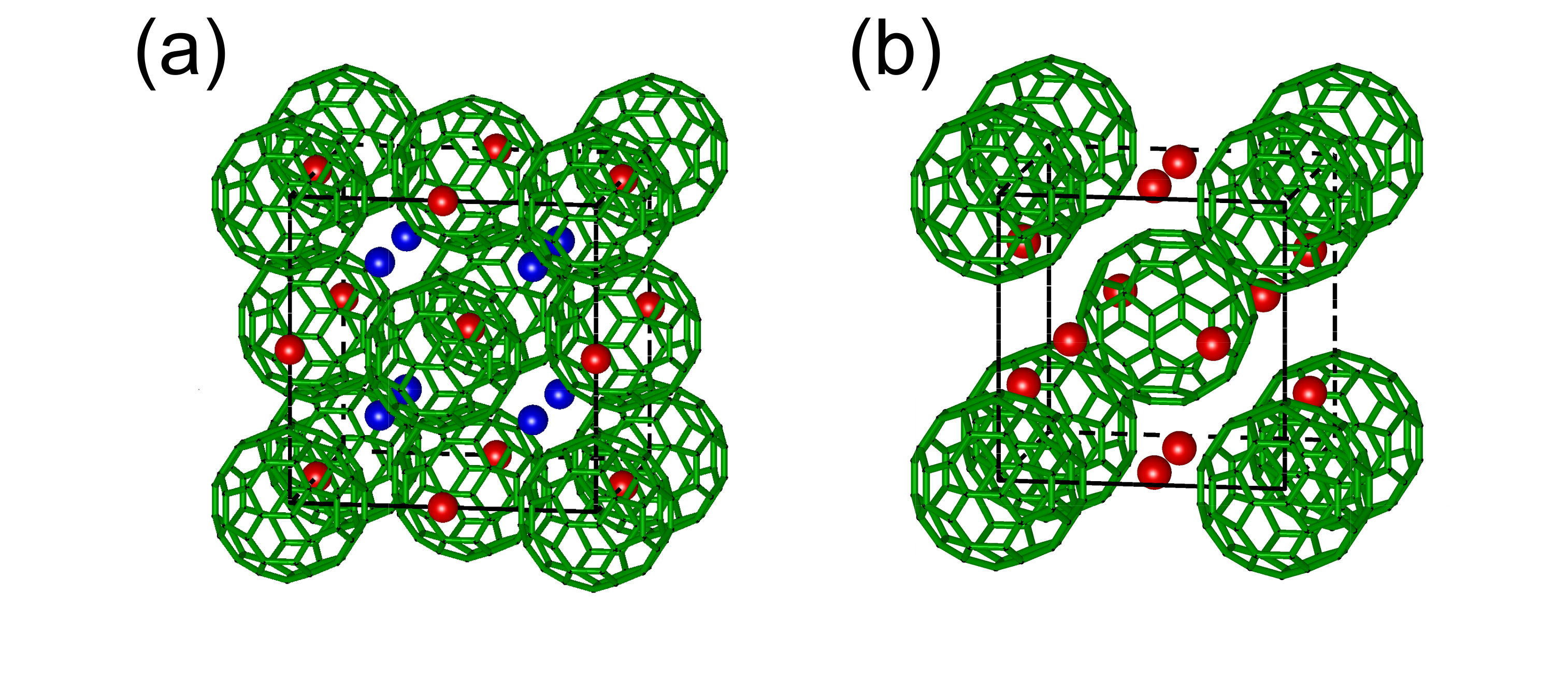}
    \caption{\label{fig:Struct_FCC_A15} Crystal structures of A$_{3}$C$_{60}$. C$_{60}$ molecules are depicted by the green framework. (a). Fcc unit cell; alkali atoms sitting in octahedral and tetrahedral interstitials are shown by red and blue balls. (b). Bcc-like unit cell for A15 structure; Cs atoms are indicated by red balls.}
\end{figure}

\section{VIBRONIC INTERACTION IN CUBIC FULLERIDES}
Cubic A$_{3}$C$_{60}$ crystalize into face-centered cubic (fcc) and body-centered cubic (bcc) structures (Fig. \ref{fig:Struct_FCC_A15}). 
In both types of A$_{3}$C$_{60}$, one electron from each alkali atom A transfers to the $t_{1u}$ LUMO orbitals of fullerenes, due to a very strong electronegativity of the latter, resulting in their three-fold population at each C$_{60}$ site. Since all fullerene molecules reside in cubic lattice points, their symmetry is reduced from I$_{h}$ to T$_{h}$ point group, implying that the $t_{1u}$ orbitals keep their three-fold degeneracy (belonging now to the $T_{u}$ irrep of the T$_{h}$ group). At the same time the icosahedral irrep $H_{g}$ splits under this symmetry reduction into $E_{g}$ and $T_{g}$ \cite{Altmann1994}.

In the following we consider the simplest situation of fully localized LUMO electrons at sites.

\subsection{The Jahn-Teller Hamiltonian for localized LUMO electrons at C$_{60}$ sites}
The JT Hamiltonian for A$_{3}$C$_{60}$ reads as follows:
\begin{equation}
\hat{H}_\text{JT} = \hat{H}_{\text{ph}} + \sum_{\mathbf{n}} \Big( \hat{H}_\text{U}^{\mathbf{n}} + \hat{H}_\text{H}^{\mathbf{n}} + \hat{V}_\text{a}^{\mathbf{n}} + \hat{V}_\text{e}^{\mathbf{n}} +
\hat{V}_\text{t}^{\mathbf{n}} \Big) .
\label{Eq:H_JT}
\end{equation}
The first term is the phonon Hamiltonian of the crystal,
\begin{equation}
\hat{H}_{\text{ph}} = \sum_{\kappa \mathbf{k}} \Big( \frac{1}{2} \hat{P}_{\kappa \mathbf{k}}^2 + \frac{1}{2} \omega_{\kappa \mathbf{k}}^2 Q_{\kappa \mathbf{k}}^2 \Big) ,
\label{Eq:H_ph}
\end{equation}
expressed via phonon coordinates $Q_{\kappa \mathbf{k}}$ characterized by the wave vector $\mathbf{k}$ and the branch $\kappa$. The other terms in (\ref{Eq:H_JT}) are one-site contributions with $\mathbf{n}$ denoting the unit cell which is supposed to include one A$_{3}$C$_{60}$ formula unit. $\hat{H}_\text{U}^{\mathbf{n}}$ and $\hat{H}_\text{H}^{\mathbf{n}}$ describe bielectronic interactions in the LUMO orbitals of $\mathbf{n}$-th fullerene molecule: 
\begin{eqnarray}
\hat{H}_\text{U}^{\mathbf{n}} &=& \sum_{\alpha} \sideset{}{^\prime}\sum_{\beta (\ne \alpha)} \sum_\sigma \frac{U_{\perp}}{2} 
\hat{n}_{\alpha \sigma} \hat{n}_{\beta \sigma} + \sum_{\alpha \sigma} \frac{U_{\perp}}{2} \hat{n}_{\alpha \sigma} \hat{n}_{\alpha -\sigma} \nonumber\\
&=& \frac{U_{\perp}}{2} \left( \hat{N}^2 - \hat{N} \right) , 
\nonumber\\
\hat{H}_\text{H}^{\mathbf{n}} &=& \sum_{\alpha} \sideset{}{^\prime}\sum_{\beta (\ne \alpha)} \sum_\sigma 
 \frac{J_\text{H}}{2} \left[ - \hat{n}_{\alpha \sigma} \hat{n}_{\beta \sigma}
 + \hat{c}_{\alpha \sigma}^\dagger \hat{c}_{\beta \sigma} \hat{c}_{\alpha -\sigma}^\dagger \hat{c}_{\beta -\sigma}\right. \nonumber\\
 && \left.+ \hat{c}_{\alpha \sigma}^\dagger \hat{c}_{\beta \sigma} \hat{c}_{\alpha -\sigma}^\dagger \hat{c}_{\beta -\sigma}
   \right]
 + \sum_{\alpha \sigma} J_\text{H} \hat{n}_{\alpha \sigma} \hat{n}_{\alpha -\sigma}, 
\label{Eq:H_bi}
\end{eqnarray}
where the second-quantization operators correspond to the LUMO orbitals of the corresponding fullerene site, $\alpha, \beta =x,y,z$, and $\hat{N} = \sum_{\alpha \sigma} \hat{n}_{\alpha \sigma}$ is the operator of total number of electrons in the $t_{1u}$ shell. The first term is the Coulomb repulsion of LUMO electrons, the parameter $U_{\perp}$ describing the repulsion of electrons in different orbitals within the same $t_{1u}$ shell. Since this contribution depends only on the total number of LUMO electrons on site considered constant in the present work, it gives only a constant energy shift and is dropped in the subsequent treatment. The term $\hat{H}_\text{H}^{\mathbf{n}}$ describes orbitally-specific electronic interactions which, due to the isomorphism of cubic $t_{1u}$ and atomic $p$ shells, depend on one Hund's parameter $J_\text{H}$. Besides exchange (Hund's rule) coupling and electron pairs transfer between different $t_{1u}$ orbitals, this operator also includes (through the last term) the difference between the electron repulsion in the same and different orbitals, $U_{\parallel} - U_{\perp} =2J_\text{H}$.

The last three terms in the sum in Eq.(\ref{Eq:H_JT}) describe the vibronic interaction with three types of nuclear distortions at a given site, with irreps (of the T$_h$ group) contained in the symmetric square $\left[T_{u}^2\right]=A_{g}\oplus E_{g}\oplus T_{g}$\cite{bersuker2006JTeffect},   
\begin{widetext}
\begin{eqnarray}
\hat{V}_\text{A}^{\mathbf{n}} &=& \sum_{\mu} V_{\mu A} q_{\mu A}^{\mathbf{n}} \hat{N} , \nonumber
\\
\hat{V}_\text{E}^{\mathbf{n}} &=& \sum_\sigma \sum_{\mu} V_{\mu E} 
 \left(\hat{c}_{x\sigma}^\dagger, \hat{c}_{y\sigma}^\dagger, \hat{c}_{z\sigma}^\dagger \right)
 \begin{pmatrix}
  \frac{1}{2}q_{\mu\theta}^{\mathbf{n}} - \frac{\sqrt{3}}{2} q_{\mu\epsilon}^{\mathbf{n}} & 0  & 0 \\
  0 & \frac{1}{2}q_{\mu\theta}^{\mathbf{n}} + \frac{\sqrt{3}}{2} q_{\mu\epsilon}^{\mathbf{n}} & 0 \\
  0 & 0 & -q_\theta^{\mathbf{n}} 
 \end{pmatrix}
 \begin{pmatrix}
  \hat{c}_{x\sigma}\\
  \hat{c}_{y\sigma}\\
  \hat{c}_{z\sigma}
 \end{pmatrix}, \nonumber
\\ 
\hat{V}_\text{T}^{\mathbf{n}} &=& \sum_\sigma \sum_{\mu} V_{\mu T} 
 \left(\hat{c}_{x\sigma}^\dagger, \hat{c}_{y\sigma}^\dagger, \hat{c}_{z\sigma}^\dagger \right)
 \begin{pmatrix}
  0 & \frac{\sqrt{3}}{2} q_{\mu\zeta}^{\mathbf{n}} & \frac{\sqrt{3}}{2} q_{\mu\eta}^{\mathbf{n}} \\
  \frac{\sqrt{3}}{2} q_{\mu\zeta}^{\mathbf{n}} & 0 & \frac{\sqrt{3}}{2} q_{\mu\xi}^{\mathbf{n}} \\
  \frac{\sqrt{3}}{2} q_{\mu\eta}^{\mathbf{n}} & \frac{\sqrt{3}}{2} q_{\mu\xi}^{\mathbf{n}} & 0 
 \end{pmatrix}
 \begin{pmatrix}
  \hat{c}_{x\sigma}\\
  \hat{c}_{y\sigma}\\
  \hat{c}_{z\sigma}
 \end{pmatrix},
\label{Eq:V_JT}
\end{eqnarray}
\end{widetext}
where $q_{\mu\gamma}^{\mathbf{n}}$ are symmetrized distortions at the $\mathbf{n}$-th fullerene site transforming after the row $\gamma$ of the corresponding irrep; $\mu$ counts the repeating distortions of a given type. The latter include both distortions of the fullerene itself and displacements of surrounding atoms.    

The operator $\hat{V}_\text{A}$ depends on the total population number, i.e., a constant in the electronic space. It doesn't lead to the splitting of the $t_{1u}$ orbitals and merely represents a constant force acting on the environment of the $\mathbf{n}$-th site in a totally symmetric fashion, i.e. without destroying its cubic site symmetry. Summed up over all sites, the effect of such terms results in the reoptimization of the structure of the cubic crystal, which is assumed to be already done during quantum chemistry calculations. 

The other two operators describe the JT coupling to the local distortions at the $\mathbf{n}$-th site. Under the symmetry reduction I$_{h}\rightarrow$T$_{h}$, two of five symmetrized functions of the irrep $H_{g}$, $\gamma =\theta ,\epsilon$, form the basis of the irrep $E_{g}$, while the other three, $\gamma =\xi , \eta ,\zeta$, - of the irrep $T_{g}$. Then the Clebsh-Gordan coefficients [the numerical coefficients in two matrices in (\ref{Eq:V_JT})] coincide for the two groups: $\langle t\alpha |\Gamma\gamma \;t\beta\rangle =\langle t_{1}\alpha |H \gamma \;t_{1}\beta\rangle $, where $\Gamma= E, T$ and $\alpha ,\beta =x,y,z$. Therefore, for equal vibronic coupling constants of two types, $V_{\mu E} =V_{\mu T}$, the matrix describing the JT coupling with five symmetrized nuclear distortions in (\ref{Eq:V_JT}) coincides with the one for the $t_{1u} \otimes H_g$ JT problem \cite{OBrien1969a, Chancey1997}.

The assumption of localized electrons in the LUMO orbitals of fullerene sites is valid for insulating cubic fullerides Cs$_{3}$C$_{60}$ in the A15 structure (bcc) and fcc lattice at ambient pressure, and in all expanded fullerides such as, e.g., 
Li$_{3}$ NH$_3$C$_{60}$ \cite{durand2003mott, Chibotaru2005a}.
It is a reasonable approximation in metallic fullerides which are not far from Mott-Hubbard metal-insulator transition. These are the fcc Cs$_{3}$C$_{60}$ under pressure, Rb$_{3}$C$_{60}$ and several A$_x$A'$_{3-x}$C$_{60}$ compounds with different alkali atoms A and A'. As a matter of fact, an experimental proof for (dynamical) JT effect in Rb$_{3}$C$_{60}$ is the detection of a spin gap in its NMR spectrum \cite{PhysRevB.66.155124}.   

Another approximation, of a single C$_{60}$ per unit cell, was adopted to reduce the unit cell to one formula unit because of complications with the calculation of phonon spectrum. This situation is strictly realized only in mixed compounds AA'$_2$C$_{60}$. All other fullerides are characterized by different forms of merohedral arrangement of C$_{60}$'s (their C$_2$ rotations around a cubic axis). Thus the A15 Cs$_{3}$C$_{60}$ is characterized by the merohedral order (two C$_{60}$'s in a unit cell), while the three other fcc compounds investigated here are subject to merohedral disorder \cite{Stephens1991structure, Teslic1995shortrange, Mazin1993orientational}. 
We believe that the neglect of merohedral arrangement will not affect the main conclusions of this study, the main reason being the high symmetry of the C$_{60}$ balls.         

\subsection{Calculation of orbital vibronic coupling constants}

\begin{figure}[tb]
\begin{center}
\begin{tabular}{ll}
(a)~~~~~~~~~~~~~~~~~~~~~~~~~~~~~~~~~~~ & (b) \\
\multicolumn{2}{c}{\includegraphics[width=1.0\linewidth,height=4.0cm]{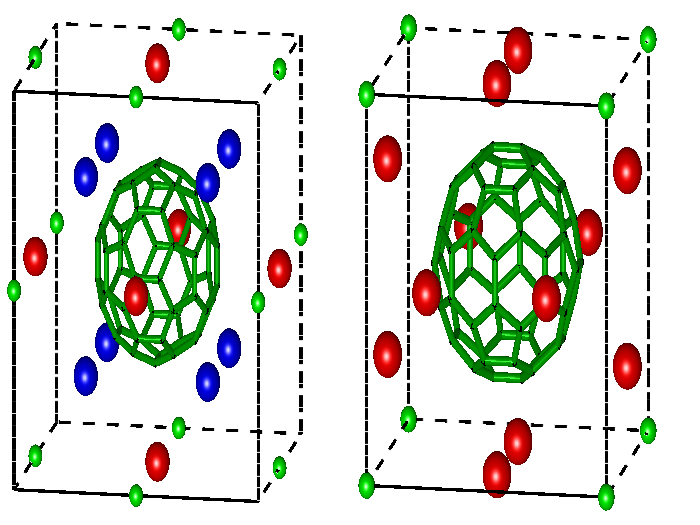}}
\end{tabular}
\end{center}
    \caption{ A$_{n}$C$_{60}$ fragments used for the calculation of vibronic coupling constants.
(a) Fcc unit cell; octahedral and cubic alkali atoms are shown by red and blue balls.
(b). Bcc-like unit cell (A15 structure); Cs atoms are indicated by red balls. Nearest neighbor C$_{60}$ molecules contributing to interfullerene active JT modes are shown by green balls.
}
\label{Fig:VCCsStructFCCA15}
\end{figure}

The vibronic coupling constants at fullerene sites have been extracted from the splitting of three $t_{1u}$ LUMO orbitals in function of the amplitude of active local nuclear distortions
\footnote{
They correspond to expectation values of the first derivatives of the Hamiltonian with respect to the mass weighted normal coordinates of the corresponding nuclear modes \cite{bersuker2006JTeffect} 
and involve the $t_{1u}$ orbitals only because the contributions from doubly occupied orbitals are zero due to symmetry [see for detailed discussion in Ref. \cite{Sato2006}]. 
In the following we use the atomic units, $E_h/\sqrt{m_e} a_0$, where $E_h$ is Hartree, $m_e$ the electron mass, and $a_0$ the Bohr radius (1 a.u.=2.73 10$^{11}$ dyn/g$^{1/2})$. 
}.
To this end, DFT calculations of A$_{n}$C$_{60}$ clusters including nearest and next-nearest neighbor alkaline atoms and point charges replacing other alkaline atoms and fullerenes (Fig. \ref{Fig:VCCsStructFCCA15}) have been done with Gaussian09 package \cite{g09}. Hybrid B3LYP exchange correlation functional \cite{Becke1993a} and triple-zeta basis set (6-311G(d)) were employed. The three LUMO orbitals have been populated by three electrons with parallel spins. For given local nuclear distortions, ten calculations corresponding to their different amplitude have been performed. From them the slopes of LUMO orbital energies in vicinity of equilibrium point have been extracted via an interpolation procedure and vibronic coupling constants derived for a number of active JT modes making use of vibronic matrices from Eq. (\ref{Eq:V_JT}). 

\subsubsection{Intrafullerene modes}
As active JT modes, the normal vibrational H$_{g}$ modes of isolated C$_{60}$ \cite{Liu2018b} have been chosen, which merely subdivide into E$_g$ and T$_{g}$ modes in cubic fullerides. We also calculated the vibronic coupling constants for intrafullerene A$_{g}$ modes. The explicit form of the corresponding nuclear displacements is given in the Suplementary Material (SM) \cite{SM}. 
We note that these modes are not vibrational eigenmodes in fullerides and have been chosen for the sake of comparison with isolated C$_{60}^{3-}$. We could have considered arbitrary combinations of them as well because the resulting electron-phonon Hamiltonian is invariant with respect to any such choice. 

The calculated vibronic coupling constants are shown in Table \ref{tab:VCC_constants_M}. 

Despite the molecular character of fulleride crystals, 
the difference between vibronic coupling constants for E$_g$ and T$_{g}$ modes appears to be non-negligible and these constants also deffer from corresponding vibronic coupling constants of isolated fullerene ion (last column in Table \ref{tab:VCC_constants_M}).
We did not consider the coupling to intrafullerene modes of other I$_h$ genealogy (G$_g$, T$_{1g}$ and T$_{2g}$), which according to the selection rules for the T$_h$ group also become active in fullerides. Because the coupling to these modes is absent for isolated fullerene anions, we expect it to be negligible in fullerides. 
\begin{table}[tb]
    \caption{\label{tab:VCC_constants_M}
 Vibronic coupling constants for E$_g$ and T$_{g}$ intrafullerene distortions of H$_g$ genealogy compared to similar coupling constants in isolated C$_{60}^{-}$ (in 10$^{-6}~ a. u.$.).
   }
    \begin{ruledtabular}
    \begin{tabular}{cccccccccc}
        &\multicolumn{6}{c}{fcc}    & \multicolumn{2}{c}{A15} &  \\
        \cline{2-7} \cline{8-9}
        & \multicolumn{2}{c}{K$_{3}$C$_{60}$} &  \multicolumn{2}{c}{Rb$_{3}$C$_{60}$}&\multicolumn{2}{c}{Cs$_{3}$C$_{60}$} & \multicolumn{2}{c}{Cs$_{3}$C$_{60}$}& C$_{60}^-$ \\
    \cline{2-7} \cline{8-9}
        &E$_{g}$ &T$_{g} $ &E$_{g}$ &T$_{g} $ &E$_{g}$ &T$_{g} $ &E$_{g}$ &T$_{g}$  &H$_g$\\
\hline
H$_{g}$(1)&18.9   &19.9   &19.2   &19.8   &19.4   &19.4   &19.0   &19.7  & 19.2  \\
H$_{g}$(2)&42.4   &39.9   &43.4   &41.0   &44.4   &42.3   &44.4   &43.0  & 45.0  \\
H$_{g}$(3)&80.1   &78.7   &80.2   &78.9   &80.4   &79.5   &81.0   &79.0  & 75.4  \\
H$_{g}$(4)&56.3   &53.0   &56.6   &53.3   &56.7   &53.9   &54.8   &55.6  & 55.4  \\
H$_{g}$(5)&77.5   &77.2   &76.1   &76.8   &74.8   &76.3   &75.1   &75.3  & 76.6  \\
H$_{g}$(6)&58.1   &58.5   &58.6   &58.9   &58.9   &59.4   &58.1   &59.8  & 57.8  \\
H$_{g}$(7)&200.4   &202.3   &199.9   &203.7   &299.7   &204.7   &201.2   &204.0  & 209.9  \\
H$_{g}$(8)&206.1   &201.5   &205.4   &202.4   &204.7   &203.3   &205.0   &203.5  & 204.3  \\
\hline
A$_{g}(1)$&\multicolumn{2}{c}{66.7} &\multicolumn{2}{c}{67.8} &\multicolumn{2}{c}{70.2} &\multicolumn{2}{c}{70.0}& 26.4\\
A$_{g}(2)$&\multicolumn{2}{c}{311.8} &\multicolumn{2}{c}{310.5} &\multicolumn{2}{c}{308.7} &\multicolumn{2}{c}{309.5}& 238.0
    \end{tabular}
    \end{ruledtabular}
\end{table}

\subsubsection{Alkali modes}
In fcc A$_{3}$C$_{60}$ each fullerene is surrounded by a cube of nearest neighbor and an octahedron of next-nearest neighbor alkali atoms (blue and red balls in Fig. \ref{fig:Struct_FCC_A15}a, respectively). The closest alkali environment contributes with one E$_g$ and two T$_{g}$ modes of symmetrized distortions, while the octahedral environment with one E$_g$ and one T$_{g}$ modes. The expressions of these modes via atomic displacements are given in SM \cite{SM} and also are depicted in various textbooks (see, e.g. \cite{bersuker2006JTeffect}). 
In the case of A15 fulleride, the closest alkali atoms form a pseudo octahedron (two atoms at each face of surrounding cube) as shown in Fig. \ref{fig:Struct_FCC_A15}b. The active distortions include two E$_g$ and two T$_{g}$ modes whose form is given in SM \cite{SM}.  

\begin{table}[tb]
\caption{\label{tab:VCC_constants_A_FCC_A15}
  Vibronic coupling constants for symmetrized distortions of cubic (cub), octahedal (oct) frame of surrounding alkali atoms in fcc A$_{3}$C$_{60}$, and pseudo-octahedral (p-oct) frame in A15 (in 10$^{-6}~ a. u.$)
}
\begin{ruledtabular}
\begin{tabular}{cccccccc}
        &\multicolumn{6}{c}{fcc}    & A15  \\
\cline{2-7} 
        & \multicolumn{2}{c}{K$_{3}$C$_{60}$} &  \multicolumn{2}{c}{Rb$_{3}$C$_{60}$}&\multicolumn{2}{c}{Cs$_{3}$C$_{60}$} & Cs$_{3}$C$_{60}$ \\
\cline{2-7} 
        &   cub     &   oct     &   cub     &   oct     &   cub     &   oct    & p-oct \\
\hline
A$_{g}$(1) & 88.8   & 68.2      & 60.6      & 44.6      & 48.4      & 39.1     &  4.1 \\
E$_{g}$(1) & 2.6    & 2.4       & 1.6       & 1.5       & 1.0       & 1.1      &  1.8 \\
T$_{2g}$(1)& 2.7    & 0.8       & 2.0       & 0.5       & 1.7       & 0.4      &  0.9 \\
T$_{2g}$(2)& 0.2    &    -      & 0.2       &    -      & 0.1       &    -     &  0.9 \\
A$_{g}$(2) &   -    &    -      &    -      &    -      &    -      &    -     &  57.0 \\
E$_{g}$(2) &   -    &    -      &    -      &    -      &    -      &    -     &  0.8 \\
T$_{g}$(3) &   -    &    -      &    -      &    -      &    -      &    -     &  0.8 \\
\end{tabular}
\end{ruledtabular}
\end{table}

The calculated vibronic coupling constants are given in Table \ref{tab:VCC_constants_A_FCC_A15}.
As we can see, they are smaller than the vibronic coupling constants for intrafullerene modes by ca one order of magnitude (an exception are alkali A$_{g}$ modes for fcc A$_{3}$C$_{60}$). 
Nevertheless these modes are related to acoustic and low-frequency optical phonons  (see the next section), therefore their contribution to JT stabilization cannot be neglected from the start. 
Moreover, as will be seen below, their contribution is crucial for the warping of one-site APES and for the interaction of JT distortions at different sites.

\subsubsection{Interfullerene modes}
Each fullerene molecule in fcc A$_{3}$C$_{60}$ is surrounded by twelve C$_{60}$'s forming a cub-octahedron (Fig. 2a). The latter yields two E$_g$ and two T$_{g}$ modes of symmetrized distortions. In A15 fulleride the nearest alkali atoms form a cube (Fig. 2b) which gives one E$_g$ and two T$_{g}$ modes. The expressions of these modes via atomic displacements are given in SM \cite{SM}. Given a large spacing between fullerene molecules, only the electrostatic interaction between them is taken into account in the calculation of vibronic coupling constants.

\begin{table}[tb]
    \caption{\label{tab:VCC_constants_A_interC60}
 Vibronic coupling constants for active symmetrized distortions of fullerene molecules surrounding a given C$_{60}$ in fcc and A15 fullerides (in 10$^{-6}~ a. u.$).
}
    \begin{ruledtabular}
    \begin{tabular}{ccccc}
        &\multicolumn{3}{c}{fcc}    & A15  \\
\cline{2-4} 
        & K$_{3}$C$_{60}$ & Rb$_{3}$C$_{60}$&Cs$_{3}$C$_{60}$ & Cs$_{3}$C$_{60}$ \\
\cline{2-4} 
    \hline
A$_{g}$    & 35.1 & 34.3      & 32.5     & 27.7\\
E$_{g}$(1) & 0.2  & 0.1      & 0.1      & 0.2\\
E$_{g}$(2) & 0.2  & 0.2      & 0.2      & - \\
T$_{2g}$(1)& 0.2  & 0.2      & 0.2      & 0.3\\
T$_{2g}$(2)& 0.3  & 0.3      & 0.2      & 0.1\\
    \end{tabular}
    \end{ruledtabular}
\end{table}

The calculated vibronic coupling constants are given in Table \ref{tab:VCC_constants_A_interC60}.
They are obtained one order of magnitude smaller than the vibronic coupling constants for alkali modes (Table \ref{tab:VCC_constants_A_FCC_A15}). This is because the interfullerene vibronic coupling can be seen as an electrostatic interaction of a set of electric dipoles (arising from shifted charges $q=3 e$) with the quadrupolar distribution of $t_{1u}$ LUMO electrons, scaling as $R^{-4}$ with a distance $R$ to the center of C$_{60}$. On the other hand, there is an additional covalent contribution to vibronic coupling constants for alkali modes. 
%
%
\section{THE PHONON SPECTRUM OF CUBIC FULLERIDES}
In order to take into account exactly the effect of the environment on the JT effect at fullerene sites and the intersite interaction of their active JT modes, the precise knowledge on phonon modes of A$_3$C$_{60}$ is decisive. 
Many attempts have been undertaken to calculate the phonon dispersion. 
The first calculations by Varma {\it et al.} [Science 1991] and  You {\it et al.}. \cite{Phon_wholeC60} in the beginning of 1990s were semiempirical but reflected qualitatively the basic features of phonon spectrum in A$_{3}$C$_{60}$ fullerides. In spite of that, they could not achieve a correct description of intra-fullerene vibronic interaction. 
Later works \cite{IONIZED_PP, Phonon_imaginary_Nomura} with first principles methods also could not provide accurate information on the low-energy phonon modes: In these calculations, the low-energy frequecies become imaginary. 
In this work, this issue is solved by employing new ionized pseudopotentials for alkali atoms. 
%

\subsection{Calculation details}
%
To reduce the calculational load, in both fcc and A15 fullerides the fullerenes were supposed to be completely ordered.  
Experimental lattice constants were used for the starting crystal structure \cite{Structure_nature_K_Rb, Ganin2008a}, then they are fully relaxed in order to get good phonon calulations. 
%
%
The phonon calculations have been done by density functional perturbation theory (DFPT) \cite{dfpt} with exchange correlation funcionals of LDA and PBE types \cite{PP_PZ, PBE-1}.
Given large unit cells in A$_{3}$C$_{60}$, soft pseudopotentials (PPs) have been employed. 
To this end, following Akashi and Arias\cite{IONIZED_PP}, we chose the configuration 
(3p)$^{6.0}$(4s)$^{0.0}$(3d)$^{0.0}$, (4p)$^{6.0}$(5s)$^{0.0}$(4d)$^{0.0}$ and (5p)$^{6.0}$(6s)$^{0.0}$(5d)$^{0.0}$ for K, Rb, and Cs, respectively, 
with the nonlinear core correction \cite{Nonlinear_core_correction}. 
The relativistic effects in alkali atoms were considered within scalar relativistic approximation \cite{Scalar_relativistic}. 
All the PPs were generated by Trolled-Martins\cite{TM-method} method using {\tt atomic} code within the plane wave based package Quantum Espresso \cite{PWscfcode}.  
For the LDA functional the parametrization of Perdew and Zunger (PZ) \cite{PP_PZ} was used and 
for the GGA functional the parametrization of Perdew, Burke, and Ernzerhof (PBE) \cite{PBE-1} was employed. 
The details of generation of ionized PPs and the test calculations are given in SM \cite{SM}.

The plane-wave kinetic energy cutoff was set to 60 Ry with the density cut-off of 240 Ry, 
and shifted $4\times4\times4$ Monkhorst-Pack meshes were used to perform Brillouin zone integration in order to ensure the convergence of the results. 
The convergence of the total energy was set to be better than 10$^{-14}$ Hartree and forces on the atoms were limited within 10$^{-5}$ Ry/a.u. 

%
\subsection{The phonon dispersion}
\begin{figure}[tb]
\includegraphics [height=1.0\linewidth,angle=-90]{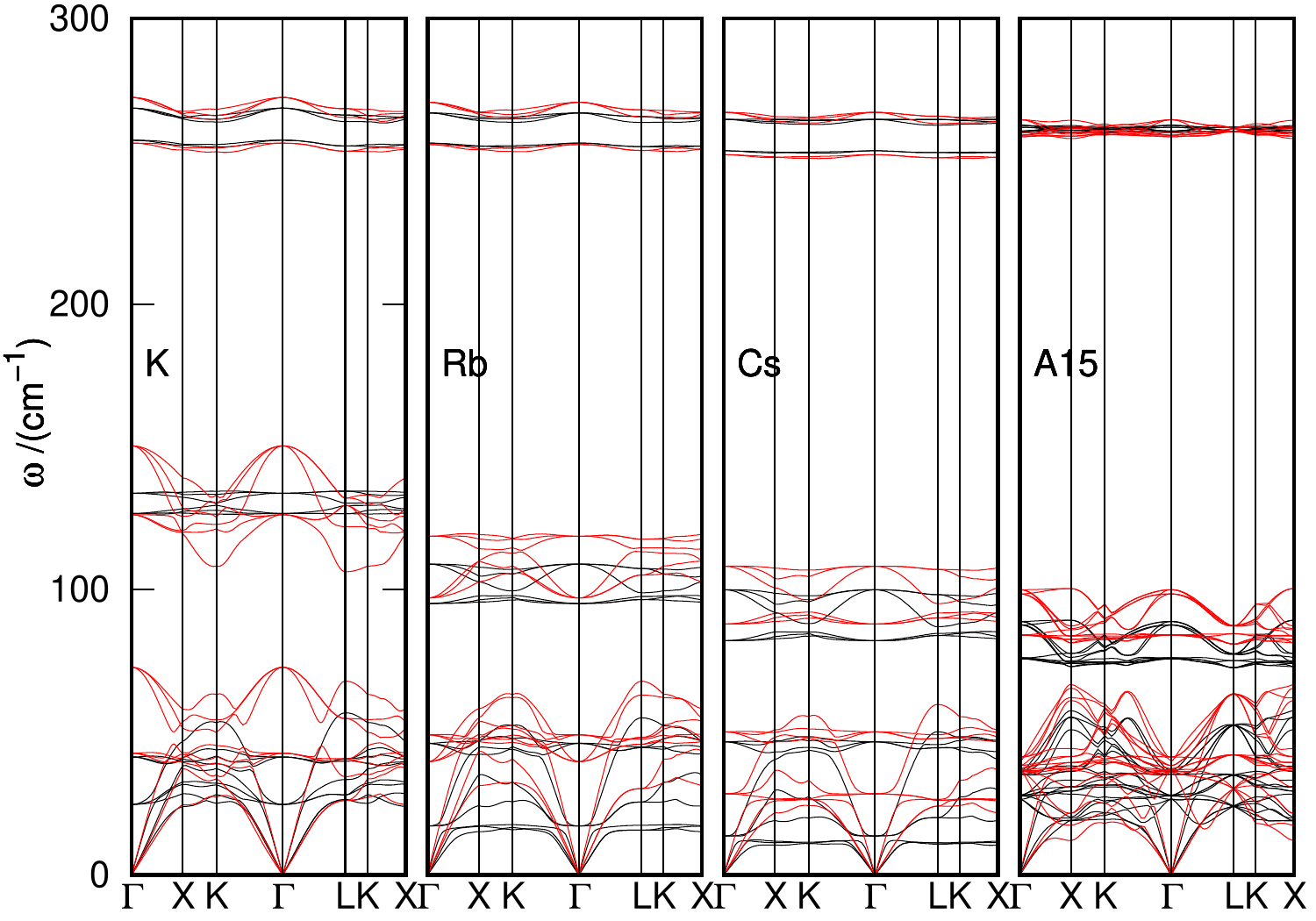}
    \caption{\label{Fig:Phon_dis_300} Phonon dispersion for A$_{3}$C$_{60}$, including A15. Red and black lines correspond to PBE and LDA, respectively.}
\end{figure}

The phonon dispersion for A$_{3}$C$_{60}$ is shown in Fig. \ref{Fig:Phon_dis_300}, where the frequency range was restricted to 300 cm$^{-1}$ for visibility (see the SM \cite{SM} for the dispersion of all phonon bands). 
It is evident that no imaginary modes appear anymore in the present calculations with both PBE and LDA functionals, contrary to the previous calculations \cite{IONIZED_PP, Phonon_imaginary_Nomura}. 

Visualization of polarization vectors at the $\Gamma$ point allows us to conclude that 
the lowest three optical branches, 4-6, correspond to opposite displacements of fullerene and a next nearest alkali atom, siting in the octahedral interstitial, along three Cartesian axes. The next branches, 7-9, are three librational (pure rotational) modes of fullerene molecules. The next branches, 10-12, correspond to displacements of two nearest alkali atoms hybridized with $H_{g}\gamma $(1) vibrations ($\gamma= \xi, \eta, \zeta$) of C$_{60}$ cage. Finally the upper mixed branches, 7-15, represent opposite translations of two nearest alkali atoms and fullerene. All higher branches arise from  almost pure intrafullerene vibrations. Thus the highest five bands in Fig. 3 correspond to $H_{g}$(1) intrafullerene vibrations, split at the $\Gamma$ point into $E_g$ and $T_g$ degenerate phonons according to cubic symmetry of the lattice.  

Above 150 cm$^{-1}$,  
PBE gives slightly smaller frequencies than LDA albeit displaying similar dispersion, 
while in the lower frequency range, 0$\div$150 cm$^{-1}$, the results of PBE and LDA disagree significantly. 
%
The calculated frequencies of the H$_{g}$ intrafullerene phonons at the $\Gamma$ point are tabulated in Table S2. 
One can see that the frequencies calculated by Nomura and Arita \cite{Phonon_imaginary_Nomura} with constrained DFPT within LDA are close to our results, 
and both are in a good agreement with the experimental Raman data \cite{Zhou1992, Zhou1993, Mitch1992, Mitch1993}.
\section{THE LOWEST ADIABATIC POTENTIAL ENERGY SURFACE}
The cooperative Jahn-Teller dynamics in fullerides is highly complex and has never been assessed even for the simplest, insulating compounds. The character of JT dynamics can be understood by analyzing the lowest adiabatic potential energy surface (APES) of the crystal, when all JT centers (fullerenes) are in the ground electronic state for given distortions of the lattice. 
\subsection{Static JT effect in terms of electronic vectors at C$_{60}$ sites}
The potential energy operator for A$_3$C$_{60}$ is obtained by dropping the kinetic energy of phonons (we also neglect some other contributions mentioned in Sect.IIA) from the full Jahn-Teller Hamiltonian in Eq. (\ref{Eq:H_JT}):
\begin{eqnarray}
\hat{U}_\text{JT} &=& \sum_{\kappa \mathbf{k}} \frac{1}{2} \omega_{\kappa \mathbf{k}}^2 Q_{\kappa \mathbf{k}}^2
 + \sum_{\mathbf{n}} \Big( \hat{H}_\text{H}^{\mathbf{n}} +\sum_{\Gamma =E,T}\sum_{\mu} V_{\mu\Gamma} \nonumber \\
&&\times \sum_{\gamma (\in \Gamma)} q_{\mu\gamma}^{\mathbf{n}}\sum_{\alpha ,\beta}\sum_{\sigma} 
\langle t_{1}\alpha |H \gamma \;t_{1}\beta\rangle \;
\hat{c}_{\alpha\sigma}^{\mathbf{n}\dagger} \hat{c}_{\beta\sigma}^{\mathbf{n}}
\Big) ,
\label{Eq:U_JT}
\end{eqnarray}
where the operator of JT coupling is expressed through Clebsh-Gordan coefficients for the icosahedral group [see the discussion after Eq. (\ref{Eq:V_JT})].
The lowest APES is obtained by diagonalizing the electronic operator in (\ref{Eq:U_JT}) corresponding to each site $\mathbf{n}$ and considering its lowest eigenvalues in function of local JT distortions {$q_{\mu\gamma}^{\mathbf{n}}$}. Thus obtained function of local JT distortions of all sites of the crystal is further investigated for extremes. 

For the investigation of the extremes of APES there exists a more convenient approach proposed by \"{O}pie and Pryce for molecular JT problems \cite{Opik1957}, which we extend here. It essentially exploits the existence of a bijective relation (one-to-one correspondence) between the extremes of an APES and the electronic function corresponding to nuclear distortions in these extremes. 
Basing on this property, \"{O}pik and Pryce proposed to find first the equilibrium nuclear coordinates for an arbitrary form of the electronic wave function (expressed via arbitrary fixed parameters - adiabatic coordinates), and then to investigate the extremes of the obtained energy functional in the space of these adiabatic coordinates. 
These adiabatic coordinates can be viewed as directional cosines of the vector representing an arbitrary wave function in the functional space of electronic basis functions involved in the Jahn-Teller effect. The representation via such vectors (electronic pseudospins) was widely used for the investigation of static JT effect in molecular systems and cooperative JT effect in solids for simple JT interaction on sites. 
It was extended for molecular systems with multimode vibronic coupling \cite{Ceulemans1996isostationary}.
Here we further extend this approach over multimode JT coupling in crystals involving multielectric JT sites, a situation realized in our fullerides.

The basis of electronic wave functions involved in JT effect at a C$_{60}^{3-}$ ion includes spin doublet terms of the $t_{1u}^3$ electronic configuration, the atomic-like $^{2}$P and $^{2}$D molecular terms amounting to eight electronic wave functions for a given projection of the total spin $S=1/2$ \cite{Auerbach1994, OBrien1996}. For arbitrary distortions, in the presence of multiplet splitting operator $\hat{H}_\text{H}^{\mathbf{n}}$ all these eight electronic states (equivalently eight spin-doublet Slater determinants) are generally admixed to the ground adiabatic wave function of C$_{60}^{3-}$. 
However, it was shown \cite{StructureK3C60} that in the case of multiply charged fullerene anions, with $n=2,3,4$, 
the ground adiabatic multielectronic wave function corresponds in a good approximation to the lowest eigenvalue of the operator of JT coupling only. 
This means that the effect of multiplet splitting operator $\hat{H}_\text{H}^{\mathbf{n}}$ can be taken into account in the lowest order of perturbation theory. This is done in Sect. IV B.1.

\begin{figure}[tb]
\begin{center}
\begin{tabular}{ll}
(a) & (b) \\
\includegraphics[width=0.4\linewidth]{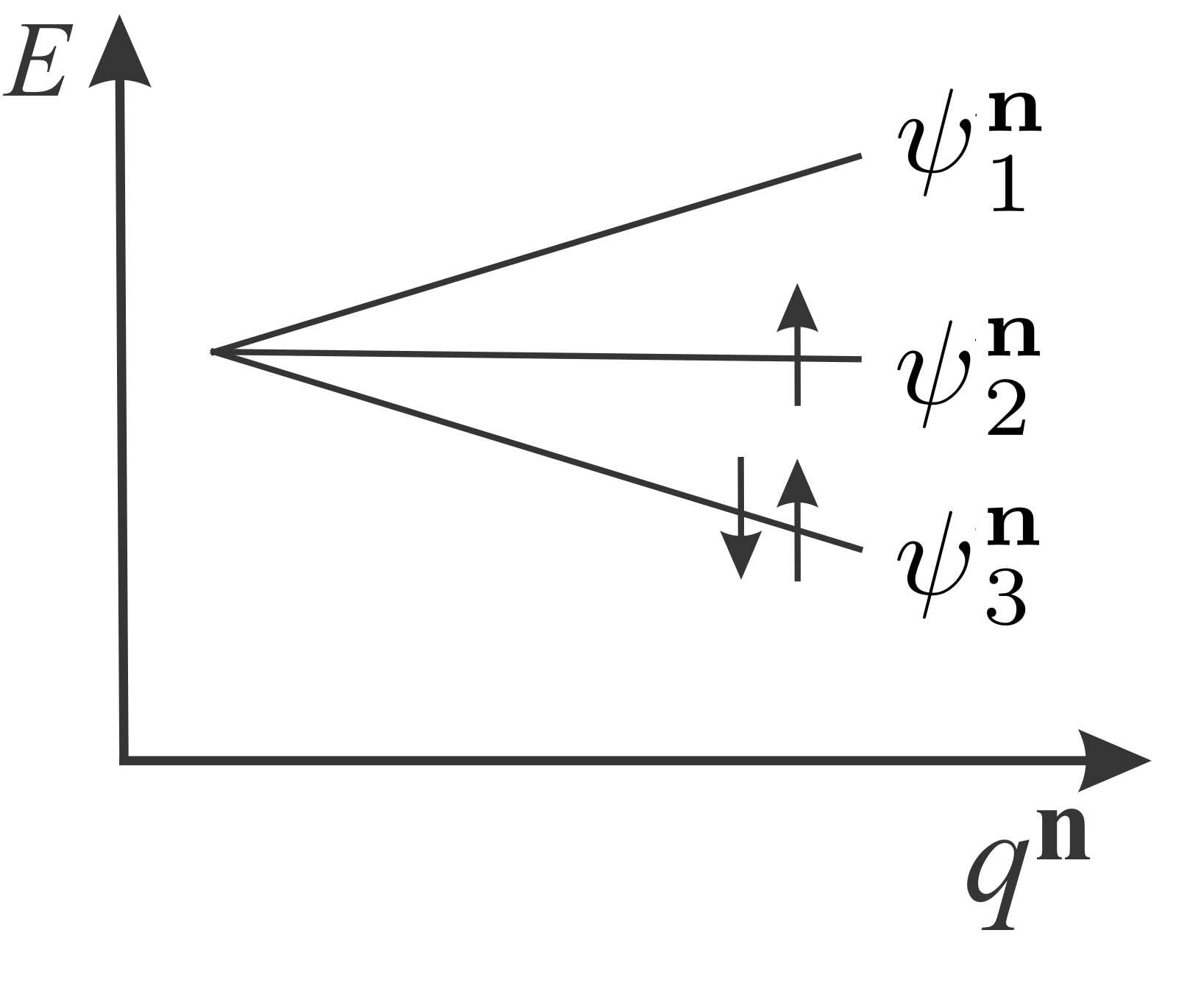}
&
\includegraphics[width=0.6\linewidth]{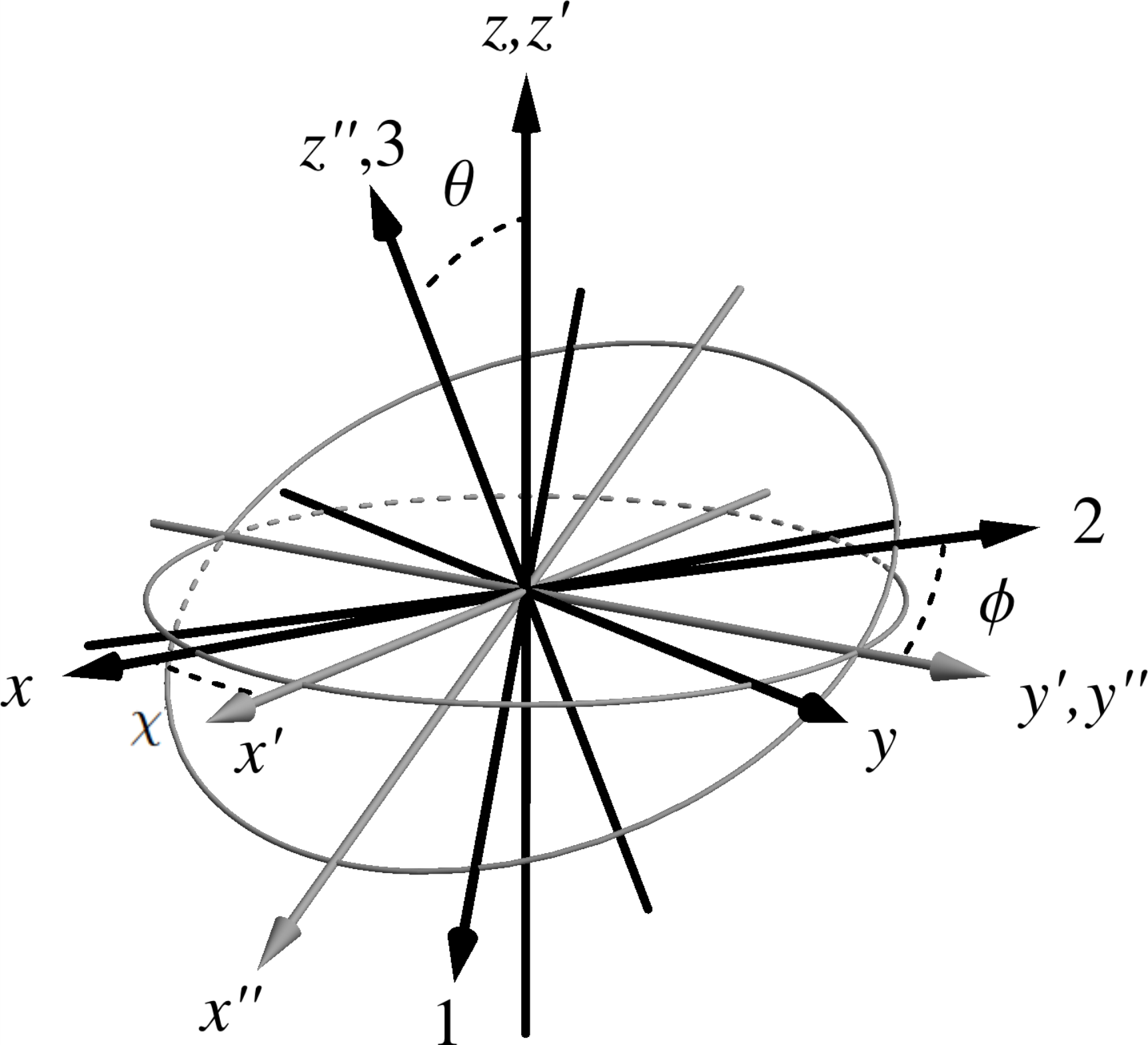}
\end{tabular}
\end{center}
\caption{Adiabatic orbitals at C$_{60}^{3-}$ sites. (a) Splitting and population of adiabatic orbitals in the ground adiabatic electronic state $\Psi_{\uparrow}^{\mathbf{n}}$. (b) Adiabatic orbitals 1,2,3 obtained from the rotations of orthorombic LUMO orbitals $x,y,z$ (corresponding to $t_{1u}\alpha$, $\alpha =x,y,z$) by three Euler angles.}
\label{Fig:Orbitsplitingcoordinates}
\end{figure}

Given the JT coupling at each site is described by one-electron operators, Eq. (\ref{Eq:U_JT}), its ground-state multielectronic eigenfunction ($\Psi^{\mathbf{n}}_{\sigma}$) is a Slater determinant of the lowest occupied eigenorbitals (adiabatic orbitals) $\psi_{i\sigma}^{\mathbf{n}} = \hat{a}_{i\sigma}^{\mathbf{n}\dagger} |0\rangle$, (Fig. \ref{Fig:Orbitsplitingcoordinates}(a)): 
\begin{eqnarray}
\Psi^{\mathbf{n}}_{\sigma} &=&  \hat{a}_{2\sigma}^{\mathbf{n}\dagger} \hat{a}_{3\uparrow}^{\mathbf{n}\dagger} \hat{a}_{3\downarrow}^{\mathbf{n}\dagger} |0\rangle ,
\nonumber\\
\hat{a}_{3\sigma}^{\mathbf{n}\dagger} &=& x_{\mathbf{n}} \hat{c}_{x\sigma}^{\mathbf{n}\dagger} + y_{\mathbf{n}} \hat{c}_{y\sigma}^{\mathbf{n}\dagger} + z_{\mathbf{n}} \hat{c}_{z\sigma}^{\mathbf{n}\dagger} , \nonumber\\
\hat{a}_{2\sigma}^{\mathbf{n}\dagger} &=& \bar{x}_{\mathbf{n}} \hat{c}_{x\sigma}^{\mathbf{n}\dagger} + \bar{y}_{\mathbf{n}} \hat{c}_{y\sigma}^{\mathbf{n}\dagger} + \bar{z}_{\mathbf{n}} \hat{c}_{z\sigma}^{\mathbf{n}\dagger} ,
\label{Eq:Psi_n}
\end{eqnarray}
where the adiabatic coordinates $x,y,z$ and $\bar{x},\bar{y},\bar{z}$ are directional cosines of the doubly occupied and the half filled adiabatic orbitals (Fig. \ref{Fig:Orbitsplitingcoordinates}(a)) w.r.t. three Cartesian axes representing the reference t$_{1u}$ orbitals (Fig. \ref{Fig:Orbitsplitingcoordinates}(b)). The unit vectors $(x,y,z)$ and $(\bar{x},\bar{y},\bar{z})$ are obviously orthogonal. Due to the electronic independence of the sites, the ground-state adiabatic wave function of the whole crystal $\Phi$ is merely a direct product 
of adiabatic wave functions $\Psi^{\mathbf{n}}_{\sigma}$ at different sites,
\begin{equation}
\Phi = \prod_{\mathbf{n}} \hat{a}_{2\sigma}^{\mathbf{n}\dagger} \hat{a}_{3\uparrow}^{\mathbf{n}\dagger} \hat{a}_{3\downarrow}^{\mathbf{n}\dagger} |0\rangle .
\label{Eq:Phi}
\end{equation}

Averaging the potential energy operator (\ref{Eq:U_JT}) on the ground-state adiabatic wave function (\ref{Eq:Phi}), we obtain:
\begin{eqnarray}
\langle U\rangle &&\equiv  \langle \Phi |\hat{U}_\text{JT}|\Phi\rangle = 
\sum_{\kappa \mathbf{k}} \frac{1}{2} \omega_{\kappa \mathbf{k}}^2 Q_{\kappa \mathbf{k}}^2 
 + \sum_{\mathbf{n}} \Big[ \langle \Psi^{\mathbf{n}}_{\sigma}| \hat{H}_\text{H}^{\mathbf{n}}|\Psi^{\mathbf{n}}_{\sigma}\rangle  \nonumber\\
&& +\sum_{\Gamma =E,T}\sum_{\mu} V_{\mu\Gamma} \sum_{\gamma (\in \Gamma)} q_{\mu\gamma}^{\mathbf{n}}
\big( 2R_{\gamma}^{\mathbf{n}} + \bar{R}_{\gamma}^{\mathbf{n}} \big) \big] ,
\label{Eq:<E>}
\end{eqnarray}
where $R$ and $\bar{R}$ are tensorial combinations of adiabatic coordinates:
\begin{eqnarray}
R_{\gamma}^{\mathbf{n}} &=& \sum_{\alpha ,\beta} \langle t_{1}\alpha |H \gamma \;t_{1}\beta\rangle \alpha^{\mathbf{n}} \beta^{\mathbf{n}} , \nonumber\\
\bar{R}_{\gamma}^{\mathbf{n}} &=& \sum_{\alpha ,\beta} \langle t_{1}\alpha |H \gamma \;t_{1}\beta\rangle \bar{\alpha}^{\mathbf{n}} \bar{\beta}^{\mathbf{n}} . 
\label{Eq:R}
\end{eqnarray}
Next we minimize the expression (\ref{Eq:<E>}) w.r.t. nuclear coordinates. Having in mind that the phonon coordinates form a complete, linearly independent set, we first expand the local JT distortions on sites through the latter (the Van Vleck expansion):
\begin{equation}
q_{\mu\gamma}^{\mathbf{n}} = \sum_{\kappa \mathbf{k}} a_{\mu\gamma}^{\mathbf{n}}(\kappa \mathbf{k}) \; Q_{\kappa \mathbf{k}} , 
\label{Eq:VV}
\end{equation}
where $a_{\mu\gamma}^{\mathbf{n}}(\kappa \mathbf{k})$ are Van Vleck coefficients \cite{VanVleck1940}. They are obtained by decomposition of $q_{\mu\gamma}^{\mathbf{n}}$ into the displacements of involved atoms, and the latter into phonon coordinates using the calculated phonon frequencies and polarization vectors in Sect III. Substituting (\ref{Eq:VV}) into (\ref{Eq:<E>}) and minimizing the obtained expression after phonon coordinates, we obtain the equilibrium value of the latter,
\begin{equation}
Q_{\kappa \mathbf{k}}^{(0)}= -\frac{1}{\omega_{\kappa \mathbf{k}}^2} \sum_{\mathbf{n}} 
\sum_{\Gamma =E,T}\sum_{\mu} V_{\mu\Gamma} \sum_{\gamma (\in \Gamma)} a_{\mu\gamma}^{\mathbf{n}}(\kappa \mathbf{k})
\big( 2R_{\gamma}^{\mathbf{n}} + \bar{R}_{\gamma}^{\mathbf{n}} \big) , 
\label{Eq:Q^0}
\end{equation}
in terms of adiabatic coordinates on sites. Substituting the equilibrium coordinates (\ref{Eq:Q^0}) back into the potential energy expression (\ref{Eq:<E>}) we obtain its equilibrium (extremal after $Q_{\kappa \mathbf{k}}$) form in terms of adiabatic coordinates only:
\begin{eqnarray} 
\langle U\rangle^{(0)} &&= \sum_{\mathbf{n}} \langle \Psi^{\mathbf{n}}_{\sigma}| \hat{H}_\text{H}^{\mathbf{n}}|\Psi^{\mathbf{n}}_{\sigma}\rangle  -
 \frac{1}{2} \sum_{\mathbf{n}_1} \sum_{\mathbf{n}_2} \sum_{\mu_1 \Gamma_1} \sum_{\mu_2 \Gamma_2} V_{\mu_1 \Gamma_1} V_{\mu_2 \Gamma_2} \nonumber \\
&&\times \sum_{\gamma_1 (\in \Gamma_1 )} \sum_{\gamma_2 (\in \Gamma_2 )} \zeta_{\mu_1 \gamma_1}^{\mu_2 \gamma_2} (\mathbf{n}_2 -\mathbf{n}_1 ) \nonumber \\
&&\times \big( 2R_{\gamma_1}^{\mathbf{n}_1} + \bar{R}_{\gamma_1}^{\mathbf{n}_1} \big)
\big( 2R_{\gamma_2}^{\mathbf{n}_2} + \bar{R}_{\gamma_2}^{\mathbf{n}_2} \big) ,
\label{Eq:U^0}
\end{eqnarray}
where the introduced parameters,
\begin{equation}
\zeta_{\mu_1 \gamma_1}^{\mu_2 \gamma_2} (\mathbf{n}_2 -\mathbf{n}_1 ) = 
\sum_{\kappa \mathbf{k}} \frac{a_{\mu_1\gamma_1}^{\mathbf{n}_1}(\kappa \mathbf{k}) a_{\mu_2\gamma_2}^{\mathbf{n}_2}(\kappa \mathbf{k})}{\omega_{\kappa \mathbf{k}}^2},
\label{Eq:zeta}
\end{equation}
describe the relaxation of the lattice along the local nuclear coordinates $q_{\mu_1\gamma_1}^{\mathbf{n}_1}$ in response to local JT distortions $q_{\mu_2\gamma_2}^{\mathbf{n}_2}$ of unity amplitude. Due to translation symmetry of fullerene sites, these parameters depend only on the lattice vector connecting the positions of two local distortions. It is evident from Eqs. (\ref{Eq:zeta}) and (\ref{Eq:VV}), that the knowledge of the phonon polarization vectors and frequencies allows to define completely these parameters.

The tensors $R_\gamma$ are obtained from Eq. (\ref{Eq:R}) using explicit JT matrices in (\ref{Eq:V_JT}):
\begin{eqnarray} 
\Gamma = E: && \nonumber\\
&& R_\theta = \frac{1}{2} \big( x^2 +y^2 \big) -z^2 , \;\; R_\epsilon =-\frac{\sqrt{3}}{2} \big( x^2 -y^2 \big) , \nonumber\\
\Gamma = T: && \nonumber\\
&& R_\xi = \sqrt{3} yz , \;\; R_\eta = \sqrt{3} xz , \;\;R_\zeta = \sqrt{3} xy , \nonumber\\
\label{Eq:R_explicit}
\end{eqnarray}
and similar expressions (in terms of $\bar{x}$, $\bar{y}$ and $\bar{z}$) hold for tensors $\bar{R}_\gamma$ (we dropped for simplicity all indices $\mathbf{n}$). 

It is convenient to present the APES from Eq. (\ref{Eq:U^0}) as a sum of one-site and two-sites contributions,
\begin{equation}
\langle U\rangle^{(0)} = \sum_{\mathbf{n}} W^{(1)}_{\mathbf{n}} + \sum_{\mathbf{n}_1 < \mathbf{n}_2} W^{(2)}_{\mathbf{n}_1 , \mathbf{n}_2} .
\label{Eq:Two_contributions}
\end{equation}
where $W^{(1)}$ and $W^{(2)}$ depend on the adiabatic coordinates of one center and two centers, respectively. 
In the following we calculate and analyze these quantities for different A$_3$C$_{60}$. 

\subsection{APES at individual JT sites}
The contributions $W^{(1)}_{\mathbf{n}}$ (terms $\mathbf{n}_1 = \mathbf{n}_2$ in (\ref{Eq:U^0})) involve response parameters (\ref{Eq:zeta}) obeying the following relations \cite{Chibotaru1994}:
\begin{equation} 
\zeta_{\mu_1 \gamma_1}^{\mu_2 \gamma_2} (\mathbf{0})  = \zeta_{\mu_1 \gamma_1}^{\mu_2 \gamma_1}(\mathbf{0}) \delta_{\gamma_1 ,\gamma_2} 
 \equiv \zeta_{\mu_1 \mu_2}^{ \Gamma} \delta_{\gamma_1 ,\gamma_2} , 
\label{Eq:zeta_symmetry}
\end{equation}
where $\zeta_{\mu_1 \mu_2}^{ \Gamma}$ is common for all $\gamma\in \Gamma$. Then the contribution to the APES from a given JT site can be written as follows (we drop hereafter the index of the site):
\begin{eqnarray}
W^{(1)} &&=  \langle |\Psi_{\sigma}| \hat{H}_\text{H}|\Psi_{\sigma}\rangle -\frac{1}{2} \sum_{\Gamma =E,T} \Big[ \sum_{\mu_1 \mu_2} V_{\mu_1 \Gamma} V_{\mu_2 \Gamma}
\zeta_{\mu_1 \mu_2}^{ \Gamma} \Big] \nonumber \\
&&\times \sum_{\gamma (\in \Gamma )} \big( 2R_{\gamma} + \bar{R}_{\gamma} \big)^2 .
\label{Eq:W1_general}
\end{eqnarray}

The first term is the averaged multiplet splitting interaction in the adiabatic electronic state $\Psi_{\sigma}$.
Given the adiabatic orbitals can be seen as rotated reference electronic orbitals (Fig. \ref{Fig:Orbitsplitingcoordinates}(b)) in virtue of $t_{1u} - p$ isomorphism, the operator $\hat{H}_\text{H}$, being invariant under rotations of coordinate system, could be written in the basis of adiabatic orbitals from the beginning.
That is the $\hat{c}_{\alpha\sigma}$, ($\alpha =x, y, z$) operators in Eq. (\ref{Eq:H_bi}) can be replaced by the $\hat{a}_{i\sigma}$ ($i=1,2,3$) operators, Eq. (\ref{Eq:Psi_n}) (the lacking operator $\hat{a}_{1\sigma}$ is uniquely defined for given $\hat{a}_{2\sigma}$ and $\hat{a}_{3\sigma}$). Therefore, the matrix element of $\hat{H}_\text{H}$ will not depend on the adiabatic coordinates, i.e., will be a constant,
\begin{equation}
\langle \Psi_{\sigma}| \hat{H}_\text{H}|\Psi_{\sigma}\rangle = E_\text{H} ,
\label{Eq:E_H}
\end{equation}
linearly scaling with $J_\text{H}$.

The JT part in (\ref{Eq:W1_general}) can be conveniently rewritten as follows. Consider first equal quantities in the square brackets for both $\Gamma$. Then the last summation in (\ref{Eq:W1_general}) can be extended over all $\gamma \in H$ which, after substituting Eqs. (\ref{Eq:R_explicit}), gives the following equalities:
\begin{eqnarray} 
&& \sum_{\gamma (\in H )} R_{\gamma}^2 = \sum_{\gamma (\in H )} \bar{R}_{\gamma}^2 = 1 ,  \nonumber\\
&& \sum_{\gamma (\in H )} R_{\gamma} \bar{R}_{\gamma} = -1/2 .
\label{Eq:sum_RR}
\end{eqnarray}
The second relation becomes evident if one passes to a coordinate system $XYZ$ where one adiabatic vector ($x,y,z$) is directed along $Z$ and the other, ($\bar{x},\bar{y},\bar{z}$), lies in the $XY$ plane. The obtained relations show that the one-site APES is independent from electronic coordinates in the considered case. Given that the latter are parameterized by three Euler angles (Fig. \ref{Fig:Orbitsplitingcoordinates}(b)), we conclude that a three-dimensional continuum of equipotential minima (a three-dimensional trough) is realized at the bottom of lowest APES in this approximation. The motion at the bottom of this trough is isomorphic with the rotation of a rigid body which determines the structure of low-lying vibronic levels in isolated C$_{60}^{3-}$ ions \cite{Auerbach1994, OBrien1996}.

Using the relations (\ref{Eq:sum_RR}), Eq. (\ref{Eq:W1_general}) can be given in two equivalent forms:
\begin{equation}
W^{(1)} = E_\text{H} -3 E_{\text{JT}}^{E} + 
\big( E_{\text{JT}}^{E}  - E_{\text{JT}}^{T} \big) 
 \sum_{\gamma (\in T )} \big( 2R_{\gamma} + \bar{R}_{\gamma} \big)^2 ,
\label{Eq:W1_e}
\end{equation}
or
\begin{equation}
W^{(1)} = E_\text{H} -3 E_{\text{JT}}^{T} + 
\big( E_{\text{JT}}^{T}  - E_{\text{JT}}^{E} \big) 
 \sum_{\gamma (\in E )} \big( 2R_{\gamma} + \bar{R}_{\gamma} \big)^2 ,
\label{Eq:W1_t}
\end{equation}
where the parameters
\begin{eqnarray}
E_{\text{JT}}^{E} = && \frac{1}{2} \sum_{\mu_1 \mu_2} V_{\mu_1 E} V_{\mu_2 E} \zeta_{\mu_1 \mu_2}^{ E} , \nonumber\\
E_{\text{JT}}^{T} = && \frac{1}{2} \sum_{\mu_1 \mu_2} V_{\mu_1 T} V_{\mu_2 T} \zeta_{\mu_1 \mu_2}^{ T}  ,
\label{Eq:E_JT^E+T}
\end{eqnarray}
are JT stabilization energies after the distortions of $E$ and $T$ type, respectively, in the case of a single electron presenting in the $t_{1u}$ shell. The last term in (\ref{Eq:W1_e}) and (\ref{Eq:W1_t}) describes the warping of the bottom of the trough in terms of adiabatic coordinates of two occupied adiabatic orbitals (equivalently three Euler angles). The parameter defining the amplitude of the warping scales with the difference of energies of JT stabilization after the distortions of E and T type. When these two stabilization energies are equal, the warping contribution disappears and we end up with a three-dimensional trough similarly to an isolated C$_{60}^{3-}$ ion. Given the complexity of the systems (multicenter and multimode JT effect) the expression for $W^{(1)}$ looks remarkably simple. In the following we analyze the contributions to $W^{(1)}$ in different cubic fullerides using the calculated vibronic constants and phonon spectra.  

\subsubsection{Contributions to the static JT stabilization}
A consistent definition of JT stabilzation energy in the presence of warping is obtained via averaging the second term in Eq. (\ref{Eq:W1_general}) over all points of the trough. This is equivalent with the averaging over adiabatic coordinates (Euler angles) of the sums involving $R_{\gamma}$ and $\bar{R}_{\gamma}$ for $\gamma \in E $ and $\gamma \in T $, respectively. Integration over all Euler angles gives for these terms the weights $2/5$ and $3/5$, respectively. With them we can write for the Jahn-Teller stabilization energy,
\begin{equation}
E_{\text{JT}} = 3 \left( \frac{2}{5} E_{\text{JT}}^{E} + \frac{3}{5} E_{\text{JT}}^{T} \right) .
\label{Eq:E_JT}
\end{equation}
In the limit of an isolated C$_{60}^{3-}$, when the coupling to the vibrational eigenmodes is considered from the beginning, the Van Vleck coefficients in Eq. (\ref{Eq:VV}) become elements of a unity matrix, 
so that the response matrix $\mathbf{\zeta}$ becomes diagonal too, $\zeta_{\mu_1 \mu_2}^{ H} = \delta_{\mu_1 ,\mu_2}/\omega_{\mu H}^2$. Then we recover the usual expression for the multimode $t\otimes H$ JT problem involving three electrons, 
$E_{\text{JT}} = (3/2) \sum_{\mu} V_{\mu H}^2 /\omega_{\mu H}^2$.

The first term in (\ref{Eq:W1_general}) can be evaluated more accurately than suggested in Eq. (\ref{Eq:E_H}), by applying a second order perturbation theory after $\hat{H}_\text{H}$ (see the SM \cite{SM}), yielding
\begin{equation}
E_\text{H}^{(2)} = J_\text{H}-\frac{J_\text{H}^2}{4E_{\text{JT}}}
\label{Eq:E_H}
\end{equation}
The first term here is the destabilization energy [the only one given by Eq. (\ref{Eq:E_H}) ],
while the second term represents a small correction due to a small value of Hund's coupling parameter (ca 40 meV) compared to the E$_{\text{JT}}$ of ca 150 meV \cite{Iwahara2013, Liu2018Dynamical}. Note that the stabilization energy is further increased in C$_{60}^{3-}$ by ca 90 meV due to a dynamical delocalization of JT distortions in the trough \cite{Iwahara2013, Liu2018Dynamical}, which diminishes further this correction thus enhancing the criterion of applicability of single-determinant adiabatic wave function (\ref{Eq:Psi_n}). In the following we neglect the contribution (\ref{Eq:E_H}) which only gives a constant shift of energy on sites.

This elastic response parameters $\zeta_{\mu_1 \mu_2}^{ \Gamma}$ can be expressed through the lattice Green's functions \cite{Chibotaru1994, Chibotaru2003}. Evaluating the latter via the integration over the Brillouin zone of the crystal for atomic displacements of different pairs of atoms we obtain the elastic response parameters for all relevant local modes and evaluate their contribution to $E_{\text{JT}}$. 

Table \ref{tab:EJT_all_contrib} shows the calculated $E_{\text{JT}}$ as well as its contributions from different modes. The components F, A and FF are separated contributions from intrafullerene, alkali  and interfullerene modes, respectively. 
We can see that the off-diagonal contributions after $\mu$ (the interference terms) are important for intrafullerene $H_g$ modes because the latter are not vibrational eigenmodes in fullerides. 
The next 
rows, 
A-F, F-FF and A-FF, represent the contributions from alkali-intrafullerene, intrafullerene-interfullerene and interfullerene-alkali interference terms. We may conclude that the intrafullerene modes give the major contribution to JT stabilization as expected, while interaction with alkali mode increases $E_{\text{JT}}$ by few percents in LDA calculations. At the same time the effect of interfullerene modes is negligible.

\begin{table}[tb]
    \caption{\label{tab:EJT_all_contrib}
The JT stabilization energy and its contributions (in meV).
The JT stabilization energy of isolated C$_{60}^{3-}$ is -150.9 meV.
  }
\begin{ruledtabular}
    \begin{tabular}{ccccccccc}
       &\multicolumn{6}{c}{fcc}    & \multicolumn{2}{c}{A15}  \\
        \cline{2-9} 
        & \multicolumn{2}{c}{K$_{3}$C$_{60}$} &  \multicolumn{2}{c}{Rb$_{3}$C$_{60}$}&\multicolumn{2}{c}{Cs$_{3}$C$_{60}$} & \multicolumn{2}{c}{Cs$_{3}$C$_{60}$}  \\
    \cline{2-9} 
&LDA &PBE& LDA &PBE &LDA &PBE &LDA &PBE \\
\hline
\multirow{2}{*}{F}\footnotemark[1]& -144.9 &-153.6 &-148.9 &-153.4 &-150.1 &-155.3 &-150.4 &-156.7  \\
         &-171.1 &-179.5 &-176.7 &-181.7 &-184.6 &-190.4 &-178.0 &-187.3  \\
\hline
\multirow{2}{*}{A}\footnotemark[1]&  -16.1 &  -7.2 & -17.6 &  -2.5 & -19.0 &  -3.9 &  -3.0 &  -2.7  \\
         & -15.7 &  -6.8 & -17.3 &  -2.4 & -18.7 &  -3.9 &  -3.0 &  -2.7  \\
\hline
\multirow{2}{*}{FF}\footnotemark[1]&   -0.2 &  -0.2 &  -0.2 &  -0.1 &  -0.2 &  -0.1 &  -0.2 &  -0.1 \\
        &   -0.1 &  -0.1 &  -0.1 &  -0.1 &  -0.1 &  -0.1 &  -0.2 &  -0.1  \\
\hline
      A-F&  -0.2&  -0.2&  -0.4&  -0.1&  -0.6&  -0.2&   0.6&   0.7 \\
\hline
     F-FF&   0.0&   0.0&   0.0&   0.0&   0.1&   0.1&   0.0&  -0.0 \\
\hline
     A-FF&  -0.0&  -0.1&  -0.0&  -0.0&  -0.0&  -0.0&   0.0&   0.0 \\
\hline
 $E_{JT}$&-161.4&-161.2&-167.1&-156.1&-169.8&-159.5&-152.9&-158.8 \\
\end{tabular}
\end{ruledtabular}
\footnotemark[1]{The data in the second row correspond to the neglect of off-diagonal contributions after $\mu$.}
\end{table}

To get further insight into the origin of these contributions we inspect the elastic response parameters entering Eqs. (\ref{Eq:E_JT^E+T}), which have the meaning of inverse effective force constants (rigidity) w.r.t. the corresponding active distortion mode. 
The inverse square root of $\zeta_{\mu\mu}^{\Gamma}$ which corresponds to the effective frequency of the corresponding mode can be referred in SM \cite{SM}. We can see that in the case of intrafullerene modes these agree well with the corresponding vibrational eigenmodes of isolated fullerene ion supporting the molecular crystal character of fullerides. At the same time, the effective frequencies for alkali and interfullerene modes are much lower amounting to few tens of wave numbers for some of them. This explains the obtained important contribution of alkali modes to the JT stabilization despite the vibronic coupling constants are one order of magnitude smaller than for intrafullerene modes. Partly the small value of $\big( \zeta_{\mu\mu}^{\Gamma} \big)^{-1/2}$ for alkali modes is explained by the existence of low-frequency optical phonons involving these modes as it is evident from the phonon dispersion (Fig. 3). Another contribution comes from the acoustic phonons as can be seen from Table \ref{tab:EJT_all_contrib_AC}.
\subsubsection{Warping of the APES}
Weighting the expressions (\ref{Eq:W1_e}) and (\ref{Eq:W1_t}) in a similar way as in Eq. (\ref{Eq:E_JT}), we obtain:
\begin{equation}
W^{(1)} = E_\text{H} -E_{\text{JT}} + W_{\text{warp}} , 
\label{Eq:W1_final}
\end{equation}
where the warping term is given by
\begin{eqnarray}
W_{\text{warp}} &&= 
\frac{1}{2}\big( E_{\text{JT}}^{E}  - E_{\text{JT}}^{T} \big) 
\Big[ \frac{2}{5} \sum_{\gamma (\in T )} \big( 2R_{\gamma} + \bar{R}_{\gamma} \big)^2 \nonumber \\
&&- \frac{3}{5} \sum_{\gamma (\in E )} \big( 2R_{\gamma} + \bar{R}_{\gamma} \big)^2 \Big] .
\label{Eq:W_warp}
\end{eqnarray}
Note that the warping term now averages to zero when integrated over the trough.

Using the equality
\begin{equation} 
 \sum_{\gamma (\in H )} \Big( 2R_{\gamma} - \bar{R}_{\gamma} \Big)^2 = 3 ,  
\label{Eq:sum_3}
\end{equation}
emerging from the relations (\ref{Eq:sum_RR}), we can represent the warping term (\ref{Eq:W_warp}) only in terms of tensors with $\gamma \in T$ only:
\begin{equation}
W_{\text{warp}} = \Delta \Big[ -\frac{9}{5} + \sum_{\gamma (\in T )} \big( 2R_{\gamma} + \bar{R}_{\gamma} \big)^2 \Big] ,
\label{Eq:warping}
\end{equation}
where the warping parameter was introduced:
\begin{equation} 
\Delta = \frac{1}{2}\big( E_{\text{JT}}^{E} - E_{\text{JT}}^{T} \big) 
\label{Eq:warping_parameter}
\end{equation}
This parameter is evaluated in a similar way as $E_{\text{JT}}$. 
Table \ref{tab:EJT_ET_all_contrib} gives the calculated $\Delta$ for the four materials, as well as its separated contributions from intrafullerene, alkali and interfullerene modes and their intereference. The notations and meaning of the columns are similar to Table \ref{tab:EJT_all_contrib}. We observe that the main contribution to the warping comes almost entirely from alkali modes in the case of fcc fullerides and from the intrafullerene modes in the case of A15 structure. 
In all compounds (except for potassium fulleride within LDA approximation) the contributions from intrafullerene and alkali atoms are of opposite sign. 
Note that now the interference of intrafullerene and alkali modes is relatively large. Table \ref{tab:EJT_ET_all_contrib_AC} shows that $\Delta$ is mostly contributed by the acoustic phonon modes in the case of fcc fullerides. On the contrary, it is contributed by optical phonon modes in the case of A15 fulleride. We stress that the warping is entirely due to the interaction of fullerene molecules with the environment. In the case of isolated fullerene ions the warping of the lowest APES doesn't arise in the approximation of linear JT coupling. On the other hand, the quadratic JT coupling was shown to have negligible effect on its warping \cite{Liu2018b}.

\begin{table*}[tb]
    \caption{\label{tab:EJT_ET_all_contrib}
The warping parameter $\Delta$ and its contributions (in meV).
 }
\begin{ruledtabular}
    \begin{tabular}{ccccccccc}
       &\multicolumn{6}{c}{fcc}    & \multicolumn{2}{c}{A15}  \\
        \cline{2-9} 
        & \multicolumn{2}{c}{K$_{3}$C$_{60}$} &  \multicolumn{2}{c}{Rb$_{3}$C$_{60}$}&\multicolumn{2}{c}{Cs$_{3}$C$_{60}$} & \multicolumn{2}{c}{Cs$_{3}$C$_{60}$}  \\
    \cline{2-9} 
&LDA &PBE& LDA &PBE &LDA &PBE &LDA &PBE \\
\hline
\multirow{2}{*}{F}\footnotemark[1]&  0.044 &-0.526 &-0.577 &-1.109 &-1.112 &-1.809 &-0.532 &-0.287 \\
         & 1.992 & 1.670 & 1.715 & 1.104 & 1.798 & 0.706 & 0.150 & 0.485 \\
\hline
\multirow{2}{*}{A}\footnotemark[1]&  4.454 & 1.548 & 4.727 & 0.402 & 4.869 & 0.668 & 0.137 & 0.211 \\
         & 4.362 & 1.464 & 4.670 & 0.380 & 4.825 & 0.643 & 0.340 & 0.371 \\
\hline
\multirow{2}{*}{FF}\footnotemark[1]& -0.018 &-0.014 &-0.019 &-0.015 &-0.025 &-0.017 &-0.003 &-0.006 \\
        & -0.014 &-0.009 &-0.015 &-0.011 &-0.018 &-0.011 & 0.003 & 0.000 \\
\hline
      A-F& 0.153 & 0.194 & 0.256 & 0.148 & 0.333 & 0.213 &-0.179 &-0.123 \\
\hline
     F-FF&-0.036 &-0.060 &-0.043 &-0.032 &-0.065 &-0.052 &-0.003 & 0.003 \\
\hline
     A-FF& 0.021 & 0.023 & 0.019 & 0.009 & 0.020 & 0.010 &-0.024 &-0.017 \\
\hline
 $\Delta$& 4.618 & 1.165 & 4.364 &-0.598 & 4.019 &-0.985 &-0.605 &-0.220 \\
\end{tabular}
\end{ruledtabular}
\footnotemark[1]{The data in the second row correspond to the neglect of off-diagonal contributions after $\mu$.}
\end{table*}

%

\begin{figure}[tb]
\begin{center}
\begin{tabular}{l}
(a) \\
\includegraphics[width=5cm,height=2.8cm]{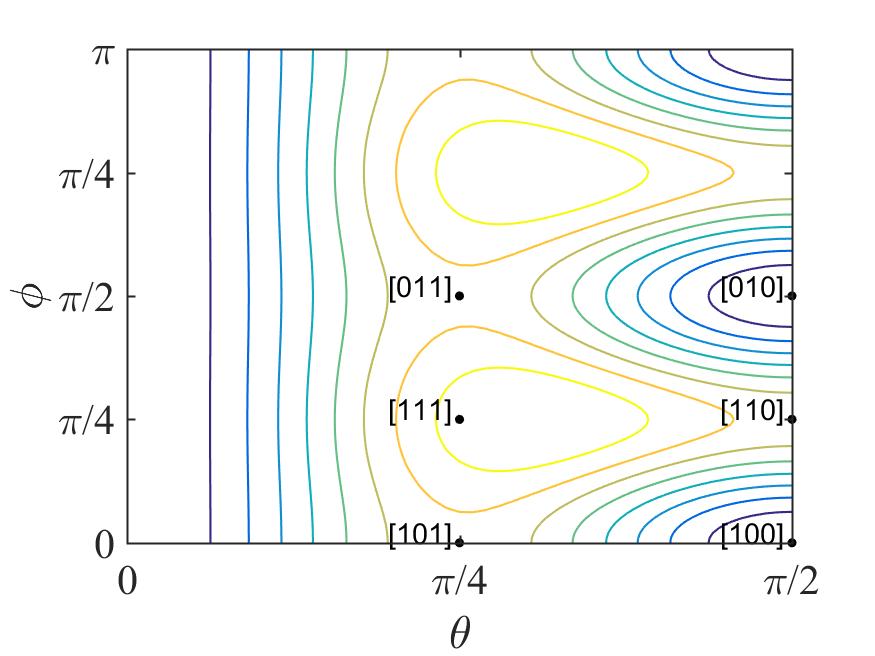} \\
(b) \\
\includegraphics[width=9cm]{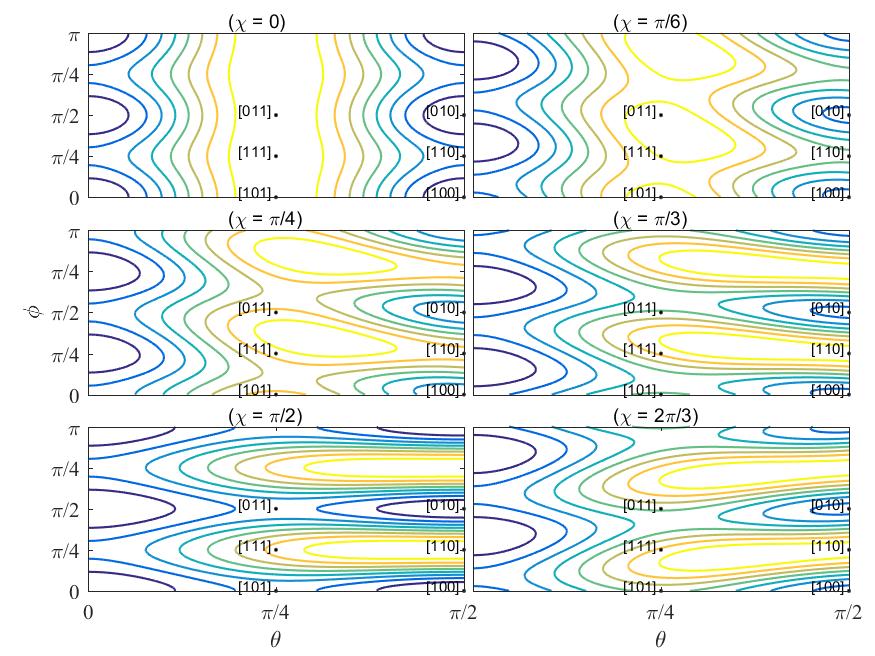}
\end{tabular}
\end{center}
\caption{Contour plots of the warping function in Eq. (\ref{Eq:warping}) for an irreducible domain of Euler angles $\theta$ and $\phi$ (a) C$_{60}^{2-}$ ion with cubic warping. (b) Cross section of a three-dimensional surface, corresponding to the APES of C$_{60}^{3-}$ site in cubic A$_3$C$_{60}$, at indicated values of Euler angle $\chi$. The lines correspond to increasing energies for colors varying from white to black.}
\label{Fig:ConterplotofAPES}
\end{figure}

The warping itself arises from the second factor in (\ref{Eq:warping}). This factor represents a warping function which depends on the adiabatic coordinates only and is universal for all investigated fullerides. For the sake of analysis, consider first the case of only two electrons in the lowest adiabatic orbital (the case of C$_{60}^{2-}$ ion). In the absence of warping this situation leads to a two-dimensional trough in the space of spherical angles $\theta$ and $\phi$ because of the constraint $x^2 +y^2 +z^2 =1$. The warping function reduces in this case to an expression depending only on the adiabatic coordinates of this adiabatic orbital, $-6[x^4 +y^4 +z^4 -(7/10)r^4 ]$, where $r=1$. The expression in the square brackets coincides up to a constant -1/10 with the fourth-order cubic invariant \cite{Abragam1970electron}. Such a function was considered phenomenologically by O'Brien in connection with the warping of the lowest APES of $d\otimes H$ JT problem \cite{OBrien1996, Chancey1997} subject to a perturbation from a cubic environment. The obtained warping function is shown in Fig. \ref{Fig:ConterplotofAPES}a and has similar extremal properties as in the case of of $d\otimes H$ JT problem. Thus two groups of extremes, three of [001] type (tetragonal) and four of [111] type (trigonal) appear, which are either minima or maxima depending on the sign of the parameter $\Delta$. On the contrary, six extremes of [011] type (rhombic) are always saddle points.     

In the case of C$_{60}^{3-}$, the warping function will depend on three additional adiabatic coordinates entering the tensors $\bar{R}_{\gamma}$ in (\ref{Eq:warping}). It is easier, therefore, to investigate it in the space of three Euler angles defining both occupied adiabatic orbitals (Fig.\ref{Fig:ConterplotofAPES}b). To this end we need to express the six adiabatic coordinates entering Eq. (\ref{Eq:warping}) via these angles \cite{Auerbach1994, OBrien1996},
\begin{eqnarray}
&& x =  \sin{\theta} \cos{\phi} ,\; y = \sin{\theta} \sin{\phi} ,\; z = \cos{\theta} , \nonumber\\
&& \bar{x} =  -\sin{\chi} \cos{\theta} \cos{\phi} - \cos{\chi} \sin{\phi} ,  \nonumber\\
&& \bar{y} = \cos{\chi} \cos{\phi}-\sin{\chi} \cos{\theta} \sin{\phi} , \nonumber\\
&& \bar{z} = \sin{\chi} \sin{\theta} .
\label{Eq:Euler}
\end{eqnarray}

The obtained $\theta -\phi$ maps of the warping function for several values of the angle $\chi$ are shown in Fig. 5b. We can see that the extremal structure of the lowest APES has the same morphology as in the case of C$_{60}^{2-}$ (Fig. \ref{Fig:ConterplotofAPES}a). The new feature is the cyclic shift of the landscape in function of $\chi$ with a period of $\pi$. For $\Delta >0$ the minima are of trigonal type and the lowest energy path connecting pairs of these minima, e.g., [111] and [1-11] (not shown in Fig. \ref{Fig:ConterplotofAPES}), is via the saddle points ([101]). For $\Delta <0$ the minima are of tetragonal type (e.g., [001]) and the lowest-energy path connecting minima of these type are again saddle points albeit now the path passes through them in a perpendicular direction. 

\begin{figure}[tb]
\begin{center}
\begin{tabular}{ll}
(a)~~~~~~~~~~~~~~~~~~~~~~~~~~~~~~~~~~~~~ & (b) \\
\multicolumn{2}{c}{\includegraphics[width=1.0\linewidth]{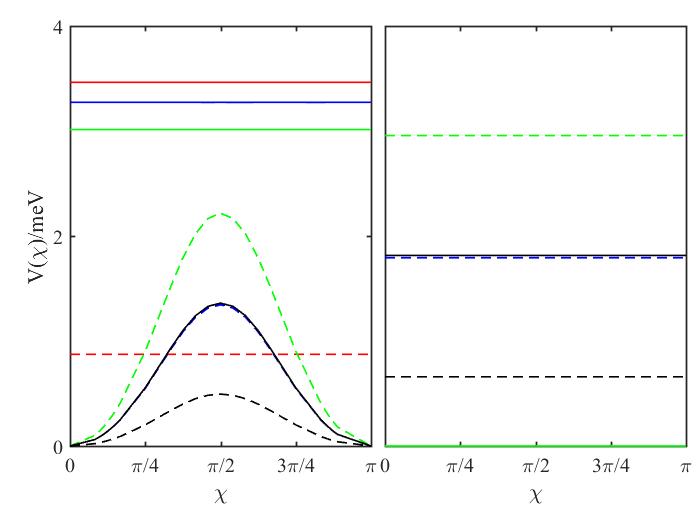}}
\end{tabular}
\end{center}
\caption{Calculated barriers (a) and energies of the minimum (b) of the lowest APES at a C$_{60}^{3-}$ in cubic fullerides in function of the Euler angle $\chi$. The energy of the lowest minimum among the investigated systems is taken as zero in (b).
Solid and dashed lines are the results of LDA and PBE calculations; red, blue, green, and black lines correspond to fcc K$_3$C$_{60}$, Rb$_3$C$_{60}$, Cs$_3$C$_{60}$ and A15 Cs$_3$C$_{60}$ fullerides. In (b), the solid red and blue lines are hidden behind the solid green line; the dashed red line is hidden behind the solid green line.
}
\label{Fig:PLOTbarrierandmin}
\end{figure}

To establish the character of JT dynamics, it is necessary to know the height of the barrier (the energy of the saddle point measured from the minimum) separating the minima. 
Fig. \ref{Fig:PLOTbarrierandmin}a shows the dependence of the barriers on the angle $\chi$. One can see that the barrier doesn't depend on $\chi$ in the case of trigonal minima, whereas its height oscillates periodically in the case of tetragonal minima. In all systems and for all $\chi$ the height of the barrier is smaller than 4 meV which, at its turn, is smaller than the rotational quantum in the trough for an isolated C$_{60}^{3-}$ (ca 8 meV) and much smaller than the kinetic delocalization energy of JT deformations in this anion (ca 90 meV) \cite{Iwahara2013, Liu2018Dynamical}. 
This precludes the localization JT deformation after $\theta$ and $\phi$ in vicinity of the minima. At the same time, the energy of the minima is independent from $\chi$ irrespective of their type (Fig. \ref{Fig:PLOTbarrierandmin}b). This means that a one-dimensional trough after the coordinate $0<\chi\leq 2\pi$ will be preserved for the warping of any amplitude, a situation not realized in the case of the cubic warping of a two-dimensional trough as Fig. \ref{Fig:ConterplotofAPES}a shows.

On the basis of these results we conclude that the JT dynamics at one site in the A$_3$C$_{60}$ cubic fullerides corresponds to weakly hindered rotations of JT deformations in the three-dimensional trough.      

\subsection{Intersite interaction of JT centers}
The two-site contributions to the lowest APES,  $W^{(2)}_{\mathbf{n}_1 , \mathbf{n}_2}$, can be extracted from Eq. (\ref{Eq:U^0}) in the following form:
\begin{eqnarray} 
 W^{(2)}_{\mathbf{n}_1 , \mathbf{n}_2} &&=    
\sum_{\Gamma_1} \sum_{\Gamma_2} 
\sum_{\gamma_1 (\in \Gamma_1 )} \sum_{\gamma_2 (\in \Gamma_2 )} B_{\gamma_1}^{\gamma_2} (\mathbf{n}_2 -\mathbf{n}_1 ) \nonumber \\ 
&& \times \big( 2R_{\gamma_1}^{\mathbf{n}_1} + \bar{R}_{\gamma_1}^{\mathbf{n}_1} \big)
\big( 2R_{\gamma_2}^{\mathbf{n}_2} + \bar{R}_{\gamma_2}^{\mathbf{n}_2} \big) , \nonumber\\
 B_{\gamma_1}^{\gamma_2} (\mathbf{n}_2 -\mathbf{n}_1 ) &&\equiv  - \frac{1}{2} 
\sum_{\mu_1 ( \Gamma_1 )} \sum_{\mu_2 ( \Gamma_2 )} V_{\mu_1 \Gamma_1} V_{\mu_2 \Gamma_2} \nonumber \\
&&\times \zeta_{\mu_1 \gamma_1}^{\mu_2 \gamma_2} (\mathbf{n}_2 -\mathbf{n}_1 ) .
\label{Eq:W^2}
\end{eqnarray}
We calculated the interaction parameters $B_{\gamma_1}^{\gamma_2} (\Delta\mathbf{n} )$ for nearest neighbor and next nearest neighbor fullerene pairs. In the former case $\Delta\mathbf{n} \parallel [101]$ for fcc and $\parallel [111]$ for bcc fullerides, respectively (Fig. \ref{fig:Struct_FCC_A15}). The next nearest neighbors are located along cubic axes for both kind of lattices ($\Delta\mathbf{n} \parallel [001]$). The results are given in Tables VII and IX. 

One can see that the interaction parameters for the A15 fulleride are much smaller than in fcc compounds. A similar situation was also found for the warping parameter (Table VII) pointing to a generally weaker symmetry lowering effect in bcc lattices compared to fcc ones (we remind that in the present calculations the merohedral order in the A15 compound was replaced by a full order of fullerene molecules). In fcc fullerides the interaction parameters corresponding to different pairs $\gamma_1 \gamma_2$ is highly selective. Actually, it is non-negligible only for $\theta\theta$, $\theta\xi$ and $\xi\xi$ coupling in nearest neighbors and only for $\theta\theta$ coupling in next nearest neighbor pairs. Remarkably, the latter is significantly stronger than all coupling in the nearest neighbor pairs. Even so it doesn't exceed 4 meV (LDA result for Cs$_3$C$_{60}$). However, the interaction with more distant fullerenes is negligible. We expect that the interaction of these pairs is mainly governed by acoustic phonon modes due to their larger dispersion (Fig. 2), which implies its $R^{-3}$ dependence on the interfullerene separation $R$.

We may conclude from this study that the interaction of JT distortions on different fullerene sites is too weak to quench their dynamics and localize them at some points in the troughs. There can be however some correlation in the rotations of these distortions. For instance, for the strongest $\theta\theta$ coupling of next nearest neighbor pairs the maximal negative value of the corresponding tensorial factor in Eq. (\ref{Eq:W^2}) reaches -9/4 for $\theta$ distortions on the two sites of opposite sign. For fcc Cs$_3$C$_{60}$ this gives a correlation energy amounting to half of rotational quantum (in the LDA calculation). We can speculate that this interaction can give rise at low temperature to an antiferroditorsive ordering of $\theta$ distortions at next nearest neighbor fullerene sites. These static distortions are expected to have a small amplitude and will coexist with the rotation of JT deformations of much larger amplitude at each C$_{60}$. This scenario corroborate the low-temperature NMR data for insulating fcc Cs$_3$C$_{60}$ which evidence weak static distortions of fullerenes gradually disappearing with the rise of temperature \cite{Potocnik2014}.

Finally we would like to point out that the results of the present and previous subsections are strongly dependent on the type of exchange correlation functional used in the DFT calculations. While it is not clear for the moment what functional from the two employed here is to be preferred for phonon calculations (there is no clear answer from the literature), we notice that the use of more involved hybrid functionals is prohibited for the present systems due to a large number of atoms in the unit cell (even for the highest degree of ordering of fullerene molecules in the lattice). 
To unambiguously determine the low-energy phonon dispersion, it is desired that such calculations will become feasible. 
\begin{table}[tb]
    \caption{\label{tab:EJT_all_contrib_TERM3_N}
The interaction parameter $B_{\gamma_1}^{\gamma_2} (\Delta\mathbf{n} )$ for nearest neighbor fullerene sites (in meV), calculated by LDA and PBE (parentheses).
 }
\begin{ruledtabular}
    \begin{tabular}{cccccc}
     &   &\multicolumn{3}{c}{fcc}    & A15  \\
\cline{3-5} 
     &   & K$_{3}$C$_{60}$ & Rb$_{3}$C$_{60}$&Cs$_{3}$C$_{60}$ & Cs$_{3}$C$_{60}$ \\
\cline{3-5} 
\multirow{5}{*}{E$_\theta$} &E$_\theta$ & -0.55( -0.17)& -0.52( -0.44)& -0.36( -0.39)& -0.01( -0.02) \\
&E$_\epsilon$ & -0.06(  0.09)& -0.03(  0.02)& -0.02(  0.01)&  0.15(  0.18) \\
&T$_\xi$ & -0.24( -0.30)& -0.26( -0.22)& -0.26( -0.24)&  0.02(  0.03) \\
&T$_\eta$ &  0.02(  0.04)&  0.03(  0.01)&  0.03(  0.02)&  0.30(  0.27) \\
&T$_\zeta$ & -0.01(  0.07)& -0.01(  0.01)& -0.02(  0.01)& -0.08( -0.13) \\
\hline
\multirow{4}{*}{E$_\epsilon$} &E$_\epsilon$ & -0.06(  0.09)& -0.03(  0.02)& -0.02(  0.01)&  0.15(  0.18) \\
&T$_\xi$ & -0.06( -0.11)& -0.06( -0.04)& -0.03( -0.04)&  0.01(  0.01) \\
&T$_\eta$ & -0.02( -0.04)& -0.03( -0.01)& -0.03( -0.02)& -0.06( -0.09) \\
&T$_\zeta$ & -0.03( -0.09)& -0.03( -0.01)& -0.05( -0.01)& -0.05( -0.05) \\
\hline
\multirow{3}{*}{T$_\xi$} & T$_\xi$ & -0.50( -0.56)& -0.65( -0.41)& -1.04( -0.73)&  0.02( -0.00) \\
&T$_\eta$ &  0.10(  0.12)&  0.13(  0.12)&  0.15(  0.15)& -0.07( -0.05) \\
&T$_\zeta$ & -0.00(  0.02)&  0.01(  0.06)&  0.00(  0.04)& -0.06( -0.04) \\
\hline
\multirow{2}{*}{T$_\eta$} &T$_\eta$ & -0.07( -0.10)& -0.09( -0.08)& -0.10( -0.10)& -0.18( -0.14) \\
&T$_\zeta$ & -0.01(  0.05)& -0.01( -0.00)& -0.01(  0.01)& -0.01(  0.01) \\
\hline
T$_\zeta$ &T$_\zeta$ & -0.11(  0.08)& -0.05( -0.09)&  0.06( -0.06)& -0.15( -0.09) \\
\end{tabular}
\end{ruledtabular}
\end{table}

\begin{table}[tb]
    \caption{\label{tab:EJT_all_contrib_TERM3_NN}
The interaction parameter $B_{\gamma_1}^{\gamma_2} (\Delta\mathbf{n} )$ for next nearest neighbor fullerene sites (in meV), calculated by LDA and PBE (parentheses).
 }
\begin{ruledtabular}
    \begin{tabular}{cccccccccc}
     &   &\multicolumn{3}{c}{fcc}    & A15  \\
\cline{3-5} 
     &   & K$_{3}$C$_{60}$ & Rb$_{3}$C$_{60}$&Cs$_{3}$C$_{60}$ & Cs$_{3}$C$_{60}$ \\
\cline{3-5} 
\multirow{5}{*}{E$_\theta$} &E$_\theta$ &  2.49(  0.79)&  3.25(  0.50)&  4.20(  1.07)&  0.04(  0.04) \\
&E$_\epsilon$ &  0.03(  0.07)&  0.17(  0.05)&  0.35(  0.12)&  0.00(  0.02) \\
&T$_\xi$ & -0.00( -0.00)& -0.00( -0.00)& -0.00( -0.00)&  0.03(  0.05) \\
&T$_\eta$ & -0.04( -0.04)& -0.04( -0.01)& -0.06( -0.01)&  0.03(  0.06) \\
&T$_\zeta$ &  0.00(  0.00)&  0.00(  0.00)& -0.00( -0.00)&  0.01(  0.02) \\
\hline
\multirow{4}{*}{E$_\epsilon$} &E$_\epsilon$ &  0.03(  0.07)&  0.17(  0.05)&  0.35(  0.12)&  0.00(  0.02) \\
&T$_\xi$ &  0.00(  0.00)&  0.00(  0.00)&  0.00(  0.00)&  0.01(  0.00) \\
&T$_\eta$ & -0.05( -0.08)& -0.05( -0.00)& -0.08( -0.01)&  0.02(  0.02) \\
&T$_\zeta$ & -0.00( -0.00)& -0.00( -0.00)& -0.00( -0.00)&  0.02(  0.05) \\
\hline
\multirow{3}{*}{T$_\xi$} & T$_\xi$ & -0.02(  0.08)&  0.04( -0.05)&  0.10( -0.00)& -0.09( -0.07) \\
&T$_\eta$ & -0.00( -0.01)& -0.01( -0.00)& -0.02( -0.01)& -0.05( -0.03) \\
&T$_\zeta$ &  0.05(  0.05)&  0.00( -0.01)& -0.01( -0.03)& -0.08( -0.07) \\
\hline
\multirow{2}{*}{T$_\eta$} &T$_\eta$ &  0.03(  0.03)&  0.04(  0.02)&  0.03(  0.03)& -0.02( -0.02) \\
&T$_\zeta$ &  0.00(  0.00)&  0.00(  0.00)&  0.00(  0.00)& -0.03( -0.02) \\
\hline
T$_\zeta$ &T$_\zeta$ & -0.02(  0.01)&  0.00( -0.02)&  0.02( -0.03)&  0.03(  0.04) \\
\end{tabular}
\end{ruledtabular}
\end{table}

\section{DISCUSSION AND CONCLUSIONS}
In this paper, we investigate for the first time the modification of JT effect on fullerene anions C$_{60}^{3-}$ when they are incorporated in the cubic lattices of A$_{3}$C$_{60}$ fullerides. The interaction of each fullerene molecule with the environment leads to modification of JT stabilization energy and to the warping of the trough at each fullerene site, as well as to the interaction of JT distortions at different sites. We studied these effects in three fcc fullerides with A=K,Rb,Cs and in Cs$_3$C$_{60}$ with bcc (A15) structure by using the results of DFT calculations of orbital vibronic coupling constants at C$_{60}$ sites and of phonon spectra of these materials. The key quantities defining the character of JT effect in these crystals are the elastic response parameters for local JT distortions, which are evaluated on the basis of phonon calculations. Using these response parameters and the vibronic coupling constants the lowest APES has been calculated and analyzed in these materials. To this end, an expression for the lowest APES in function of trough coordinates at the fullerene sites has been derived. We found that the JT stabilization energy increases by few percents compared to isolated C$_{60}^{3-}$ and a warping of the trough amounting to few meV occurs in all investigated compounds. The interaction of JT distortions on nearest- and next-nearest-neighbor fullerene sites is quite selective w.r.t. to their symmetry and turns out to be of similar order of magnitude as the warping of the APES on individual sites. All these effects are mostly due to the interaction of C$_{60}$ ions with the displacements of neighbor alkali atoms. 

The obtained results concerning the APES of A$_3$C$_{60}$ are sufficient for the description of eventual static cooperative JT effect in such systems, i.e., the distorted equilibrium geometry and its evolution with the temperature, including the structural phase transition \cite{Kaplan_book, bersuker2006JTeffect}. However, in cubic fullerides the dynamical JT effect on fullerene sites, destroying any ordering of static JT deformations is the most probable scenario \cite{StructureK3C60, Iwahara2016} because the energy of kinetic delocalization of JT deformations along the trough is of the order of ca 90 meV \cite{Iwahara2013}, which exceeds by two orders of magnitude the interaction parameters for JT active distortions on nearest-neighbor fullerene sites (Table VI). Given this situation, the dynamical cooperative JT effect could be considered within an approach using a single set of $H_g$ modes (effective modes) on each fullerene site \cite{Iwahara2013, StructureK3C60} with JT, warping and interaction parameters derived in the present work. One should note, however, that the latter will by further modified when the JT dynamics on sites is considered in a pretty similar way as the isotope substitution leads to localization of JT distortions in perfectly equipotential troughs \cite{Iwahara2012}. However, such modifications are not expected to change the order of magnitude of the calculated parameters corresponding to static JT distortions. We may conclude, therefore, that the present study supports the picture of weakly hindered independent rotations of JT deformations at C$_{60}$ sites in all A$_3$C$_{60}$.

The results obtained in this work are relevant to a number of the observable properties 
of insulating cubic fullerides such as the NMR \cite{Potocnik2014} and infrared \cite{Klupp2012, Kamaras2013} spectra. 
For instance, a weak localization of JT distortions on fullerene sites due to their elastic coupling could slightly lower the symmetry of C$_{60}$ anions, which can be the reason for the ``orbital glass'' behavior found in the NMR spectra of Cs$_3$C$_{60}$ \cite{Potocnik2014}. 
This static and dynamic symmetry lowering of fullerene molecules is also expected to show up in the fine structure of infrared spectra of insulating A$_3$C$_{60}$, the reason for which the latter have not been completely assigned by the simulations of isolated C$_{60}^{3-}$ units \cite{Matsuda2018}. 
Complementary information on the JT dynamics is expected to be gained from other spectroscopy as well such as optical absorption, Raman and inelastic neutron scattering.
Further application of the present results to the rationalization of various spectroscopic data of insulating fullerides will result in a thorough understanding of the physics of dynamical cooperative JT effect in these materials.

\section*{Acknowledgements}
Z.H. gratefully acknowledges funding by the China Scholarship Council.
N.I. was partly supported by the GOA program from KU Leuven and and the scientific research grant R-143-000-A80-114 of the National University of Singapore.
The computational resources and services used in this work were provided by the VSC (Flemish Supercomputer Center) funded by the Research Foundation - Flanders (FWO) and the Flemish Government - department EWI.
This work was funded by the Researchers Supporting Project Number (RSP-2020/267) King Saud University, Riyadh, Saudi Arabia.
\appendix
\section{Effect of acoustic modes on $E_{\text{JT}}$ and $\Delta$}
Considering only the lowest three phonon branches in Eq. (\ref{Eq:E_JT^E+T}), we calculate their contributions to JT energies, Eq. (\ref{Eq:E_JT}), in Table \ref{tab:EJT_all_contrib_AC}, 
and to the warping parameters, Eq. (\ref{Eq:warping_parameter}), in Table \ref{tab:EJT_ET_all_contrib_AC}. 

\begin{table}[tb]
    \caption{\label{tab:EJT_all_contrib_AC}
The contribution of acoustic modes to the total JT stabilization energy and its components (in meV).
 }
\resizebox{\linewidth}{!}{    \begin{ruledtabular}
    \begin{tabular}{ccccccccc}
&\multicolumn{2}{c}{K$_3$C$_{60}$} &\multicolumn{2}{c}{Rb$_3$C$_{60}$} &\multicolumn{2}{c}{Cs$_3$C$_{60}$} &\multicolumn{2}{c}{A15} \\
\hline
&LDA &PBE& LDA &PBE &LDA &PBE &LDA &PBE \\
\hline
\multirow{2}{*}{F}\footnotemark[1]&   -0.1 &  -0.1 &  -0.4 &  -0.1 &  -0.8 &  -0.4 &  -0.0 &  -0.0 \\
         &  -1.4 &  -0.2 &  -2.5 &  -0.1 &  -4.2 &  -0.6 &  -0.1 &  -0.3 \\
\hline
\multirow{2}{*}{A}\footnotemark[1]&  -12.6 &  -2.4 & -16.5 &  -0.4 & -18.3 &  -2.3 &  -1.0 &  -1.2 \\
         & -12.1 &  -2.3 & -16.1 &  -0.4 & -17.9 &  -2.3 &  -1.0 &  -1.3 \\
\hline
\multirow{2}{*}{FF}\footnotemark[1]&   -0.1 &  -0.1 &  -0.0 &  -0.1 &  -0.0 &  -0.1 &  -0.1 &  -0.1 \\
        &   -0.1 &  -0.1 &  -0.0 &  -0.1 &  -0.0 &  -0.1 &  -0.1 &  -0.1 \\
\hline
      A-F&  -0.1 &  -0.0 &  -0.4 &  -0.0 &  -0.5 &  -0.0 &   0.0 &   0.0 \\
\hline
     F-FF&   0.0 &   0.1 &   0.1 &   0.0 &   0.1 &   0.1 &   0.0 &   0.0 \\
\hline
     A-FF&  -0.0 &  -0.0 &  -0.0 &  -0.0 &  -0.0 &  -0.0 &  -0.0 &  -0.0 \\
\hline
 $\Delta$E$_{JT}$& -12.9 &  -2.6 & -17.3 &  -0.5 & -19.6 &  -2.8 &  -1.1 &  -1.4 \\
\end{tabular}
\end{ruledtabular}}
\footnotemark[1]{The data in the second row correspond to the neglect of off-diagonal contributions after $\mu$.}
\end{table}

\begin{table}[tb]
    \caption{\label{tab:EJT_ET_all_contrib_AC}
The contribution from acoustic modes to warping parameter, $\Delta_{ac}$, and its components (in meV).
  }
\resizebox{\linewidth}{!}{    \begin{ruledtabular}
    \begin{tabular}{ccccccccc}
       &\multicolumn{6}{c}{fcc}    & \multicolumn{2}{c}{A15}  \\
        \cline{2-9} 
        & \multicolumn{2}{c}{K$_{3}$C$_{60}$} &  \multicolumn{2}{c}{Rb$_{3}$C$_{60}$}&\multicolumn{2}{c}{Cs$_{3}$C$_{60}$} & \multicolumn{2}{c}{Cs$_{3}$C$_{60}$}  \\
    \cline{2-9} 
&LDA &PBE& LDA &PBE &LDA &PBE &LDA &PBE \\
\hline
\multirow{2}{*}{F}\footnotemark[1]& -0.048 &-0.232 & 0.107 &-0.263 & 0.256 & 0.093 &-0.031 &-0.021 \\
         & 0.392 &-0.112 & 0.815 &-0.130 & 1.379 & 0.174 & 0.008 & 0.049 \\
\hline
\multirow{2}{*}{A}\footnotemark[1]&  3.747 & 0.407 & 4.824 &-0.003 & 5.191 & 0.480 & 0.035 & 0.119 \\
         & 3.617 & 0.395 & 4.747 &-0.024 & 5.149 & 0.458 & 0.132 & 0.212 \\
\hline
\multirow{2}{*}{FF}\footnotemark[1]& -0.002 &-0.015 & 0.005 &-0.010 & 0.006 &-0.004 & 0.014 & 0.005 \\
        & -0.001 &-0.006 & 0.002 &-0.005 & 0.003 & 0.001 & 0.019 & 0.010 \\
\hline
      A-F& 0.060 & 0.034 & 0.125 & 0.016 & 0.178 & 0.014 &-0.037 & 0.012 \\
\hline
     F-FF&-0.028 &-0.059 &-0.021 &-0.035 &-0.032 &-0.051 &-0.003 & 0.001 \\
\hline
     A-FF& 0.013 & 0.020 & 0.011 & 0.008 & 0.010 & 0.010 &-0.017 &-0.016 \\
\hline
 $\Delta_{ac}$& 3.741 & 0.155 & 5.051 &-0.286 & 5.610 & 0.542 &-0.040 & 0.100 \\
\end{tabular}
\end{ruledtabular}}
\footnotemark[1]{The data in the second row correspond to the neglect of off-diagonal contributions after $\mu$.}
\end{table}

\bibliography{BIB_A3C60_summary}

\begin{thebibliography}{61}%
\makeatletter
\providecommand \@ifxundefined [1]{%
 \@ifx{#1\undefined}
}%
\providecommand \@ifnum [1]{%
 \ifnum #1\expandafter \@firstoftwo
 \else \expandafter \@secondoftwo
 \fi
}%
\providecommand \@ifx [1]{%
 \ifx #1\expandafter \@firstoftwo
 \else \expandafter \@secondoftwo
 \fi
}%
\providecommand \natexlab [1]{#1}%
\providecommand \enquote  [1]{``#1''}%
\providecommand \bibnamefont  [1]{#1}%
\providecommand \bibfnamefont [1]{#1}%
\providecommand \citenamefont [1]{#1}%
\providecommand \href@noop [0]{\@secondoftwo}%
\providecommand \href [0]{\begingroup \@sanitize@url \@href}%
\providecommand \@href[1]{\@@startlink{#1}\@@href}%
\providecommand \@@href[1]{\endgroup#1\@@endlink}%
\providecommand \@sanitize@url [0]{\catcode `\\12\catcode `\$12\catcode
  `\&12\catcode `\#12\catcode `\^12\catcode `\_12\catcode `\%12\relax}%
\providecommand \@@startlink[1]{}%
\providecommand \@@endlink[0]{}%
\providecommand \url  [0]{\begingroup\@sanitize@url \@url }%
\providecommand \@url [1]{\endgroup\@href {#1}{\urlprefix }}%
\providecommand \urlprefix  [0]{URL }%
\providecommand \Eprint [0]{\href }%
\providecommand \doibase [0]{http://dx.doi.org/}%
\providecommand \selectlanguage [0]{\@gobble}%
\providecommand \bibinfo  [0]{\@secondoftwo}%
\providecommand \bibfield  [0]{\@secondoftwo}%
\providecommand \translation [1]{[#1]}%
\providecommand \BibitemOpen [0]{}%
\providecommand \bibitemStop [0]{}%
\providecommand \bibitemNoStop [0]{.\EOS\space}%
\providecommand \EOS [0]{\spacefactor3000\relax}%
\providecommand \BibitemShut  [1]{\csname bibitem#1\endcsname}%
\let\auto@bib@innerbib\@empty
\bibitem [{\citenamefont {Gunnarsson}(1997)}]{Gunnarsson1997a}%
  \BibitemOpen
  \bibfield  {author} {\bibinfo {author} {\bibfnamefont {O.}~\bibnamefont
  {Gunnarsson}},\ }\bibfield  {title} {\enquote {\bibinfo {title}
  {Superconductivity in fullerides},}\ }\href
  {http://link.aps.org/doi/10.1103/RevModPhys.69.575} {\bibfield  {journal}
  {\bibinfo  {journal} {Reviews of Modern Physics}\ }\textbf {\bibinfo {volume}
  {69}},\ \bibinfo {pages} {575--606} (\bibinfo {year} {1997})}\BibitemShut
  {NoStop}%
\bibitem [{\citenamefont {Tanigaki}\ \emph {et~al.}(1992)\citenamefont
  {Tanigaki}, \citenamefont {Hirosawa}, \citenamefont {Ebbesen}, \citenamefont
  {Mizuki}, \citenamefont {Shimakawa}, \citenamefont {Kubo}, \citenamefont
  {Tsai},\ and\ \citenamefont {Kuroshima}}]{Structure_nature_K_Rb}%
  \BibitemOpen
  \bibfield  {author} {\bibinfo {author} {\bibfnamefont {K.}~\bibnamefont
  {Tanigaki}}, \bibinfo {author} {\bibfnamefont {I.}~\bibnamefont {Hirosawa}},
  \bibinfo {author} {\bibfnamefont {T.~W.}\ \bibnamefont {Ebbesen}}, \bibinfo
  {author} {\bibfnamefont {J.}~\bibnamefont {Mizuki}}, \bibinfo {author}
  {\bibnamefont {Shimakawa}}, \bibinfo {author} {\bibfnamefont
  {Y.}~\bibnamefont {Kubo}}, \bibinfo {author} {\bibfnamefont {J.~S.}\
  \bibnamefont {Tsai}}, \ and\ \bibinfo {author} {\bibfnamefont
  {S.}~\bibnamefont {Kuroshima}},\ }\bibfield  {title} {\enquote {\bibinfo
  {title} {Superconductivity in sodium-and lithium-containing alkali-metal
  fullerides},}\ }\href {http://dx.doi.org/10.1038/356419a0} {\bibfield
  {journal} {\bibinfo  {journal} {Nature}\ }\textbf {\bibinfo {volume} {356}},\
  \bibinfo {pages} {419--421} (\bibinfo {year} {1992})}\BibitemShut {NoStop}%
\bibitem [{\citenamefont {Ganin}\ \emph {et~al.}(2008)\citenamefont {Ganin},
  \citenamefont {Takabayashi}, \citenamefont {Khimyak}, \citenamefont
  {Margadonna}, \citenamefont {Tamai}, \citenamefont {Rosseinsky},\ and\
  \citenamefont {Prassides}}]{Ganin2008a}%
  \BibitemOpen
  \bibfield  {author} {\bibinfo {author} {\bibfnamefont {A.~Y.}\ \bibnamefont
  {Ganin}}, \bibinfo {author} {\bibfnamefont {Y.}~\bibnamefont {Takabayashi}},
  \bibinfo {author} {\bibfnamefont {Y.~Z.}\ \bibnamefont {Khimyak}}, \bibinfo
  {author} {\bibfnamefont {S.}~\bibnamefont {Margadonna}}, \bibinfo {author}
  {\bibfnamefont {A.}~\bibnamefont {Tamai}}, \bibinfo {author} {\bibfnamefont
  {M.~J.}\ \bibnamefont {Rosseinsky}}, \ and\ \bibinfo {author} {\bibfnamefont
  {K.}~\bibnamefont {Prassides}},\ }\bibfield  {title} {\enquote {\bibinfo
  {title} {Bulk superconductivity at 38 k in a molecular system},}\ }\href
  {\doibase 10.1038/nmat2179} {\bibfield  {journal} {\bibinfo  {journal} {Nat
  Mater}\ }\textbf {\bibinfo {volume} {7}},\ \bibinfo {pages} {367--71}
  (\bibinfo {year} {2008})}\BibitemShut {NoStop}%
\bibitem [{\citenamefont {Takabayashi}\ \emph {et~al.}(2009)\citenamefont
  {Takabayashi}, \citenamefont {Ganin}, \citenamefont {Jegli\v{c}},
  \citenamefont {Ar\v{c}on}, \citenamefont {Takano}, \citenamefont {Iwasa},
  \citenamefont {Ohishi}, \citenamefont {Takata}, \citenamefont {Takeshita},
  \citenamefont {Prassides},\ and\ \citenamefont
  {Rosseinsky}}]{Takabayashi2009}%
  \BibitemOpen
  \bibfield  {author} {\bibinfo {author} {\bibfnamefont {Y.}~\bibnamefont
  {Takabayashi}}, \bibinfo {author} {\bibfnamefont {A.~Y.}\ \bibnamefont
  {Ganin}}, \bibinfo {author} {\bibfnamefont {P.}~\bibnamefont {Jegli\v{c}}},
  \bibinfo {author} {\bibfnamefont {D.}~\bibnamefont {Ar\v{c}on}}, \bibinfo
  {author} {\bibfnamefont {T.}~\bibnamefont {Takano}}, \bibinfo {author}
  {\bibfnamefont {Y.}~\bibnamefont {Iwasa}}, \bibinfo {author} {\bibfnamefont
  {Y.}~\bibnamefont {Ohishi}}, \bibinfo {author} {\bibfnamefont
  {M.}~\bibnamefont {Takata}}, \bibinfo {author} {\bibfnamefont
  {N.}~\bibnamefont {Takeshita}}, \bibinfo {author} {\bibfnamefont
  {K.}~\bibnamefont {Prassides}}, \ and\ \bibinfo {author} {\bibfnamefont
  {M.~J.}\ \bibnamefont {Rosseinsky}},\ }\bibfield  {title} {\enquote {\bibinfo
  {title} {{The Disorder-Free Non-BCS Superconductor Cs$_3$C$_{60}$ Emerges
  from an Antiferromagnetic Insulator Parent State}},}\ }\href {\doibase
  10.1126/science.1169163} {\bibfield  {journal} {\bibinfo  {journal}
  {Science}\ }\textbf {\bibinfo {volume} {323}},\ \bibinfo {pages} {1585--1590}
  (\bibinfo {year} {2009})}\BibitemShut {NoStop}%
\bibitem [{\citenamefont {Ganin}\ \emph {et~al.}(2010)\citenamefont {Ganin},
  \citenamefont {Takabayashi}, \citenamefont {Jegli¿}, \citenamefont {Ar¿on},
  \citenamefont {Poto¿nik}, \citenamefont {Baker}, \citenamefont {Ohishi},
  \citenamefont {McDonald}, \citenamefont {Tzirakis}, \citenamefont {McLennan},
  \citenamefont {Darling}, \citenamefont {Takata}, \citenamefont {Rosseinsky},\
  and\ \citenamefont {Prassides}}]{Ganin2010a}%
  \BibitemOpen
  \bibfield  {author} {\bibinfo {author} {\bibfnamefont {Alexey~Y.}\
  \bibnamefont {Ganin}}, \bibinfo {author} {\bibfnamefont {Yasuhiro}\
  \bibnamefont {Takabayashi}}, \bibinfo {author} {\bibfnamefont {Peter}\
  \bibnamefont {Jegli¿}}, \bibinfo {author} {\bibfnamefont {Denis}\
  \bibnamefont {Ar¿on}}, \bibinfo {author} {\bibfnamefont {Anton}\ \bibnamefont
  {Poto¿nik}}, \bibinfo {author} {\bibfnamefont {Peter~J.}\ \bibnamefont
  {Baker}}, \bibinfo {author} {\bibfnamefont {Yasuo}\ \bibnamefont {Ohishi}},
  \bibinfo {author} {\bibfnamefont {Martin~T.}\ \bibnamefont {McDonald}},
  \bibinfo {author} {\bibfnamefont {Manolis~D.}\ \bibnamefont {Tzirakis}},
  \bibinfo {author} {\bibfnamefont {Alec}\ \bibnamefont {McLennan}}, \bibinfo
  {author} {\bibfnamefont {George~R.}\ \bibnamefont {Darling}}, \bibinfo
  {author} {\bibfnamefont {Masaki}\ \bibnamefont {Takata}}, \bibinfo {author}
  {\bibfnamefont {Matthew~J.}\ \bibnamefont {Rosseinsky}}, \ and\ \bibinfo
  {author} {\bibfnamefont {Kosmas}\ \bibnamefont {Prassides}},\ }\bibfield
  {title} {\enquote {\bibinfo {title} {{Polymorphism control of
  superconductivity and magnetism in Cs$_3$C$_{60}$ close to the Mott
  transition}},}\ }\href {\doibase 10.1038/nature09120} {\bibfield  {journal}
  {\bibinfo  {journal} {Nature}\ }\textbf {\bibinfo {volume} {466}},\ \bibinfo
  {pages} {221--225} (\bibinfo {year} {2010})}\BibitemShut {NoStop}%
\bibitem [{\citenamefont {Mitrano}\ \emph {et~al.}(2016)\citenamefont
  {Mitrano}, \citenamefont {Cantaluppi}, \citenamefont {Nicoletti},
  \citenamefont {Kaiser}, \citenamefont {Perucchi}, \citenamefont {Lupi},
  \citenamefont {Pietro}, \citenamefont {Pontiroli}, \citenamefont {Ricc\'{o}},
  \citenamefont {Clark}, \citenamefont {Jaksch},\ and\ \citenamefont
  {Cavalleri}}]{Mitrano2016}%
  \BibitemOpen
  \bibfield  {author} {\bibinfo {author} {\bibfnamefont {M.}~\bibnamefont
  {Mitrano}}, \bibinfo {author} {\bibfnamefont {A.}~\bibnamefont {Cantaluppi}},
  \bibinfo {author} {\bibfnamefont {D.}~\bibnamefont {Nicoletti}}, \bibinfo
  {author} {\bibfnamefont {S.}~\bibnamefont {Kaiser}}, \bibinfo {author}
  {\bibfnamefont {A.}~\bibnamefont {Perucchi}}, \bibinfo {author}
  {\bibfnamefont {S.}~\bibnamefont {Lupi}}, \bibinfo {author} {\bibfnamefont
  {P.~Di}\ \bibnamefont {Pietro}}, \bibinfo {author} {\bibfnamefont
  {D.}~\bibnamefont {Pontiroli}}, \bibinfo {author} {\bibfnamefont
  {M.}~\bibnamefont {Ricc\'{o}}}, \bibinfo {author} {\bibfnamefont {S.~R.}\
  \bibnamefont {Clark}}, \bibinfo {author} {\bibfnamefont {D.}~\bibnamefont
  {Jaksch}}, \ and\ \bibinfo {author} {\bibfnamefont {A.}~\bibnamefont
  {Cavalleri}},\ }\bibfield  {title} {\enquote {\bibinfo {title} {{Possible
  light-induced superconductivity in K$_3$C$_{60}$ at high temperature}},}\
  }\href {https://www.nature.com/articles/nature16522} {\bibfield  {journal}
  {\bibinfo  {journal} {Nature}\ }\textbf {\bibinfo {volume} {530}},\ \bibinfo
  {pages} {461} (\bibinfo {year} {2016})}\BibitemShut {NoStop}%
\bibitem [{\citenamefont {Cantaluppi}\ \emph {et~al.}(2018)\citenamefont
  {Cantaluppi}, \citenamefont {Buzzi}, \citenamefont {Jotzu}, \citenamefont
  {Nicoletti}, \citenamefont {Mitrano}, \citenamefont {Pontiroli},
  \citenamefont {Ricc\'{o}}, \citenamefont {Perucchi}, \citenamefont {Pietro},\
  and\ \citenamefont {Cavalleri}}]{Cantaluppi2017}%
  \BibitemOpen
  \bibfield  {author} {\bibinfo {author} {\bibfnamefont {A.}~\bibnamefont
  {Cantaluppi}}, \bibinfo {author} {\bibfnamefont {M.}~\bibnamefont {Buzzi}},
  \bibinfo {author} {\bibfnamefont {G.}~\bibnamefont {Jotzu}}, \bibinfo
  {author} {\bibfnamefont {D.}~\bibnamefont {Nicoletti}}, \bibinfo {author}
  {\bibfnamefont {M.}~\bibnamefont {Mitrano}}, \bibinfo {author} {\bibfnamefont
  {D.}~\bibnamefont {Pontiroli}}, \bibinfo {author} {\bibfnamefont
  {M.}~\bibnamefont {Ricc\'{o}}}, \bibinfo {author} {\bibfnamefont
  {A.}~\bibnamefont {Perucchi}}, \bibinfo {author} {\bibfnamefont {P.~Di}\
  \bibnamefont {Pietro}}, \ and\ \bibinfo {author} {\bibfnamefont
  {A.}~\bibnamefont {Cavalleri}},\ }\bibfield  {title} {\enquote {\bibinfo
  {title} {{Pressure tuning of light-induced superconductivity in
  K$_3$C$_{60}$}},}\ }\href {\doibase 10.1038/s41567-018-0134-8} {\bibfield
  {journal} {\bibinfo  {journal} {Nat. Phys.}\ }\textbf {\bibinfo {volume}
  {14}},\ \bibinfo {pages} {837} (\bibinfo {year} {2018})}\BibitemShut
  {NoStop}%
\bibitem [{\citenamefont {Yildirim}\ \emph {et~al.}(1996)\citenamefont
  {Yildirim}, \citenamefont {Barbedette}, \citenamefont {Fischer},
  \citenamefont {Lin}, \citenamefont {Robert}, \citenamefont {Petit},\ and\
  \citenamefont {Palstra}}]{Yildirim1996}%
  \BibitemOpen
  \bibfield  {author} {\bibinfo {author} {\bibfnamefont {T.}~\bibnamefont
  {Yildirim}}, \bibinfo {author} {\bibfnamefont {L.}~\bibnamefont
  {Barbedette}}, \bibinfo {author} {\bibfnamefont {J.~E.}\ \bibnamefont
  {Fischer}}, \bibinfo {author} {\bibfnamefont {C.~L.}\ \bibnamefont {Lin}},
  \bibinfo {author} {\bibfnamefont {J.}~\bibnamefont {Robert}}, \bibinfo
  {author} {\bibfnamefont {P.}~\bibnamefont {Petit}}, \ and\ \bibinfo {author}
  {\bibfnamefont {T.~T.~M.}\ \bibnamefont {Palstra}},\ }\bibfield  {title}
  {\enquote {\bibinfo {title} {${T}_{c}$ vs carrier concentration in cubic
  fulleride superconductors},}\ }\href {\doibase 10.1103/PhysRevLett.77.167}
  {\bibfield  {journal} {\bibinfo  {journal} {Physical Review Letters}\
  }\textbf {\bibinfo {volume} {77}},\ \bibinfo {pages} {167--170} (\bibinfo
  {year} {1996})}\BibitemShut {NoStop}%
\bibitem [{\citenamefont {Kerkoud}\ \emph {et~al.}(1996)\citenamefont
  {Kerkoud}, \citenamefont {Auban-Senzier}, \citenamefont {Jérome},
  \citenamefont {Brazovskii}, \citenamefont {Luk'Yanchuk}, \citenamefont
  {Kirova}, \citenamefont {Rachdi},\ and\ \citenamefont {Goze}}]{Kerkoud1996}%
  \BibitemOpen
  \bibfield  {author} {\bibinfo {author} {\bibfnamefont {R.}~\bibnamefont
  {Kerkoud}}, \bibinfo {author} {\bibfnamefont {P.}~\bibnamefont
  {Auban-Senzier}}, \bibinfo {author} {\bibfnamefont {D.}~\bibnamefont
  {Jérome}}, \bibinfo {author} {\bibfnamefont {S.}~\bibnamefont {Brazovskii}},
  \bibinfo {author} {\bibfnamefont {I.}~\bibnamefont {Luk'Yanchuk}}, \bibinfo
  {author} {\bibfnamefont {N.}~\bibnamefont {Kirova}}, \bibinfo {author}
  {\bibfnamefont {F.}~\bibnamefont {Rachdi}}, \ and\ \bibinfo {author}
  {\bibfnamefont {C.}~\bibnamefont {Goze}},\ }\bibfield  {title} {\enquote
  {\bibinfo {title} {{Insulator-metal transition in Rb$_{4}$C$_{60}$ under
  pressure from 13C-NMR}},}\ }\href {\doibase
  https://doi.org/10.1016/0022-3697(95)00113-1} {\bibfield  {journal} {\bibinfo
   {journal} {Journal of Physics and Chemistry of Solids}\ }\textbf {\bibinfo
  {volume} {57}},\ \bibinfo {pages} {143--152} (\bibinfo {year}
  {1996})}\BibitemShut {NoStop}%
\bibitem [{\citenamefont {Ihara}\ \emph {et~al.}(2010)\citenamefont {Ihara},
  \citenamefont {Alloul}, \citenamefont {Wzietek}, \citenamefont {Pontiroli},
  \citenamefont {Mazzani},\ and\ \citenamefont {Ricc\`o}}]{Ihara2010}%
  \BibitemOpen
  \bibfield  {author} {\bibinfo {author} {\bibfnamefont {Y.}~\bibnamefont
  {Ihara}}, \bibinfo {author} {\bibfnamefont {H.}~\bibnamefont {Alloul}},
  \bibinfo {author} {\bibfnamefont {P.}~\bibnamefont {Wzietek}}, \bibinfo
  {author} {\bibfnamefont {D.}~\bibnamefont {Pontiroli}}, \bibinfo {author}
  {\bibfnamefont {M.}~\bibnamefont {Mazzani}}, \ and\ \bibinfo {author}
  {\bibfnamefont {M.}~\bibnamefont {Ricc\`o}},\ }\bibfield  {title} {\enquote
  {\bibinfo {title} {{NMR Study of the Mott Transitions to Superconductivity in
  the Two ${\mathrm{Cs}}_{3}{\mathrm{C}}_{60}$ Phases}},}\ }\href {\doibase
  10.1103/PhysRevLett.104.256402} {\bibfield  {journal} {\bibinfo  {journal}
  {Phys. Rev. Lett.}\ }\textbf {\bibinfo {volume} {104}},\ \bibinfo {pages}
  {256402} (\bibinfo {year} {2010})}\BibitemShut {NoStop}%
\bibitem [{\citenamefont {Nava}\ \emph {et~al.}(2018)\citenamefont {Nava},
  \citenamefont {Giannetti}, \citenamefont {Georges}, \citenamefont {Tosatti},\
  and\ \citenamefont {Fabrizio}}]{Nava2018cooling}%
  \BibitemOpen
  \bibfield  {author} {\bibinfo {author} {\bibfnamefont {A.}~\bibnamefont
  {Nava}}, \bibinfo {author} {\bibfnamefont {C.}~\bibnamefont {Giannetti}},
  \bibinfo {author} {\bibfnamefont {A.}~\bibnamefont {Georges}}, \bibinfo
  {author} {\bibfnamefont {E.}~\bibnamefont {Tosatti}}, \ and\ \bibinfo
  {author} {\bibfnamefont {M.}~\bibnamefont {Fabrizio}},\ }\bibfield  {title}
  {\enquote {\bibinfo {title} {{Cooling quasiparticles in A$_{3}$C$_{60}$
  fullerides by excitonic mid-infrared absorption}},}\ }\href {\doibase
  10.1038/nphys4288} {\bibfield  {journal} {\bibinfo  {journal} {Nature
  Physics}\ }\textbf {\bibinfo {volume} {14}},\ \bibinfo {pages} {154--159}
  (\bibinfo {year} {2018})}\BibitemShut {NoStop}%
\bibitem [{\citenamefont {Zadik}\ \emph {et~al.}(2015)\citenamefont {Zadik},
  \citenamefont {Takabayashi}, \citenamefont {Klupp}, \citenamefont {Colman},
  \citenamefont {Ganin}, \citenamefont {{Poto\v{c}nik}}, \citenamefont
  {{Jegli\v{c}}}, \citenamefont {{Ar\v{c}on}}, \citenamefont {Matus},
  \citenamefont {{Kamar\'{a}s}}, \citenamefont {Kasahara}, \citenamefont
  {Iwasa}, \citenamefont {Fitch}, \citenamefont {Ohishi}, \citenamefont
  {Garbarino}, \citenamefont {Kato}, \citenamefont {Rosseinsky},\ and\
  \citenamefont {Prassides}}]{Zadik2015}%
  \BibitemOpen
  \bibfield  {author} {\bibinfo {author} {\bibfnamefont {R.~H.}\ \bibnamefont
  {Zadik}}, \bibinfo {author} {\bibfnamefont {Y.}~\bibnamefont {Takabayashi}},
  \bibinfo {author} {\bibfnamefont {G.}~\bibnamefont {Klupp}}, \bibinfo
  {author} {\bibfnamefont {R.~H.}\ \bibnamefont {Colman}}, \bibinfo {author}
  {\bibfnamefont {A.~Y.}\ \bibnamefont {Ganin}}, \bibinfo {author}
  {\bibfnamefont {A.}~\bibnamefont {{Poto\v{c}nik}}}, \bibinfo {author}
  {\bibfnamefont {P.}~\bibnamefont {{Jegli\v{c}}}}, \bibinfo {author}
  {\bibfnamefont {D.}~\bibnamefont {{Ar\v{c}on}}}, \bibinfo {author}
  {\bibfnamefont {P.}~\bibnamefont {Matus}}, \bibinfo {author} {\bibfnamefont
  {K.}~\bibnamefont {{Kamar\'{a}s}}}, \bibinfo {author} {\bibfnamefont
  {Y.}~\bibnamefont {Kasahara}}, \bibinfo {author} {\bibfnamefont
  {Y.}~\bibnamefont {Iwasa}}, \bibinfo {author} {\bibfnamefont {A.~N.}\
  \bibnamefont {Fitch}}, \bibinfo {author} {\bibfnamefont {Y.}~\bibnamefont
  {Ohishi}}, \bibinfo {author} {\bibfnamefont {G.}~\bibnamefont {Garbarino}},
  \bibinfo {author} {\bibfnamefont {K.}~\bibnamefont {Kato}}, \bibinfo {author}
  {\bibfnamefont {M.~J.}\ \bibnamefont {Rosseinsky}}, \ and\ \bibinfo {author}
  {\bibfnamefont {K.}~\bibnamefont {Prassides}},\ }\bibfield  {title} {\enquote
  {\bibinfo {title} {{Optimized unconventional superconductivity in a molecular
  Jahn-Teller metal}},}\ }\href
  {http://advances.sciencemag.org/content/1/3/e1500059} {\bibfield  {journal}
  {\bibinfo  {journal} {Sci. Adv.}\ }\textbf {\bibinfo {volume} {1}},\ \bibinfo
  {pages} {e1500059} (\bibinfo {year} {2015})}\BibitemShut {NoStop}%
\bibitem [{\citenamefont {Iwahara}\ and\ \citenamefont
  {Chibotaru}(2015)}]{StructureK3C60}%
  \BibitemOpen
  \bibfield  {author} {\bibinfo {author} {\bibfnamefont {N.}~\bibnamefont
  {Iwahara}}\ and\ \bibinfo {author} {\bibfnamefont {L.~F.}\ \bibnamefont
  {Chibotaru}},\ }\bibfield  {title} {\enquote {\bibinfo {title} {{Dynamical
  Jahn-Teller instability in metallic fullerides}},}\ }\href {\doibase
  10.1103/PhysRevB.91.035109} {\bibfield  {journal} {\bibinfo  {journal} {Phys.
  Rev. B}\ }\textbf {\bibinfo {volume} {91}},\ \bibinfo {pages} {035109}
  (\bibinfo {year} {2015})}\BibitemShut {NoStop}%
\bibitem [{\citenamefont {Iwahara}\ \emph {et~al.}(2010)\citenamefont
  {Iwahara}, \citenamefont {Sato}, \citenamefont {Tanaka},\ and\ \citenamefont
  {Chibotaru}}]{Iwahara2010a}%
  \BibitemOpen
  \bibfield  {author} {\bibinfo {author} {\bibfnamefont {N.}~\bibnamefont
  {Iwahara}}, \bibinfo {author} {\bibfnamefont {T.}~\bibnamefont {Sato}},
  \bibinfo {author} {\bibfnamefont {K.}~\bibnamefont {Tanaka}}, \ and\ \bibinfo
  {author} {\bibfnamefont {L.~F.}\ \bibnamefont {Chibotaru}},\ }\bibfield
  {title} {\enquote {\bibinfo {title} {{Vibronic coupling in C$_{60}^-$ anion
  revisited: Derivations from photoelectron spectra and DFT calculations}},}\
  }\href@noop {} {\bibfield  {journal} {\bibinfo  {journal} {Phys. Rev. B}\
  }\textbf {\bibinfo {volume} {82}},\ \bibinfo {pages} {245409} (\bibinfo
  {year} {2010})}\BibitemShut {NoStop}%
\bibitem [{\citenamefont {Huang}\ and\ \citenamefont
  {Liu}(2020)}]{Zhishuo2020}%
  \BibitemOpen
  \bibfield  {author} {\bibinfo {author} {\bibfnamefont {Z.}~\bibnamefont
  {Huang}}\ and\ \bibinfo {author} {\bibfnamefont {D.}~\bibnamefont {Liu}},\
  }\bibfield  {title} {\enquote {\bibinfo {title} {{Dynamical Jahn-Teller
  effect in the first excited C$_{60}^-$}},}\ }\href
  {https://onlinelibrary.wiley.com/doi/abs/10.1002/qua.26148} {\bibfield
  {journal} {\bibinfo  {journal} {International Journal of Quantum Chemistry}\
  }\textbf {\bibinfo {volume} {120}},\ \bibinfo {pages} {e26148} (\bibinfo
  {year} {2020})},\ \Eprint
  {http://arxiv.org/abs/https://onlinelibrary.wiley.com/doi/pdf/10.1002/qua.26148}
  {https://onlinelibrary.wiley.com/doi/pdf/10.1002/qua.26148} \BibitemShut
  {NoStop}%
\bibitem [{\citenamefont {Altmann}\ and\ \citenamefont
  {Herzig}(1994)}]{Altmann1994}%
  \BibitemOpen
  \bibfield  {author} {\bibinfo {author} {\bibfnamefont {S.~L.}\ \bibnamefont
  {Altmann}}\ and\ \bibinfo {author} {\bibfnamefont {P.}~\bibnamefont
  {Herzig}},\ }\href {https://phaidra.univie.ac.at/view/o:104731} {\emph
  {\bibinfo {title} {{Point-Group Theory Tables}}}}\ (\bibinfo  {publisher}
  {Claredon Press},\ \bibinfo {address} {Oxford},\ \bibinfo {year}
  {1994})\BibitemShut {NoStop}%
\bibitem [{\citenamefont {Klupp}\ \emph {et~al.}(2012)\citenamefont {Klupp},
  \citenamefont {Matus}, \citenamefont {{Kamar\'{a}s}}, \citenamefont {Ganin},
  \citenamefont {McLennan}, \citenamefont {Rosseinsky}, \citenamefont
  {Takabayashi}, \citenamefont {McDonald},\ and\ \citenamefont
  {Prassides}}]{Klupp2012}%
  \BibitemOpen
  \bibfield  {author} {\bibinfo {author} {\bibfnamefont {G.}~\bibnamefont
  {Klupp}}, \bibinfo {author} {\bibfnamefont {P.}~\bibnamefont {Matus}},
  \bibinfo {author} {\bibfnamefont {K.}~\bibnamefont {{Kamar\'{a}s}}}, \bibinfo
  {author} {\bibfnamefont {A.~Y.}\ \bibnamefont {Ganin}}, \bibinfo {author}
  {\bibfnamefont {A.}~\bibnamefont {McLennan}}, \bibinfo {author}
  {\bibfnamefont {M.~J.}\ \bibnamefont {Rosseinsky}}, \bibinfo {author}
  {\bibfnamefont {Y.}~\bibnamefont {Takabayashi}}, \bibinfo {author}
  {\bibfnamefont {M.~T.}\ \bibnamefont {McDonald}}, \ and\ \bibinfo {author}
  {\bibfnamefont {K.}~\bibnamefont {Prassides}},\ }\bibfield  {title} {\enquote
  {\bibinfo {title} {{Dynamic Jahn-Teller effect in the parent insulating state
  of the molecular superconductor Cs$_3$C$_{60}$}},}\ }\href
  {https://www.nature.com/articles/ncomms1910} {\bibfield  {journal} {\bibinfo
  {journal} {Nat. Commun.}\ }\textbf {\bibinfo {volume} {3}},\ \bibinfo {pages}
  {912} (\bibinfo {year} {2012})}\BibitemShut {NoStop}%
\bibitem [{\citenamefont {Kamar\'{a}s}\ \emph {et~al.}(2013)\citenamefont
  {Kamar\'{a}s}, \citenamefont {Klupp}, \citenamefont {Matus}, \citenamefont
  {Ganin}, \citenamefont {McLennan}, \citenamefont {Rosseinsky}, \citenamefont
  {Takabayashi}, \citenamefont {McDonald},\ and\ \citenamefont
  {Prassides}}]{Kamaras2013}%
  \BibitemOpen
  \bibfield  {author} {\bibinfo {author} {\bibfnamefont {K.}~\bibnamefont
  {Kamar\'{a}s}}, \bibinfo {author} {\bibfnamefont {G.}~\bibnamefont {Klupp}},
  \bibinfo {author} {\bibfnamefont {P.}~\bibnamefont {Matus}}, \bibinfo
  {author} {\bibfnamefont {A.~Y.}\ \bibnamefont {Ganin}}, \bibinfo {author}
  {\bibfnamefont {A.}~\bibnamefont {McLennan}}, \bibinfo {author}
  {\bibfnamefont {M.~J.}\ \bibnamefont {Rosseinsky}}, \bibinfo {author}
  {\bibfnamefont {Y.}~\bibnamefont {Takabayashi}}, \bibinfo {author}
  {\bibfnamefont {M.~T.}\ \bibnamefont {McDonald}}, \ and\ \bibinfo {author}
  {\bibfnamefont {K.}~\bibnamefont {Prassides}},\ }\bibfield  {title} {\enquote
  {\bibinfo {title} {{Mott localization in the correlated superconductor
  Cs$_3$C$_{60}$ resulting from the molecular Jahn-Teller effect}},}\ }\href
  {http://stacks.iop.org/1742-6596/428/i=1/a=012002} {\bibfield  {journal}
  {\bibinfo  {journal} {J. Phys.: Conf. Ser.}\ }\textbf {\bibinfo {volume}
  {428}},\ \bibinfo {pages} {012002} (\bibinfo {year} {2013})}\BibitemShut
  {NoStop}%
\bibitem [{\citenamefont {Poto{\v{c}}nik}\ \emph {et~al.}(2014)\citenamefont
  {Poto{\v{c}}nik}, \citenamefont {Ganin}, \citenamefont {Takabayashi},
  \citenamefont {McDonald}, \citenamefont {Heinmaa}, \citenamefont
  {Jegli{\v{c}}}, \citenamefont {Stern}, \citenamefont {Rosseinsky},
  \citenamefont {Prassides},\ and\ \citenamefont {Ar{\v{c}}on}}]{Potocnik2014}%
  \BibitemOpen
  \bibfield  {author} {\bibinfo {author} {\bibfnamefont {A.}~\bibnamefont
  {Poto{\v{c}}nik}}, \bibinfo {author} {\bibfnamefont {A.~Y.}\ \bibnamefont
  {Ganin}}, \bibinfo {author} {\bibfnamefont {Y.}~\bibnamefont {Takabayashi}},
  \bibinfo {author} {\bibfnamefont {M.~T.}\ \bibnamefont {McDonald}}, \bibinfo
  {author} {\bibfnamefont {I.}~\bibnamefont {Heinmaa}}, \bibinfo {author}
  {\bibfnamefont {P.}~\bibnamefont {Jegli{\v{c}}}}, \bibinfo {author}
  {\bibfnamefont {R.}~\bibnamefont {Stern}}, \bibinfo {author} {\bibfnamefont
  {M.~J.}\ \bibnamefont {Rosseinsky}}, \bibinfo {author} {\bibfnamefont
  {K.}~\bibnamefont {Prassides}}, \ and\ \bibinfo {author} {\bibfnamefont
  {D.}~\bibnamefont {Ar{\v{c}}on}},\ }\bibfield  {title} {\enquote {\bibinfo
  {title} {{Jahn-Teller orbital glass state in the expanded fcc
  Cs$_{3}$C$_{60}$ fulleride}},}\ }\href@noop {} {\bibfield  {journal}
  {\bibinfo  {journal} {Chem. Sci.}\ }\textbf {\bibinfo {volume} {5}},\
  \bibinfo {pages} {3008} (\bibinfo {year} {2014})}\BibitemShut {NoStop}%
\bibitem [{\citenamefont {Bersuker}(2006)}]{bersuker2006JTeffect}%
  \BibitemOpen
  \bibfield  {author} {\bibinfo {author} {\bibfnamefont {I.}~\bibnamefont
  {Bersuker}},\ }\href {\doibase 10.1017/CBO9780511524769} {\emph {\bibinfo
  {title} {The Jahn-Teller Effect}}}\ (\bibinfo  {publisher} {Cambridge
  University Press},\ \bibinfo {year} {2006})\BibitemShut {NoStop}%
\bibitem [{\citenamefont {O'Brien}(1969)}]{OBrien1969a}%
  \BibitemOpen
  \bibfield  {author} {\bibinfo {author} {\bibfnamefont {M.~C.~M.}\
  \bibnamefont {O'Brien}},\ }\bibfield  {title} {\enquote {\bibinfo {title}
  {{Dynamic Jahn--Teller Effect in an Orbital Triplet State Coupled to Both
  $E_g$ and $T_{2g}$ Vibrations}},}\ }\href@noop {} {\bibfield  {journal}
  {\bibinfo  {journal} {Phys. Rev.}\ }\textbf {\bibinfo {volume} {187}},\
  \bibinfo {pages} {407} (\bibinfo {year} {1969})}\BibitemShut {NoStop}%
\bibitem [{\citenamefont {Chancey}\ and\ \citenamefont
  {O'Brien}(1997)}]{Chancey1997}%
  \BibitemOpen
  \bibfield  {author} {\bibinfo {author} {\bibfnamefont {C.~C.}\ \bibnamefont
  {Chancey}}\ and\ \bibinfo {author} {\bibfnamefont {M.~C.~M.}\ \bibnamefont
  {O'Brien}},\ }\href {https://press.princeton.edu/titles/6243.html} {\emph
  {\bibinfo {title} {The Jahn--Teller Effect in C$_{60}$ and Other Icosahedral
  Complexes}}}\ (\bibinfo  {publisher} {Princeton University Press},\ \bibinfo
  {address} {Princeton},\ \bibinfo {year} {1997})\BibitemShut {NoStop}%
\bibitem [{\citenamefont {Durand}\ \emph {et~al.}(2003)\citenamefont {Durand},
  \citenamefont {Darling}, \citenamefont {Dubitsky}, \citenamefont {Zaopo},\
  and\ \citenamefont {Rosseinsky}}]{durand2003mott}%
  \BibitemOpen
  \bibfield  {author} {\bibinfo {author} {\bibfnamefont {P.}~\bibnamefont
  {Durand}}, \bibinfo {author} {\bibfnamefont {G.~R.}\ \bibnamefont {Darling}},
  \bibinfo {author} {\bibfnamefont {Y.}~\bibnamefont {Dubitsky}}, \bibinfo
  {author} {\bibfnamefont {A.}~\bibnamefont {Zaopo}}, \ and\ \bibinfo {author}
  {\bibfnamefont {M.~J.}\ \bibnamefont {Rosseinsky}},\ }\bibfield  {title}
  {\enquote {\bibinfo {title} {{The Mott--Hubbard insulating state and orbital
  degeneracy in the superconducting C$_{60}^{3-}$ fulleride family}},}\
  }\href@noop {} {\bibfield  {journal} {\bibinfo  {journal} {Nature materials}\
  }\textbf {\bibinfo {volume} {2}},\ \bibinfo {pages} {605--610} (\bibinfo
  {year} {2003})}\BibitemShut {NoStop}%
\bibitem [{\citenamefont {Chibotaru}(2005)}]{Chibotaru2005a}%
  \BibitemOpen
  \bibfield  {author} {\bibinfo {author} {\bibfnamefont {L.~F.}\ \bibnamefont
  {Chibotaru}},\ }\bibfield  {title} {\enquote {\bibinfo {title}
  {{Spin-Vibronic Superexchange in Mott-Hubbard Fullerides}},}\ }\href
  {\doibase 10.1103/PhysRevLett.94.186405} {\bibfield  {journal} {\bibinfo
  {journal} {Phys. Rev. Lett.}\ }\textbf {\bibinfo {volume} {94}},\ \bibinfo
  {pages} {186405} (\bibinfo {year} {2005})}\BibitemShut {NoStop}%
\bibitem [{\citenamefont {Brouet}\ \emph {et~al.}(2002)\citenamefont {Brouet},
  \citenamefont {Alloul}, \citenamefont {Garaj},\ and\ \citenamefont
  {Forr\'o}}]{PhysRevB.66.155124}%
  \BibitemOpen
  \bibfield  {author} {\bibinfo {author} {\bibfnamefont {V.}~\bibnamefont
  {Brouet}}, \bibinfo {author} {\bibfnamefont {H.}~\bibnamefont {Alloul}},
  \bibinfo {author} {\bibfnamefont {S.}~\bibnamefont {Garaj}}, \ and\ \bibinfo
  {author} {\bibfnamefont {L.}~\bibnamefont {Forr\'o}},\ }\bibfield  {title}
  {\enquote {\bibinfo {title} {Persistence of molecular excitations in metallic
  fullerides and their role in a possible metal to insulator transition at high
  temperatures},}\ }\href {\doibase 10.1103/PhysRevB.66.155124} {\bibfield
  {journal} {\bibinfo  {journal} {Phys. Rev. B}\ }\textbf {\bibinfo {volume}
  {66}},\ \bibinfo {pages} {155124} (\bibinfo {year} {2002})}\BibitemShut
  {NoStop}%
\bibitem [{\citenamefont {Stephens}\ \emph {et~al.}(1991)\citenamefont
  {Stephens}, \citenamefont {Mihaly}, \citenamefont {Lee}, \citenamefont
  {Whetten}, \citenamefont {Huang}, \citenamefont {Kaner}, \citenamefont
  {Deiderich},\ and\ \citenamefont {Holczer}}]{Stephens1991structure}%
  \BibitemOpen
  \bibfield  {author} {\bibinfo {author} {\bibfnamefont {P.~W.}\ \bibnamefont
  {Stephens}}, \bibinfo {author} {\bibfnamefont {L.}~\bibnamefont {Mihaly}},
  \bibinfo {author} {\bibfnamefont {P.~L.}\ \bibnamefont {Lee}}, \bibinfo
  {author} {\bibfnamefont {R.~L.}\ \bibnamefont {Whetten}}, \bibinfo {author}
  {\bibfnamefont {S.~M.}\ \bibnamefont {Huang}}, \bibinfo {author}
  {\bibfnamefont {R.}~\bibnamefont {Kaner}}, \bibinfo {author} {\bibfnamefont
  {F.}~\bibnamefont {Deiderich}}, \ and\ \bibinfo {author} {\bibfnamefont
  {K.}~\bibnamefont {Holczer}},\ }\bibfield  {title} {\enquote {\bibinfo
  {title} {{Structure of single-phase superconducting K$_3$C$_{60}$}},}\ }\href
  {http://dx.doi.org/10.1038/351632a0} {\bibfield  {journal} {\bibinfo
  {journal} {Nature}\ }\textbf {\bibinfo {volume} {351}},\ \bibinfo {pages}
  {632--634} (\bibinfo {year} {1991})}\BibitemShut {NoStop}%
\bibitem [{\citenamefont {Teslic}\ \emph {et~al.}(1995)\citenamefont {Teslic},
  \citenamefont {Egami},\ and\ \citenamefont {Fischer}}]{Teslic1995shortrange}%
  \BibitemOpen
  \bibfield  {author} {\bibinfo {author} {\bibfnamefont {S.}~\bibnamefont
  {Teslic}}, \bibinfo {author} {\bibfnamefont {T.}~\bibnamefont {Egami}}, \
  and\ \bibinfo {author} {\bibfnamefont {J.~E.}\ \bibnamefont {Fischer}},\
  }\bibfield  {title} {\enquote {\bibinfo {title} {{Short-range
  antiferromagnetic orientational correlations in Rb$_3$C$_{60}$}},}\ }\href
  {\doibase 10.1103/PhysRevB.51.5973} {\bibfield  {journal} {\bibinfo
  {journal} {Physical Review B}\ }\textbf {\bibinfo {volume} {51}},\ \bibinfo
  {pages} {5973--5976} (\bibinfo {year} {1995})}\BibitemShut {NoStop}%
\bibitem [{\citenamefont {Mazin}\ \emph {et~al.}(1993)\citenamefont {Mazin},
  \citenamefont {Liechtenstein}, \citenamefont {Gunnarsson}, \citenamefont
  {Andersen}, \citenamefont {Antropov},\ and\ \citenamefont
  {Burkov}}]{Mazin1993orientational}%
  \BibitemOpen
  \bibfield  {author} {\bibinfo {author} {\bibfnamefont {II}~\bibnamefont
  {Mazin}, \bibfnamefont {I.~I.}}, \bibinfo {author} {\bibfnamefont {A.~I.}\
  \bibnamefont {Liechtenstein}}, \bibinfo {author} {\bibfnamefont
  {O.}~\bibnamefont {Gunnarsson}}, \bibinfo {author} {\bibfnamefont {O.~K.}\
  \bibnamefont {Andersen}}, \bibinfo {author} {\bibfnamefont {V.~P.}\
  \bibnamefont {Antropov}}, \ and\ \bibinfo {author} {\bibfnamefont {S.~E.}\
  \bibnamefont {Burkov}},\ }\bibfield  {title} {\enquote {\bibinfo {title}
  {{Orientational order in A$_{3}$C$_{60}$: Antiferromagnetic Ising model for
  the fcc lattice}},}\ }\href {\doibase 10.1103/PhysRevLett.70.4142} {\bibfield
   {journal} {\bibinfo  {journal} {Phys Rev Lett}\ }\textbf {\bibinfo {volume}
  {70}},\ \bibinfo {pages} {4142--4145} (\bibinfo {year} {1993})}\BibitemShut
  {NoStop}%
\bibitem [{\citenamefont {Sato}\ \emph {et~al.}(2006)\citenamefont {Sato},
  \citenamefont {Tokunaga},\ and\ \citenamefont {Tanaka}}]{Sato2006}%
  \BibitemOpen
  \bibfield  {author} {\bibinfo {author} {\bibfnamefont {Tohru}\ \bibnamefont
  {Sato}}, \bibinfo {author} {\bibfnamefont {Ken}\ \bibnamefont {Tokunaga}}, \
  and\ \bibinfo {author} {\bibfnamefont {Kazuyoshi}\ \bibnamefont {Tanaka}},\
  }\bibfield  {title} {\enquote {\bibinfo {title} {{Vibronic coupling in
  cyclopentadienyl radical: A method for calculation of vibronic coupling
  constant and vibronic coupling density analysis}},}\ }\href {\doibase
  10.1063/1.2150816} {\bibfield  {journal} {\bibinfo  {journal} {The Journal of
  Chemical Physics}\ }\textbf {\bibinfo {volume} {124}},\ \bibinfo {pages}
  {024314} (\bibinfo {year} {2006})}\BibitemShut {NoStop}%
\bibitem [{\citenamefont {Frisch}\ \emph {et~al.}(2009)\citenamefont {Frisch},
  \citenamefont {Trucks}, \citenamefont {Schlegel}, \citenamefont {Scuseria},
  \citenamefont {Robb}, \citenamefont {Cheeseman}, \citenamefont {Scalmani},
  \citenamefont {Barone}, \citenamefont {Petersson}, \citenamefont {Nakatsuji},
  \citenamefont {Li}, \citenamefont {Caricato}, \citenamefont {Marenich},
  \citenamefont {Bloino}, \citenamefont {Janesko}, \citenamefont {Gomperts},
  \citenamefont {Mennucci}, \citenamefont {Hratchian}, \citenamefont {Ortiz},
  \citenamefont {Izmaylov}, \citenamefont {Sonnenberg}, \citenamefont
  {Williams-Young}, \citenamefont {Ding}, \citenamefont {Lipparini},
  \citenamefont {Egidi}, \citenamefont {Goings}, \citenamefont {Peng},
  \citenamefont {Petrone}, \citenamefont {Henderson}, \citenamefont
  {Ranasinghe}, \citenamefont {Zakrzewski}, \citenamefont {Gao}, \citenamefont
  {Rega}, \citenamefont {Zheng}, \citenamefont {Liang}, \citenamefont {Hada},
  \citenamefont {Ehara}, \citenamefont {Toyota}, \citenamefont {Fukuda},
  \citenamefont {Hasegawa}, \citenamefont {Ishida}, \citenamefont {Nakajima},
  \citenamefont {Honda}, \citenamefont {Kitao}, \citenamefont {Nakai},
  \citenamefont {Vreven}, \citenamefont {Throssell}, \citenamefont
  {Montgomery}, \citenamefont {Jr.}, \citenamefont {Ogliaro}, \citenamefont
  {Bearpark}, \citenamefont {Heyd}, \citenamefont {Brothers}, \citenamefont
  {Kudin}, \citenamefont {Staroverov}, \citenamefont {Keith}, \citenamefont
  {Kobayashi}, \citenamefont {Normand}, \citenamefont {Raghavachari},
  \citenamefont {Rendell}, \citenamefont {Burant}, \citenamefont {Iyengar},
  \citenamefont {Tomasi}, \citenamefont {Cossi}, \citenamefont {Millam},
  \citenamefont {Klene}, \citenamefont {Adamo}, \citenamefont {Cammi},
  \citenamefont {Ochterski}, \citenamefont {Martin}, \citenamefont {Morokuma},
  \citenamefont {Farkas}, \citenamefont {Foresman},\ and\ \citenamefont
  {Fox}}]{g09}%
  \BibitemOpen
  \bibfield  {author} {\bibinfo {author} {\bibfnamefont {M.~J.}\ \bibnamefont
  {Frisch}}, \bibinfo {author} {\bibfnamefont {G.~W.}\ \bibnamefont {Trucks}},
  \bibinfo {author} {\bibfnamefont {H.~B.}\ \bibnamefont {Schlegel}}, \bibinfo
  {author} {\bibfnamefont {G.~E.}\ \bibnamefont {Scuseria}}, \bibinfo {author}
  {\bibfnamefont {M.~A.}\ \bibnamefont {Robb}}, \bibinfo {author}
  {\bibfnamefont {J.~R.}\ \bibnamefont {Cheeseman}}, \bibinfo {author}
  {\bibfnamefont {G.}~\bibnamefont {Scalmani}}, \bibinfo {author}
  {\bibfnamefont {V.}~\bibnamefont {Barone}}, \bibinfo {author} {\bibfnamefont
  {G.~A.}\ \bibnamefont {Petersson}}, \bibinfo {author} {\bibfnamefont
  {H.}~\bibnamefont {Nakatsuji}}, \bibinfo {author} {\bibfnamefont
  {X.}~\bibnamefont {Li}}, \bibinfo {author} {\bibfnamefont {M.}~\bibnamefont
  {Caricato}}, \bibinfo {author} {\bibfnamefont {A.}~\bibnamefont {Marenich}},
  \bibinfo {author} {\bibfnamefont {J.}~\bibnamefont {Bloino}}, \bibinfo
  {author} {\bibfnamefont {B.~G.}\ \bibnamefont {Janesko}}, \bibinfo {author}
  {\bibfnamefont {R.}~\bibnamefont {Gomperts}}, \bibinfo {author}
  {\bibfnamefont {B.}~\bibnamefont {Mennucci}}, \bibinfo {author}
  {\bibfnamefont {H.~P.}\ \bibnamefont {Hratchian}}, \bibinfo {author}
  {\bibfnamefont {J.~V.}\ \bibnamefont {Ortiz}}, \bibinfo {author}
  {\bibfnamefont {A.~F.}\ \bibnamefont {Izmaylov}}, \bibinfo {author}
  {\bibfnamefont {J.~L.}\ \bibnamefont {Sonnenberg}}, \bibinfo {author}
  {\bibfnamefont {D.}~\bibnamefont {Williams-Young}}, \bibinfo {author}
  {\bibfnamefont {F.}~\bibnamefont {Ding}}, \bibinfo {author} {\bibfnamefont
  {F.}~\bibnamefont {Lipparini}}, \bibinfo {author} {\bibfnamefont
  {F.}~\bibnamefont {Egidi}}, \bibinfo {author} {\bibfnamefont
  {J.}~\bibnamefont {Goings}}, \bibinfo {author} {\bibfnamefont
  {B.}~\bibnamefont {Peng}}, \bibinfo {author} {\bibfnamefont {A.}~\bibnamefont
  {Petrone}}, \bibinfo {author} {\bibfnamefont {T.}~\bibnamefont {Henderson}},
  \bibinfo {author} {\bibfnamefont {D.}~\bibnamefont {Ranasinghe}}, \bibinfo
  {author} {\bibfnamefont {V.~G.}\ \bibnamefont {Zakrzewski}}, \bibinfo
  {author} {\bibfnamefont {J.}~\bibnamefont {Gao}}, \bibinfo {author}
  {\bibfnamefont {N.}~\bibnamefont {Rega}}, \bibinfo {author} {\bibfnamefont
  {G.}~\bibnamefont {Zheng}}, \bibinfo {author} {\bibfnamefont
  {W.}~\bibnamefont {Liang}}, \bibinfo {author} {\bibfnamefont
  {M.}~\bibnamefont {Hada}}, \bibinfo {author} {\bibfnamefont {M.}~\bibnamefont
  {Ehara}}, \bibinfo {author} {\bibfnamefont {K.}~\bibnamefont {Toyota}},
  \bibinfo {author} {\bibfnamefont {R.}~\bibnamefont {Fukuda}}, \bibinfo
  {author} {\bibfnamefont {J.}~\bibnamefont {Hasegawa}}, \bibinfo {author}
  {\bibfnamefont {M.}~\bibnamefont {Ishida}}, \bibinfo {author} {\bibfnamefont
  {T.}~\bibnamefont {Nakajima}}, \bibinfo {author} {\bibfnamefont
  {Y.}~\bibnamefont {Honda}}, \bibinfo {author} {\bibfnamefont
  {O.}~\bibnamefont {Kitao}}, \bibinfo {author} {\bibfnamefont
  {H.}~\bibnamefont {Nakai}}, \bibinfo {author} {\bibfnamefont
  {T.}~\bibnamefont {Vreven}}, \bibinfo {author} {\bibfnamefont
  {K.}~\bibnamefont {Throssell}}, \bibinfo {author} {\bibfnamefont {J.~A.}\
  \bibnamefont {Montgomery}}, \bibinfo {author} {\bibfnamefont {J.~E.~Peralta}\
  \bibnamefont {Jr.}}, \bibinfo {author} {\bibfnamefont {F.}~\bibnamefont
  {Ogliaro}}, \bibinfo {author} {\bibfnamefont {M.}~\bibnamefont {Bearpark}},
  \bibinfo {author} {\bibfnamefont {J.~J.}\ \bibnamefont {Heyd}}, \bibinfo
  {author} {\bibfnamefont {E.}~\bibnamefont {Brothers}}, \bibinfo {author}
  {\bibfnamefont {K.~N.}\ \bibnamefont {Kudin}}, \bibinfo {author}
  {\bibfnamefont {V.~N.}\ \bibnamefont {Staroverov}}, \bibinfo {author}
  {\bibfnamefont {T.}~\bibnamefont {Keith}}, \bibinfo {author} {\bibfnamefont
  {R.}~\bibnamefont {Kobayashi}}, \bibinfo {author} {\bibfnamefont
  {J.}~\bibnamefont {Normand}}, \bibinfo {author} {\bibfnamefont
  {K.}~\bibnamefont {Raghavachari}}, \bibinfo {author} {\bibfnamefont
  {A.}~\bibnamefont {Rendell}}, \bibinfo {author} {\bibfnamefont {J.~C.}\
  \bibnamefont {Burant}}, \bibinfo {author} {\bibfnamefont {S.~S.}\
  \bibnamefont {Iyengar}}, \bibinfo {author} {\bibfnamefont {J.}~\bibnamefont
  {Tomasi}}, \bibinfo {author} {\bibfnamefont {M.}~\bibnamefont {Cossi}},
  \bibinfo {author} {\bibfnamefont {J.~M.}\ \bibnamefont {Millam}}, \bibinfo
  {author} {\bibfnamefont {M.}~\bibnamefont {Klene}}, \bibinfo {author}
  {\bibfnamefont {C.}~\bibnamefont {Adamo}}, \bibinfo {author} {\bibfnamefont
  {R.}~\bibnamefont {Cammi}}, \bibinfo {author} {\bibfnamefont {J.~W.}\
  \bibnamefont {Ochterski}}, \bibinfo {author} {\bibfnamefont {R.~L.}\
  \bibnamefont {Martin}}, \bibinfo {author} {\bibfnamefont {K.}~\bibnamefont
  {Morokuma}}, \bibinfo {author} {\bibfnamefont {O.}~\bibnamefont {Farkas}},
  \bibinfo {author} {\bibfnamefont {J.~B.}\ \bibnamefont {Foresman}}, \ and\
  \bibinfo {author} {\bibfnamefont {D.~J.}\ \bibnamefont {Fox}},\ }\href@noop
  {} {\emph {\bibinfo {title} {Gaussian 09, Revision B.01}}},\ \bibinfo
  {address} {Wallingford, CT} (\bibinfo {year} {2009})\BibitemShut {NoStop}%
\bibitem [{\citenamefont {Becke}(1993)}]{Becke1993a}%
  \BibitemOpen
  \bibfield  {author} {\bibinfo {author} {\bibfnamefont {Axel~D.}\ \bibnamefont
  {Becke}},\ }\bibfield  {title} {\enquote {\bibinfo {title}
  {{Density-functional thermochemistry. III. The role of exact exchange}},}\
  }\href {\doibase 10.1063/1.464913} {\bibfield  {journal} {\bibinfo  {journal}
  {The Journal of Chemical Physics}\ }\textbf {\bibinfo {volume} {98}},\
  \bibinfo {pages} {5648--5652} (\bibinfo {year} {1993})},\ \Eprint
  {http://arxiv.org/abs/https://doi.org/10.1063/1.464913}
  {https://doi.org/10.1063/1.464913} \BibitemShut {NoStop}%
\bibitem [{\citenamefont {Liu}\ \emph {et~al.}(2018{\natexlab{a}})\citenamefont
  {Liu}, \citenamefont {Niwa}, \citenamefont {Iwahara}, \citenamefont {Sato},\
  and\ \citenamefont {Chibotaru}}]{Liu2018b}%
  \BibitemOpen
  \bibfield  {author} {\bibinfo {author} {\bibfnamefont {D.}~\bibnamefont
  {Liu}}, \bibinfo {author} {\bibfnamefont {Y.}~\bibnamefont {Niwa}}, \bibinfo
  {author} {\bibfnamefont {N.}~\bibnamefont {Iwahara}}, \bibinfo {author}
  {\bibfnamefont {T.}~\bibnamefont {Sato}}, \ and\ \bibinfo {author}
  {\bibfnamefont {L.~F.}\ \bibnamefont {Chibotaru}},\ }\bibfield  {title}
  {\enquote {\bibinfo {title} {{Quadratic Jahn-Teller effect of fullerene
  anions}},}\ }\href {\doibase 10.1103/PhysRevB.98.035402} {\bibfield
  {journal} {\bibinfo  {journal} {Phys. Rev. B}\ }\textbf {\bibinfo {volume}
  {98}},\ \bibinfo {pages} {035402} (\bibinfo {year}
  {2018}{\natexlab{a}})}\BibitemShut {NoStop}%
\bibitem [{SM()}]{SM}%
  \BibitemOpen
  \href@noop {} {}\bibinfo {note} {The Supplementary Material contains active
  intrafullerene normal vibrational modes, alkali and interfullerene active
  displacements; details of pseudopotential generation for phonon calculations,
  full phonon dispersion; calculation of the effect of multiplet splitting on
  the adiabatic potential on one fullerene site; analysis of elastic response
  function to JT displacements at one site.}\BibitemShut {Stop}%
\bibitem [{\citenamefont {You}\ \emph {et~al.}(1993)\citenamefont {You},
  \citenamefont {Yan},\ and\ \citenamefont {Yan}}]{Phon_wholeC60}%
  \BibitemOpen
  \bibfield  {author} {\bibinfo {author} {\bibfnamefont {J.~Q.}\ \bibnamefont
  {You}}, \bibinfo {author} {\bibfnamefont {J.~R.}\ \bibnamefont {Yan}}, \ and\
  \bibinfo {author} {\bibfnamefont {X.~H.}\ \bibnamefont {Yan}},\ }\bibfield
  {title} {\enquote {\bibinfo {title} {{Lattice dynamics of A$_3$C$_{60}$
  fullerides, where the A are alkali-metal elements}},}\ }\href
  {http://stacks.iop.org/0953-8984/5/i=46/a=001} {\bibfield  {journal}
  {\bibinfo  {journal} {Journal of Physics: Condensed Matter}\ }\textbf
  {\bibinfo {volume} {5}},\ \bibinfo {pages} {L591} (\bibinfo {year}
  {1993})}\BibitemShut {NoStop}%
\bibitem [{\citenamefont {Akashi}\ and\ \citenamefont
  {Arita}(2013)}]{IONIZED_PP}%
  \BibitemOpen
  \bibfield  {author} {\bibinfo {author} {\bibfnamefont {R.}~\bibnamefont
  {Akashi}}\ and\ \bibinfo {author} {\bibfnamefont {R.}~\bibnamefont {Arita}},\
  }\bibfield  {title} {\enquote {\bibinfo {title} {Nonempirical study of
  superconductivity in alkali-doped fullerides based on density functional
  theory for superconductors},}\ }\href@noop {} {\bibfield  {journal} {\bibinfo
   {journal} {Physical Review B}\ }\textbf {\bibinfo {volume} {88}},\ \bibinfo
  {pages} {054510} (\bibinfo {year} {2013})}\BibitemShut {NoStop}%
\bibitem [{\citenamefont {Nomura}\ and\ \citenamefont
  {Arita}(2015)}]{Phonon_imaginary_Nomura}%
  \BibitemOpen
  \bibfield  {author} {\bibinfo {author} {\bibfnamefont {Y.}~\bibnamefont
  {Nomura}}\ and\ \bibinfo {author} {\bibfnamefont {R.}~\bibnamefont {Arita}},\
  }\bibfield  {title} {\enquote {\bibinfo {title} {{Ab initio downfolding for
  electron-phonon-coupled systems: Constrained density-functional perturbation
  theory}},}\ }\href {\doibase 10.1103/PhysRevB.92.245108} {\bibfield
  {journal} {\bibinfo  {journal} {Physical Review B}\ }\textbf {\bibinfo
  {volume} {92}} (\bibinfo {year} {2015}),\
  10.1103/PhysRevB.92.245108}\BibitemShut {NoStop}%
\bibitem [{\citenamefont {Baroni}\ \emph {et~al.}(2001)\citenamefont {Baroni},
  \citenamefont {de~Gironcoli}, \citenamefont {Dal~Corso},\ and\ \citenamefont
  {Giannozzi}}]{dfpt}%
  \BibitemOpen
  \bibfield  {author} {\bibinfo {author} {\bibfnamefont {S.}~\bibnamefont
  {Baroni}}, \bibinfo {author} {\bibfnamefont {S.}~\bibnamefont
  {de~Gironcoli}}, \bibinfo {author} {\bibfnamefont {A.}~\bibnamefont
  {Dal~Corso}}, \ and\ \bibinfo {author} {\bibfnamefont {P.}~\bibnamefont
  {Giannozzi}},\ }\bibfield  {title} {\enquote {\bibinfo {title} {Phonons and
  related crystal properties from density-functional perturbation theory},}\
  }\href {\doibase 10.1103/RevModPhys.73.515} {\bibfield  {journal} {\bibinfo
  {journal} {Rev. Mod. Phys.}\ }\textbf {\bibinfo {volume} {73}},\ \bibinfo
  {pages} {515--562} (\bibinfo {year} {2001})}\BibitemShut {NoStop}%
\bibitem [{\citenamefont {Perdew}\ and\ \citenamefont {Zunger}(1981)}]{PP_PZ}%
  \BibitemOpen
  \bibfield  {author} {\bibinfo {author} {\bibfnamefont {J.~P.}\ \bibnamefont
  {Perdew}}\ and\ \bibinfo {author} {\bibfnamefont {Alex}\ \bibnamefont
  {Zunger}},\ }\bibfield  {title} {\enquote {\bibinfo {title} {Self-interaction
  correction to density-functional approximations for many-electron systems},}\
  }\href {\doibase 10.1103/PhysRevB.23.5048} {\bibfield  {journal} {\bibinfo
  {journal} {Phys. Rev. B}\ }\textbf {\bibinfo {volume} {23}},\ \bibinfo
  {pages} {5048--5079} (\bibinfo {year} {1981})}\BibitemShut {NoStop}%
\bibitem [{\citenamefont {Perdew}\ \emph {et~al.}(1996)\citenamefont {Perdew},
  \citenamefont {Burke},\ and\ \citenamefont {Ernzerhof}}]{PBE-1}%
  \BibitemOpen
  \bibfield  {author} {\bibinfo {author} {\bibfnamefont {J.~P.}\ \bibnamefont
  {Perdew}}, \bibinfo {author} {\bibfnamefont {K.}~\bibnamefont {Burke}}, \
  and\ \bibinfo {author} {\bibfnamefont {M.}~\bibnamefont {Ernzerhof}},\
  }\bibfield  {title} {\enquote {\bibinfo {title} {{Generalized Gradient
  Approximation Made Simple}},}\ }\href
  {http://link.aps.org/doi/10.1103/PhysRevLett.77.3865} {\bibfield  {journal}
  {\bibinfo  {journal} {Physical Review Letters}\ }\textbf {\bibinfo {volume}
  {77}},\ \bibinfo {pages} {3865--3868} (\bibinfo {year} {1996})}\BibitemShut
  {NoStop}%
\bibitem [{\citenamefont {Louie}\ \emph {et~al.}(1982)\citenamefont {Louie},
  \citenamefont {Froyen},\ and\ \citenamefont
  {Cohen}}]{Nonlinear_core_correction}%
  \BibitemOpen
  \bibfield  {author} {\bibinfo {author} {\bibfnamefont {S.~G.}\ \bibnamefont
  {Louie}}, \bibinfo {author} {\bibfnamefont {S.}~\bibnamefont {Froyen}}, \
  and\ \bibinfo {author} {\bibfnamefont {M.~L.}\ \bibnamefont {Cohen}},\
  }\bibfield  {title} {\enquote {\bibinfo {title} {Nonlinear ionic
  pseudopotentials in spin-density-functional calculations},}\ }\href {\doibase
  10.1103/PhysRevB.26.1738} {\bibfield  {journal} {\bibinfo  {journal}
  {Physical Review B}\ }\textbf {\bibinfo {volume} {26}},\ \bibinfo {pages}
  {1738--1742} (\bibinfo {year} {1982})}\BibitemShut {NoStop}%
\bibitem [{\citenamefont {Koelling}\ and\ \citenamefont
  {Harmon}(1977)}]{Scalar_relativistic}%
  \BibitemOpen
  \bibfield  {author} {\bibinfo {author} {\bibfnamefont {D~D}\ \bibnamefont
  {Koelling}}\ and\ \bibinfo {author} {\bibfnamefont {B~N}\ \bibnamefont
  {Harmon}},\ }\bibfield  {title} {\enquote {\bibinfo {title} {A technique for
  relativistic spin-polarised calculations},}\ }\href
  {http://stacks.iop.org/0022-3719/10/i=16/a=019} {\bibfield  {journal}
  {\bibinfo  {journal} {Journal of Physics C: Solid State Physics}\ }\textbf
  {\bibinfo {volume} {10}},\ \bibinfo {pages} {3107} (\bibinfo {year}
  {1977})}\BibitemShut {NoStop}%
\bibitem [{\citenamefont {Troullier}\ and\ \citenamefont
  {Martins}(1991)}]{TM-method}%
  \BibitemOpen
  \bibfield  {author} {\bibinfo {author} {\bibfnamefont {N.}~\bibnamefont
  {Troullier}}\ and\ \bibinfo {author} {\bibfnamefont {José~Luriaas}\
  \bibnamefont {Martins}},\ }\bibfield  {title} {\enquote {\bibinfo {title}
  {Efficient pseudopotentials for plane-wave calculations},}\ }\href {\doibase
  10.1103/PhysRevB.43.1993} {\bibfield  {journal} {\bibinfo  {journal}
  {Physical Review B}\ }\textbf {\bibinfo {volume} {43}},\ \bibinfo {pages}
  {1993--2006} (\bibinfo {year} {1991})}\BibitemShut {NoStop}%
\bibitem [{\citenamefont {Giannozzi}\ \emph {et~al.}(2009)\citenamefont
  {Giannozzi}, \citenamefont {Baroni}, \citenamefont {Bonini}, \citenamefont
  {Calandra}, \citenamefont {Car}, \citenamefont {Cavazzoni}, \citenamefont
  {Ceresoli}, \citenamefont {Chiarotti}, \citenamefont {Cococcioni},
  \citenamefont {Dabo}, \citenamefont {Corso}, \citenamefont {de~Gironcoli},
  \citenamefont {Fabris}, \citenamefont {Fratesi}, \citenamefont {Gebauer},
  \citenamefont {Gerstmann}, \citenamefont {Gougoussis}, \citenamefont
  {Kokalj}, \citenamefont {Lazzeri}, \citenamefont {Martin-Samos},
  \citenamefont {Marzari}, \citenamefont {Mauri}, \citenamefont {Mazzarello},
  \citenamefont {Paolini}, \citenamefont {Pasquarello}, \citenamefont
  {Paulatto}, \citenamefont {Sbraccia}, \citenamefont {Scandolo}, \citenamefont
  {Sclauzero}, \citenamefont {Seitsonen}, \citenamefont {Smogunov},
  \citenamefont {Umari},\ and\ \citenamefont {Wentzcovitch}}]{PWscfcode}%
  \BibitemOpen
  \bibfield  {author} {\bibinfo {author} {\bibfnamefont {P.}~\bibnamefont
  {Giannozzi}}, \bibinfo {author} {\bibfnamefont {S.}~\bibnamefont {Baroni}},
  \bibinfo {author} {\bibfnamefont {N.}~\bibnamefont {Bonini}}, \bibinfo
  {author} {\bibfnamefont {M.}~\bibnamefont {Calandra}}, \bibinfo {author}
  {\bibfnamefont {R.}~\bibnamefont {Car}}, \bibinfo {author} {\bibfnamefont
  {C.}~\bibnamefont {Cavazzoni}}, \bibinfo {author} {\bibfnamefont
  {D.}~\bibnamefont {Ceresoli}}, \bibinfo {author} {\bibfnamefont {G.~L}\
  \bibnamefont {Chiarotti}}, \bibinfo {author} {\bibfnamefont {M.}~\bibnamefont
  {Cococcioni}}, \bibinfo {author} {\bibfnamefont {I.}~\bibnamefont {Dabo}},
  \bibinfo {author} {\bibfnamefont {A.~Dal}\ \bibnamefont {Corso}}, \bibinfo
  {author} {\bibfnamefont {S.}~\bibnamefont {de~Gironcoli}}, \bibinfo {author}
  {\bibfnamefont {S.}~\bibnamefont {Fabris}}, \bibinfo {author} {\bibfnamefont
  {G.}~\bibnamefont {Fratesi}}, \bibinfo {author} {\bibfnamefont
  {R.}~\bibnamefont {Gebauer}}, \bibinfo {author} {\bibfnamefont
  {U.}~\bibnamefont {Gerstmann}}, \bibinfo {author} {\bibfnamefont
  {C.}~\bibnamefont {Gougoussis}}, \bibinfo {author} {\bibfnamefont
  {A.}~\bibnamefont {Kokalj}}, \bibinfo {author} {\bibfnamefont
  {M.}~\bibnamefont {Lazzeri}}, \bibinfo {author} {\bibfnamefont
  {L.}~\bibnamefont {Martin-Samos}}, \bibinfo {author} {\bibfnamefont
  {N.}~\bibnamefont {Marzari}}, \bibinfo {author} {\bibfnamefont
  {F.}~\bibnamefont {Mauri}}, \bibinfo {author} {\bibfnamefont
  {R.}~\bibnamefont {Mazzarello}}, \bibinfo {author} {\bibfnamefont
  {S.}~\bibnamefont {Paolini}}, \bibinfo {author} {\bibfnamefont
  {A.}~\bibnamefont {Pasquarello}}, \bibinfo {author} {\bibfnamefont
  {L.}~\bibnamefont {Paulatto}}, \bibinfo {author} {\bibfnamefont
  {C.}~\bibnamefont {Sbraccia}}, \bibinfo {author} {\bibfnamefont
  {S.}~\bibnamefont {Scandolo}}, \bibinfo {author} {\bibfnamefont
  {G.}~\bibnamefont {Sclauzero}}, \bibinfo {author} {\bibfnamefont {A.~P}\
  \bibnamefont {Seitsonen}}, \bibinfo {author} {\bibfnamefont {A.}~\bibnamefont
  {Smogunov}}, \bibinfo {author} {\bibfnamefont {P.}~\bibnamefont {Umari}}, \
  and\ \bibinfo {author} {\bibfnamefont {R.~M}\ \bibnamefont {Wentzcovitch}},\
  }\bibfield  {title} {\enquote {\bibinfo {title} {{QUANTUM ESPRESSO: a modular
  and open-source software project for quantum simulations of materials}},}\
  }\href {http://stacks.iop.org/0953-8984/21/i=39/a=395502} {\bibfield
  {journal} {\bibinfo  {journal} {Journal of Physics: Condensed Matter}\
  }\textbf {\bibinfo {volume} {21}},\ \bibinfo {pages} {395502} (\bibinfo
  {year} {2009})}\BibitemShut {NoStop}%
\bibitem [{\citenamefont {Zhou}\ \emph {et~al.}(1992)\citenamefont {Zhou},
  \citenamefont {Wang}, \citenamefont {Rao}, \citenamefont {Eklund},
  \citenamefont {Dresselhaus},\ and\ \citenamefont {Dresselhaus}}]{Zhou1992}%
  \BibitemOpen
  \bibfield  {author} {\bibinfo {author} {\bibfnamefont {P.}~\bibnamefont
  {Zhou}}, \bibinfo {author} {\bibfnamefont {K.}~\bibnamefont {Wang}}, \bibinfo
  {author} {\bibfnamefont {A.~M.}\ \bibnamefont {Rao}}, \bibinfo {author}
  {\bibfnamefont {P.~C.}\ \bibnamefont {Eklund}}, \bibinfo {author}
  {\bibfnamefont {G.}~\bibnamefont {Dresselhaus}}, \ and\ \bibinfo {author}
  {\bibfnamefont {M.~S.}\ \bibnamefont {Dresselhaus}},\ }\bibfield  {title}
  {\enquote {\bibinfo {title} {{Raman-scattering studies of homogeneous
  K$_3$C$_{60}$ films}},}\ }\href {\doibase 10.1103/PhysRevB.45.10838}
  {\bibfield  {journal} {\bibinfo  {journal} {Physical Review B}\ }\textbf
  {\bibinfo {volume} {45}},\ \bibinfo {pages} {10838--10840} (\bibinfo {year}
  {1992})}\BibitemShut {NoStop}%
\bibitem [{\citenamefont {Zhou}\ \emph {et~al.}(1993)\citenamefont {Zhou},
  \citenamefont {Wang}, \citenamefont {Eklund}, \citenamefont {Dresselhaus},\
  and\ \citenamefont {Dresselhaus}}]{Zhou1993}%
  \BibitemOpen
  \bibfield  {author} {\bibinfo {author} {\bibfnamefont {P.}~\bibnamefont
  {Zhou}}, \bibinfo {author} {\bibfnamefont {K.}~\bibnamefont {Wang}}, \bibinfo
  {author} {\bibfnamefont {P.~C.}\ \bibnamefont {Eklund}}, \bibinfo {author}
  {\bibfnamefont {G.}~\bibnamefont {Dresselhaus}}, \ and\ \bibinfo {author}
  {\bibfnamefont {M.~S.}\ \bibnamefont {Dresselhaus}},\ }\bibfield  {title}
  {\enquote {\bibinfo {title} {{Raman-scattering study of the electron-phonon
  interaction in ${\mathit{M}}_{3}$${\mathrm{C}}_{60}$ (M=K,Rb)}},}\ }\href
  {\doibase 10.1103/PhysRevB.48.8412} {\bibfield  {journal} {\bibinfo
  {journal} {Physical Review B}\ }\textbf {\bibinfo {volume} {48}},\ \bibinfo
  {pages} {8412--8417} (\bibinfo {year} {1993})}\BibitemShut {NoStop}%
\bibitem [{\citenamefont {Mitch}\ \emph {et~al.}(1992)\citenamefont {Mitch},
  \citenamefont {Chase},\ and\ \citenamefont {Lannin}}]{Mitch1992}%
  \BibitemOpen
  \bibfield  {author} {\bibinfo {author} {\bibfnamefont {M.G.}\ \bibnamefont
  {Mitch}}, \bibinfo {author} {\bibfnamefont {S.J.}\ \bibnamefont {Chase}}, \
  and\ \bibinfo {author} {\bibfnamefont {J.S.}\ \bibnamefont {Lannin}},\
  }\bibfield  {title} {\enquote {\bibinfo {title} {{Raman scattering and
  superconductivity of A$_x$C$_{60}$}},}\ }\href {\doibase
  10.1142/s0217979292002115} {\bibfield  {journal} {\bibinfo  {journal}
  {International Journal of Modern Physics B}\ }\textbf {\bibinfo {volume}
  {06}},\ \bibinfo {pages} {4013--4018} (\bibinfo {year} {1992})}\BibitemShut
  {NoStop}%
\bibitem [{\citenamefont {Mitch}\ and\ \citenamefont
  {Lannin}(1993)}]{Mitch1993}%
  \BibitemOpen
  \bibfield  {author} {\bibinfo {author} {\bibfnamefont {M.~G.}\ \bibnamefont
  {Mitch}}\ and\ \bibinfo {author} {\bibfnamefont {J.~S.}\ \bibnamefont
  {Lannin}},\ }\bibfield  {title} {\enquote {\bibinfo {title} {{Raman
  scattering and electron-phonon coupling in A$_3$C$_{60}$}},}\ }\href
  {\doibase https://doi.org/10.1016/0022-3697(93)90293-Z} {\bibfield  {journal}
  {\bibinfo  {journal} {Journal of Physics and Chemistry of Solids}\ }\textbf
  {\bibinfo {volume} {54}},\ \bibinfo {pages} {1801--1816} (\bibinfo {year}
  {1993})}\BibitemShut {NoStop}%
\bibitem [{\citenamefont {\"{O}pik}\ and\ \citenamefont
  {Pryce}(1957)}]{Opik1957}%
  \BibitemOpen
  \bibfield  {author} {\bibinfo {author} {\bibfnamefont {U.}~\bibnamefont
  {\"{O}pik}}\ and\ \bibinfo {author} {\bibfnamefont {Maurice Henry~Lecorney}\
  \bibnamefont {Pryce}},\ }\bibfield  {title} {\enquote {\bibinfo {title}
  {{Studies of the Jahn-Teller effect. I. A survey of the static problem}},}\
  }\href@noop {} {\bibfield  {journal} {\bibinfo  {journal} {Proceedings of the
  Royal Society of London. Series A. Mathematical and Physical Sciences}\
  }\textbf {\bibinfo {volume} {238}},\ \bibinfo {pages} {425--447} (\bibinfo
  {year} {1957})}\BibitemShut {NoStop}%
\bibitem [{\citenamefont {Ceulemans}\ and\ \citenamefont
  {Chibotaru}(1996)}]{Ceulemans1996isostationary}%
  \BibitemOpen
  \bibfield  {author} {\bibinfo {author} {\bibfnamefont {A.}~\bibnamefont
  {Ceulemans}}\ and\ \bibinfo {author} {\bibfnamefont {L.~F.}\ \bibnamefont
  {Chibotaru}},\ }\bibfield  {title} {\enquote {\bibinfo {title}
  {{Isostationary functions for multimode and multilevel Jahn-Teller
  systems}},}\ }\href {\doibase 10.1007/BF00186442} {\bibfield  {journal}
  {\bibinfo  {journal} {Theoretica chimica acta}\ }\textbf {\bibinfo {volume}
  {94}},\ \bibinfo {pages} {205--212} (\bibinfo {year} {1996})}\BibitemShut
  {NoStop}%
\bibitem [{\citenamefont {Auerbach}\ \emph {et~al.}(1994)\citenamefont
  {Auerbach}, \citenamefont {Manini},\ and\ \citenamefont
  {Tosatti}}]{Auerbach1994}%
  \BibitemOpen
  \bibfield  {author} {\bibinfo {author} {\bibfnamefont {A.}~\bibnamefont
  {Auerbach}}, \bibinfo {author} {\bibfnamefont {N.}~\bibnamefont {Manini}}, \
  and\ \bibinfo {author} {\bibfnamefont {E.}~\bibnamefont {Tosatti}},\
  }\bibfield  {title} {\enquote {\bibinfo {title} {{Electron-vibron
  interactions in charged fullerenes. I. Berry phases}},}\ }\href {\doibase
  10.1103/PhysRevB.49.12998} {\bibfield  {journal} {\bibinfo  {journal} {Phys.
  Rev. B}\ }\textbf {\bibinfo {volume} {49}},\ \bibinfo {pages} {12998}
  (\bibinfo {year} {1994})}\BibitemShut {NoStop}%
\bibitem [{\citenamefont {O'Brien}(1996)}]{OBrien1996}%
  \BibitemOpen
  \bibfield  {author} {\bibinfo {author} {\bibfnamefont {M.~C.~M.}\
  \bibnamefont {O'Brien}},\ }\bibfield  {title} {\enquote {\bibinfo {title}
  {{Vibronic energies in
  ${\mathrm{C}}_{60}$${\mathrm{}}^{\mathit{n}\mathrm{-}}$ and the Jahn-Teller
  effect}},}\ }\href {\doibase 10.1103/PhysRevB.53.3775} {\bibfield  {journal}
  {\bibinfo  {journal} {Phys. Rev. B}\ }\textbf {\bibinfo {volume} {53}},\
  \bibinfo {pages} {3775} (\bibinfo {year} {1996})}\BibitemShut {NoStop}%
\bibitem [{\citenamefont {Van~Vleck}(1940)}]{VanVleck1940}%
  \BibitemOpen
  \bibfield  {author} {\bibinfo {author} {\bibfnamefont {J.~H.}\ \bibnamefont
  {Van~Vleck}},\ }\bibfield  {title} {\enquote {\bibinfo {title} {Paramagnetic
  relaxation times for titanium and chrome alum},}\ }\href {\doibase
  10.1103/PhysRev.57.426} {\bibfield  {journal} {\bibinfo  {journal} {Phys.
  Rev.}\ }\textbf {\bibinfo {volume} {57}},\ \bibinfo {pages} {426--447}
  (\bibinfo {year} {1940})}\BibitemShut {NoStop}%
\bibitem [{\citenamefont {Chibotaru}(1994)}]{Chibotaru1994}%
  \BibitemOpen
  \bibfield  {author} {\bibinfo {author} {\bibfnamefont {L.}~\bibnamefont
  {Chibotaru}},\ }\bibfield  {title} {\enquote {\bibinfo {title} {{Effective
  Hamiltonian for multimode static Jahn-Teller effect in polycentre vibronic
  systems}},}\ }\href {http://stacks.iop.org/0305-4470/27/i=20/a=027}
  {\bibfield  {journal} {\bibinfo  {journal} {Journal of Physics A:
  Mathematical and General}\ }\textbf {\bibinfo {volume} {27}},\ \bibinfo
  {pages} {6919} (\bibinfo {year} {1994})}\BibitemShut {NoStop}%
\bibitem [{\citenamefont {Iwahara}\ and\ \citenamefont
  {Chibotaru}(2013)}]{Iwahara2013}%
  \BibitemOpen
  \bibfield  {author} {\bibinfo {author} {\bibfnamefont {N.}~\bibnamefont
  {Iwahara}}\ and\ \bibinfo {author} {\bibfnamefont {L.~F.}\ \bibnamefont
  {Chibotaru}},\ }\bibfield  {title} {\enquote {\bibinfo {title} {{Dynamical
  Jahn-Teller Effect and Antiferromagnetism in
  ${\mathrm{Cs}}_{3}{\mathrm{C}}_{60}$}},}\ }\href {\doibase
  10.1103/PhysRevLett.111.056401} {\bibfield  {journal} {\bibinfo  {journal}
  {Phys. Rev. Lett.}\ }\textbf {\bibinfo {volume} {111}},\ \bibinfo {pages}
  {056401} (\bibinfo {year} {2013})}\BibitemShut {NoStop}%
\bibitem [{\citenamefont {Liu}\ \emph {et~al.}(2018{\natexlab{b}})\citenamefont
  {Liu}, \citenamefont {Iwahara},\ and\ \citenamefont
  {Chibotaru}}]{Liu2018Dynamical}%
  \BibitemOpen
  \bibfield  {author} {\bibinfo {author} {\bibfnamefont {D.}~\bibnamefont
  {Liu}}, \bibinfo {author} {\bibfnamefont {N.}~\bibnamefont {Iwahara}}, \ and\
  \bibinfo {author} {\bibfnamefont {L.~F.}\ \bibnamefont {Chibotaru}},\
  }\bibfield  {title} {\enquote {\bibinfo {title} {{Dynamical Jahn-Teller
  effect of fullerene anions}},}\ }\href {\doibase 10.1103/PhysRevB.97.115412}
  {\bibfield  {journal} {\bibinfo  {journal} {Physical Review B}\ }\textbf
  {\bibinfo {volume} {97}} (\bibinfo {year} {2018}{\natexlab{b}}),\
  10.1103/PhysRevB.97.115412}\BibitemShut {NoStop}%
\bibitem [{\citenamefont {Chibotaru}(2003)}]{Chibotaru2003}%
  \BibitemOpen
  \bibfield  {author} {\bibinfo {author} {\bibfnamefont {L.F.}\ \bibnamefont
  {Chibotaru}},\ }\bibfield  {title} {\enquote {\bibinfo {title} {{Microscopic
  Approach to Cooperative Jahn-Teller Effect in Crystals with Strong Intra-Site
  Vibronic Coupling}},}\ }in\ \href {\doibase
  https://doi.org/10.1016/S0065-3276(03)44043-4} {\emph {\bibinfo {booktitle}
  {Advances in Quantum Chemistry}}},\ Vol.~\bibinfo {volume} {44}\ (\bibinfo
  {publisher} {Academic Press},\ \bibinfo {year} {2003})\ pp.\ \bibinfo {pages}
  {649 -- 667}\BibitemShut {NoStop}%
\bibitem [{\citenamefont {Abragam}\ and\ \citenamefont
  {Bleaney}(1970)}]{Abragam1970electron}%
  \BibitemOpen
  \bibfield  {author} {\bibinfo {author} {\bibfnamefont {A.}~\bibnamefont
  {Abragam}}\ and\ \bibinfo {author} {\bibfnamefont {B.}~\bibnamefont
  {Bleaney}},\ }\href {http://lib.ugent.be/catalog/rug01:000477992} {\emph
  {\bibinfo {title} {Electron Paramagnetic Resonance of Transition Ions}}}\
  (\bibinfo  {publisher} {Oxford : Clarendon press},\ \bibinfo {year}
  {1970})\BibitemShut {NoStop}%
\bibitem [{\citenamefont {Kaplan}\ and\ \citenamefont
  {Vehter}(1995)}]{Kaplan_book}%
  \BibitemOpen
  \bibfield  {author} {\bibinfo {author} {\bibfnamefont {M.D}\ \bibnamefont
  {Kaplan}}\ and\ \bibinfo {author} {\bibfnamefont {V.~G.}\ \bibnamefont
  {Vehter}},\ }\href {http://lib.ugent.be/catalog/rug01:000477992} {\emph
  {\bibinfo {title} {Cooperative Phenomena in Jahn-Teller Crystals}}}\
  (\bibinfo  {publisher} {Springer US, New York},\ \bibinfo {year}
  {1995})\BibitemShut {NoStop}%
\bibitem [{\citenamefont {Iwahara}\ and\ \citenamefont
  {Chibotaru}(2016)}]{Iwahara2016}%
  \BibitemOpen
  \bibfield  {author} {\bibinfo {author} {\bibfnamefont {Naoya}\ \bibnamefont
  {Iwahara}}\ and\ \bibinfo {author} {\bibfnamefont {Liviu~F.}\ \bibnamefont
  {Chibotaru}},\ }\bibfield  {title} {\enquote {\bibinfo {title} {Orbital
  disproportionation of electronic density is a universal feature of
  alkali-doped fullerides},}\ }\href {\doibase 10.1038/ncomms13093} {\bibfield
  {journal} {\bibinfo  {journal} {Nature Communications}\ }\textbf {\bibinfo
  {volume} {7}},\ \bibinfo {pages} {13093} (\bibinfo {year}
  {2016})}\BibitemShut {NoStop}%
\bibitem [{\citenamefont {Iwahara}\ \emph {et~al.}(2012)\citenamefont
  {Iwahara}, \citenamefont {Sato}, \citenamefont {Tanaka},\ and\ \citenamefont
  {Chibotaru}}]{Iwahara2012}%
  \BibitemOpen
  \bibfield  {author} {\bibinfo {author} {\bibfnamefont {Naoya}\ \bibnamefont
  {Iwahara}}, \bibinfo {author} {\bibfnamefont {Tohru}\ \bibnamefont {Sato}},
  \bibinfo {author} {\bibfnamefont {Kazuyoshi}\ \bibnamefont {Tanaka}}, \ and\
  \bibinfo {author} {\bibfnamefont {Liviu~F.}\ \bibnamefont {Chibotaru}},\
  }\bibfield  {title} {\enquote {\bibinfo {title} {{Mechanisms of localization
  in isotope-substituted dynamical Jahn-Teller systems}},}\ }\href {\doibase
  10.1209/0295-5075/100/43001} {\bibfield  {journal} {\bibinfo  {journal}
  {{EPL} (Europhysics Letters)}\ }\textbf {\bibinfo {volume} {100}},\ \bibinfo
  {pages} {43001} (\bibinfo {year} {2012})}\BibitemShut {NoStop}%
\bibitem [{\citenamefont {Matsuda}\ \emph {et~al.}(2018)\citenamefont
  {Matsuda}, \citenamefont {Iwahara}, \citenamefont {Tanigaki},\ and\
  \citenamefont {Chibotaru}}]{Matsuda2018}%
  \BibitemOpen
  \bibfield  {author} {\bibinfo {author} {\bibfnamefont {Yuki}\ \bibnamefont
  {Matsuda}}, \bibinfo {author} {\bibfnamefont {Naoya}\ \bibnamefont
  {Iwahara}}, \bibinfo {author} {\bibfnamefont {Katsumi}\ \bibnamefont
  {Tanigaki}}, \ and\ \bibinfo {author} {\bibfnamefont {Liviu~F.}\ \bibnamefont
  {Chibotaru}},\ }\bibfield  {title} {\enquote {\bibinfo {title} {Manifestation
  of vibronic dynamics in infrared spectra of mott insulating fullerides},}\
  }\href {\doibase 10.1103/PhysRevB.98.165410} {\bibfield  {journal} {\bibinfo
  {journal} {Phys. Rev. B}\ }\textbf {\bibinfo {volume} {98}},\ \bibinfo
  {pages} {165410} (\bibinfo {year} {2018})}\BibitemShut {NoStop}%
\end{thebibliography}%


\begin{thebibliography}{9}%
\makeatletter
\providecommand \@ifxundefined [1]{%
 \@ifx{#1\undefined}
}%
\providecommand \@ifnum [1]{%
 \ifnum #1\expandafter \@firstoftwo
 \else \expandafter \@secondoftwo
 \fi
}%
\providecommand \@ifx [1]{%
 \ifx #1\expandafter \@firstoftwo
 \else \expandafter \@secondoftwo
 \fi
}%
\providecommand \natexlab [1]{#1}%
\providecommand \enquote  [1]{``#1''}%
\providecommand \bibnamefont  [1]{#1}%
\providecommand \bibfnamefont [1]{#1}%
\providecommand \citenamefont [1]{#1}%
\providecommand \href@noop [0]{\@secondoftwo}%
\providecommand \href [0]{\begingroup \@sanitize@url \@href}%
\providecommand \@href[1]{\@@startlink{#1}\@@href}%
\providecommand \@@href[1]{\endgroup#1\@@endlink}%
\providecommand \@sanitize@url [0]{\catcode `\\12\catcode `\$12\catcode
  `\&12\catcode `\#12\catcode `\^12\catcode `\_12\catcode `\%12\relax}%
\providecommand \@@startlink[1]{}%
\providecommand \@@endlink[0]{}%
\providecommand \url  [0]{\begingroup\@sanitize@url \@url }%
\providecommand \@url [1]{\endgroup\@href {#1}{\urlprefix }}%
\providecommand \urlprefix  [0]{URL }%
\providecommand \Eprint [0]{\href }%
\providecommand \doibase [0]{http://dx.doi.org/}%
\providecommand \selectlanguage [0]{\@gobble}%
\providecommand \bibinfo  [0]{\@secondoftwo}%
\providecommand \bibfield  [0]{\@secondoftwo}%
\providecommand \translation [1]{[#1]}%
\providecommand \BibitemOpen [0]{}%
\providecommand \bibitemStop [0]{}%
\providecommand \bibitemNoStop [0]{.\EOS\space}%
\providecommand \EOS [0]{\spacefactor3000\relax}%
\providecommand \BibitemShut  [1]{\csname bibitem#1\endcsname}%
\let\auto@bib@innerbib\@empty
\bibitem [{\citenamefont {Tanigaki}\ \emph {et~al.}(1992)\citenamefont
  {Tanigaki}, \citenamefont {Hirosawa}, \citenamefont {Ebbesen}, \citenamefont
  {Mizuki}, \citenamefont {Shimakawa}, \citenamefont {Kubo}, \citenamefont
  {Tsai},\ and\ \citenamefont {Kuroshima}}]{Structure_nature_K_Rb}%
  \BibitemOpen
  \bibfield  {author} {\bibinfo {author} {\bibfnamefont {K.}~\bibnamefont
  {Tanigaki}}, \bibinfo {author} {\bibfnamefont {I.}~\bibnamefont {Hirosawa}},
  \bibinfo {author} {\bibfnamefont {T.~W.}\ \bibnamefont {Ebbesen}}, \bibinfo
  {author} {\bibfnamefont {J.}~\bibnamefont {Mizuki}}, \bibinfo {author}
  {\bibnamefont {Shimakawa}}, \bibinfo {author} {\bibfnamefont
  {Y.}~\bibnamefont {Kubo}}, \bibinfo {author} {\bibfnamefont {J.~S.}\
  \bibnamefont {Tsai}}, \ and\ \bibinfo {author} {\bibfnamefont
  {S.}~\bibnamefont {Kuroshima}},\ }\href {http://dx.doi.org/10.1038/356419a0}
  {\bibfield  {journal} {\bibinfo  {journal} {Nature}\ }\textbf {\bibinfo
  {volume} {356}},\ \bibinfo {pages} {419} (\bibinfo {year}
  {1992})}\BibitemShut {NoStop}%
\bibitem [{\citenamefont {Ganin}\ \emph {et~al.}(2008)\citenamefont {Ganin},
  \citenamefont {Takabayashi}, \citenamefont {Khimyak}, \citenamefont
  {Margadonna}, \citenamefont {Tamai}, \citenamefont {Rosseinsky},\ and\
  \citenamefont {Prassides}}]{Structure_nat-maters_Cs}%
  \BibitemOpen
  \bibfield  {author} {\bibinfo {author} {\bibfnamefont {A.~Y.}\ \bibnamefont
  {Ganin}}, \bibinfo {author} {\bibfnamefont {Y.}~\bibnamefont {Takabayashi}},
  \bibinfo {author} {\bibfnamefont {Y.~Z.}\ \bibnamefont {Khimyak}}, \bibinfo
  {author} {\bibfnamefont {S.}~\bibnamefont {Margadonna}}, \bibinfo {author}
  {\bibfnamefont {A.}~\bibnamefont {Tamai}}, \bibinfo {author} {\bibfnamefont
  {M.~J.}\ \bibnamefont {Rosseinsky}}, \ and\ \bibinfo {author} {\bibfnamefont
  {K.}~\bibnamefont {Prassides}},\ }\href {\doibase 10.1038/nmat2179}
  {\bibfield  {journal} {\bibinfo  {journal} {Nat Mater}\ }\textbf {\bibinfo
  {volume} {7}},\ \bibinfo {pages} {367} (\bibinfo {year} {2008})}\BibitemShut
  {NoStop}%
\bibitem [{\citenamefont {Giannozzi}\ \emph {et~al.}(2009)\citenamefont
  {Giannozzi}, \citenamefont {Baroni}, \citenamefont {Bonini}, \citenamefont
  {Calandra}, \citenamefont {Car}, \citenamefont {Cavazzoni}, \citenamefont
  {Ceresoli}, \citenamefont {Chiarotti}, \citenamefont {Cococcioni},
  \citenamefont {Dabo}, \citenamefont {Corso}, \citenamefont {de~Gironcoli},
  \citenamefont {Fabris}, \citenamefont {Fratesi}, \citenamefont {Gebauer},
  \citenamefont {Gerstmann}, \citenamefont {Gougoussis}, \citenamefont
  {Kokalj}, \citenamefont {Lazzeri}, \citenamefont {Martin-Samos},
  \citenamefont {Marzari}, \citenamefont {Mauri}, \citenamefont {Mazzarello},
  \citenamefont {Paolini}, \citenamefont {Pasquarello}, \citenamefont
  {Paulatto}, \citenamefont {Sbraccia}, \citenamefont {Scandolo}, \citenamefont
  {Sclauzero}, \citenamefont {Seitsonen}, \citenamefont {Smogunov},
  \citenamefont {Umari},\ and\ \citenamefont {Wentzcovitch}}]{PWscfcode}%
  \BibitemOpen
  \bibfield  {author} {\bibinfo {author} {\bibfnamefont {P.}~\bibnamefont
  {Giannozzi}}, \bibinfo {author} {\bibfnamefont {S.}~\bibnamefont {Baroni}},
  \bibinfo {author} {\bibfnamefont {N.}~\bibnamefont {Bonini}}, \bibinfo
  {author} {\bibfnamefont {M.}~\bibnamefont {Calandra}}, \bibinfo {author}
  {\bibfnamefont {R.}~\bibnamefont {Car}}, \bibinfo {author} {\bibfnamefont
  {C.}~\bibnamefont {Cavazzoni}}, \bibinfo {author} {\bibfnamefont
  {D.}~\bibnamefont {Ceresoli}}, \bibinfo {author} {\bibfnamefont {G.~L.}\
  \bibnamefont {Chiarotti}}, \bibinfo {author} {\bibfnamefont {M.}~\bibnamefont
  {Cococcioni}}, \bibinfo {author} {\bibfnamefont {I.}~\bibnamefont {Dabo}},
  \bibinfo {author} {\bibfnamefont {A.~D.}\ \bibnamefont {Corso}}, \bibinfo
  {author} {\bibfnamefont {S.}~\bibnamefont {de~Gironcoli}}, \bibinfo {author}
  {\bibfnamefont {S.}~\bibnamefont {Fabris}}, \bibinfo {author} {\bibfnamefont
  {G.}~\bibnamefont {Fratesi}}, \bibinfo {author} {\bibfnamefont
  {R.}~\bibnamefont {Gebauer}}, \bibinfo {author} {\bibfnamefont
  {U.}~\bibnamefont {Gerstmann}}, \bibinfo {author} {\bibfnamefont
  {C.}~\bibnamefont {Gougoussis}}, \bibinfo {author} {\bibfnamefont
  {A.}~\bibnamefont {Kokalj}}, \bibinfo {author} {\bibfnamefont
  {M.}~\bibnamefont {Lazzeri}}, \bibinfo {author} {\bibfnamefont
  {L.}~\bibnamefont {Martin-Samos}}, \bibinfo {author} {\bibfnamefont
  {N.}~\bibnamefont {Marzari}}, \bibinfo {author} {\bibfnamefont
  {F.}~\bibnamefont {Mauri}}, \bibinfo {author} {\bibfnamefont
  {R.}~\bibnamefont {Mazzarello}}, \bibinfo {author} {\bibfnamefont
  {S.}~\bibnamefont {Paolini}}, \bibinfo {author} {\bibfnamefont
  {A.}~\bibnamefont {Pasquarello}}, \bibinfo {author} {\bibfnamefont
  {L.}~\bibnamefont {Paulatto}}, \bibinfo {author} {\bibfnamefont
  {C.}~\bibnamefont {Sbraccia}}, \bibinfo {author} {\bibfnamefont
  {S.}~\bibnamefont {Scandolo}}, \bibinfo {author} {\bibfnamefont
  {G.}~\bibnamefont {Sclauzero}}, \bibinfo {author} {\bibfnamefont {A.~P.}\
  \bibnamefont {Seitsonen}}, \bibinfo {author} {\bibfnamefont {A.}~\bibnamefont
  {Smogunov}}, \bibinfo {author} {\bibfnamefont {P.}~\bibnamefont {Umari}}, \
  and\ \bibinfo {author} {\bibfnamefont {R.~M.}\ \bibnamefont {Wentzcovitch}},\
  }\href {http://stacks.iop.org/0953-8984/21/i=39/a=395502} {\bibfield
  {journal} {\bibinfo  {journal} {Journal of Physics: Condensed Matter}\
  }\textbf {\bibinfo {volume} {21}},\ \bibinfo {pages} {395502} (\bibinfo
  {year} {2009})}\BibitemShut {NoStop}%
\bibitem [{\citenamefont {Nomura}\ and\ \citenamefont
  {Arita}(2015)}]{Phonon_imaginary_Nomura}%
  \BibitemOpen
  \bibfield  {author} {\bibinfo {author} {\bibfnamefont {Y.}~\bibnamefont
  {Nomura}}\ and\ \bibinfo {author} {\bibfnamefont {R.}~\bibnamefont {Arita}},\
  }\href {\doibase 10.1103/PhysRevB.92.245108} {\bibfield  {journal} {\bibinfo
  {journal} {Physical Review B}\ }\textbf {\bibinfo {volume} {92}} (\bibinfo
  {year} {2015}),\ 10.1103/PhysRevB.92.245108}\BibitemShut {NoStop}%
\bibitem [{\citenamefont {Akashi}\ and\ \citenamefont
  {Arita}(2013)}]{IONIZED_PP}%
  \BibitemOpen
  \bibfield  {author} {\bibinfo {author} {\bibfnamefont {R.}~\bibnamefont
  {Akashi}}\ and\ \bibinfo {author} {\bibfnamefont {R.}~\bibnamefont {Arita}},\
  }\href@noop {} {\bibfield  {journal} {\bibinfo  {journal} {Physical Review
  B}\ }\textbf {\bibinfo {volume} {88}},\ \bibinfo {pages} {054510} (\bibinfo
  {year} {2013})}\BibitemShut {NoStop}%
\bibitem [{\citenamefont {Zhou}\ \emph {et~al.}(1992)\citenamefont {Zhou},
  \citenamefont {Wang}, \citenamefont {Rao}, \citenamefont {Eklund},
  \citenamefont {Dresselhaus},\ and\ \citenamefont {Dresselhaus}}]{Zhou1992}%
  \BibitemOpen
  \bibfield  {author} {\bibinfo {author} {\bibfnamefont {P.}~\bibnamefont
  {Zhou}}, \bibinfo {author} {\bibfnamefont {K.}~\bibnamefont {Wang}}, \bibinfo
  {author} {\bibfnamefont {A.~M.}\ \bibnamefont {Rao}}, \bibinfo {author}
  {\bibfnamefont {P.~C.}\ \bibnamefont {Eklund}}, \bibinfo {author}
  {\bibfnamefont {G.}~\bibnamefont {Dresselhaus}}, \ and\ \bibinfo {author}
  {\bibfnamefont {M.~S.}\ \bibnamefont {Dresselhaus}},\ }\href {\doibase
  10.1103/PhysRevB.45.10838} {\bibfield  {journal} {\bibinfo  {journal}
  {Physical Review B}\ }\textbf {\bibinfo {volume} {45}},\ \bibinfo {pages}
  {10838} (\bibinfo {year} {1992})}\BibitemShut {NoStop}%
\bibitem [{\citenamefont {Mitch}\ \emph {et~al.}(1992)\citenamefont {Mitch},
  \citenamefont {Chase},\ and\ \citenamefont {Lannin}}]{Mitch1992}%
  \BibitemOpen
  \bibfield  {author} {\bibinfo {author} {\bibfnamefont {M.}~\bibnamefont
  {Mitch}}, \bibinfo {author} {\bibfnamefont {S.}~\bibnamefont {Chase}}, \ and\
  \bibinfo {author} {\bibfnamefont {J.}~\bibnamefont {Lannin}},\ }\href
  {\doibase 10.1142/s0217979292002115} {\bibfield  {journal} {\bibinfo
  {journal} {International Journal of Modern Physics B}\ }\textbf {\bibinfo
  {volume} {06}},\ \bibinfo {pages} {4013} (\bibinfo {year}
  {1992})}\BibitemShut {NoStop}%
\bibitem [{\citenamefont {Mitch}\ and\ \citenamefont
  {Lannin}(1993)}]{Mitch1993}%
  \BibitemOpen
  \bibfield  {author} {\bibinfo {author} {\bibfnamefont {M.~G.}\ \bibnamefont
  {Mitch}}\ and\ \bibinfo {author} {\bibfnamefont {J.~S.}\ \bibnamefont
  {Lannin}},\ }\href {\doibase https://doi.org/10.1016/0022-3697(93)90293-Z}
  {\bibfield  {journal} {\bibinfo  {journal} {Journal of Physics and Chemistry
  of Solids}\ }\textbf {\bibinfo {volume} {54}},\ \bibinfo {pages} {1801}
  (\bibinfo {year} {1993})}\BibitemShut {NoStop}%
\bibitem [{\citenamefont {Zhou}\ \emph {et~al.}(1993)\citenamefont {Zhou},
  \citenamefont {Wang}, \citenamefont {Eklund}, \citenamefont {Dresselhaus},\
  and\ \citenamefont {Dresselhaus}}]{Zhou1993}%
  \BibitemOpen
  \bibfield  {author} {\bibinfo {author} {\bibfnamefont {P.}~\bibnamefont
  {Zhou}}, \bibinfo {author} {\bibfnamefont {K.}~\bibnamefont {Wang}}, \bibinfo
  {author} {\bibfnamefont {P.~C.}\ \bibnamefont {Eklund}}, \bibinfo {author}
  {\bibfnamefont {G.}~\bibnamefont {Dresselhaus}}, \ and\ \bibinfo {author}
  {\bibfnamefont {M.~S.}\ \bibnamefont {Dresselhaus}},\ }\href {\doibase
  10.1103/PhysRevB.48.8412} {\bibfield  {journal} {\bibinfo  {journal}
  {Physical Review B}\ }\textbf {\bibinfo {volume} {48}},\ \bibinfo {pages}
  {8412} (\bibinfo {year} {1993})}\BibitemShut {NoStop}%
\end{thebibliography}%

\end{document}


\title{Supplementary Material: Jahn-Teller effect in cubic fullerides A$_{3}$C$_{60}$}

\author{Zhishuo Huang}
\affiliation{Theory of Nanomaterials Group, KU Leuven, Celestijnenlaan 200F, B-3001 Leuven, Belgium}
\author{Munirah D. Albaqami}
\affiliation{Chemistry Department, College of Science, King Saud University, P.O. Box 2455, Riyadh 11451, Saudi Arabia}
\author{Tohru Sato}
\affiliation{Fukui Institute for Fundamental Chemistry, Kyoto University, Takano Nishihiraki-cho 34-4, Sakyo-ku, Kyoto 6068103, Japan}
\author{Naoya Iwahara}
\email[]{naoya.iwahara@gmail.com}
\affiliation{Department of Chemistry, National University of Singapore, Block S8 Level 3, 3 Science Drive 3, 117543, Singapore}
\affiliation{Theory of Nanomaterials Group, KU Leuven, Celestijnenlaan 200F, B-3001 Leuven, Belgium}
\affiliation{Graduate School of Engineering, Chiba University, 1-33 Yayoi-cho, Inage-ku, Chiba 263-8522, Japan}
\author{Liviu F. Chibotaru}
\email[]{liviu.chibotaru@kuleuven.be}
\affiliation{Theory of Nanomaterials Group, KU Leuven, Celestijnenlaan 200F, B-3001 Leuven, Belgium}
\date{\today}

\date{\today}
\maketitle
\section{Crystal contants}
\begin{table}[H]
    \caption{\label{tab:Struct_constant}
        Lattice constants $a$ of A$_{3}$C$_{60}$ used in phonon calculations.}
    \begin{ruledtabular}
    \begin{tabular}{ccccc}
        &K$_{3}$C$_{60}$\cite{Structure_nature_K_Rb}&Rb$_{3}$C$_{60}$\cite{Structure_nature_K_Rb}&Cs$_{3}$C$_{60}$\cite{Structure_nat-maters_Cs} &A15 Cs$_{3}$C$_{60}$\cite{Structure_nat-maters_Cs}\\
    \hline
        a($\overset{\circ}{A}$) & 14.240 & 14.384 & 14.793 &11.784
    \end{tabular}
    \end{ruledtabular}
\end{table}

\section{GENERATION AND TEST OF IONIZED PSEUDOPOTENTIAL}
\begin{figure}[b]
\includegraphics [angle=0,scale=0.80]{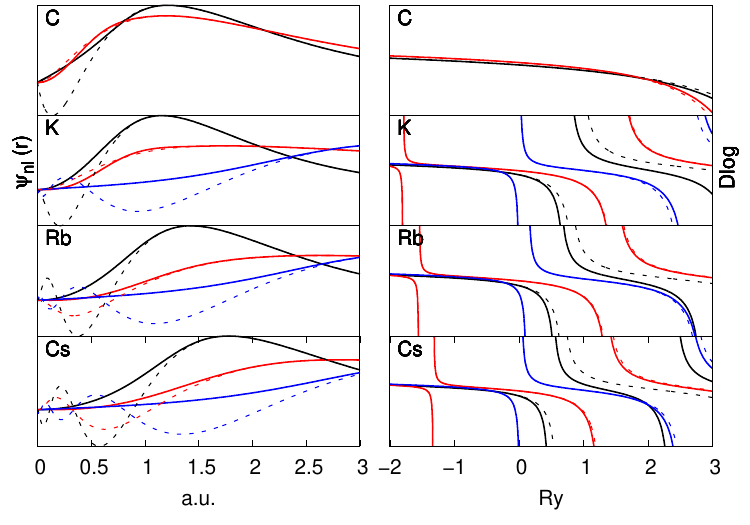}
    \caption{\label{fig:PBE_dlog_wfc} GGA-PBE PPs, the comparation of the pseudized Kohn-Sham orbitals ($\Psi_{nl}(\bold{r})$, right column) and logarithmic derivatives (Dlog, left column) for C, K, Rb, and Cs. Solid line and dashed line represented the calculation from PPs and AE, correspondingly, and black, red, and blue accounted for s, p, and d orbital, respectively.}
\end{figure}

\begin{figure}[b]
\includegraphics [angle=0,scale=0.80]{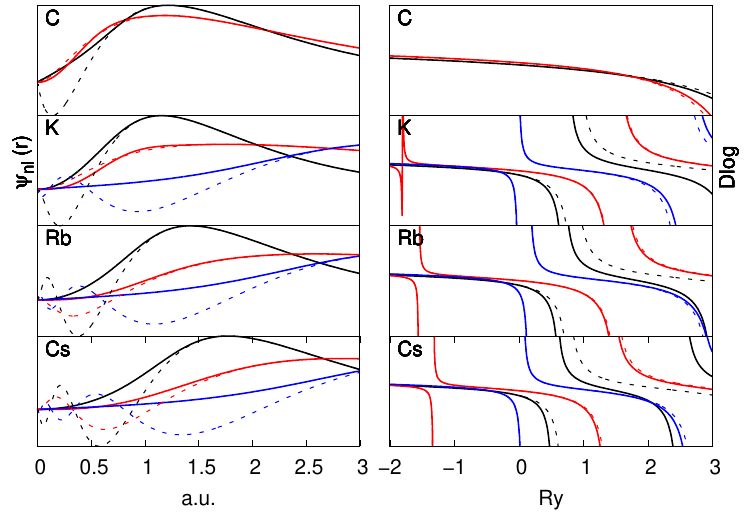}
    \caption{\label{fig:LDA_dlog_wfc} LDA-PZ PPs, the comparation of the pseudized Kohn-Sham orbitals ($\Psi_{nl}(\bold{r})$, right column) and logarithmic derivatives (Dlog, left column) for C, K, Rb, and Cs. Solid line and dashed line represented the calculation from PPs and AE, correspondingly, and black, red, and blue accounted for s, p, and d orbital, respectively.}
\end{figure}

    Energies of different elements for final consistency check were shown in Table~\ref{tab:final_consistency}, from which we could see that the energy differences of all the generated PP gave the same energy level corresponding to AE calculation, except for Rb, of which the energy difference was less than 0.1 mRy, nearly to zero. As a further check, the pseudized Kohn-Sham orbitals and the logarithmic derivatives were compared, as plotted in Fig.\ref{fig:PBE_dlog_wfc} and Fig.\ref{fig:LDA_dlog_wfc}. For both PBE and PZ PPs, the pseudized wavefunctions ($\Psi_{nl}(\bold{r})$, right column) went directly and smoothly to zero, with no wiggles or other strange features. For logarithmic derivatives (Dlog, left column), the AE orbitals and PP orbitals matched very well, and slightly deviations appeared only at relatively high ( $>$ 1Ry) energies.

    In order to confirm good the ionized PPs were good enough for use, it should be tested whether these generated PPs could give or not the nearly same information when applied in real calculations of A$_{3}$C$_{60}$. All the calculations were performed by package Quantum Espresso[\onlinecite{PWscfcode}] with the most used calculation parameters[\onlinecite{Phonon_imaginary_Nomura},\onlinecite{IONIZED_PP}]: 60 Ry for cutoff energy for the wave functions, 4$\times$4$\times$4 for Monkhorst-Pack grids of K points with 0.010 Ry for the Gaussian smearing. As we could see, the band structure of ionized PPs had a very good consistence with that of general PPs provided by Quantum Espresso PP library, with a slight dismatched at $\Gamma$ point which was only about 0.02 eV, which, to some extent, showed that the generation of ionized PPs was successful and the generated PPs behave very well in describing A$_{3}$C$_{60}$ system.

\begin{table}[b]
    \caption{\label{tab:final_consistency}
        Comparation of the final energy of all electron calculation (AE) E$_{AE}$, pseudopotential calculation (PS) E$_{PS}$, and the absolute difference between them $\Delta E$ for difference exchange-correlation functional (PBE and PZ) and difference elements (C, K, Rb, and Cs). n, l were principal quantum number and angular quantum number for pseudopotential, respectively. nl was the configuration of all-electron. Occu was the occupation of electron in the state. All the unit for energy was Ry}
    \begin{ruledtabular}
    \begin{tabular}{ccccccccc}
        &E$_{xc}$&n& l & nl &Occu& E$_{AE}$ & E$_{PS}$ & $\Delta E$ \\
    \hline
    \multirow{4}{*}{C} & \multirow{2}{*}{PBE} &1 &0 &2S &1(2.00) &-1.00980 &-1.00980 &0.00000 \\
                       &                      &2 &1 &2P &1(2.00) &-0.38872 &-0.38872 &0.00000 \\
                       \cline{2-9}
                       & \multirow{2}{*}{PZ}  &1 &0 &2S &1(2.00) &-1.00195 &-1.00195 &0.00000 \\
                       &                      &2 &1 &2P &1(2.00) &-0.39860 &-0.39860 &0.00000 \\
    \hline
    \multirow{6}{*}{K} & \multirow{3}{*}{PBE} &2 &1 &3P &1(6.00) &-1.82985 &-1.82985 &0.00000 \\
                       &                      &3 &2 &3D &1(0.00) &-0.28335 &-0.28335 &0.00000 \\
                       &                      &1 &0 &4S &1(0.00) &-0.45444 &-0.45444 &0.00000 \\

                       \cline{2-9}
                       & \multirow{3}{*}{PZ} &2 &1 &3P &1(6.00) &-1.83622 &-1.83622 &0.00000 \\
                       &                      &3 &2 &3D &1(0.00) &-0.28975 &-0.28975 &0.00000 \\
                       &                      &1 &0 &4S &1(0.00) &-0.44778 &-0.44778 &0.00000 \\
    \hline
    \multirow{6}{*}{Rb}& \multirow{3}{*}{PBE} &2 &1 &4P &1(6.00) &-1.59127 &-1.59128 &0.00001 \\
                       &                      &2 &0 &5S &1(0.00) &-0.44447 &-0.44453 &0.00006 \\
                       &                      &3 &2 &4D &1(0.00) &-0.27500 &-0.27499 &0.00001 \\
                       \cline{2-9}
                       & \multirow{3}{*}{PZ} &2 &1 &4P &1(6.00) &-1.60172 &-1.60173 &0.00001 \\
                       &                      &2 &0 &5S &1(0.00) &-0.44245 &-0.44255 &0.00009 \\
                       &                      &3 &2 &4D &1(0.00) &-0.27829 &-0.27827 &0.00001 \\
    \hline
    \multirow{6}{*}{Cs}& \multirow{3}{*}{PBE} &2 &1 &5P &1(6.00) &-1.37222 &-1.37223 &0.00000 \\
                       &                      &2 &0 &6S &1(0.00) &-0.41445 &-0.41445 &0.00000 \\
                       &                      &3 &2 &5D &1(0.00) &-0.32319 &-0.32319 &0.00000 \\
                       \cline{2-9}
                       & \multirow{3}{*}{PZ} &2 &1 &5P &1(6.00) &-1.38411 &-1.38412 &0.00000 \\
                       &                     &2 &0 &6S &1(0.00) &-0.41523 &-0.41523 &0.00000 \\
                       &                      &3 &2 &5D &1(0.00) &-0.32893 &-0.32893 &0.00000
    \end{tabular}
    \end{ruledtabular}
\end{table}

\section{FULL PHONON DISPERSION AND ELASTIC COUPLING CONSTANTS}
\begin{figure}[tb]
\includegraphics [height=\linewidth,angle=-90]{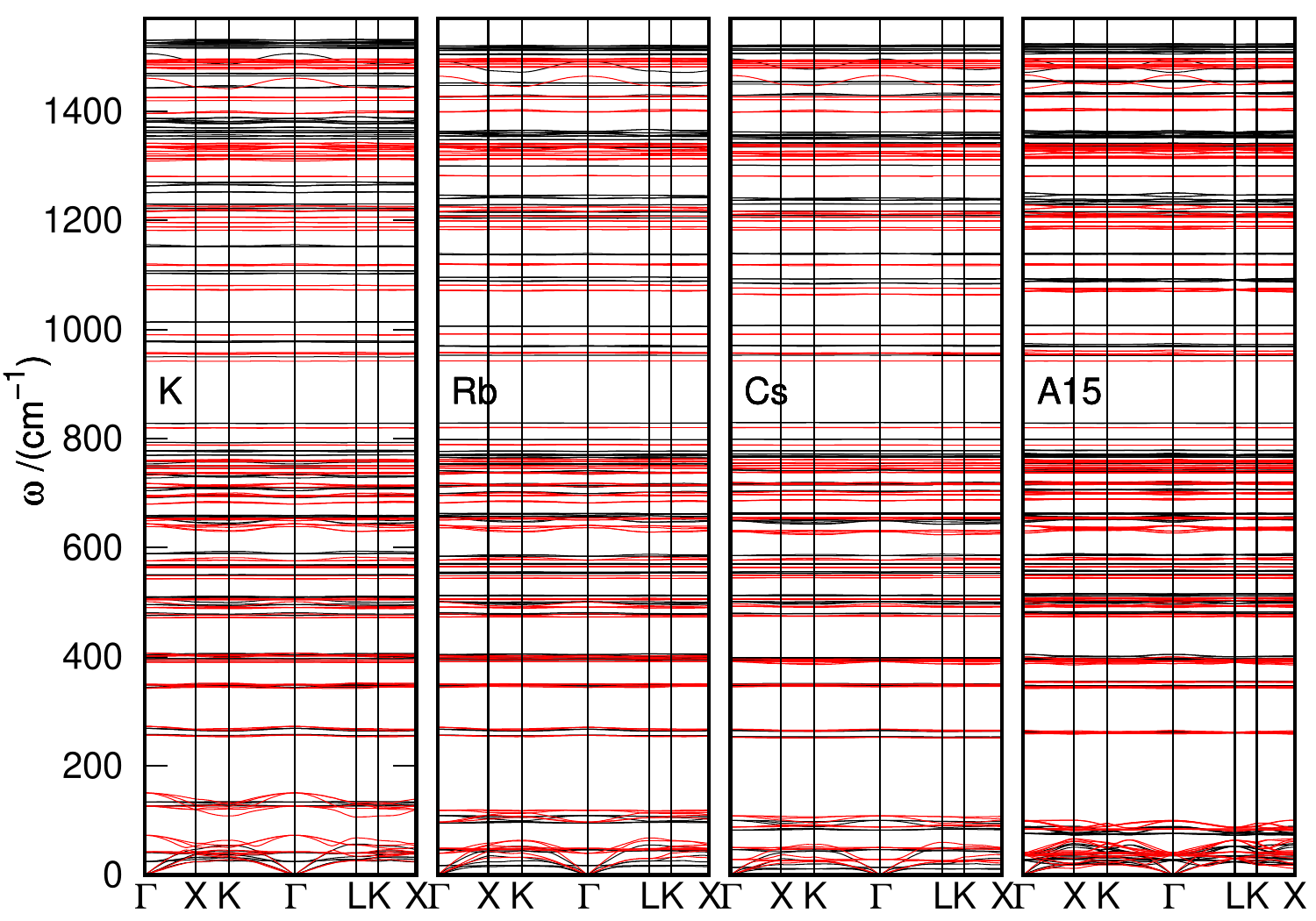}
    \caption{\label{fig:Phonon_dispersion_full} GGA-PBE PPs, the comparation of the pseudized Kohn-Sham orbitals ($\Psi_{nl}(\bold{r})$, right column) and logarithmic derivatives (Dlog, left column) for C, K, Rb, and Cs. Solid line and dashed line represented the calculation from PPs and AE, correspondingly, and black, red, and blue accounted for s, p, and d orbital, respectively.}\end{figure}

    The full phonon dispersion for A$_{3}$C$_{60}$ is shown in Fig. \ref{fig:Phonon_dispersion_full}. The acoustic and optical phonon modes were transformed into the localized modes with the similar method as the construction of the Wannier orbitals from the band orbitals. All the elaxtic coupling constant were shown in the attached txt files. Here is a short description for these attachements. The file were named by the notation of ``ECC" with respect to each system. Taking ``ECC\_K\_PB000" for an example, this means the information contained in this file is the elastic coupling constants between the local vibrational modes in the origin unit cell (labelled by $\mathbf{R}=(0,0,0)$) and that in the next unit cell (labelled by $\mathbf{R}=(0,0,0)$), while ``ECC\_K\_PB001" means the information contained in this file is the elastic coupling constants between the local vibrational modes in the origin unit cell (labelled by $\mathbf{R}=(0,0,0)$) and that in the next unit cell (labelled by $\mathbf{R}=(0,0,0)$). For the information listed in these files, the first 3 columns contain the information of local vibration modes. The first column labels the symmetry type of local vibration modes, 1 for A$_{g}$, 2 for A$_{u}$, 3 for G$_{g}$, 4 for G$_{u}$, 5 for H$_{g}$, 6 for H$_{u}$, 7 for T$_{1g}$, 8 for T$_{1u}$, 9 for T$_{2g}$, 10 for T$_{2u}$, 11 for translational modes, 12 for pure rotational modes of fullerene modes, and 13 for normal modes of alkli-atoms. For A15, there are two extra labels, which are 14 for pure rotational modes of the second fullerene molecule sitting on the center of the unit cell, and 15 for the vibronic modes between two types of fullerene. The second columns showed the $i$th order of each type of local modes. And the third column had the vibrational frequency, main fro the local modes obtained within fullerene molecule, of which we set all the frequencies for the vibrational modes of translational, pure rotational and alkli-atoms to 0. The unit for the frequencies are cm$^{-1}$, for the elastic coupling constants are Rydberg.

\begin{table}[tb]
    \caption{\label{tab:Hg_frequency}
 Phonon frequencies (cm$^{-1}$) of $H_{g}$ modes at the $\Gamma$ point calculated by density functional perturbation theory (DFPT) with LDA and PBE exchange-correlation functionals (denoted by superscipts $L$ and $P$in the first column, respectively) compared with the results of calculations by Nomura and Arita \cite{Phonon_imaginary_Nomura} and
experimental frequencies.}
    \begin{ruledtabular}
    \begin{tabular}{cccccccccccccc}
           &\multicolumn{9}{c}{DFT} & \multicolumn{3}{c}{exp.}\\
    \cline{2-11}
           &\multicolumn{6}{c}{present} &\multicolumn{3}{c}{Nomura et al} & & & \\
    \cline{2-7}\cline{8-10}\cline{11-12}
Mode&\multicolumn{2}{c}{K$_{3}$C$_{60}$}&\multicolumn{2}{c}{Rb$_{3}$C$_{60}$}&\multicolumn{2}{c}{Cs$_{3}$C$_{60}$}&K$_{3}$C$_{60}$&Rb$_{3}$C$_{60}$&Cs$_{3}$C$_{60}$&K$_{3}$C$_{60}$&Rb$_{3}$C$_{60}$&Cs$_{3}$C$_{60}$\\
    \cline{2-3}\cline{4-5}\cline{6-7}
    &E$_{g}$&T$_{g}$&E$_{g}$&T$_{g}$&E$_{g}$&T$_{g}$&     &     &    &  \\
    \hline
    $H_{g}^{L}$(1)& 257& 269& 256& 266& 254& 265& 257, 268& 255, 267& 256,277& \multirow{2}{*}{ 268\footnotemark[1]}  &\multirow{2}{*}{ 263\footnotemark[2], 265\footnotemark[3]}&\multirow{2}{*}{--} \\
    $H_{g}^{P}$(1)& 257& 274& 256& 271& 252& 267& --    &  --     &   --   &   && \\
    $H_{g}^{L}$(2)& 407& 403& 404& 403& 394& 397& 423, 425& 422, 423& 422, 425& \multirow{2}{*}{ 416\footnotemark[1]}  &\multirow{2}{*}{ 415\footnotemark[2], 395\footnotemark[3]}&\multirow{2}{*}{418\footnotemark[2]} \\
    $H_{g}^{P}$(2)& 406& 401& 399& 402& 391& 391&  --    &  --     &   --    &   && \\
    $H_{g}^{L}$(3)& 653& 657& 660& 658& 649& 650& 683, 686& 683, 686& 686, 688& \multirow{2}{*}{ 712\footnotemark[1]}  &\multirow{2}{*}{ 713\footnotemark[3]}&\multirow{2}{*}{--} \\
    $H_{g}^{P}$(3)& 637& 642& 636& 641& 629& 632&   --    &  --     &   --    &   && \\
    $H_{g}^{L}$(4)& 777& 778& 771& 769& 772& 768& 777, 778& 777, 778& 780, 788& \multirow{2}{*}{ 752\footnotemark[1]}  &\multirow{2}{*}{ 749\footnotemark[2], 743\footnotemark[3]}&\multirow{2}{*}{--} \\
    $H_{g}^{P}$(4)& 770& 768& 762& 758& 761& 755&   --    &  --     &   --    &   && \\
    $H_{g}^{L}$(5)&1107&1103&1096&1090&1093&1085&1110,1114&1110,1114&1116,1125& \multirow{2}{*}{ -- }  &\multirow{2}{*}{--}&\multirow{2}{*}{--} \\
    $H_{g}^{P}$(5)&1094&1087&1071&1081&1076&1065&   --    &  --     &   --    &   && \\
    $H_{g}^{L}$(6)&1270&1263&1246&1240&1241&1237&1267,1273&1267,1272&1277,1283& \multirow{2}{*}{ -- }  &\multirow{2}{*}{--}&\multirow{2}{*}{--} \\
    $H_{g}^{P}$(6)&1251&1240&1224&1216&1217&1211&   --    &  --     &   --    &   && \\
    $H_{g}^{L}$(7)&1382&1380&1361&1363&1357&1354&1402,1407&1403,1405&1415,1415& \multirow{2}{*}{1409\footnotemark[1]}  &\multirow{2}{*}{--}&\multirow{2}{*}{--} \\
    $H_{g}^{P}$(7)&1357&1353&1333&1334&1327&1323&   --    &  --     &   --    &   && \\
    $H_{g}^{L}$(8)&1521&1518&1512&1515&1512&1515&1531,1536&1531,1535&1541,1544& \multirow{2}{*}{1550\footnotemark[1]}  &\multirow{2}{*}{--}&\multirow{2}{*}{--} \\
    $H_{g}^{P}$(8)&1499&1497&1486&1491&1484&1488&   --    &  --     &   --    &   &&
    \end{tabular}
    \end{ruledtabular}
    \footnotetext[1]{Ref.\cite{Zhou1992}} \footnotetext[2]{Ref.\cite{Mitch1992, Mitch1993}} \footnotetext[3]{Ref.\cite{Zhou1993}}
\end{table}


\section{JT active vibration modes}
\subsection{JT active vibration modes of Fullerene}
\begin{figure}[H]
\begin{center}
\begin{tabular}{b{1.0cm} m{2.8cm} b{1.0cm} m{2.8cm}}
 $A_g$(1) &
\includegraphics[bb =  0 0 342 342, width=2.8cm]{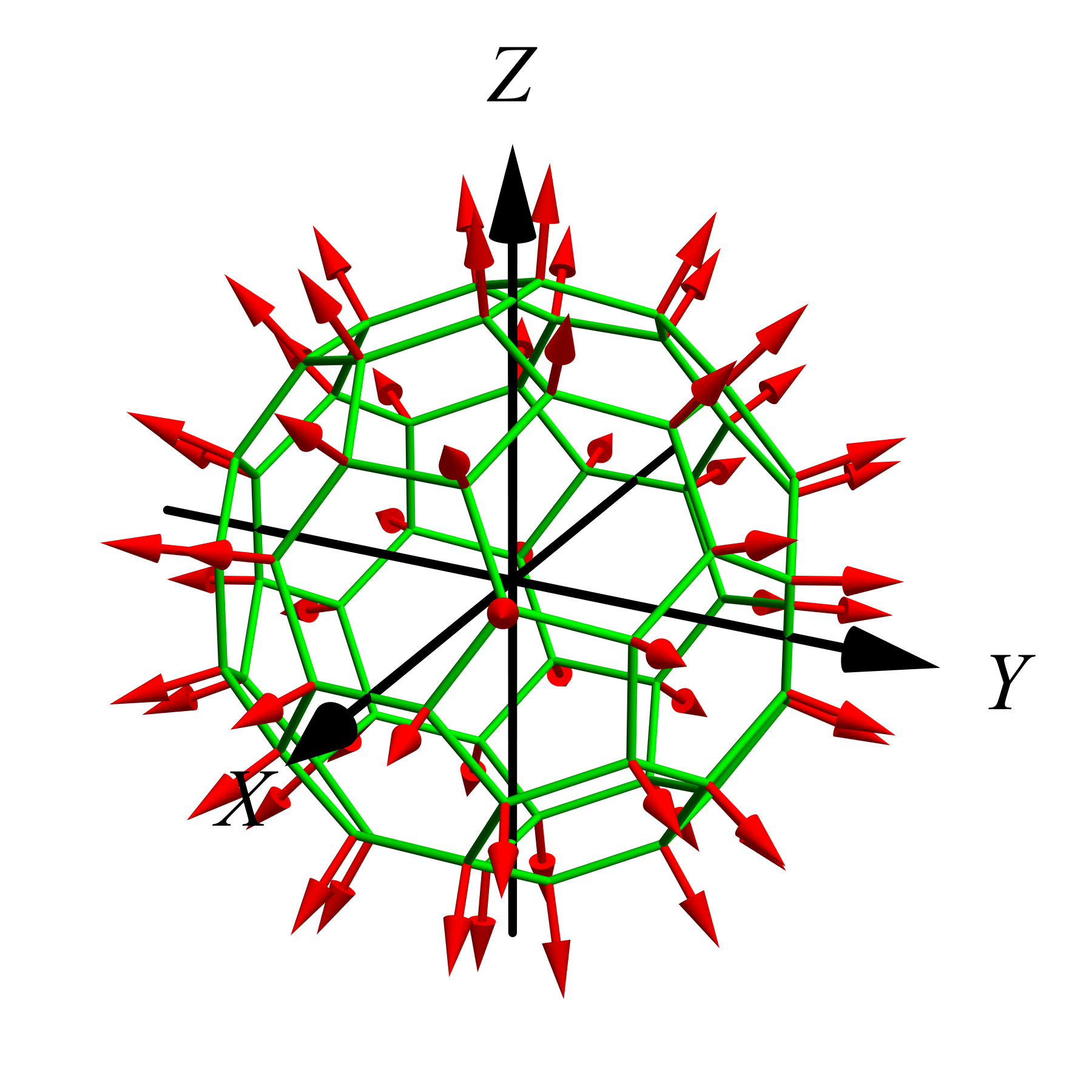} &
 $A_g$(2) &
\includegraphics[bb =  0 0 342 342, width=2.8cm]{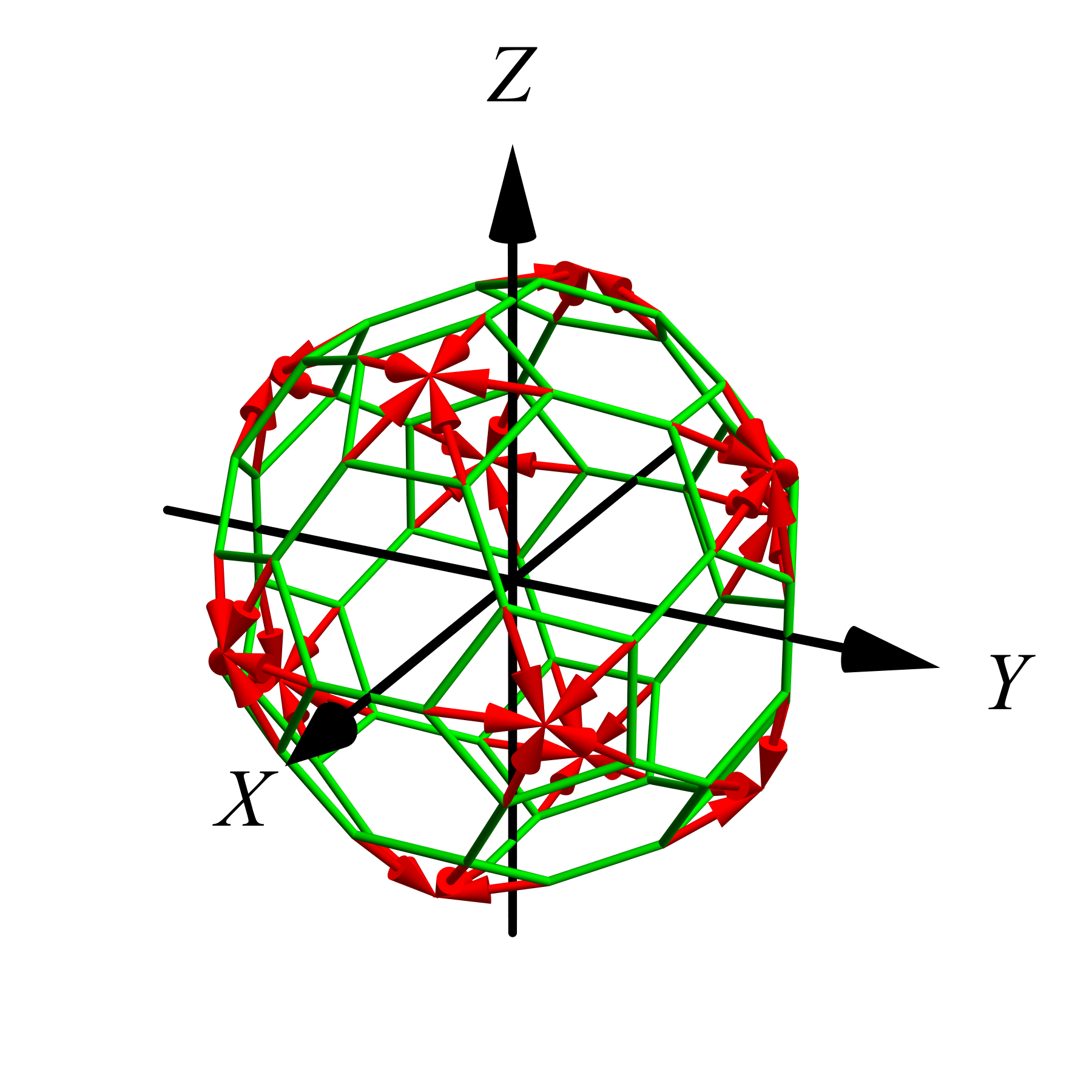}
\end{tabular}
\end{center}
\caption[Schematic representation of $A_g$ modes]{Schematic representation of $A_g$ modes}
\label{Fig:Ag}
\end{figure}

\begin{figure}[H]
\begin{center}
\begin{tabular}{b{1.0cm} m{2.7cm} m{2.7cm} m{2.7cm} m{2.7cm} m{2.7cm} }
  & \multicolumn{1}{c}{$\theta$} & \multicolumn{1}{c}{$\epsilon$} & \multicolumn{1}{c}{$\xi$} & \multicolumn{1}{c}{$\eta$} & \multicolumn{1}{c}{$\zeta$} \\
$h_g(1)$
&
\includegraphics[bb =  0 0 342 342, width=2.7cm]{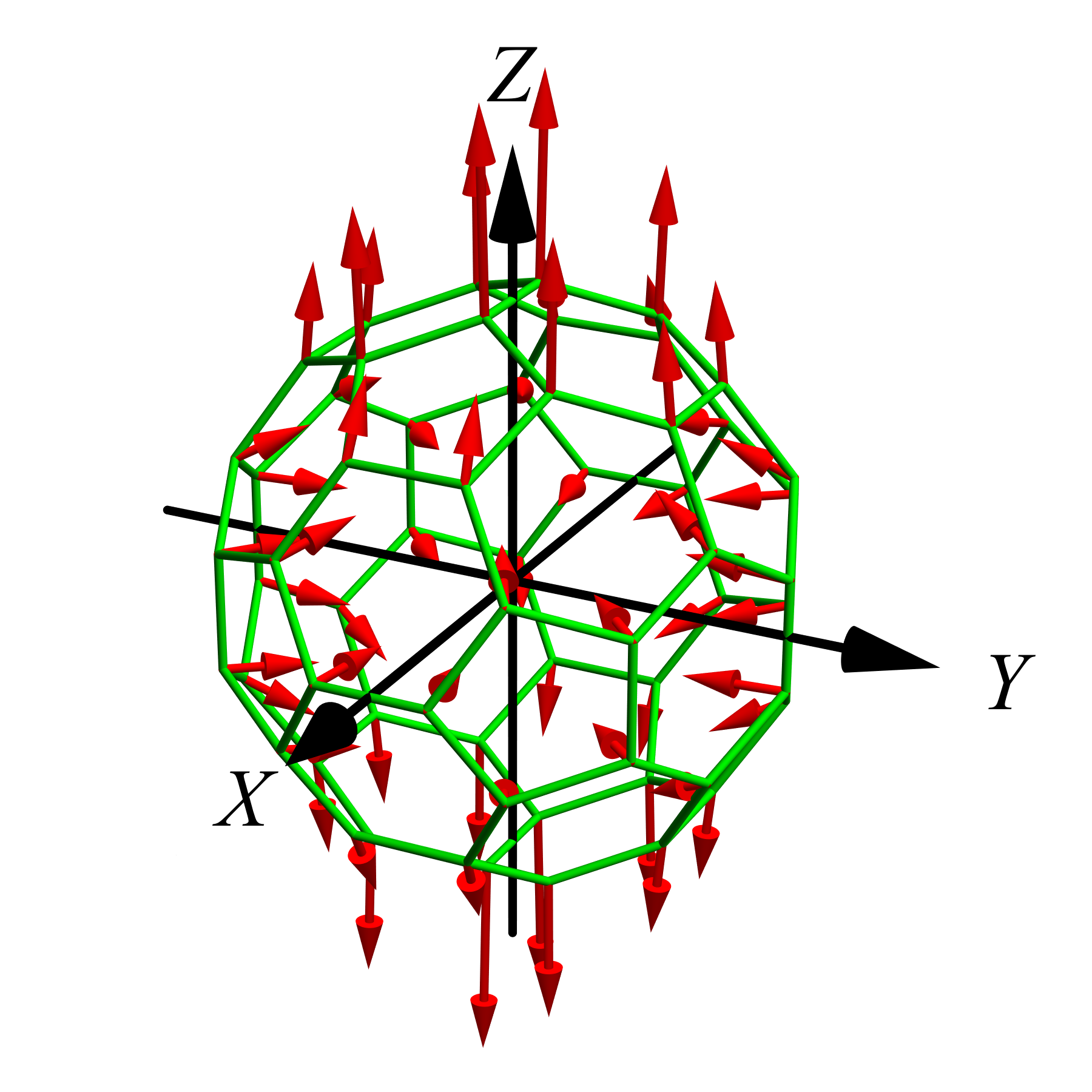}
&
\includegraphics[bb =  0 0 342 342, width=2.7cm]{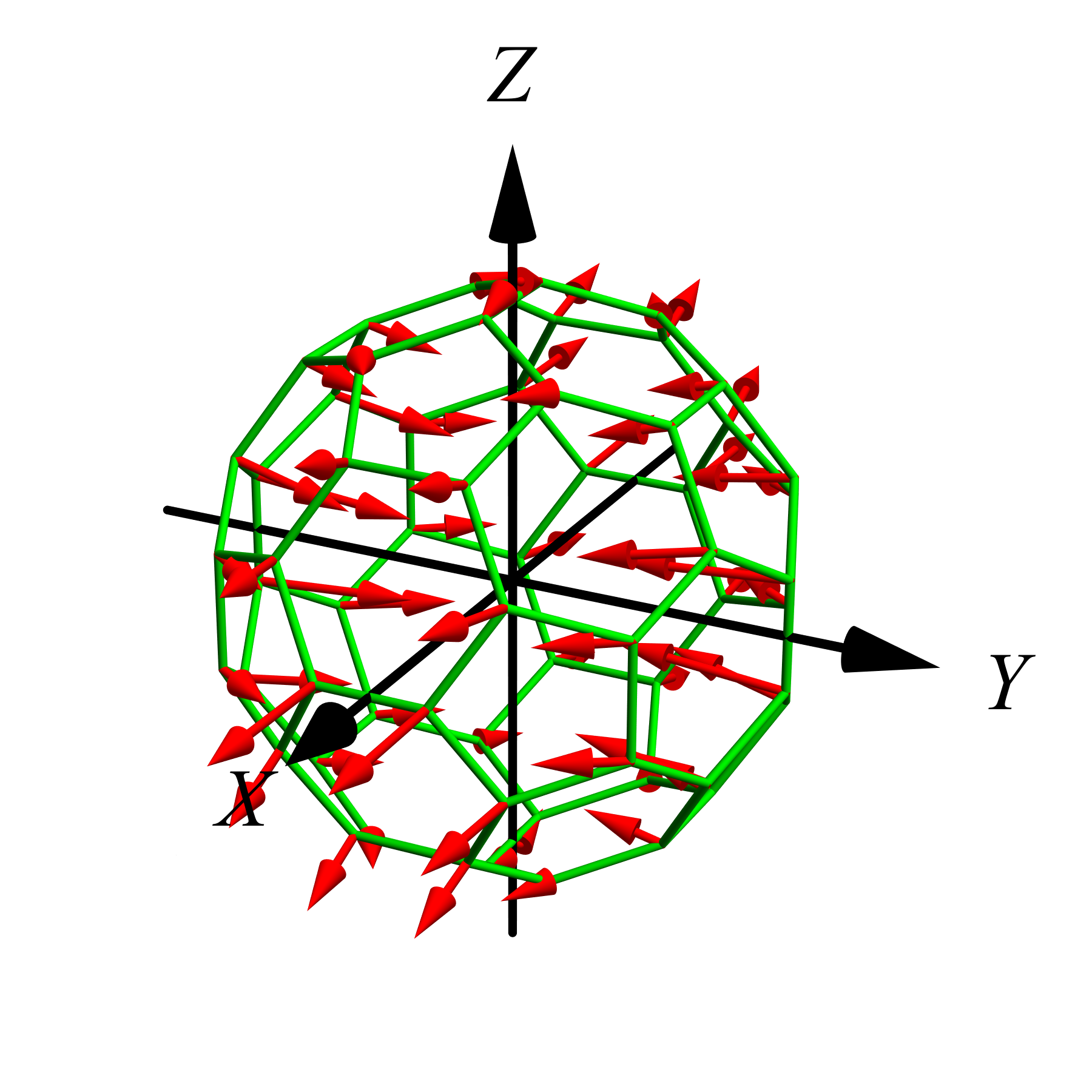}
&
\includegraphics[bb =  0 0 342 342, width=2.7cm]{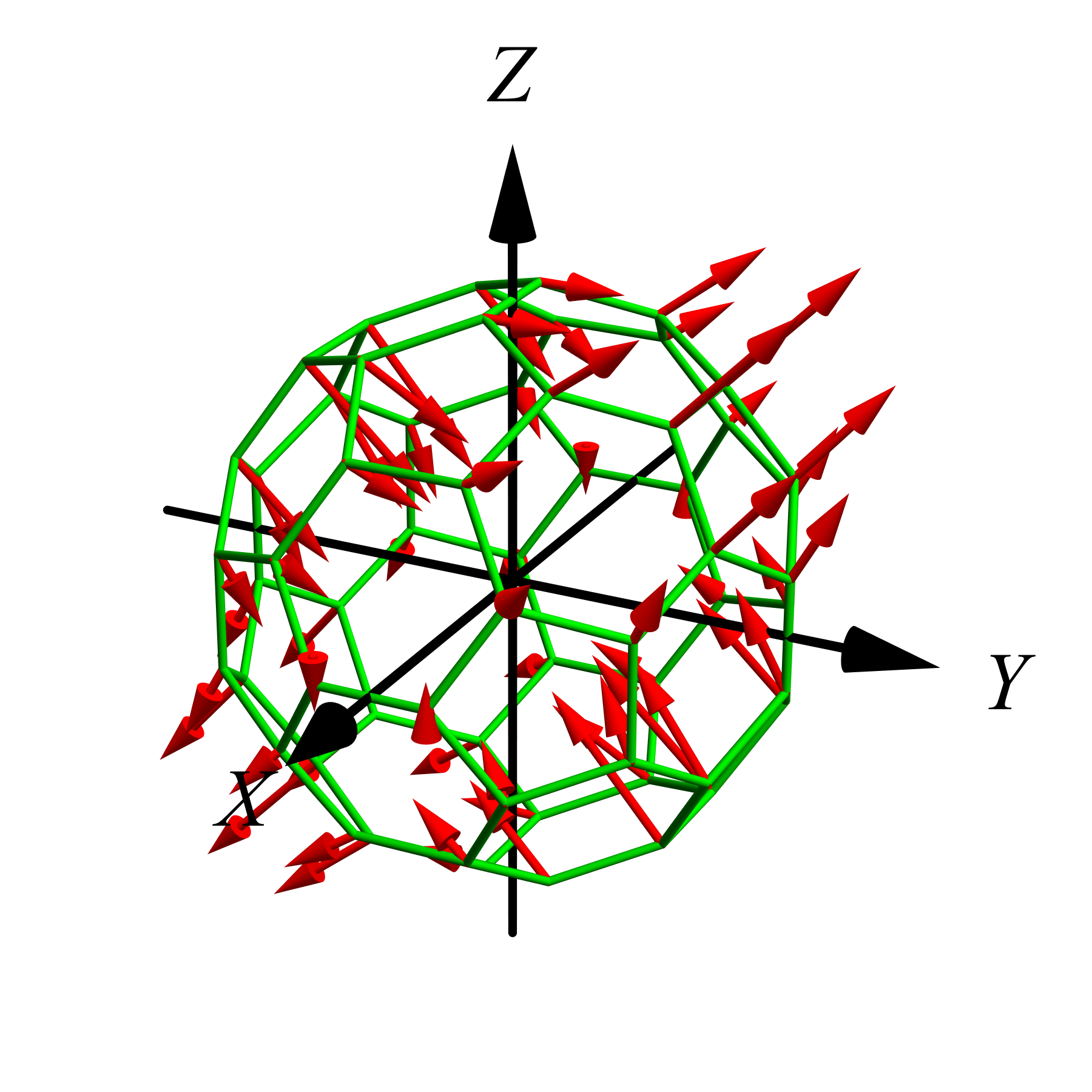}
&
\includegraphics[bb =  0 0 342 342, width=2.7cm]{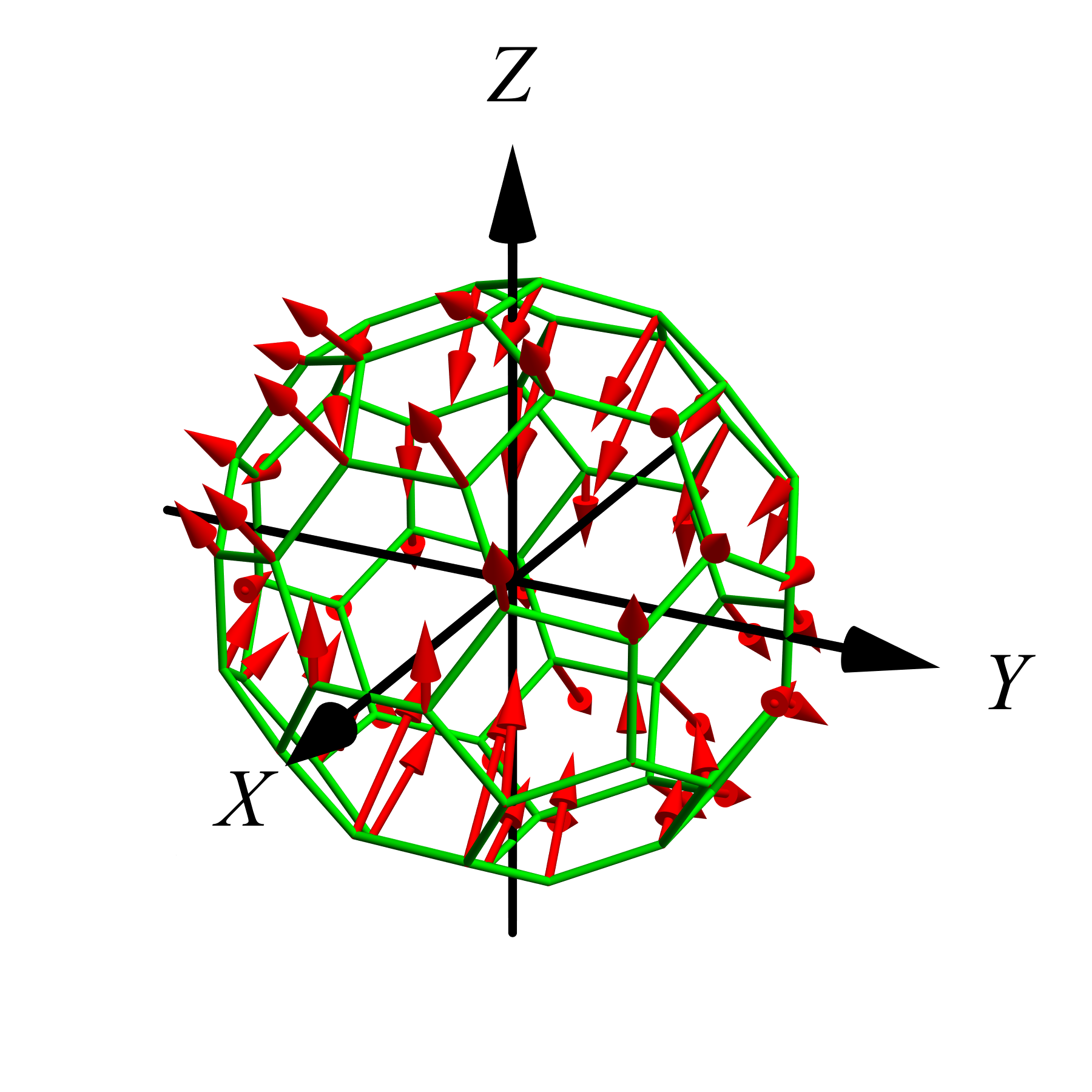}
&
\includegraphics[bb =  0 0 342 342, width=2.7cm]{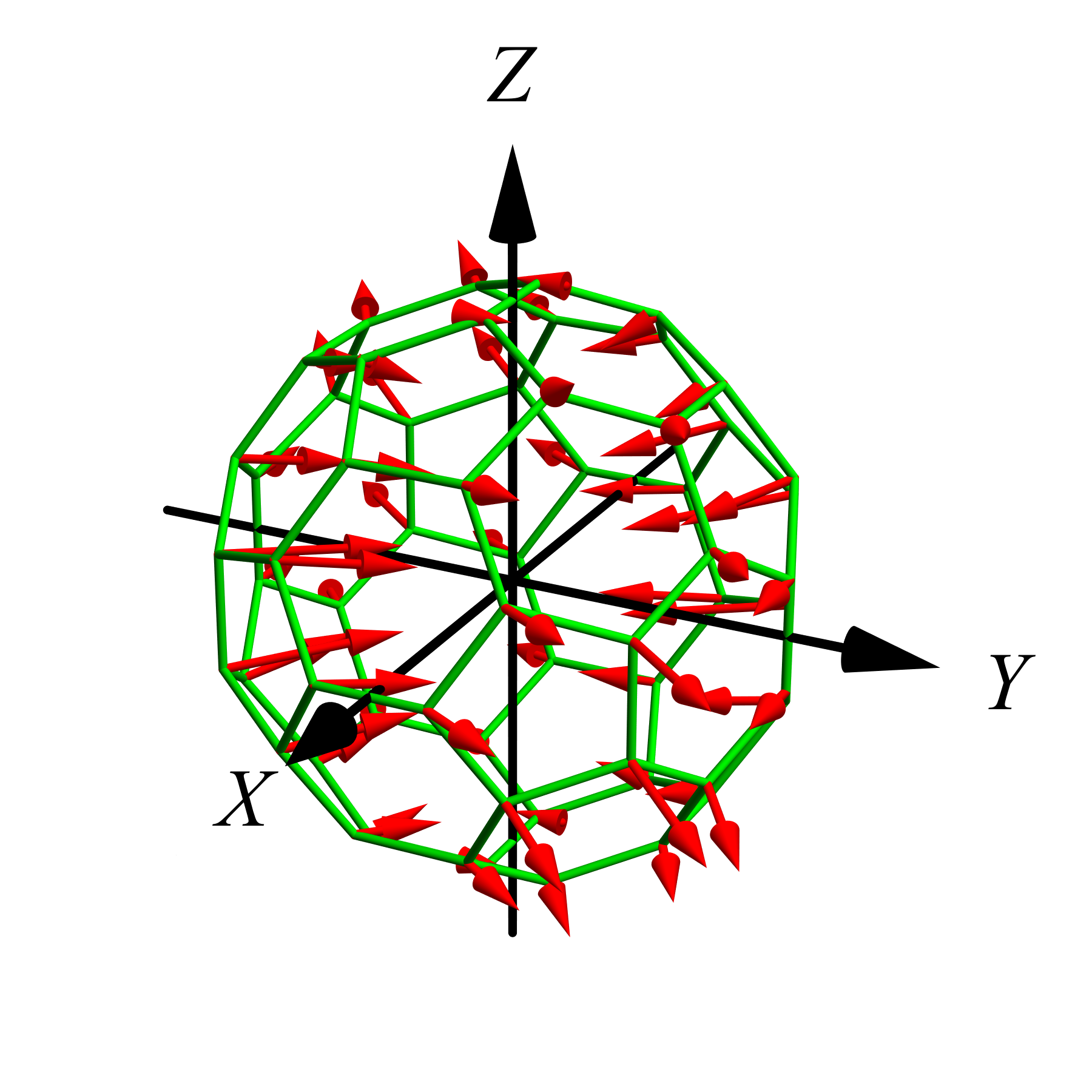}
\\
 $h_g(2)$
&
\includegraphics[bb =  0 0 342 342, width=2.7cm]{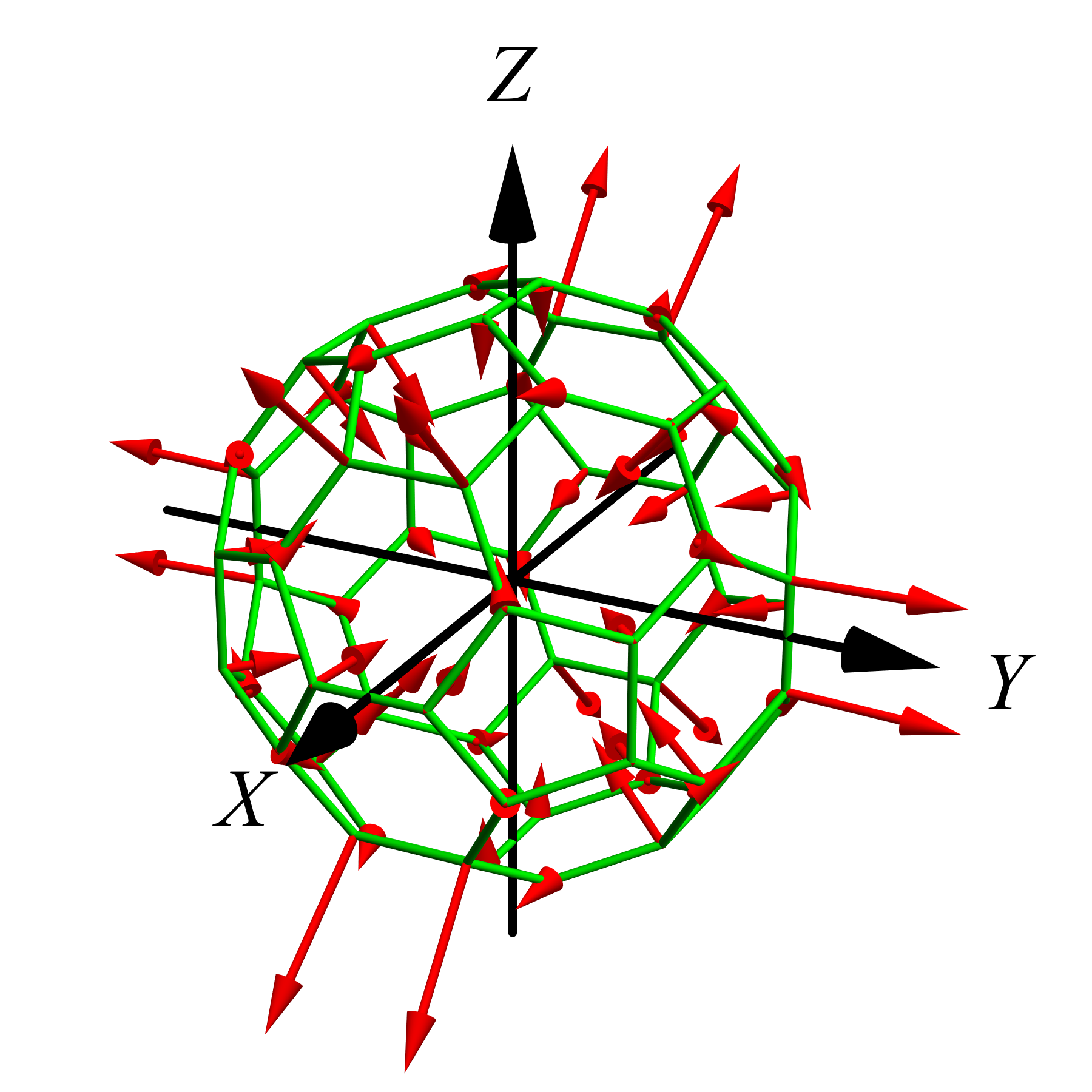}
&
\includegraphics[bb =  0 0 342 342, width=2.7cm]{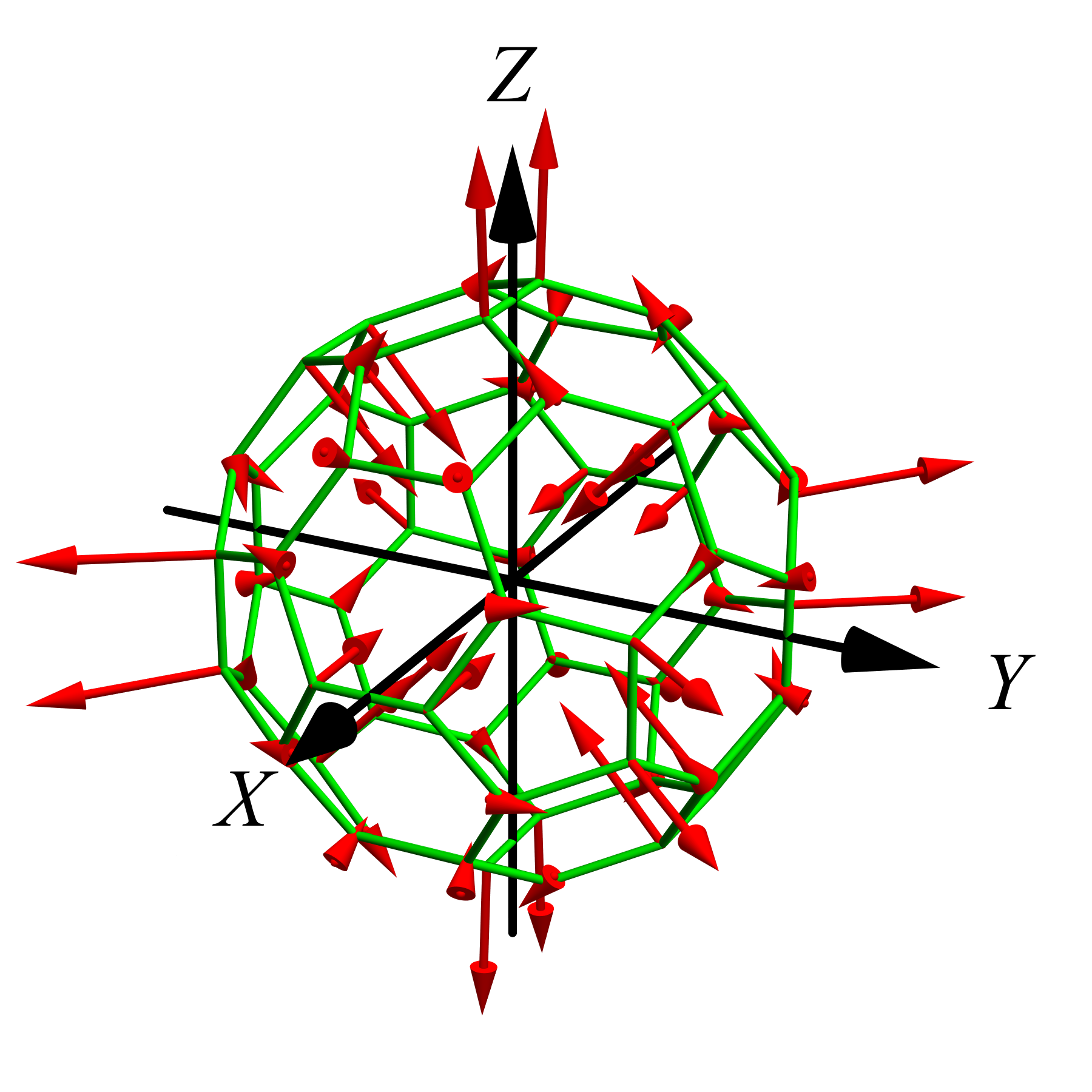}
&
\includegraphics[bb =  0 0 342 342, width=2.7cm]{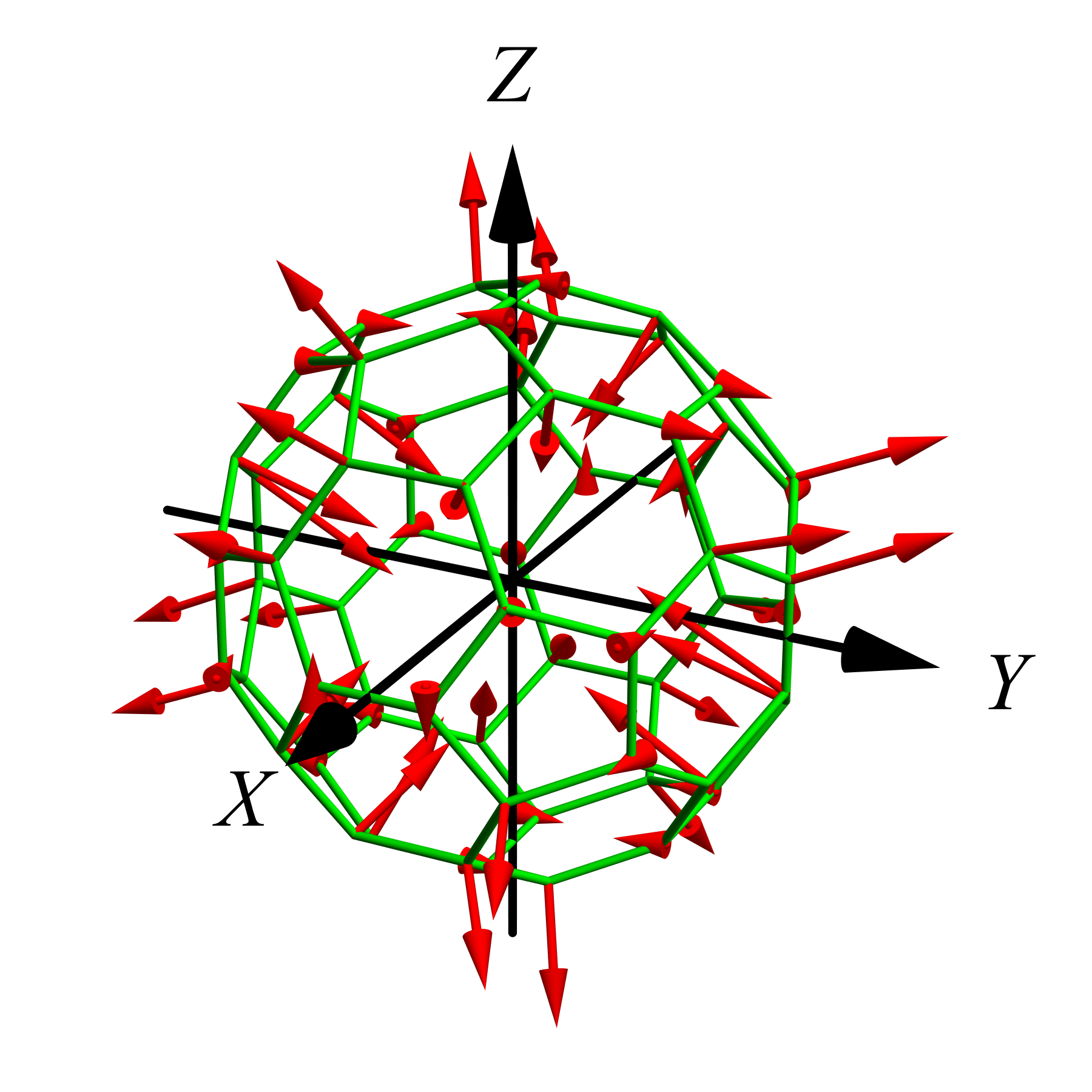}
&
\includegraphics[bb =  0 0 342 342, width=2.7cm]{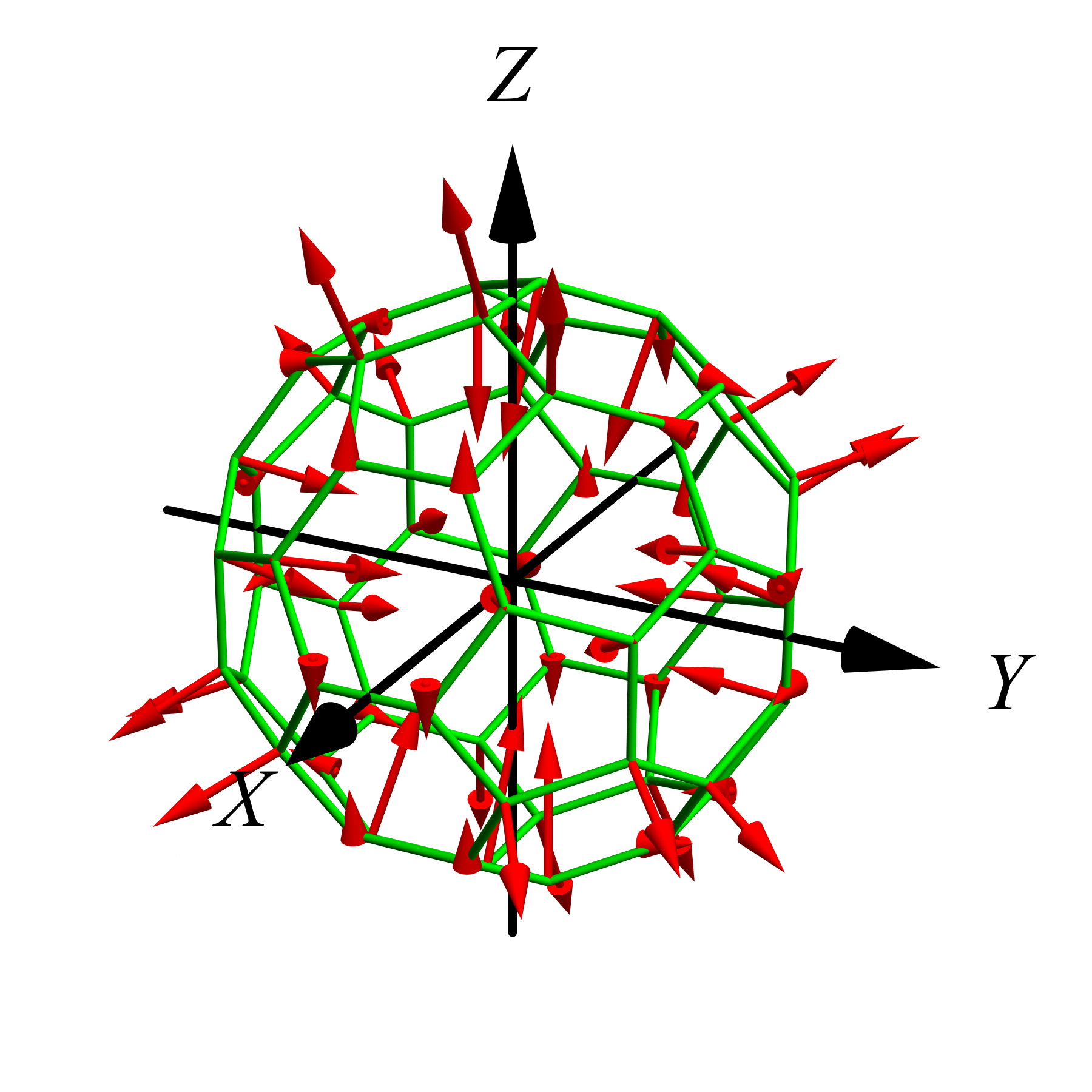}
&
\includegraphics[bb =  0 0 342 342, width=2.7cm]{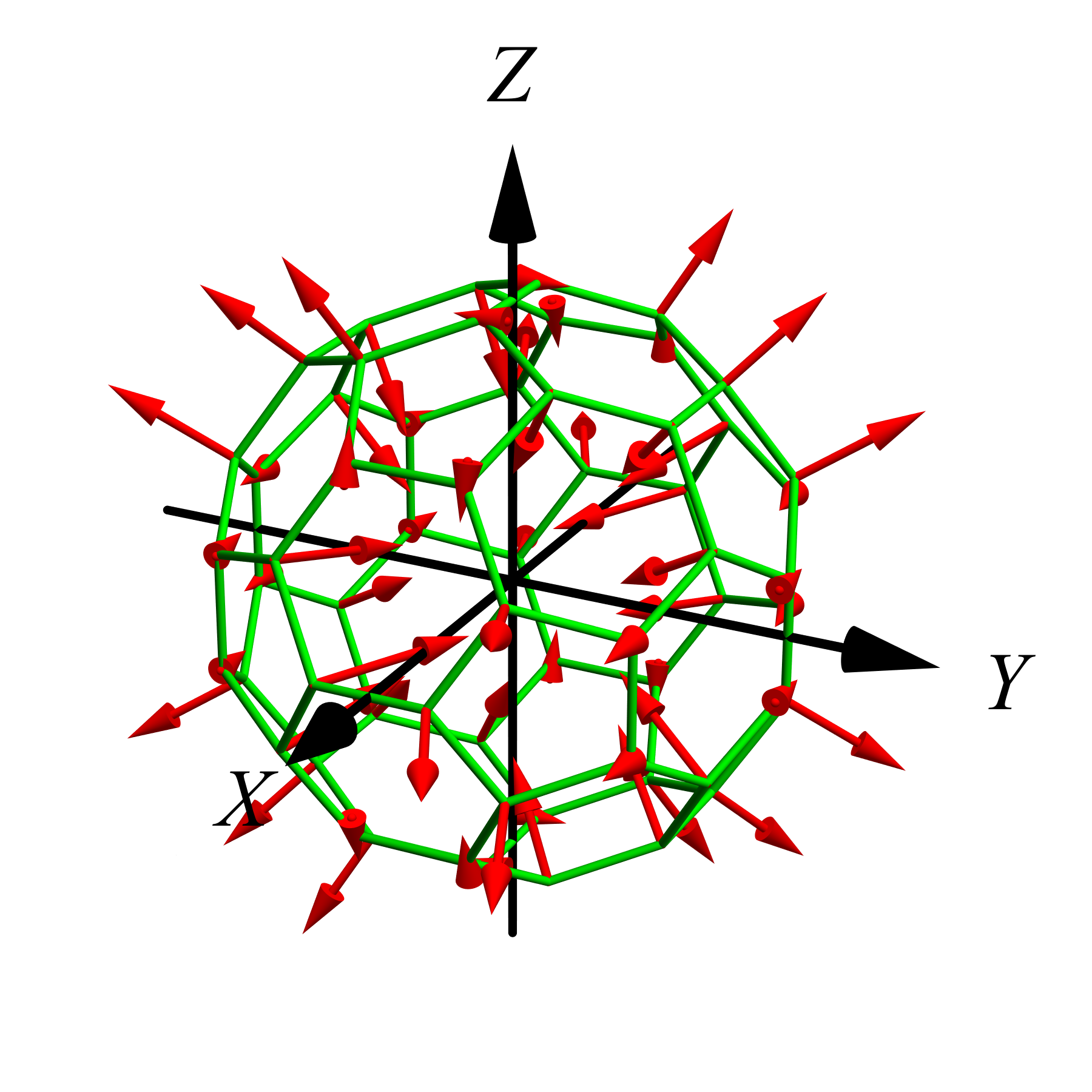}
\\
 $h_g(3)$
&
\includegraphics[bb =  0 0 342 342, width=2.7cm]{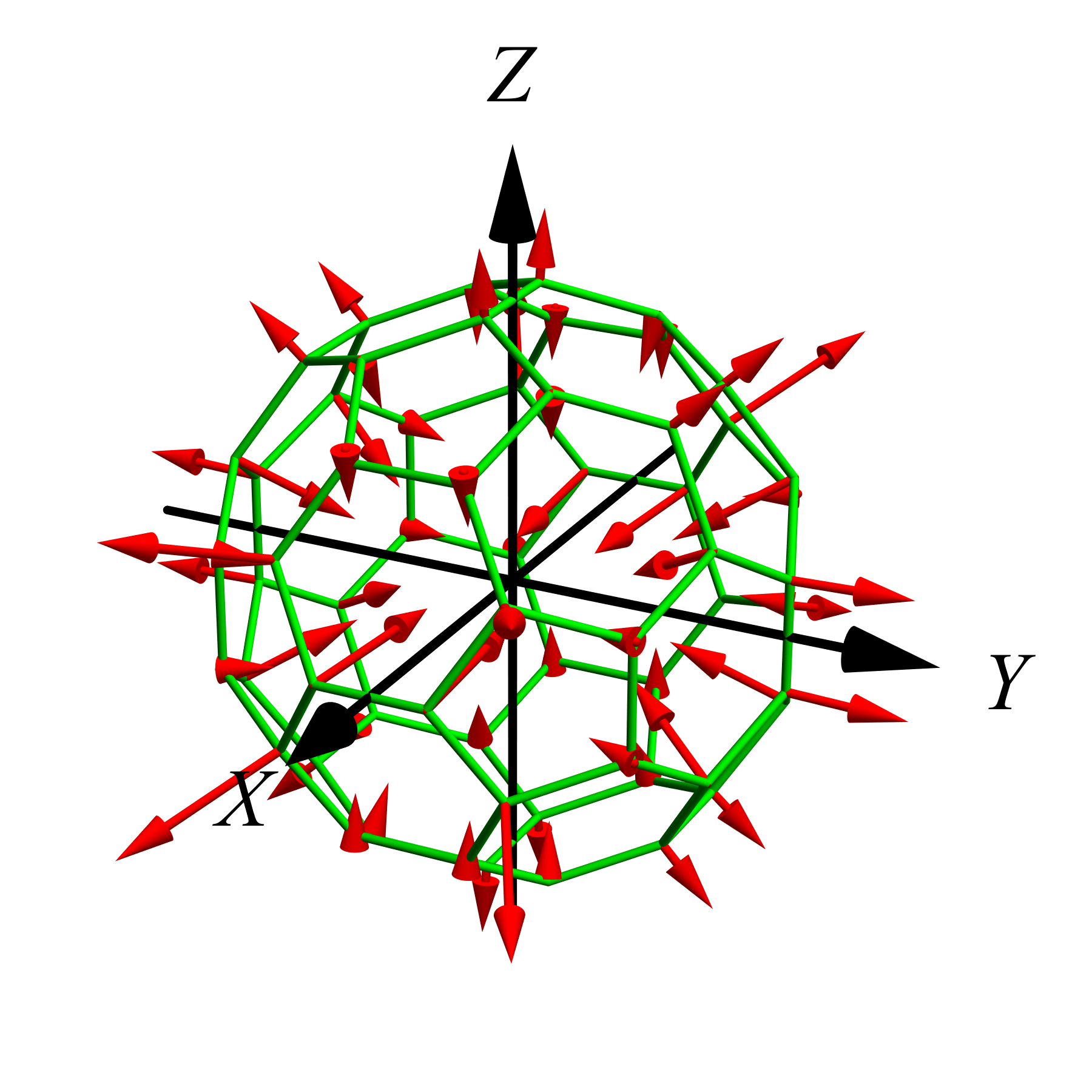}
&
\includegraphics[bb =  0 0 342 342, width=2.7cm]{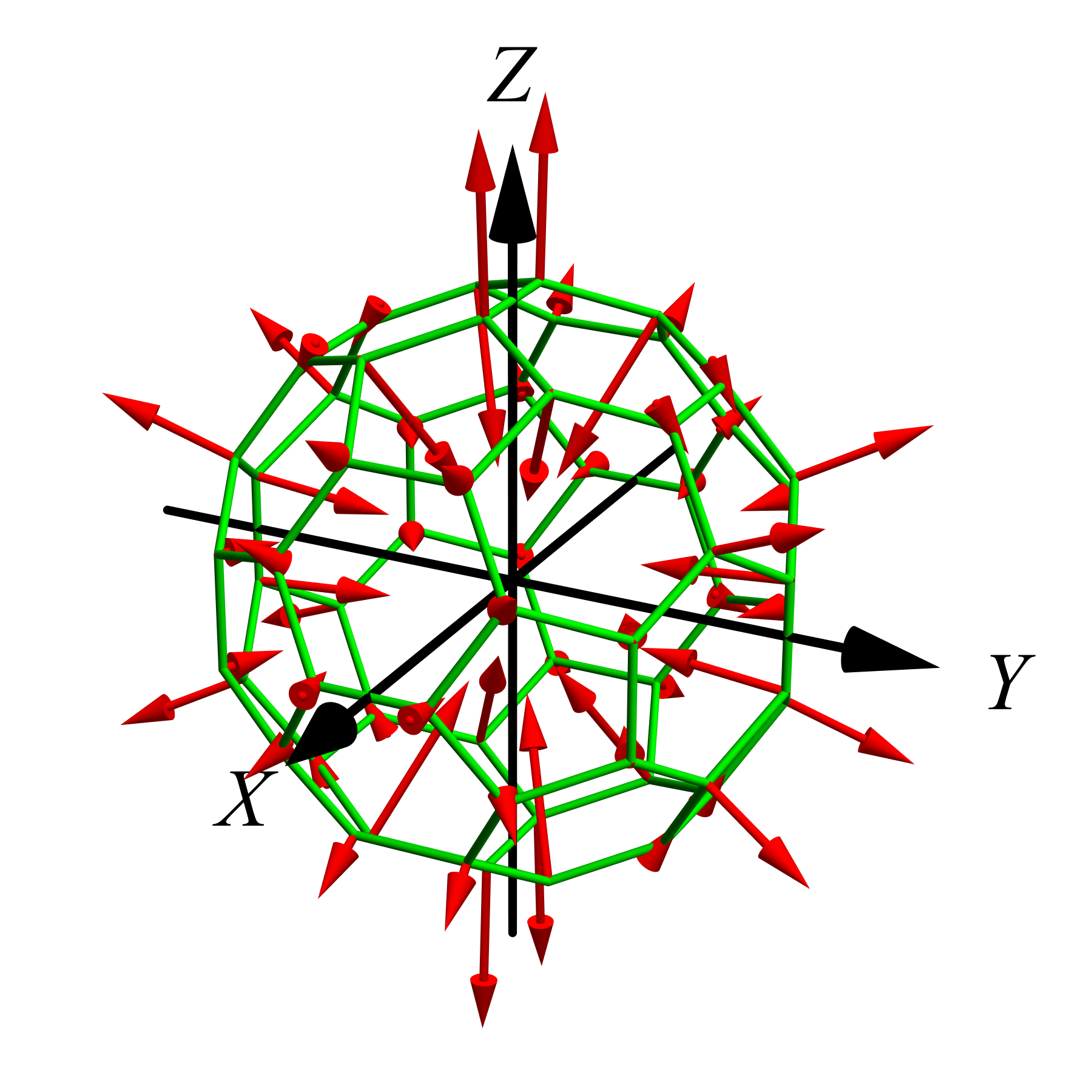}
&
\includegraphics[bb =  0 0 342 342, width=2.7cm]{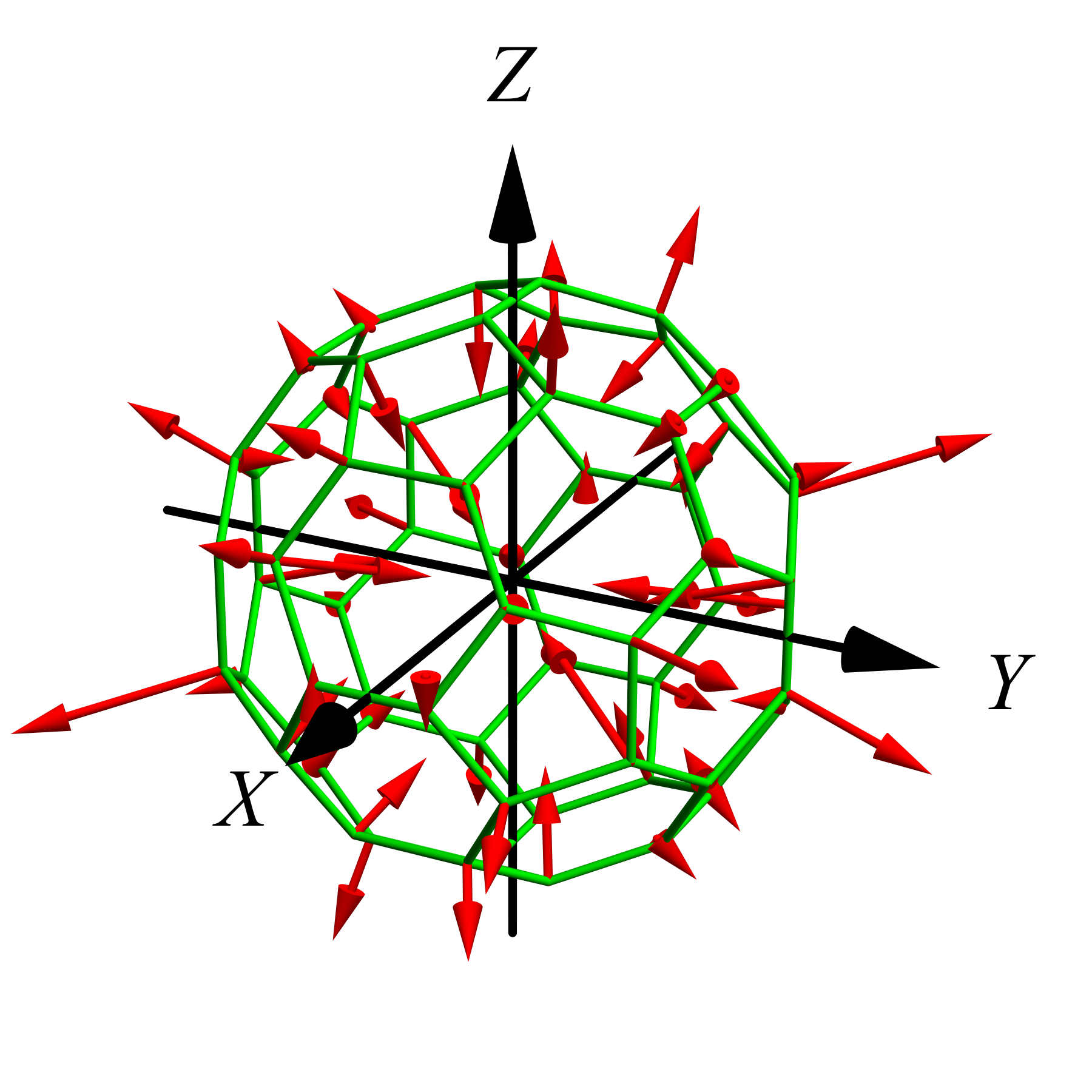}
&
\includegraphics[bb =  0 0 342 342, width=2.7cm]{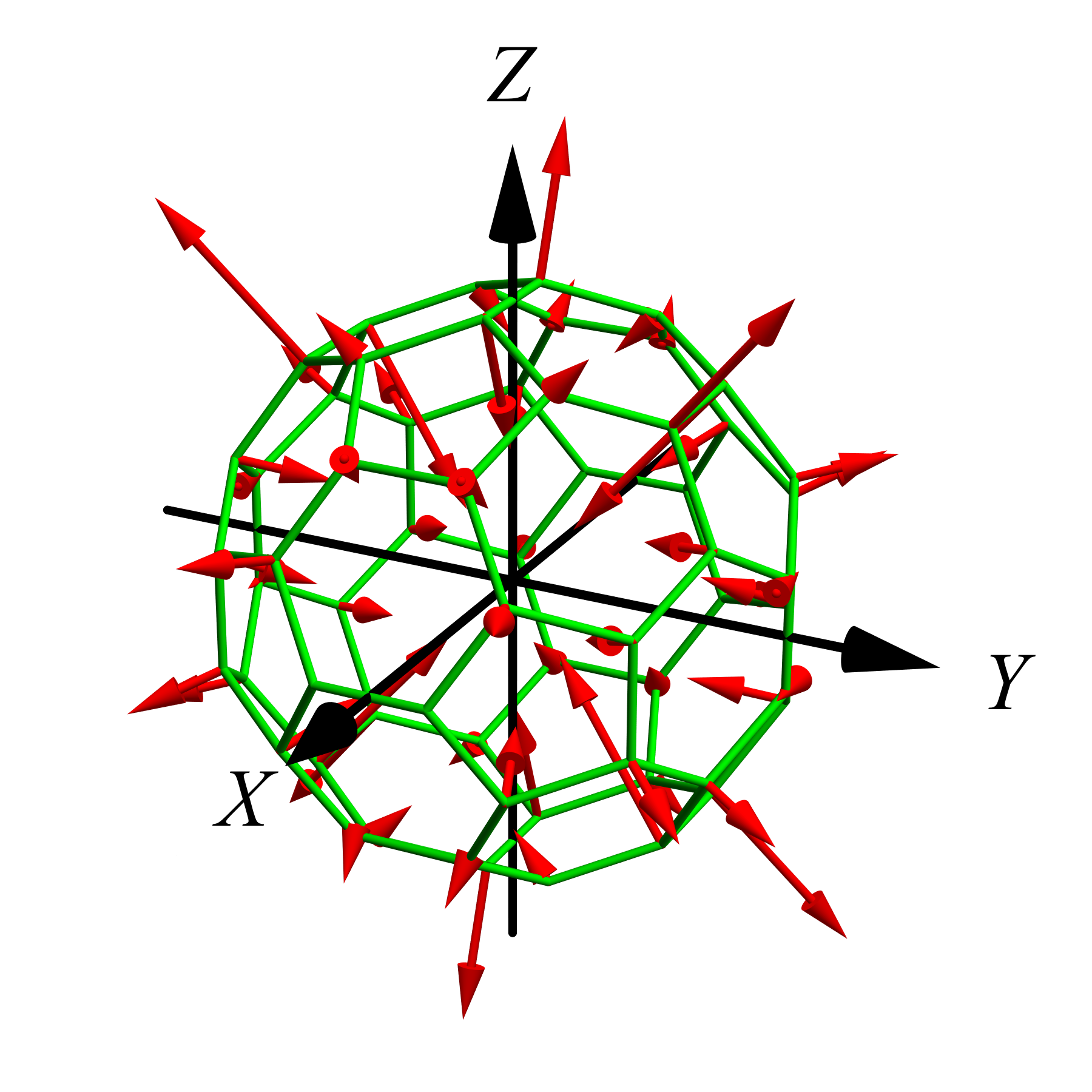}
&
\includegraphics[bb =  0 0 342 342, width=2.7cm]{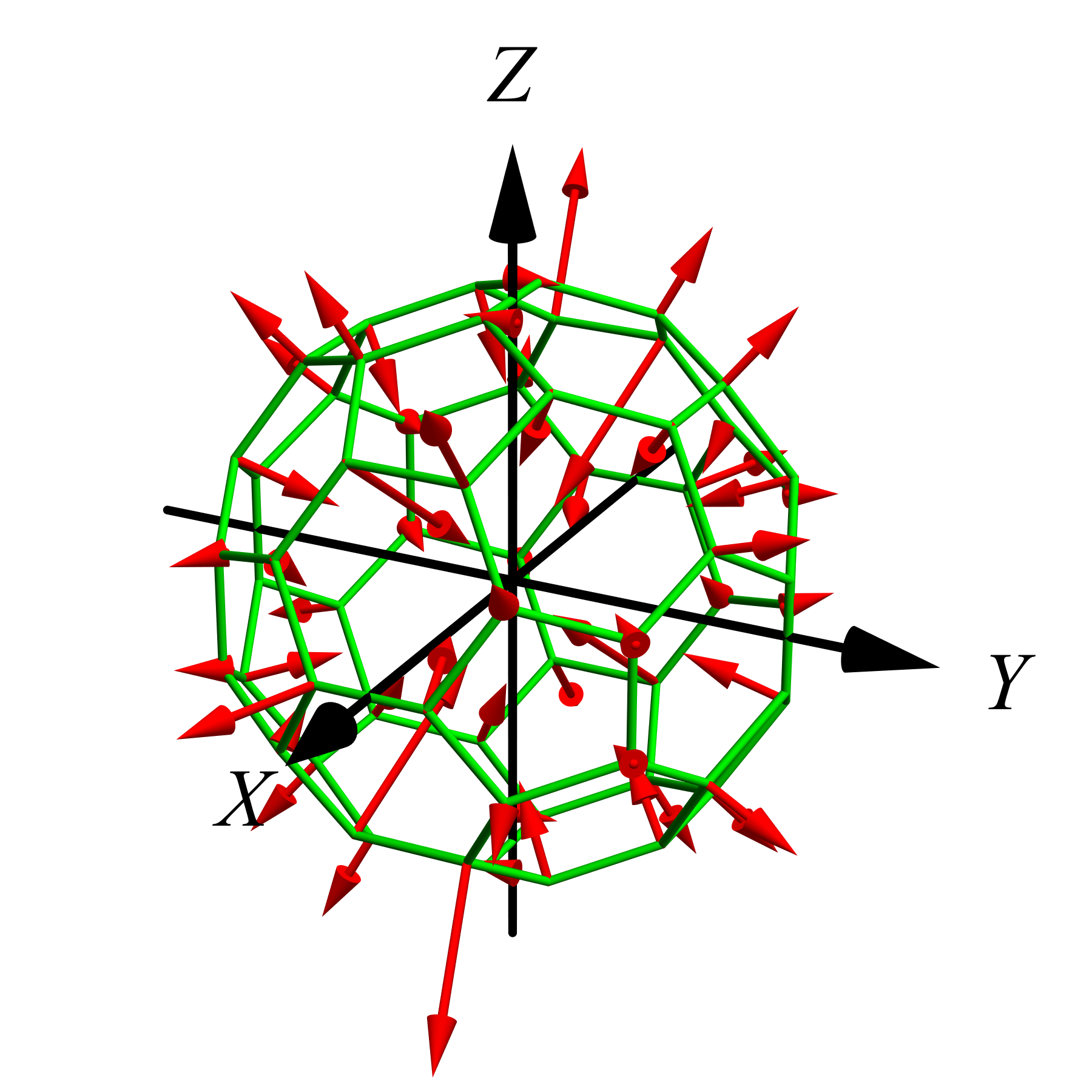}
\\
 $h_g(4)$
&
\includegraphics[bb =  0 0 342 342, width=2.7cm]{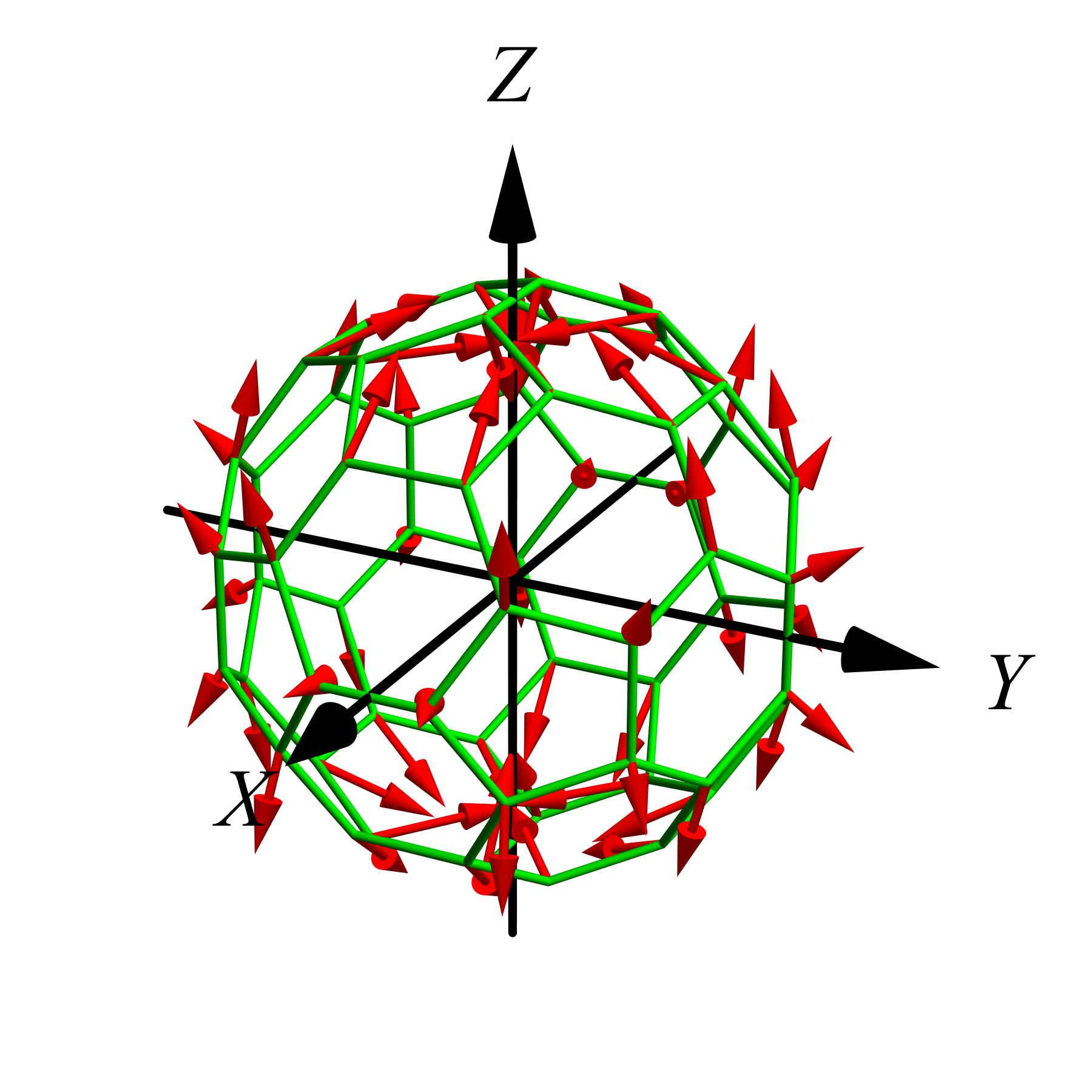}
&
\includegraphics[bb =  0 0 342 342, width=2.7cm]{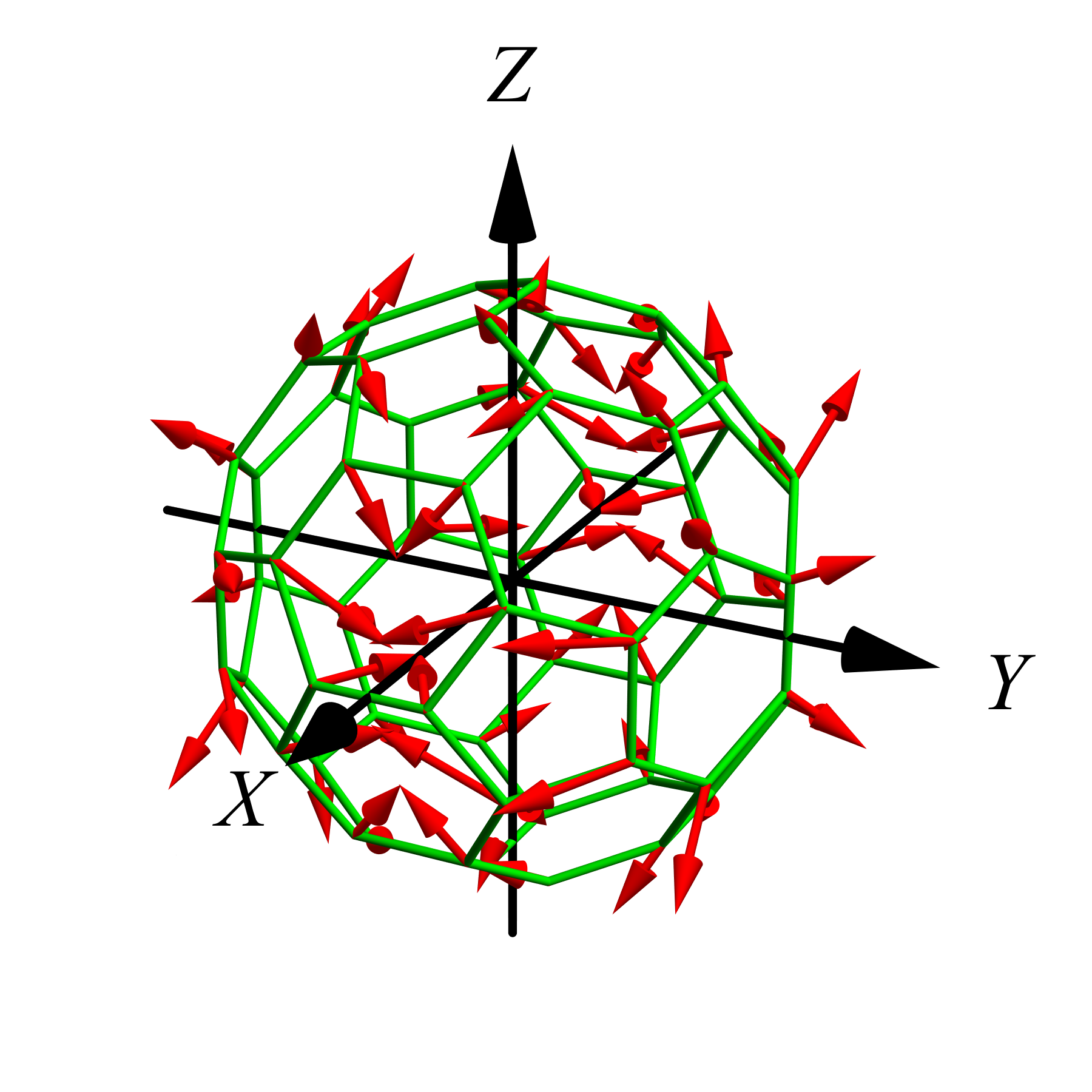}
&
\includegraphics[bb =  0 0 342 342, width=2.7cm]{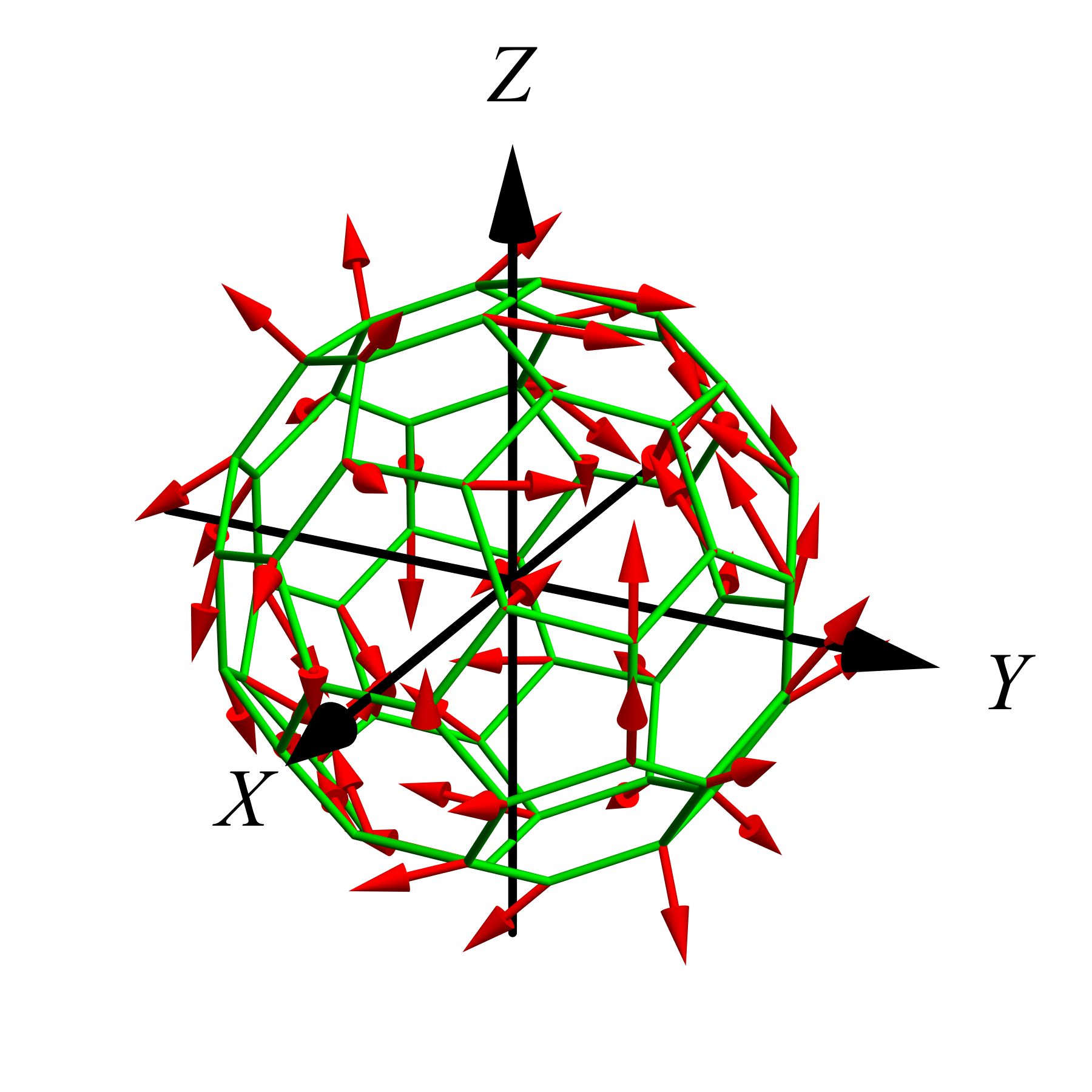}
&
\includegraphics[bb =  0 0 342 342, width=2.7cm]{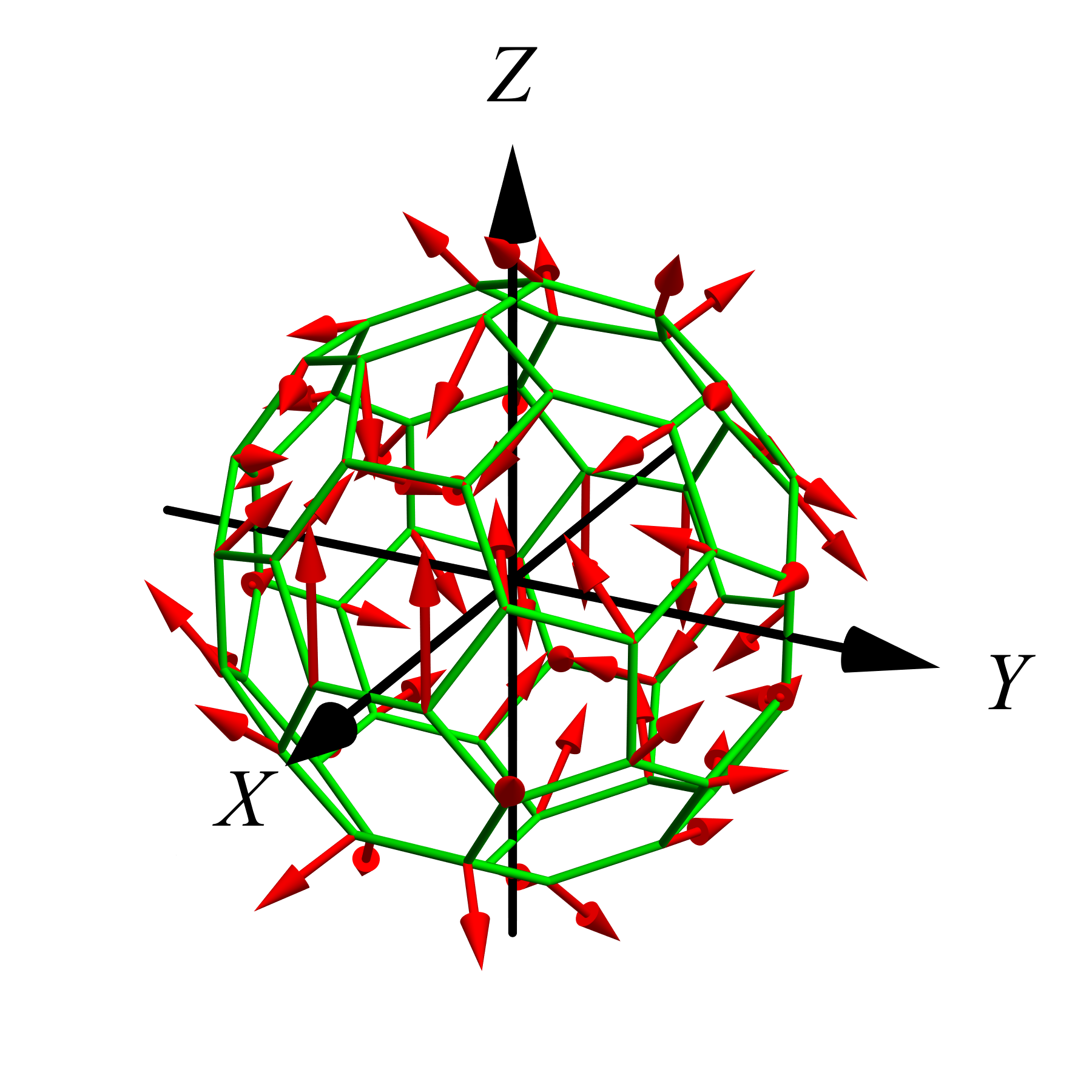}
&
\includegraphics[bb =  0 0 342 342, width=2.7cm]{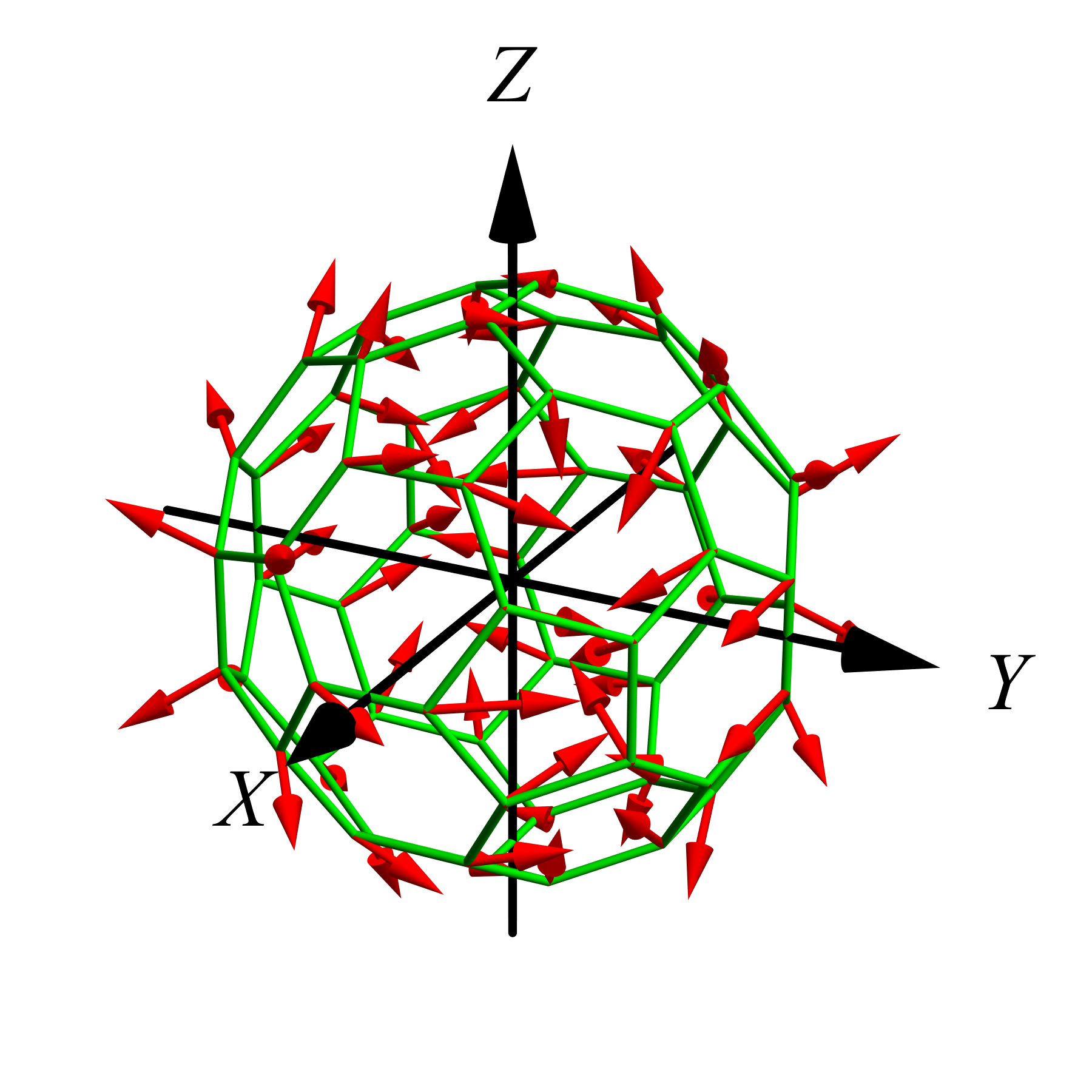}
\\
 $h_g(5)$
&
\includegraphics[bb =  0 0 342 342, width=2.7cm]{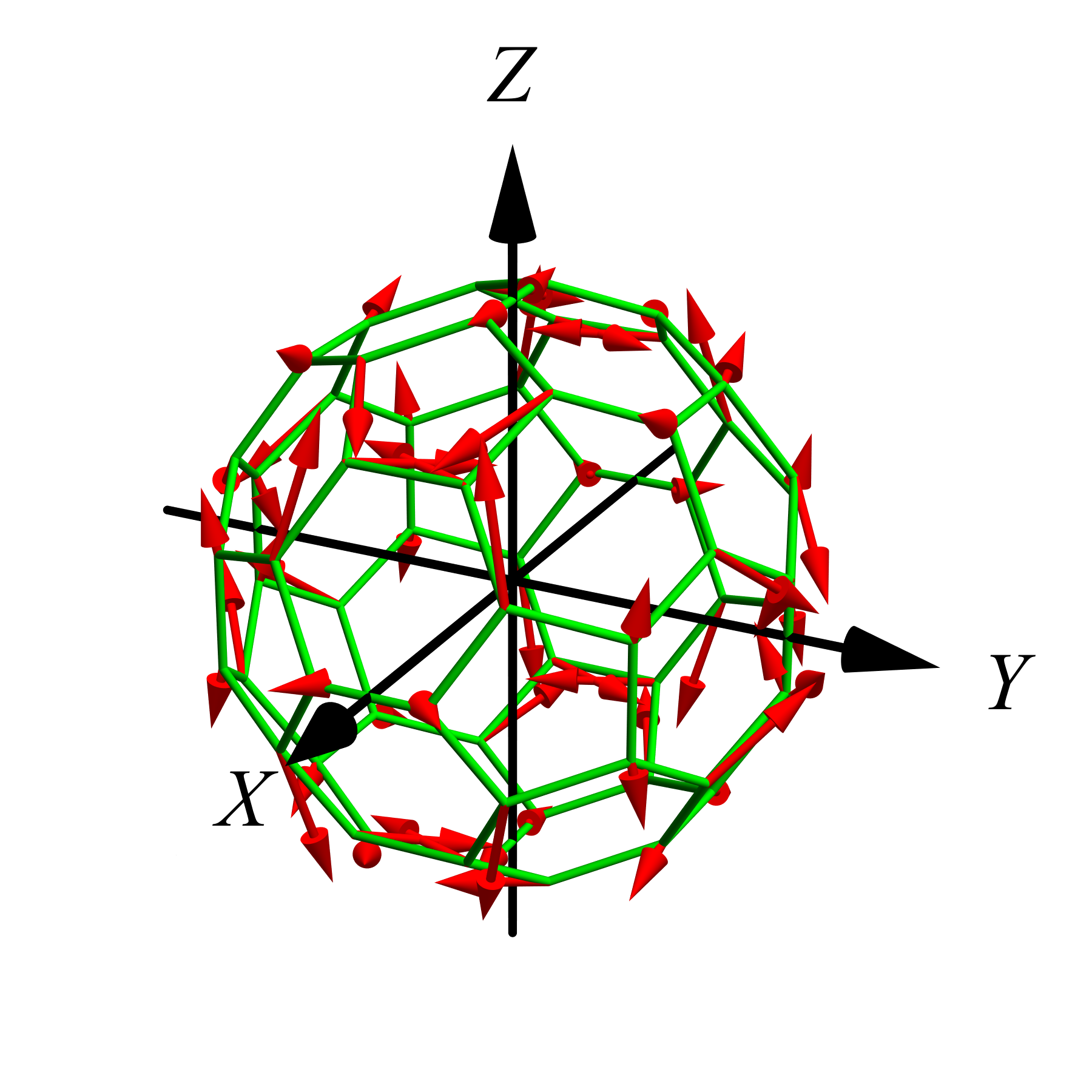}
&
\includegraphics[bb =  0 0 342 342, width=2.7cm]{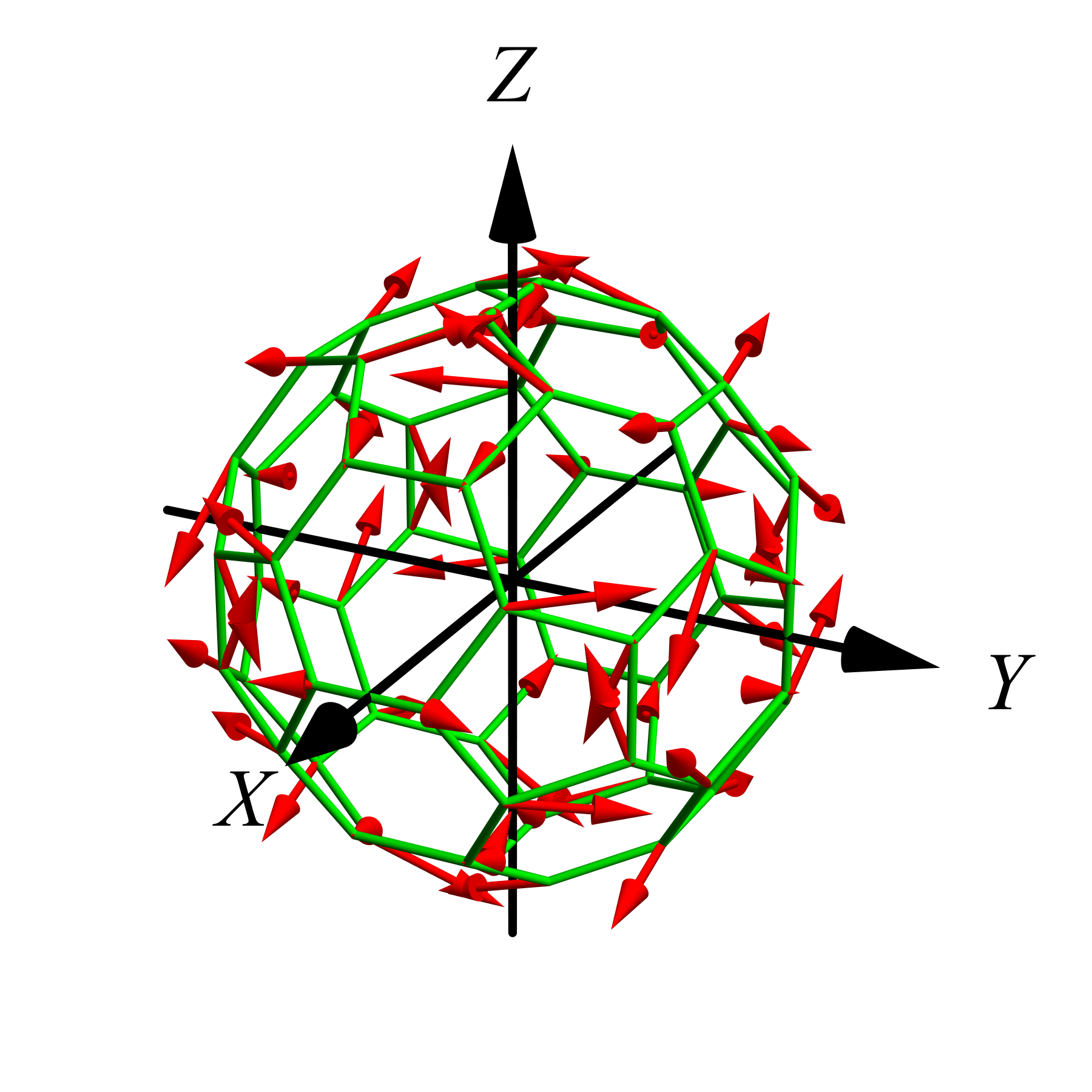}
&
\includegraphics[bb =  0 0 342 342, width=2.7cm]{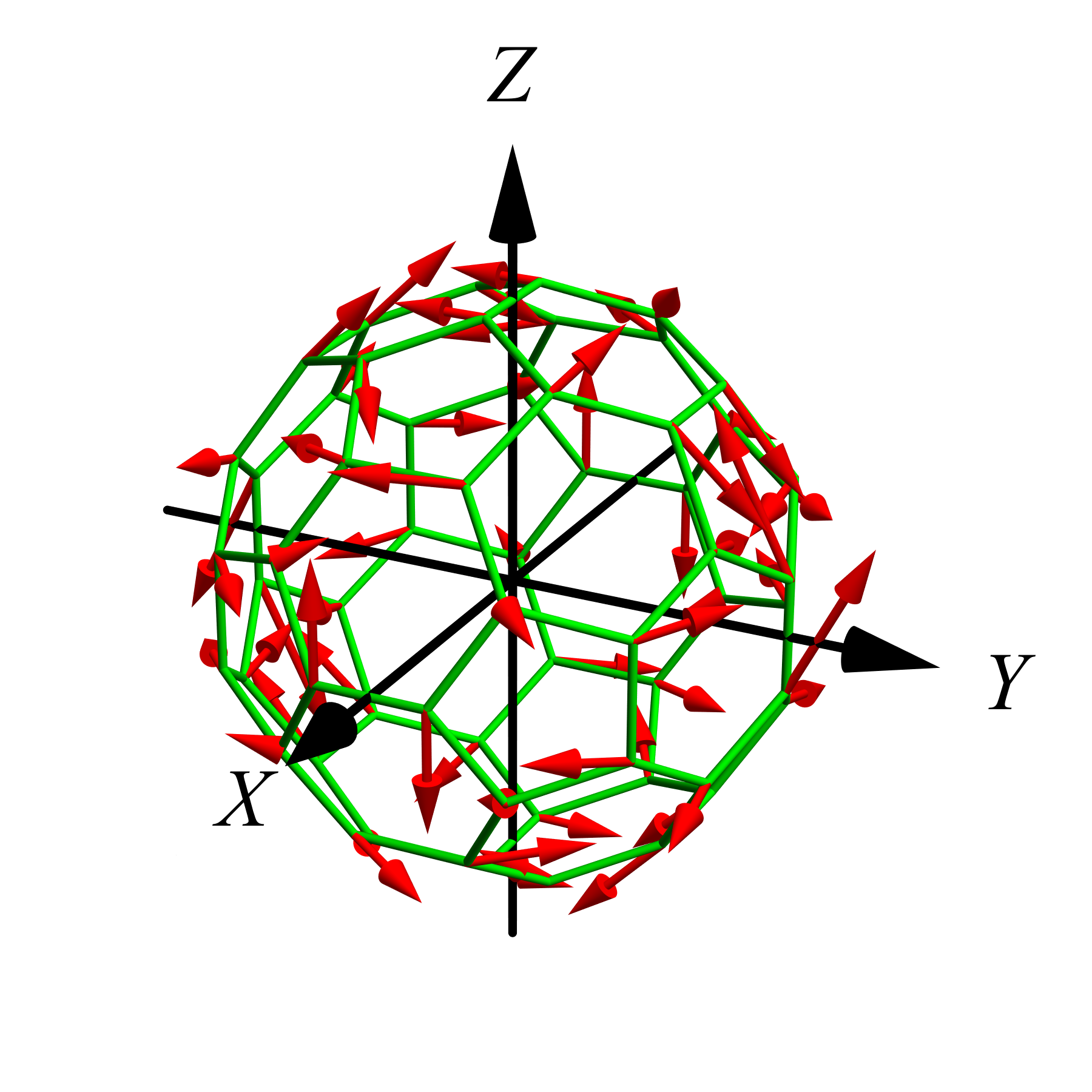}
&
\includegraphics[bb =  0 0 342 342, width=2.7cm]{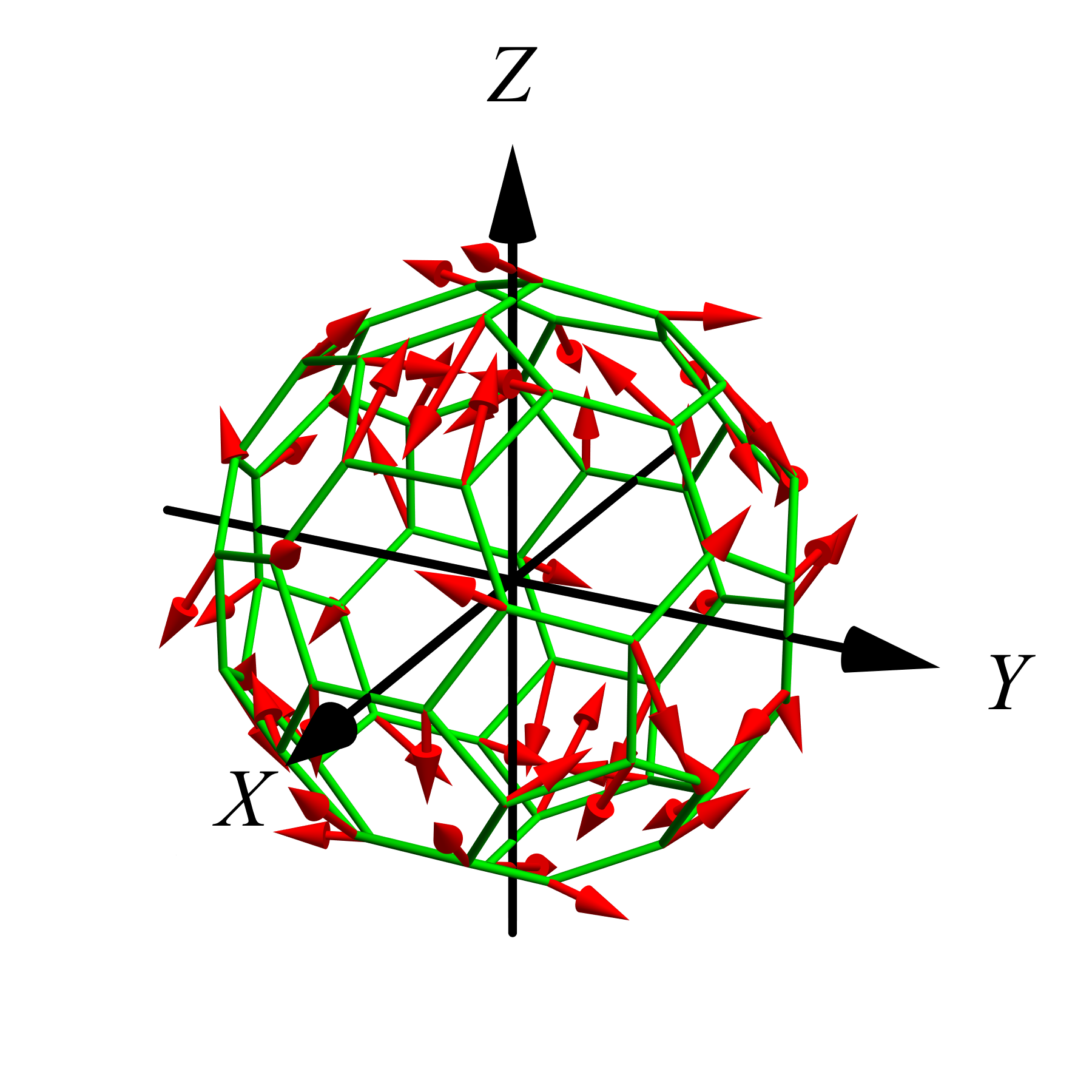}
&
\includegraphics[bb =  0 0 342 342, width=2.7cm]{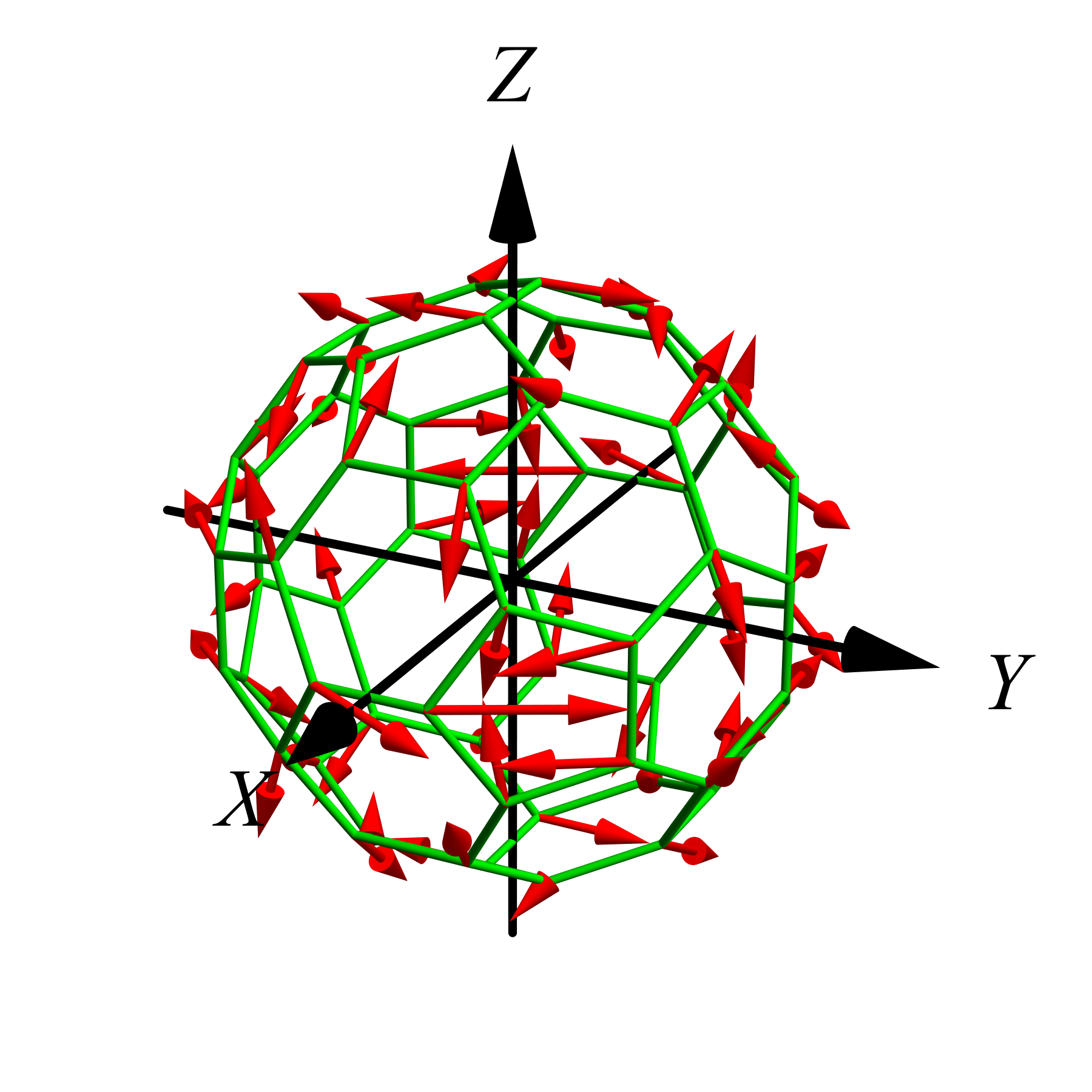}
\\
 $h_g(6)$
&
\includegraphics[bb =  0 0 342 342, width=2.7cm]{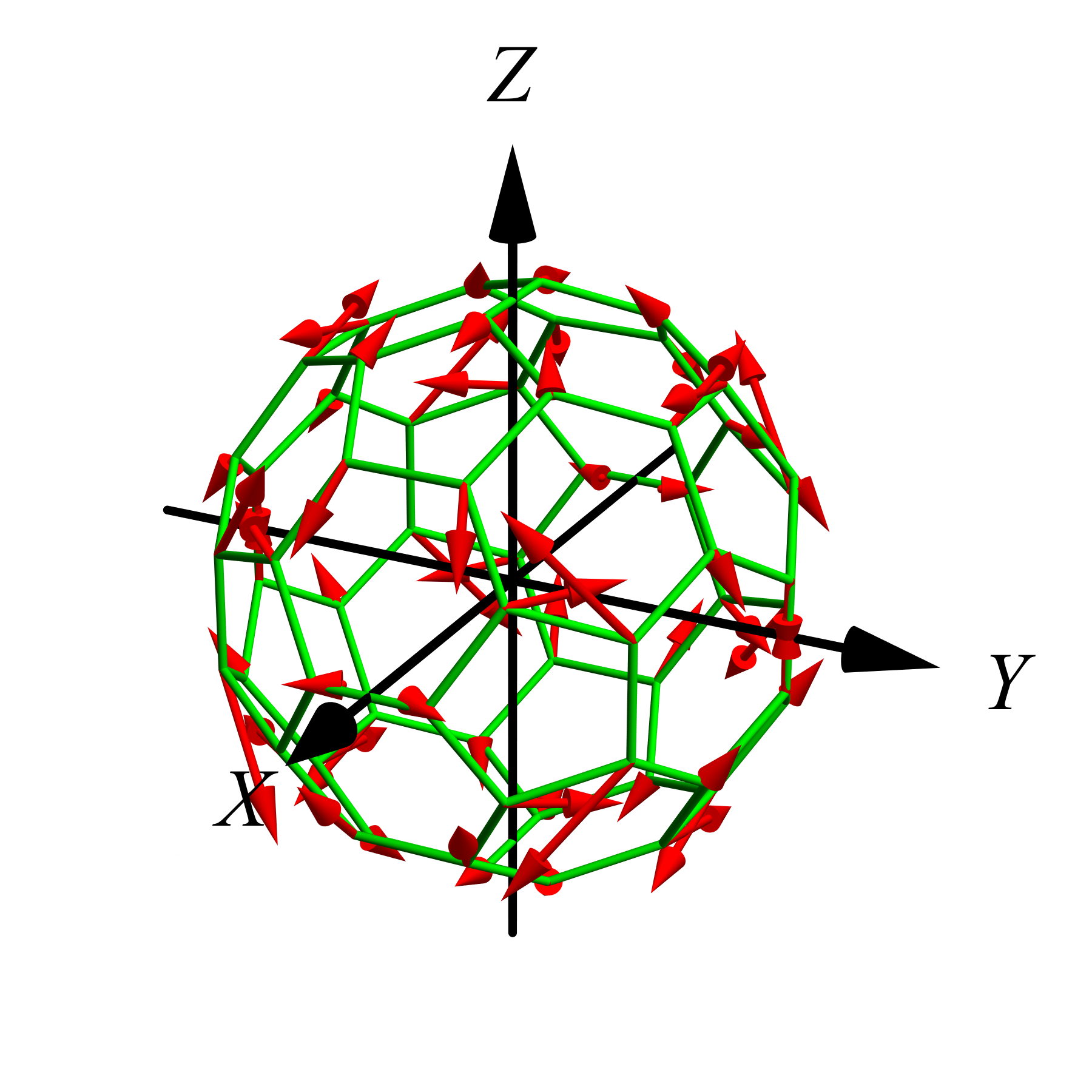}
&
\includegraphics[bb =  0 0 342 342, width=2.7cm]{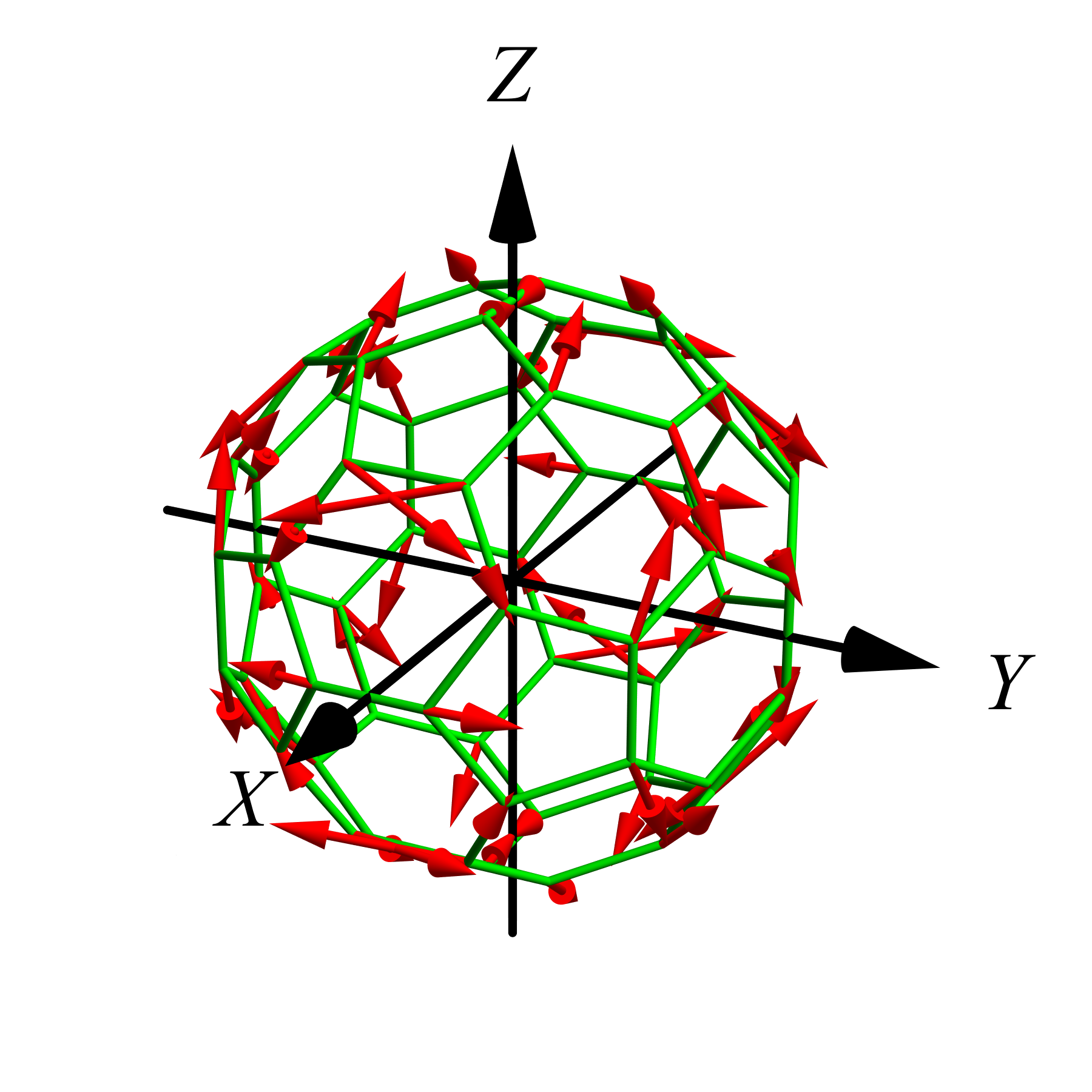}
&
\includegraphics[bb =  0 0 342 342, width=2.7cm]{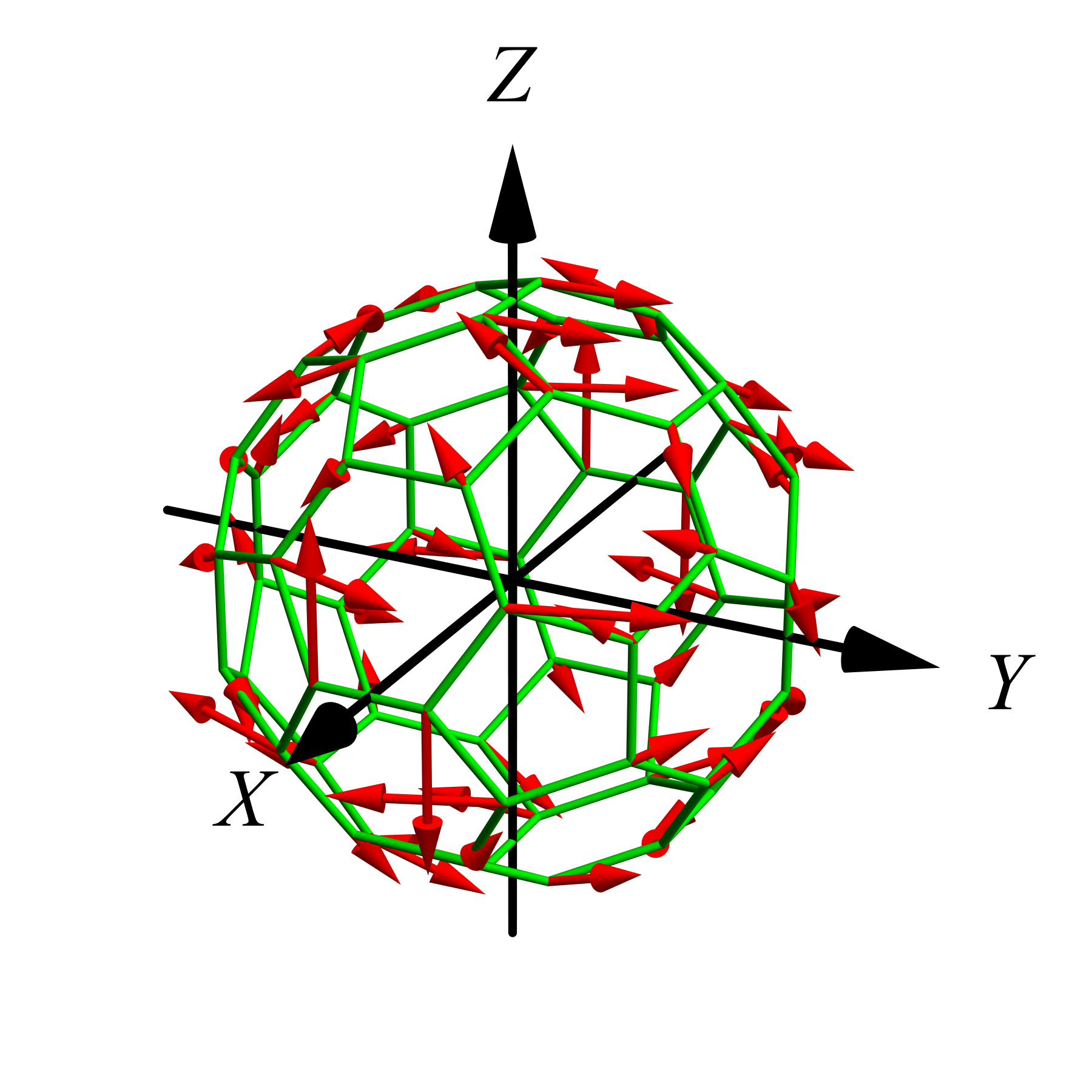}
&
\includegraphics[bb =  0 0 342 342, width=2.7cm]{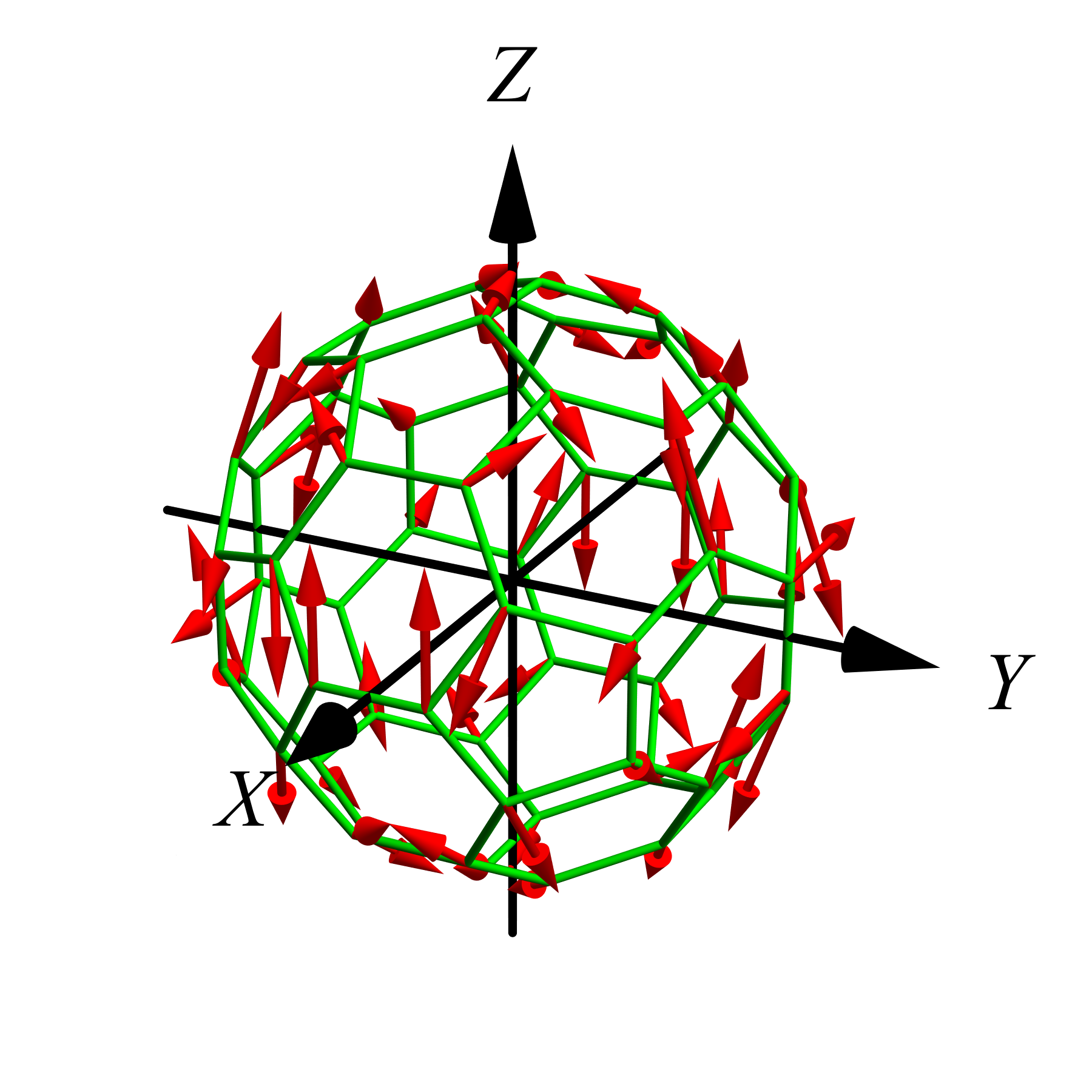}
&
\includegraphics[bb =  0 0 342 342, width=2.7cm]{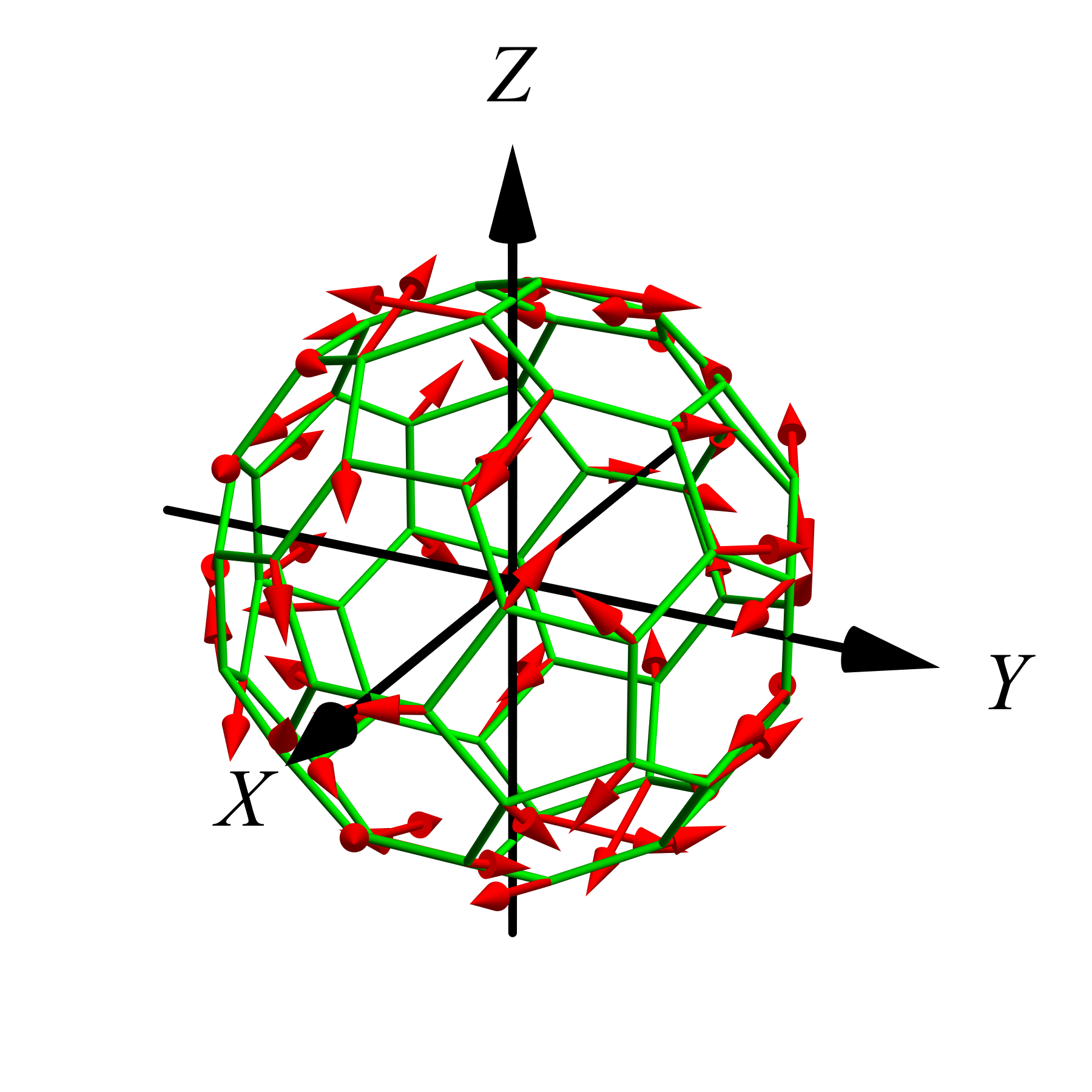}
\\
 $h_g(7)$
&
\includegraphics[bb =  0 0 342 342, width=2.7cm]{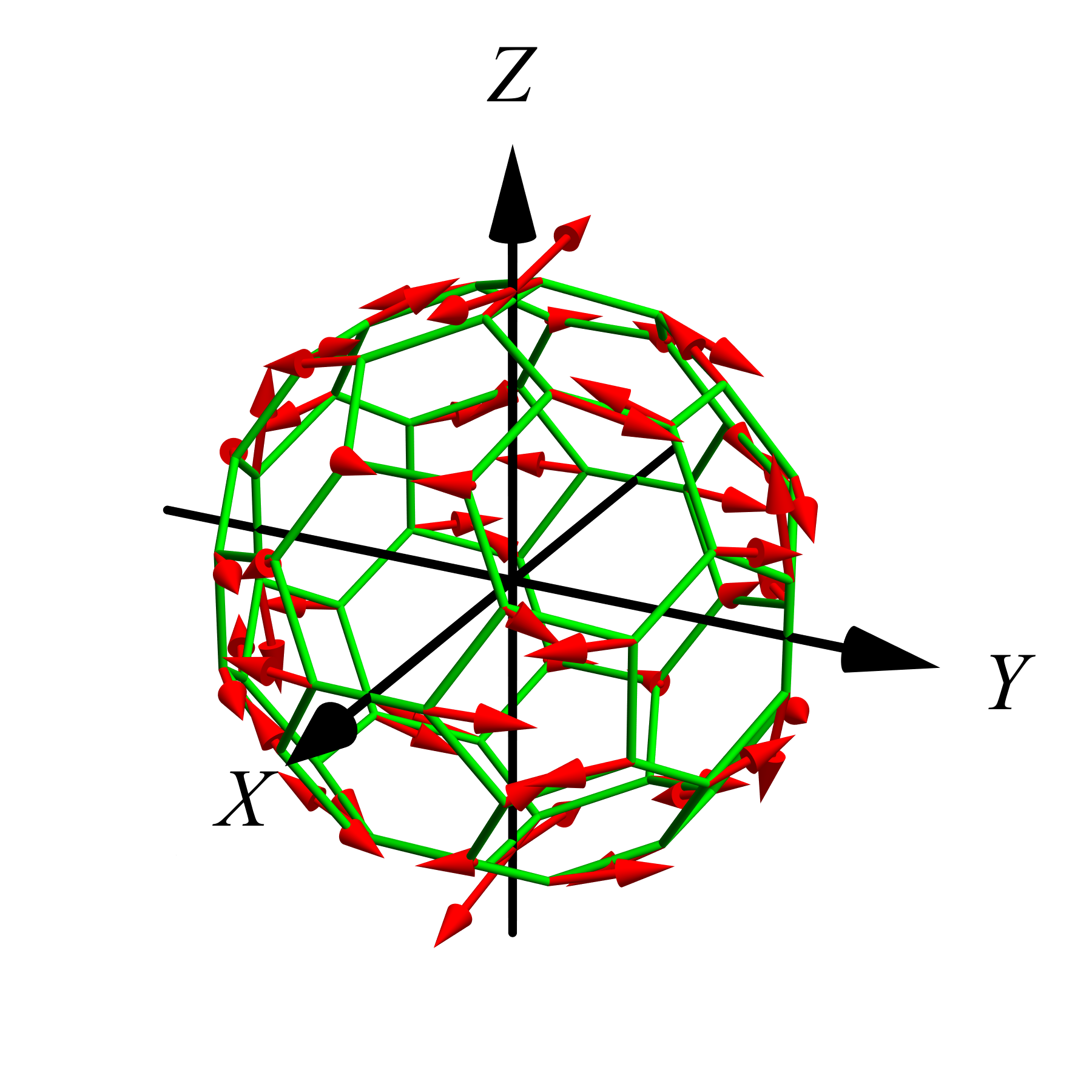}
&
\includegraphics[bb =  0 0 342 342, width=2.7cm]{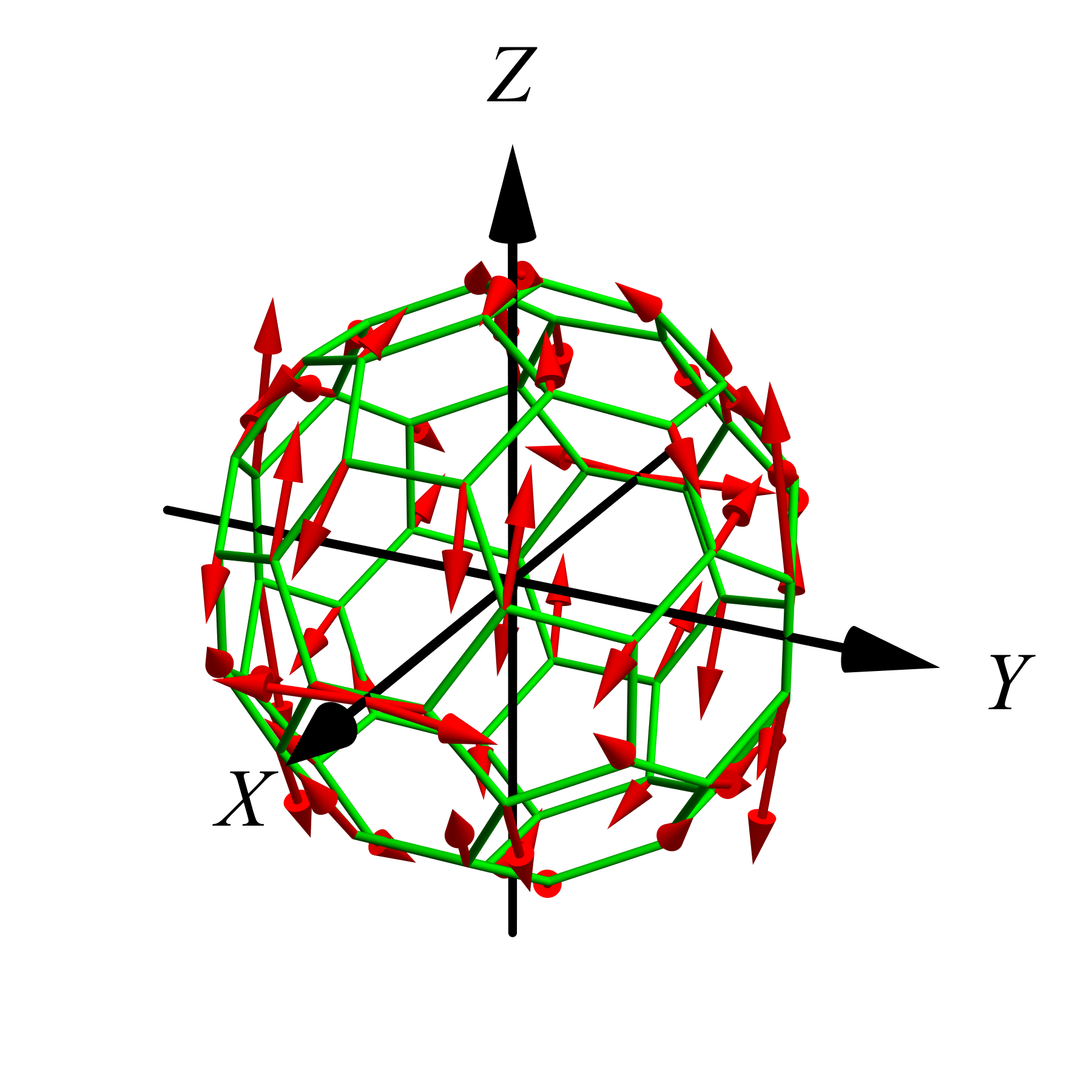}
&
\includegraphics[bb =  0 0 342 342, width=2.7cm]{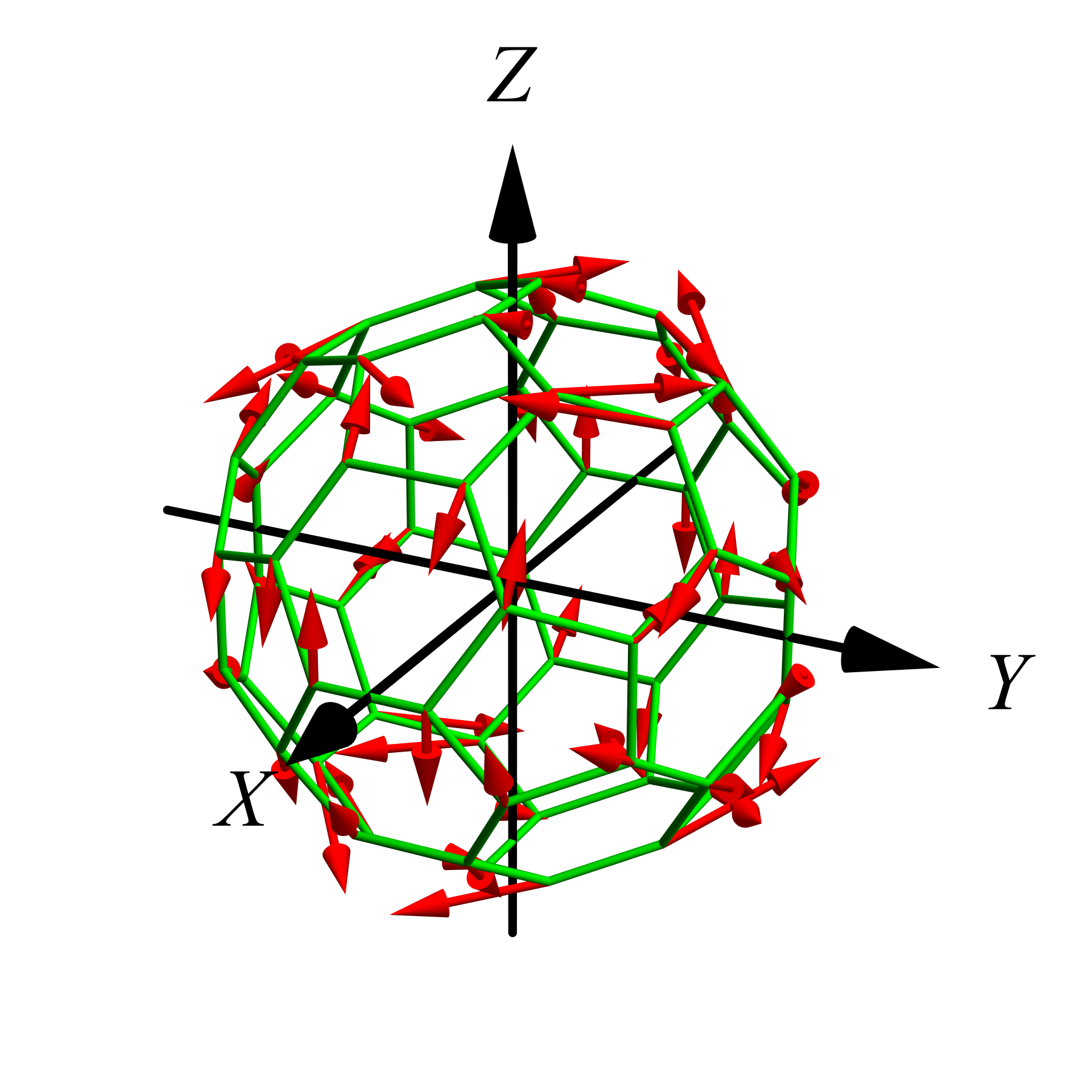}
&
\includegraphics[bb =  0 0 342 342, width=2.7cm]{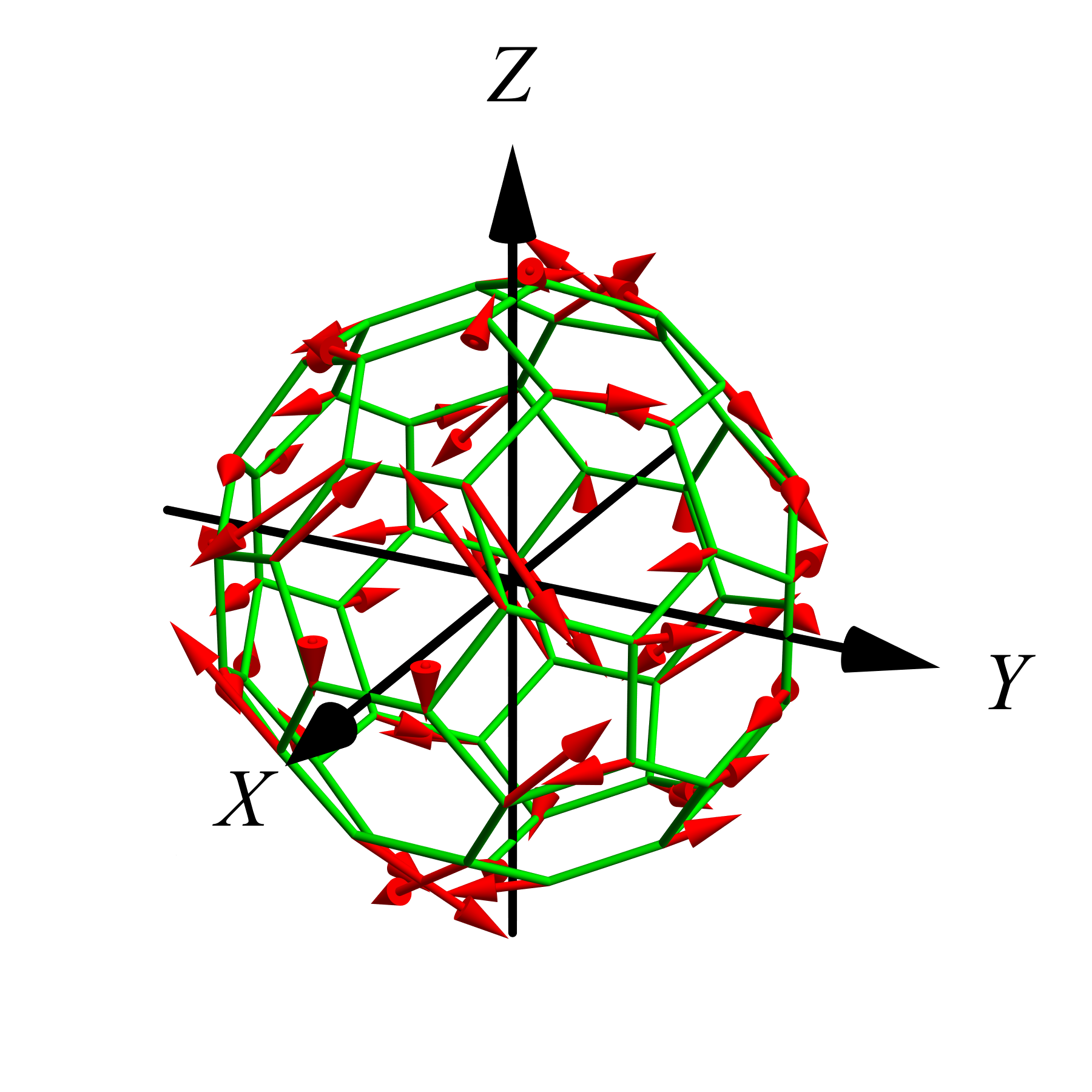}
&
\includegraphics[bb =  0 0 342 342, width=2.7cm]{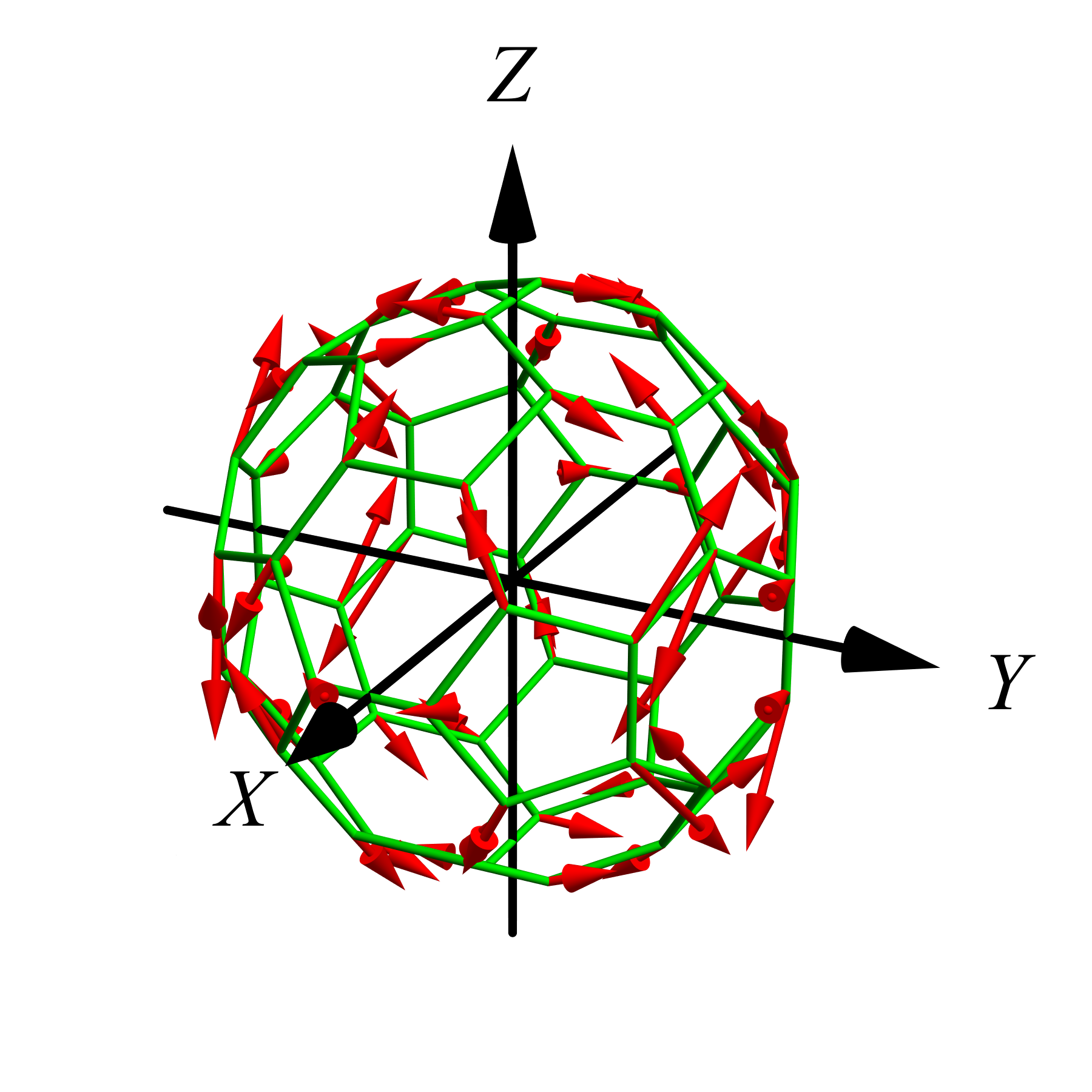}
\\
 $h_g(8)$
&
\includegraphics[bb =  0 0 342 342, width=2.7cm]{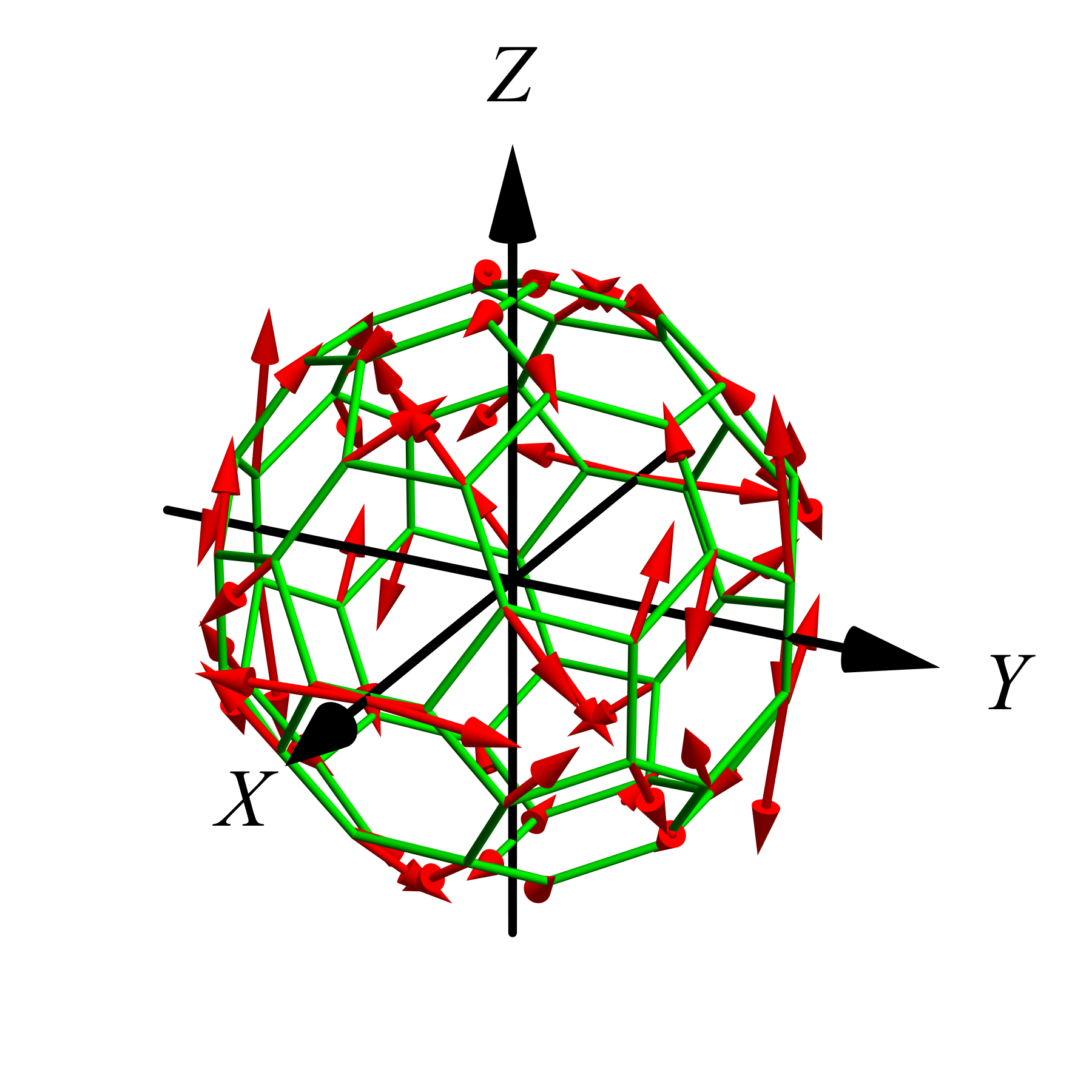}
&
\includegraphics[bb =  0 0 342 342, width=2.7cm]{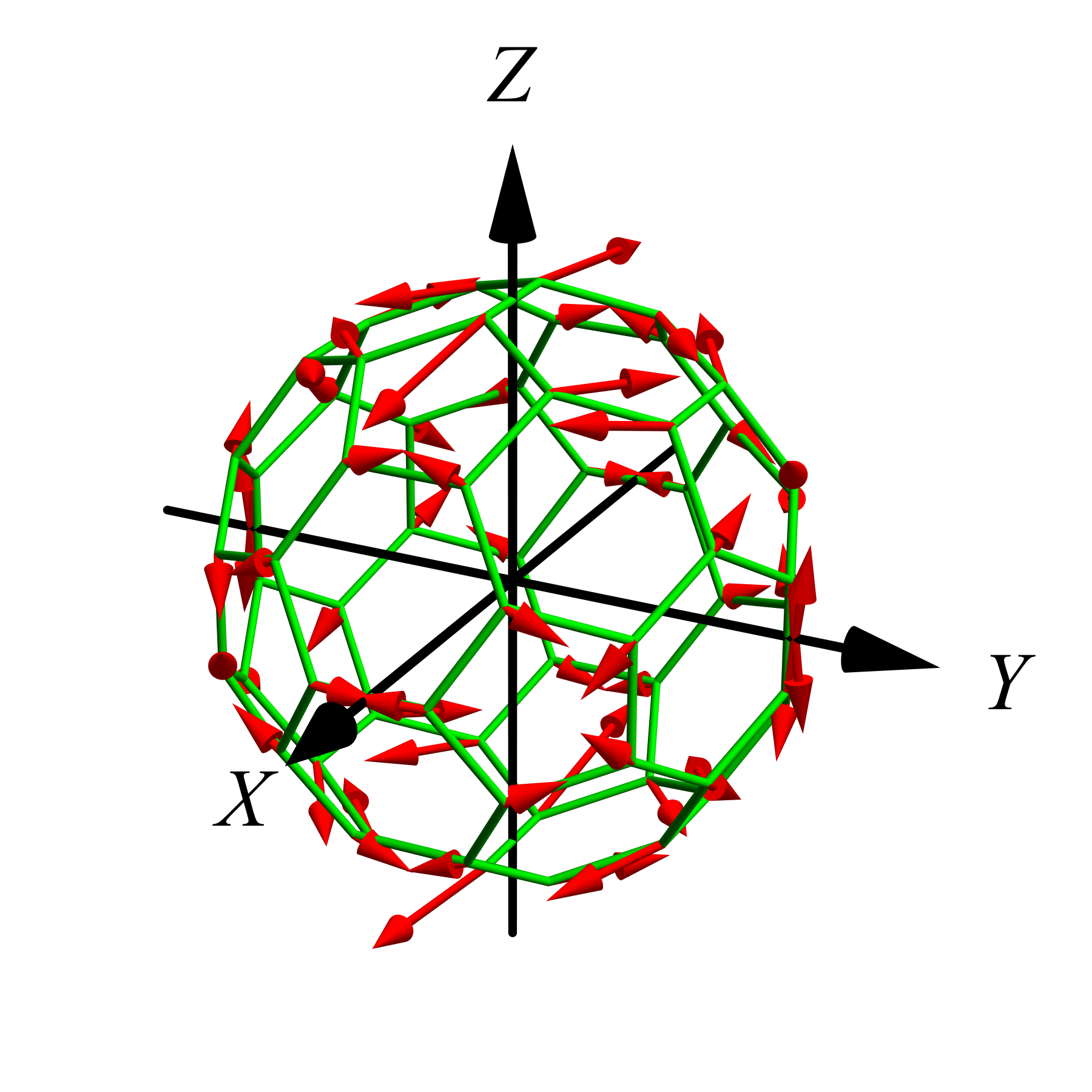}
&
\includegraphics[bb =  0 0 342 342, width=2.7cm]{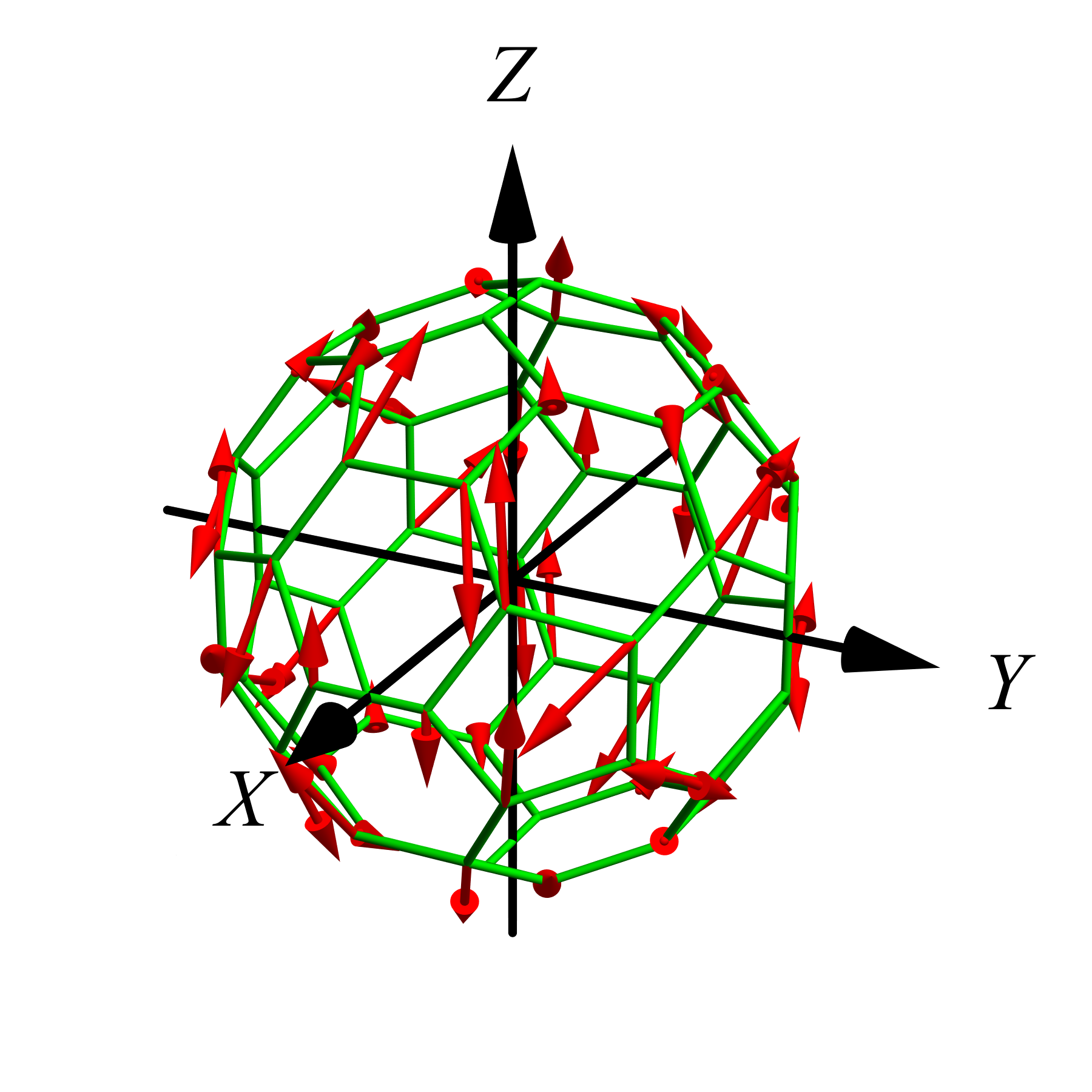}
&
\includegraphics[bb =  0 0 342 342, width=2.7cm]{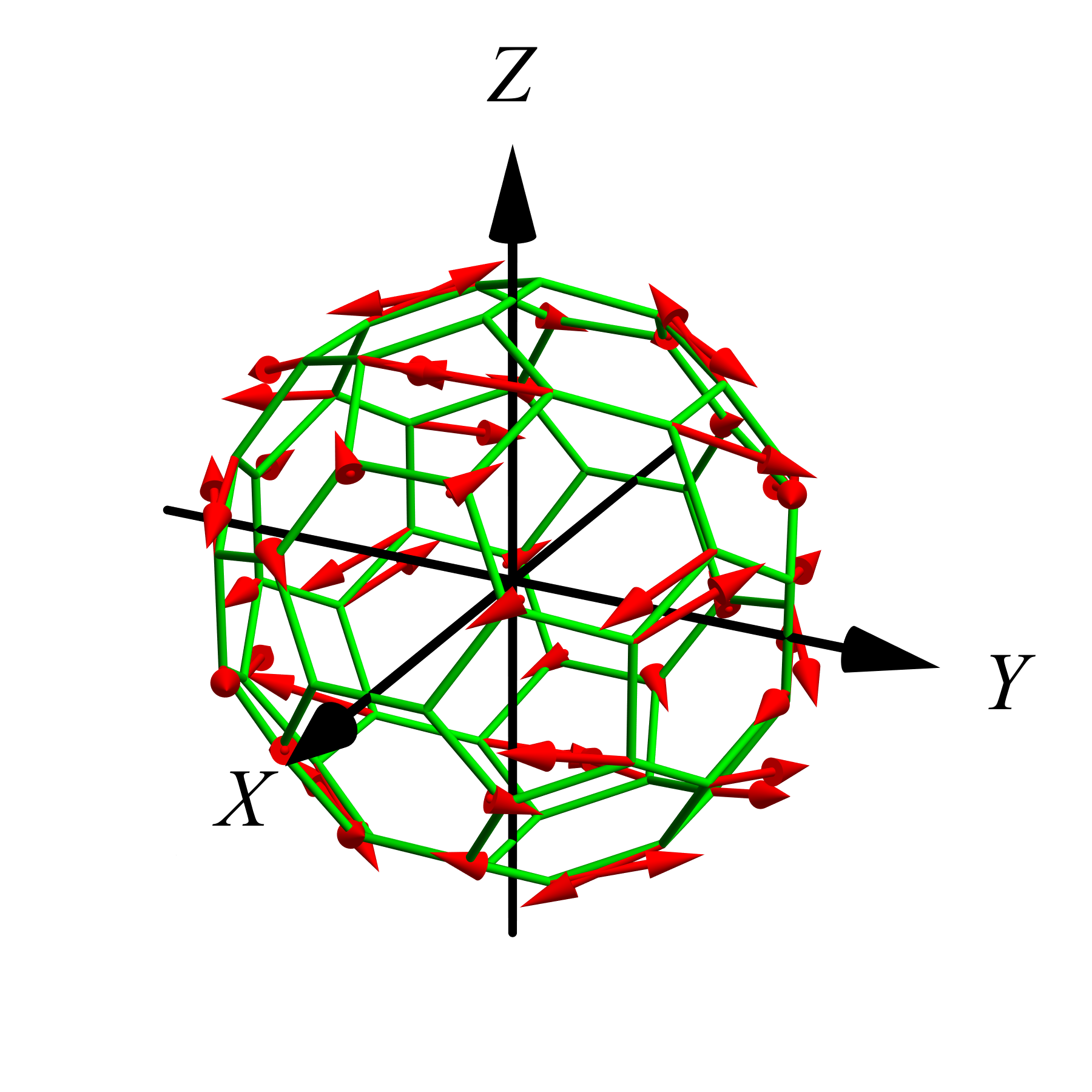}
&
\includegraphics[bb =  0 0 342 342, width=2.7cm]{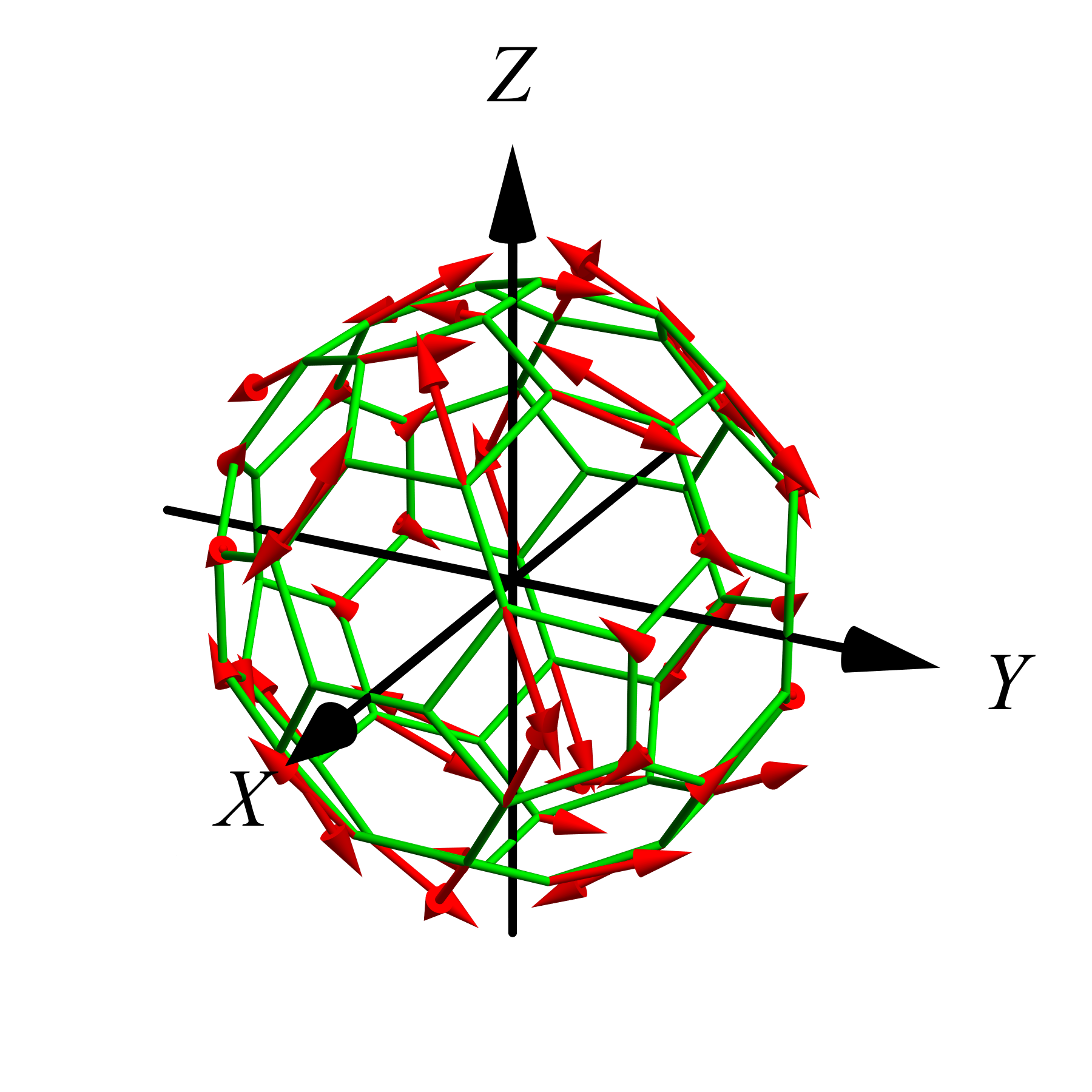}
\end{tabular}
\end{center}
\caption[Schematic representation of $H_g$ modes]{Schematic representation of $H_g$ modes}
\label{Fig:Hg}
\end{figure}

The symmetrization of the mass-weighted normal modes were done by using projection operator.
The symmetrized normal modes were shown below (Figs. ~\ref{Fig:Ag} and ~\ref{Fig:Hg}).

\subsection{Symmetry adapted alkali JT active modes}
\label{Sec:SAAJTmodes}
\begin{figure}[htb]
\begin{center}
\includegraphics[height=8cm]{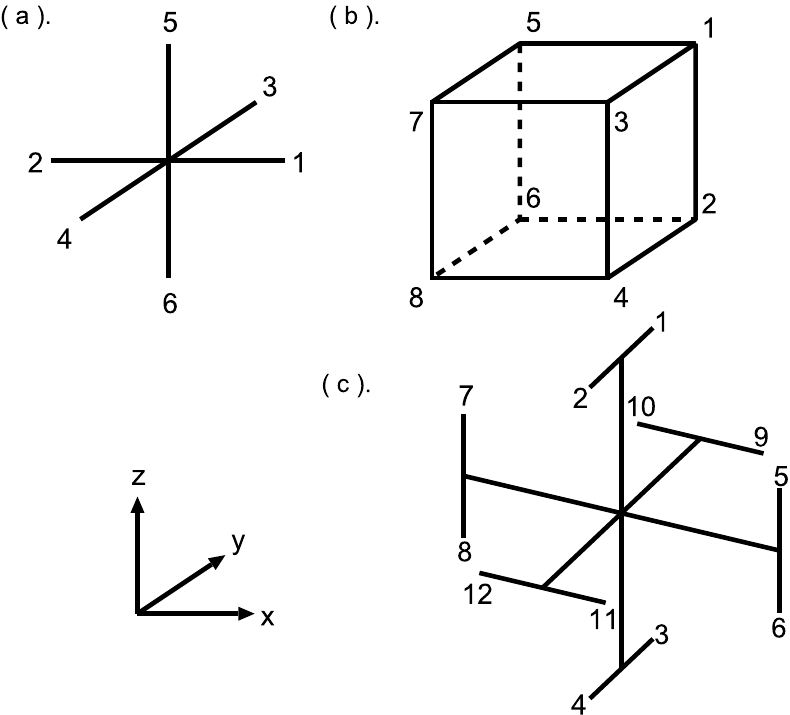}
\end{center}
\caption{Structures of alkali atom in both FCC and BCC structures. (a) and (b) correspond to octahedron and cube, respectively, and (c) represents pseudo-octahedron.}
\label{Fig:fcc_structure}
\end{figure}

Structures and coordinates of octahedron and cube are shown in Fig. \ref{Fig:fcc_structure} ((a) and (b)) and Table \ref{Table:coord_fcc}.

\begin{table}[H]
\caption{Coordinates of octahedron, cube and pseudo-octahedron Akali atoms.}
\label{Table:coord_fcc}
\begin{ruledtabular}
\begin{center}
\begin{tabular}{cccc}
   & octahedron & cube & pseudo-octahedron\\
\hline
A1 & $(1,0,0)$  & $1/2(1,1,1)$    & $1/4( 2, 1, 0)$\\
A2 & $(-1,0,0)$ & $1/2(1,1,-1)$   & $1/4( 2, 1, 0)$\\
A3 & $(0,1,0)$  & $1/2(1,-1,1)$   & $1/4(-2,-1, 0)$\\
A4 & $(0,-1,0)$ & $1/2(1,-1,-1)$  & $1/4(-2,1, 0)$\\
A5 & $(0,0,1)$  & $1/2(-1,1,1)$   & $1/4( 0,-2, 1)$\\
A6 & $(0,0,-1)$ & $1/2(-1,1,-1)$  & $1/4( 0, 2,-1)$\\
A7 & -          & $1/2(-1,-1,1)$  & $1/4( 0, 2, 1)$\\
A8 & -          & $1/2(-1,-1,-1)$ & $1/4( 0,-2,-1)$\\
A9 & -          & $1/2(-1,-1,-1)$ & $1/4(-1, 0,-2)$\\
A10 & -         & $1/2(-1,-1,-1)$ & $1/4( 1, 0,-2)$\\
A11 & -         & $1/2(-1,-1,-1)$ & $1/4(-1, 0, 2)$\\
A12 & -         & $1/2(-1,-1,-1)$ & $1/4( 1, 0, 2)$\\
\end{tabular}
\end{center}
\end{ruledtabular}
\end{table}

The irreducible representation of LUMO is $t_{1u}$. The orbital couples to $A_{1g}$, $E_g$, and $T_{2g}$ vibrational modes.
And the symmetry adapted vibrational vectors for cube, octahedron, and pseudo-octahedron are given in Tables \ref{Table:mode_oct}, \ref{Table:mode_cube}, and \ref{Table:mode_poctahedron}, respectively.

\begin{table}[H]
\caption{Symmetry adapted modes of octahedron}
\label{Table:mode_oct}
\begin{ruledtabular}
\begin{center}
\begin{tabular}{ccccc}
      & $a_{1g}$ & \multicolumn{2}{c}{$e_g$} & $t_{2g}$ \\
      &          & $\theta$ & $\epsilon$ & $\zeta$ \\
\hline
A1    & $\frac{1}{\sqrt{6}}( 1, 0, 0)$ & $\frac{1}{2\sqrt{3}}(-1, 0, 0)$ & $\frac{1}{2}(1,0,0)$  & $\frac{1}{2}(0,  1, 0)$ \\
A2    & $\frac{1}{\sqrt{6}}(-1, 0, 0)$ & $\frac{1}{2\sqrt{3}}( 1, 0, 0)$ & $\frac{1}{2}(-1,0,0)$ & $\frac{1}{2}(0, -1, 0)$ \\
A3    & $\frac{1}{\sqrt{6}}(0,  1, 0)$ & $\frac{1}{2\sqrt{3}}(0, -1, 0)$ & $\frac{1}{2}(0,-1,0)$ & $\frac{1}{2}( 1, 0, 0)$ \\
A4    & $\frac{1}{\sqrt{6}}(0, -1, 0)$ & $\frac{1}{2\sqrt{3}}(0,  1, 0)$ & $\frac{1}{2}(0,1,0)$  & $\frac{1}{2}(-1, 0, 0)$ \\
A5    & $\frac{1}{\sqrt{6}}(0, 0,  1)$ & $\frac{1}{2\sqrt{3}}(0, 0,  2)$ & $(0,0,0)$  & $(0, 0, 0)$ \\
A6    & $\frac{1}{\sqrt{6}}(0, 0, -1)$ & $\frac{1}{2\sqrt{3}}(0, 0, -2)$ & $(0,0,0)$  & $(0, 0, 0)$ \\
\end{tabular}
\end{center}
\end{ruledtabular}
\end{table}

\begin{table}[H]
\caption{Symmetry adapted modes of cube}
\label{Table:mode_cube}
\begin{ruledtabular}
\begin{center}
\begin{tabular}{cccccc}
      & $a_{1g}$ & \multicolumn{2}{c}{$e_g$} & $t_{2g}$(1) & $t_{2g}(2)$ \\
      &          & $\theta$ & $\epsilon$ & & \\
\hline
A1    & $\frac{1}{\sqrt{24}}(1,1,1)$    & $\frac{1}{\sqrt{48}}(-1,-1,2)$  & $\frac{1}{4}(1,-1,0)$ & $\frac{1}{\sqrt{24}}(1,1,1)$    & $\frac{1}{\sqrt{48}}(1,1,-2)$ \\
A2    & $\frac{1}{\sqrt{24}}(1,1,-1)$   & $\frac{1}{\sqrt{48}}(-1,-1,-2)$ & $\frac{1}{4}(1,-1,0)$ & $\frac{1}{\sqrt{24}}(1,1,-1)$   & $\frac{1}{\sqrt{48}}(1,1,2)$ \\
A3    & $\frac{1}{\sqrt{24}}(1,-1,1)$   & $\frac{1}{\sqrt{48}}(-1,1,2)$   & $\frac{1}{4}(1,1,0)$ & $\frac{1}{\sqrt{24}}(-1,1,-1)$  & $\frac{1}{\sqrt{48}}(-1,1,2)$ \\
A4    & $\frac{1}{\sqrt{24}}(1,-1,-1)$  & $\frac{1}{\sqrt{48}}(-1,1,-2)$  & $\frac{1}{4}(1,1,0)$ & $\frac{1}{\sqrt{24}}(-1,1,1)$   & $\frac{1}{\sqrt{48}}(-1,1,-2)$ \\
A5    & $\frac{1}{\sqrt{24}}(-1,1,1)$   & $\frac{1}{\sqrt{48}}(1,-1,2)$   & $\frac{1}{4}(-1,-1,0)$ & $\frac{1}{\sqrt{24}}(1,-1,-1)$  & $\frac{1}{\sqrt{48}}(1,-1,2)$ \\
A6    & $\frac{1}{\sqrt{24}}(-1,1,-1)$  & $\frac{1}{\sqrt{48}}(1,-1,-2)$  & $\frac{1}{4}(-1,-1,0)$ & $\frac{1}{\sqrt{24}}(1,-1,1)$   & $\frac{1}{\sqrt{48}}(1,-1,-2)$ \\
A7    & $\frac{1}{\sqrt{24}}(-1,-1,1)$  & $\frac{1}{\sqrt{48}}(1,1,2)$    & $\frac{1}{4}(-1,1,0)$ & $\frac{1}{\sqrt{24}}(-1,-1,1)$  & $\frac{1}{\sqrt{48}}(-1,-1,-2)$ \\
A8    & $\frac{1}{\sqrt{24}}(-1,-1,-1)$ & $\frac{1}{\sqrt{48}}(1,1,-2)$   & $\frac{1}{4}(-1,1,0)$ & $\frac{1}{\sqrt{24}}(-1,-1,-1)$ & $\frac{1}{\sqrt{48}}(-1,-1,2)$ \\
\end{tabular}
\end{center}
\end{ruledtabular}
\end{table}

\begin{table}[H]
\caption{Symmetry adapted modes of pseudo-octahedron}
\label{Table:mode_poctahedron}
\resizebox{\textwidth}{11cm}{ \begin{ruledtabular}
\begin{tabular}{cccc}
\hline
 & A$_g$(1) & E$_{g\theta}$(1) & E$_{g\epsilon}$(1) \\
\hline
A1&  0.000000   0.247154  -0.149159&  0.000000   0.149157   0.247155& -0.000000   0.009650   0.193490\\
A2&  0.000000  -0.247154  -0.149159&  0.000000  -0.149157   0.247155& -0.000000  -0.009650   0.193490\\
A3&  0.000000   0.247154   0.149159&  0.000000   0.149157  -0.247155& -0.000000   0.009660  -0.193490\\
A4&  0.000000  -0.247154   0.149159&  0.000000  -0.149157  -0.247155& -0.000000  -0.009650  -0.193490\\
A5& -0.149159   0.000000   0.247154&  0.247155   0.000000   0.149157&  0.202060  -0.000000   0.082150\\
A6& -0.149159   0.000000  -0.247154&  0.247155   0.000000  -0.149157&  0.202060  -0.000000  -0.082150\\
A7&  0.149159   0.000000   0.247154& -0.247155   0.000000   0.149157& -0.202060  -0.000000   0.082150\\
A8&  0.149159   0.000000  -0.247154& -0.247155   0.000000  -0.149157& -0.202060  -0.000000  -0.082150\\
A9&  0.247154  -0.149159   0.000000&  0.149157   0.247155   0.000000& -0.091800  -0.395550  -0.000000\\
A10& -0.247154  -0.149159   0.000000& -0.149157   0.247155   0.000000&  0.091800  -0.395550  -0.000000\\
A11&  0.247154   0.149159   0.000000&  0.149157  -0.247155   0.000000& -0.091800   0.395550  -0.000000\\
A12& -0.247154   0.149159   0.000000& -0.149157  -0.247155   0.000000&  0.091800   0.395550  -0.000000\\
\hline
 & A$_g$(2) & E$_{g\theta}$(2) & E$_{g\epsilon}$(2) \\
\hline
A1&  0.000000  -0.100430  -0.345030& -0.000000   0.126520  -0.068090&  0.000000   0.374810  -0.074450\\
A2&  0.000000   0.100430  -0.345030& -0.000000  -0.126520  -0.068090&  0.000000  -0.374810  -0.074450\\
A3&  0.000000  -0.100430   0.345030& -0.000000   0.126520   0.068090&  0.000000   0.374810   0.074450\\
A4&  0.000000   0.100430   0.345030& -0.000000  -0.126520   0.068090&  0.000000  -0.374810   0.074450\\
A5&  0.340090   0.000000   0.058580& -0.030430  -0.000000   0.261330&  0.096200   0.000000  -0.296970\\
A6&  0.340080   0.000000  -0.058580& -0.030430  -0.000000  -0.261330&  0.096200   0.000000   0.296970\\
A7& -0.340080   0.000000   0.058580&  0.030430  -0.000000   0.261330& -0.096200   0.000000  -0.296970\\
A8& -0.340080   0.000000  -0.058580&  0.030430  -0.000000  -0.261330& -0.096200   0.000000   0.296970\\
A9&  0.041860   0.004950   0.000000& -0.387850   0.098520  -0.000000& -0.077840  -0.021740   0.000000\\
A10& -0.041860   0.004950   0.000000&  0.387850   0.098530  -0.000000&  0.077830  -0.021740   0.000000\\
A11&  0.041860  -0.004950   0.000000& -0.387850  -0.098520  -0.000000& -0.077830   0.021740   0.000000\\
A12& -0.041860  -0.004950   0.000000&  0.387850  -0.098530  -0.000000&  0.077830   0.021740   0.000000\\
\hline
 & T$_{2g\xi}$(1) & T$_{2g\xi}$(2) & T$_{2g\xi}$(3) \\
\hline
A1&  0.000000  -0.183513   0.143035&  0.034419  -0.000000  -0.000000&  0.441224   0.000000   0.000000\\
A2&  0.000000  -0.183513  -0.143035&  0.034419  -0.000000  -0.000000& -0.441224   0.000000   0.000000\\
A3&  0.000000   0.183513   0.143035& -0.034419  -0.000000  -0.000000&  0.441224   0.000000   0.000000\\
A4&  0.000000   0.183513  -0.143035& -0.034419  -0.000000  -0.000000& -0.441224   0.000000   0.000000\\
A5&  0.000000   0.441225   0.000000& -0.143036  -0.000000   0.183517&  0.000000  -0.034417   0.000000\\
A6&  0.000000  -0.441225   0.000000&  0.143036  -0.000000   0.183517&  0.000000  -0.034417   0.000000\\
A7&  0.000000   0.441225   0.000000& -0.143036  -0.000000  -0.183517&  0.000000   0.034417   0.000000\\
A8&  0.000000  -0.441225   0.000000&  0.143036  -0.000000  -0.183517&  0.000000   0.034417   0.000000\\
A9&  0.000000   0.000000  -0.034417& -0.000000  -0.000000  -0.441223& -0.183514   0.143037   0.000000\\
A10&  0.000000   0.000000  -0.034417& -0.000000  -0.000000   0.441223& -0.183514  -0.143037   0.000000\\
A11&  0.000000   0.000000   0.034417& -0.000000  -0.000000  -0.441223&  0.183514   0.143037   0.000000\\
A12&  0.000000   0.000000   0.034417& -0.000000  -0.000000   0.441223&  0.183514  -0.143037   0.000000\\
\hline
\end{tabular}
\end{ruledtabular}}
\end{table}

And the symmetry adapted vibrational vectors for inter-fullerence in FCC and A15 cases are given in Tables \ref{Table:mode_FCC_inter}, \ref{Table:mode_A15_inter}, respectively.

\begin{table}[H]
\caption{Symmetry adapted modes of inter-fullerene of FCC fullerides.}
\label{Table:mode_FCC_inter}
\resizebox{\textwidth}{11cm}{\begin{ruledtabular}
\begin{tabular}{ccccc}
\hline
 & A$_g$ & E$_{g\theta}$(1) & E$_{g\epsilon}$(1)  \\
\hline
A1&  0.000000   0.204124   0.204124& -0.000000  -0.142461  -0.146206&  0.000000  -0.251075  -0.248911\\
A2&  0.000000  -0.204124   0.204124& -0.000000   0.142461  -0.146206&  0.000000   0.251075  -0.248911\\
A3&  0.000000  -0.204124  -0.204124& -0.000000   0.142461   0.146206&  0.000000   0.251075   0.248911\\
A4&  0.000000   0.204124  -0.204124& -0.000000  -0.142461   0.146206&  0.000000  -0.251075   0.248911\\
A5&  0.204124   0.000000   0.204124& -0.142461  -0.000000  -0.146206&  0.251075   0.000000   0.248911\\
A6& -0.204124   0.000000   0.204124&  0.142461  -0.000000  -0.146206& -0.251075   0.000000   0.248911\\
A7& -0.204124   0.000000  -0.204124&  0.142461  -0.000000   0.146206& -0.251075   0.000000  -0.248911\\
A8&  0.204124   0.000000  -0.204124& -0.142461  -0.000000   0.146206&  0.251075   0.000000  -0.248911\\
A9&  0.204124   0.204124   0.000000&  0.288667   0.288667  -0.000000&  0.002164  -0.002164   0.000000\\
A10& -0.204124   0.204124   0.000000& -0.288667   0.288667  -0.000000& -0.002164  -0.002164   0.000000\\
A11& -0.204124  -0.204124   0.000000& -0.288667  -0.288667  -0.000000& -0.002164   0.002164   0.000000\\
A12&  0.204124  -0.204124   0.000000&  0.288667  -0.288667  -0.000000&  0.002164   0.002164   0.000000\\
\hline
 & E$_{g\theta}$(2) & E$_{g\epsilon}$(2) & T$_{2g\xi}$(1) \\
\hline
A1& -0.000000  -0.251074   0.248911&  0.000000  -0.142457   0.146210&  0.000000   0.348929   0.348929\\
A2& -0.000000   0.251074   0.248911&  0.000000   0.142457   0.146210&  0.000000   0.348929  -0.348929\\
A3& -0.000000   0.251074  -0.248911&  0.000000   0.142457  -0.146210&  0.000000  -0.348929  -0.348929\\
A4& -0.000000  -0.251074  -0.248911&  0.000000  -0.142457  -0.146210&  0.000000  -0.348929   0.348929\\
A5& -0.251074  -0.000000   0.248911&  0.142457   0.000000  -0.146210&  0.000000   0.056999   0.000000\\
A6&  0.251074  -0.000000   0.248911& -0.142457   0.000000  -0.146210&  0.000000   0.056999   0.000000\\
A7&  0.251074  -0.000000  -0.248911& -0.142457   0.000000   0.146210&  0.000000  -0.056999   0.000000\\
A8& -0.251074  -0.000000  -0.248911&  0.142457   0.000000   0.146210&  0.000000  -0.056999   0.000000\\
A9&  0.002163   0.002163  -0.000000&  0.288667  -0.288667   0.000000&  0.000000   0.000000   0.056999\\
A10& -0.002163   0.002163  -0.000000& -0.288667  -0.288667   0.000000&  0.000000   0.000000   0.056999\\
A11& -0.002163  -0.002163  -0.000000& -0.288667   0.288667   0.000000&  0.000000   0.000000  -0.056999\\
A12&  0.002163  -0.002163  -0.000000&  0.288667   0.288667   0.000000&  0.000000   0.000000  -0.056999\\
\hline
 & T$_{2g\xi}$(2) &- & -\\
\hline
A1&  0.056997   0.000000   0.000000&-&-\\
A2&  0.056997   0.000000   0.000000&-&-\\
A3& -0.056997   0.000000   0.000000&-&-\\
A4& -0.056997   0.000000   0.000000&-&-\\
A5&  0.348929   0.000000   0.348929&-&-\\
A6&  0.348929   0.000000  -0.348929&-&-\\
A7& -0.348929   0.000000  -0.348929&-&-\\
A8& -0.348929   0.000000   0.348929&-&-\\
A9&  0.000000   0.000000   0.056997&-&-\\
A10&  0.000000   0.000000  -0.056997&-&-\\
A11&  0.000000   0.000000  -0.056997&-&-\\
A12&  0.000000   0.000000   0.056997&-&-\\
\hline
\end{tabular}
\end{ruledtabular}}
\end{table}

\begin{table}[H]
\caption{Symmetry adapted modes of inter-fullerene of A15.}
\label{Table:mode_A15_inter}
\begin{ruledtabular}
\begin{tabular}{cccc}
\hline
 & A$_g$ & E$_{g\theta}$ & E$_{g\epsilon}$ \\
\hline
A1&  0.204124   0.204124   0.204124& -0.144338  -0.144338   0.288675&  0.250000  -0.250000   0.000000\\
A2& -0.204124   0.204124   0.204124&  0.144338  -0.144338   0.288675& -0.250000  -0.250000   0.000000\\
A3&  0.204124   0.204124  -0.204124& -0.144338  -0.144338  -0.288675&  0.250000  -0.250000   0.000000\\
A4& -0.204124   0.204124  -0.204124&  0.144338  -0.144338  -0.288675& -0.250000  -0.250000   0.000000\\
A5&  0.204124  -0.204124   0.204124& -0.144338   0.144338   0.288675&  0.250000   0.250000   0.000000\\
A6& -0.204124  -0.204124   0.204124&  0.144338   0.144338   0.288675& -0.250000   0.250000   0.000000\\
A7&  0.204124  -0.204124  -0.204124& -0.144338   0.144338  -0.288675&  0.250000   0.250000   0.000000\\
A8& -0.204124  -0.204124  -0.204124&  0.144338   0.144338  -0.288675& -0.250000   0.250000   0.000000\\
\hline
 & T$_{2g\xi}$(1) & T$_{2g\xi}$(2) &- \\
\hline
A1&  0.204124   0.204124   0.204124&  0.144338   0.144338  -0.288675&-\\
A2&  0.204124  -0.204124  -0.204124&  0.144338  -0.144338   0.288675&-\\
A3&  0.204124   0.204124  -0.204124&  0.144338   0.144338   0.288675&-\\
A4&  0.204124  -0.204124   0.204124&  0.144338  -0.144338  -0.288675&-\\
A5& -0.204124   0.204124  -0.204124& -0.144338   0.144338   0.288675&-\\
A6& -0.204124  -0.204124   0.204124& -0.144338  -0.144338  -0.288675&-\\
A7& -0.204124   0.204124   0.204124& -0.144338   0.144338  -0.288675&-\\
A8& -0.204124  -0.204124  -0.204124& -0.144338  -0.144338   0.288675&-\\
\hline
\end{tabular}
\end{ruledtabular}
\end{table}

\section{Elastic coupling information}
\begin{table}[tb]
    \caption{\label{tab:Freq_from_zeta}
Effective frequencies extracted from corresponding elastic response parameters $\zeta$, compared with the frequencies of isolated C$_{60}$ (in $cm^{-1}$).
Calculations are done with LDA and PBE (in the parentheses) functionals.
 }
\resizebox{\linewidth}{10cm}{    \begin{ruledtabular}
    \begin{tabular}{ccccccc}
      &  & K$_{3}$C$_{60}$ & Rb$_{3}$C$_{60}$ & Cs$_{3}$C$_{60}$ & Cs$_{3}$C$_{60}$ (A15)& Isolated C$_{60}$ \\
\hline
\multirow{10}{*}{IntraC60}&A$_{g}$(1) &       497(       488) &       492(       490) &       491(       488) &       496(       489) &       496 \\
&A$_{g}$(2) &      1468(      1427) &      1448(      1440) &      1446(      1438) &      1472(      1444) &      1470 \\
&H$_{g}$(1) &       256(       254) &       255(       255) &       253(       251) &       254(       252) &       273 \\
&H$_{g}$(2) &       403(       402) &       402(       399) &       394(       390) &       398(       393) &       437 \\
&H$_{g}$(3) &       653(       641) &       656(       639) &       649(       630) &       656(       636) &       710 \\
&H$_{g}$(4) &       765(       748) &       755(       748) &       749(       740) &       755(       742) &       774 \\
&H$_{g}$(5) &      1104(      1078) &      1093(      1078) &      1088(      1071) &      1092(      1076) &      1099 \\
&H$_{g}$(6) &      1279(      1233) &      1256(      1234) &      1254(      1229) &      1257(      1232) &      1250 \\
&H$_{g}$(7) &      1335(      1285) &      1301(      1275) &      1270(      1239) &      1289(      1253) &      1428 \\
&H$_{g}$(8) &      1455(      1407) &      1427(      1399) &      1400(      1365) &      1418(      1380) &      1575 \\
\hline
\multirow{4}{*}{Akali:cube}&A$_{g}$ &       116(        97) &        83(        99) &        65(        74) &-&- \\
&E$_{g}$ &       111(        96) &        78(        93) &        61(        70) &-&- \\
&T$_{2g}$(1) &       104(        93) &        73(        86) &        55(        63) &-&- \\
&T$_{2g}$(2) &       107(        95) &        76(        88) &        58(        66) &-&- \\
\hline
\multirow{3}{*}{Akali:oct}&A$_{g}$ &        32(        39) &        17(        58) &        11(        27) &-&- \\
&E$_{g}$ &        28(        46) &        16(        49) &        11(        26) &-&- \\
&T$_{2g}$ &        30(        41) &        17(        54) &        12(        26) &-&- \\
\hline
\multirow{7}{*}{Akali:poct}&A$_{g}$(1) &-&-&-&        36(        37) &- \\
&A$_{g}$(2) &-&-&-&        40(        40) &- \\
&E$_{g}$(1) &-&-&-&        51(        52) &- \\
&E$_{g}$(2) &-&-&-&        32(        32) &- \\
&T$_{2g}$(1) &-&-&-&        48(        55) &- \\
&T$_{2g}$(2) &-&-&-&        43(        49) &- \\
&T$_{2g}$(3) &-&-&-&        44(        51) &- \\
\hline
\multirow{5}{*}{InerC60:FCC}&A$_{g}$(1) &        43(        49) &        43(        53) &        39(        48) &-&- \\
&E$_{g}$(1) &        35(        35) &        32(        41) &        26(        33) &-&- \\
&E$_{g}$(2) &        35(        36) &        33(        42) &        29(        35) &-&- \\
&T$_{2g}$(1) &        34(        35) &        33(        41) &        28(        34) &-&- \\
&T$_{2g}$(2) &        40(        43) &        39(        48) &        34(        42) &-&- \\
\hline
\multirow{4}{*}{InerC60:A15}&A$_{g}$(1) &-&-&-&        21(        24) &- \\
&E$_{g}$(1) &-&-&-&        21(        24) &- \\
&T$_{2g}$(1) &-&-&-&        29(        32) &- \\
&T$_{2g}$(2) &-&-&-&        31(        34) &- \\
    \end{tabular}
\end{ruledtabular}}
\end{table}

\section{adiabatic potential of isolated trivalent C$_{60}$}
The adiabatic potential energy surface of isolated C$_{60}^{3-}$ is calculated.
The model Hamiltonian consists of harmonic potential, Hund coupling $\hat{H}_\text{H}$, and linear Jahn-Teller coupling $\hat{V}_\text{h} = \hat{V}_\text{e} + \hat{V}_\text{t}$.
The explicit forms of the Hund and Jahn-Teller couplings are given in Eqs. ((3) and (4) in the main text, respectively.
The vibronic coupling parameters for the $e$ and $t_2$ modes are the same, $V_\text{e} = V_\text{t} = V$.

For the analysis of the adiabatic potential energy surface, it is convenient to use the adiabatic orbitals which diagonalizes $\hat{V}_\text{h}$:
\begin{eqnarray}
 \hat{H}_\text{JT} &=& \sum_\sigma V q 
 \left(\hat{a}_{1\sigma}^\dagger, \hat{a}_{2\sigma}^\dagger, \hat{a}_{3\sigma}^\dagger \right)
 \begin{pmatrix}
  -\cos \left(\alpha - \frac{2\pi}{3} \right) & 0 & 0 \\
  0 & -\cos \left(\alpha + \frac{2\pi}{3} \right) & 0 \\
  0 & 0 & -\cos \alpha
 \end{pmatrix}
 \begin{pmatrix}
  \hat{a}_{1\sigma}\\
  \hat{a}_{2\sigma}\\
  \hat{a}_{3\sigma}
 \end{pmatrix}.
\end{eqnarray}
$q$ is the length of the $H_g$ normal coordinates.
See for details, Refs. O'Brien 1971, Auerbach 1994, O'Brien 1996, Iwahara 2018.

When the adiabatic orbitals are populated by three electrons without Jahn-Teller deformations, the Hund coupling splits the electron configurations into $^4S$, $^2D$ and $^2P$ terms.
The Jahn-Teller effect is relevant to the doublet terms: The $^2P$ term is higher in energy by $2J_\text{H}$ than $^2D$ term.

The electronic states for the $^2D$ terms are expressed by
\begin{eqnarray}
&& \frac{1}{\sqrt{6}} \left(
 2
 \hat{a}^\dagger_{1\downarrow}
 \hat{a}^\dagger_{2\uparrow}
 \hat{a}^\dagger_{3\uparrow}
 -
 \hat{a}^\dagger_{1\uparrow}
 \hat{a}^\dagger_{2\downarrow}
 \hat{a}^\dagger_{3\uparrow}
 -
 \hat{a}^\dagger_{1\uparrow}
 \hat{a}^\dagger_{2\uparrow}
 \hat{a}^\dagger_{3\downarrow}
 \right),
\quad
 \frac{1}{\sqrt{2}} \left(
 \hat{a}^\dagger_{1\uparrow}
 \hat{a}^\dagger_{2\downarrow}
 \hat{a}^\dagger_{3\uparrow}
 -
 \hat{a}^\dagger_{1\uparrow}
 \hat{a}^\dagger_{2\uparrow}
 \hat{a}^\dagger_{3\downarrow}
 \right),
\nonumber\\
&& \frac{1}{\sqrt{2}}
 \left(
 \hat{a}^\dagger_{1\uparrow}
 \hat{a}^\dagger_{2\uparrow}
 \hat{a}^\dagger_{1\downarrow}
 -
 \hat{a}^\dagger_{2\uparrow}
 \hat{a}^\dagger_{3\uparrow}
 \hat{a}^\dagger_{3\downarrow}
 \right),
 \quad
 \frac{1}{\sqrt{2}}
 \left(
 \hat{a}^\dagger_{2\uparrow}
 \hat{a}^\dagger_{3\uparrow}
 \hat{a}^\dagger_{2\downarrow}
 -
 \hat{a}^\dagger_{3\uparrow}
 \hat{a}^\dagger_{1\uparrow}
 \hat{a}^\dagger_{1\downarrow}
 \right),
 \frac{1}{\sqrt{2}}
 \left(
 \hat{a}^\dagger_{3\uparrow}
 \hat{a}^\dagger_{1\uparrow}
 \hat{a}^\dagger_{3\downarrow}
 -
 \hat{a}^\dagger_{1\uparrow}
 \hat{a}^\dagger_{2\uparrow}
 \hat{a}^\dagger_{2\downarrow}
 \right),
\nonumber\\
\end{eqnarray}
and those for the $^2P$ term are
\begin{eqnarray}
&& \frac{1}{\sqrt{2}}
 \left(
 \hat{a}^\dagger_{3\uparrow}
 \hat{a}^\dagger_{1\uparrow}
 \hat{a}^\dagger_{3\downarrow}
+
 \hat{a}^\dagger_{1\uparrow}
 \hat{a}^\dagger_{2\uparrow}
 \hat{a}^\dagger_{2\downarrow}
 \right),
 \quad
 \frac{1}{\sqrt{2}}
 \left(
 \hat{a}^\dagger_{1\uparrow}
 \hat{a}^\dagger_{2\uparrow}
 \hat{a}^\dagger_{1\downarrow}
+
 \hat{a}^\dagger_{2\uparrow}
 \hat{a}^\dagger_{3\uparrow}
 \hat{a}^\dagger_{3\downarrow}
 \right),
 \quad
 \frac{1}{\sqrt{2}}
 \left(
 \hat{a}^\dagger_{2\uparrow}
 \hat{a}^\dagger_{3\uparrow}
 \hat{a}^\dagger_{2\downarrow}
+
 \hat{a}^\dagger_{3\uparrow}
 \hat{a}^\dagger_{1\uparrow}
 \hat{a}^\dagger_{1\downarrow}
 \right).
\nonumber\\
\label{Eq:config_LS_diag}
\end{eqnarray}
With the use of the multiplet states as the basis, the Jahn-Teller coupling matrix is written as
\begin{eqnarray}
 \left(\mathbf{V}_\text{h}\right)_{PD} &=& V q
 \left(
\begin{array}{cccccccc}
 0 & 0 & 0 & 0 & -\sin \left(\alpha +\frac{\pi }{6}\right)-\cos \alpha \\
 0 & 0 & \sin \left(-\alpha + \frac{\pi}{6} \right)+\cos \alpha & 0 & 0 \\
 0 & 0 & 0 & \sqrt{3} \sin \alpha & 0 \\
\end{array}
\right),
\nonumber\\
 \left(\mathbf{V}_\text{h}\right)_{DP} &=& \left[\left(\mathbf{V}_\text{h}\right)_{DP}\right]^T.
\end{eqnarray}
The diagonal $3\times 3$ and $5\times 5$ blocks for the $P$ and $D$ terms are zero.

The eigenvalues of $\mathbf{H}_\text{H} + \mathbf{V}_\text{h}$ for the spin doublet terms are given as
\begin{eqnarray}
&& 0, \quad
 0, \quad
 J_\text{H} \mp \sqrt{J_\text{H}^2 + 3\left[V q \sin\alpha\right]^2},
\nonumber\\
&& J_\text{H} \mp \sqrt{J_\text{H}^2 + 3\left[V q \cos\left(\alpha + \frac{\pi}{6}\right)\right]^2}, \quad
 J_\text{H} \mp \sqrt{J_\text{H}^2 + 3\left[V q \cos\left(\alpha - \frac{\pi}{6}\right)\right]^2}.
\end{eqnarray}
The minima of the APES locate at (elastic energy $\omega^2 q^2/2$ is considered too)
\begin{eqnarray}
 (q,\alpha) &=& \left(\frac{\sqrt{3}V}{\omega^2} \sqrt{1 - \left(\frac{J_\text{H}}{3 V^2/\omega^2}\right)^2}, \frac{\pi}{2}\right),
\end{eqnarray}
with the Jahn-Teller stabilization energy
\begin{eqnarray}
 E_\text{JT}^{(3)} &=& J_\text{H} -\frac{3 V^2}{2\omega^2} - \frac{J_\text{H}^2}{6V^2/\omega^2}.
\end{eqnarray}
The $^2D$ term energy is used as the origin of the energy.



\bibliography{BIB_A3C60_summary_SM}